\documentclass[12pt]{article}          

\usepackage{cite,amsmath,amssymb,isolatin1,alltt,amsxtra,epsfig}
\usepackage[mathscr]{eucal}

\numberwithin{equation}{section}

\let\ex=\times          
\def\IP{{\mathbb P}} \def\IZ{{\mathbb Z}} \def\IR{{\mathbb R}} 
\def\IC{{\mathbb C}}                    
\let\D=\Delta \let\s=\sigma \let\Th=\Theta \let\S=\Sigma  
\def\2{{1\over2}} \let\<=\langle \let\>=\rangle \def\new#1\endnew{{\bf #1}}
\def\BE{\begin{equation}}       \def\EE{\end{equation}}
\def\BEA{\begin{eqnarray}}      \def\EEA{\end{eqnarray}}
\def\ifundefined#1{\expandafter\ifx\csname#1\endcsname\relax}

\newcounter{TRefNX} \let\OLDcite=\cite  \makeatletter
\def\makeTRefs#1{\@for  \NewTRef:=#1\do{\global\makeTRef{\NewTRef}}}
\def\makeTRef#1{\ifundefined{TRef#1}\stepcounter{TRefNX}%
\expandafter\xdef\csname TRef#1\endcsname{\theTRefNX}\fi}\makeatother

\ifundefined{draftmode} 
\else    

\def\NEWcite#1{\makeTRefs{#1}\OLDcite{#1}}
\let\cite=\NEWcite 
   \newcount\HOUR  \HOUR=\the\time \divide\HOUR by 60    \multiply \HOUR by 60
   \newcount\MIN   \MIN=\the\time  \advance\MIN by -\HOUR   \divide\HOUR by 60
   \ifnum   \MIN>9  \def\printTIME{{\it\the\HOUR\,:\,\the\MIN}}
         \else   \def\printTIME{{\it\the\HOUR\,:\,0\the\MIN}} \fi 
   
   \def\LLab#1{\BP(0,0)\unitlength=1mm\put(-9,-1.6){\makebox(0,0)[cr]{\tiny #1
        \rlap{$_{_{\makeatletter\csname TRef#1\endcsname\makeatother}}$}}}\EP}
   \def\BP{\begin{picture}} \def\EP{\end{picture}} \let\OLDbib=\bibitem 
   \def\NEWbib#1{\OLDbib{#1}\LLab{#1}} \let\bibitem=\NEWbib
\fi

\let\Msize=\footnotesize        \def\VL{\;\vrule\;}     \def\v{\nu^*}
\def\BM{\Msize\begin{matrix}}           \def\EM{\end{matrix}} 
\def\MN M:#1 #2 N:#3 #4 {{(#1_{#2},#3_{#4})}}
\def\MNH M:#1 #2 N:#3 #4 H:#5,#6 [#7]{{(#1_{#2},#3_{#4})^{#5,#6}_{#7}}}

\newcommand{\um}{\phantom{-}}
\newcommand{\ds}{\displaystyle}
\newcommand{\cF}{{\cal F}}
\newcommand{\cM}{{\cal M}}

\def\dd{\mathrm{d}}

\def\cI{\mathcal{I}}
\def\tD{\mathrm{D}}
\def\tE{\mathrm{E}}
\def\tBl{\mathrm{Bl}}
\def\tdP{\mathrm{dP}}
\def\mC{\mathbb{C}}
\def\mF{\mathbb{F}}
\def\mP{\mathbb{P}}
\def\mR{\mathbb{R}}
\def\mT{\mathbb{T}}
\def\mZ{\mathbb{Z}}
\def\cO{\mathcal{O}}
\def\cL{\mathcal{L}}
\DeclareMathOperator{\rank}{rk} 
\DeclareMathOperator{\ch}{c}
\def\CY{Calabi--Yau }
\def\Vol{\mathrm{Vol}}
\def\conv{\mathrm{conv}}
\def\r{\mathrm{r}}

\begin{document}

\begin{titlepage}
\begin{center}

\begin{flushright}
MAD-TH-04-10\\
TUW-04-30\\
ESI 1519\\
hep-th/0410018\\
\vspace{1.5cm}\end{flushright}

{\LARGE {Topological String Amplitudes, Complete Intersection Calabi--Yau 
        Spaces and Threshold Corrections}}

\vspace{1cm}

{\bf A.~Klemm$^*$, M.~Kreuzer$^{\diamond}$, E.~Riegler$^{\diamond}$ and 
        E.~Scheidegger$^{\diamond}$}\footnote{\mbox{email: \tt 
aklemm@physics.wisc.edu, kreuzer, riegler \& esche@hep.itp.tuwien.ac.at}}\\
\vskip 1cm

{\em $^{*}$ UW-Madison Physics Department\\
1150 University Avenue\\
Madison, WI 53706-1390, USA

$^{\diamond}$Institut f\"ur Theoretische Physik,
Technische Universit\"at Wien \\
Wiedner Hauptstr. 8-10/136, 
A-1040 Vienna, Austria  }
\vskip .1in

\end{center}

\vspace{1.5cm}

\centerline{\bf ABSTRACT } 

We present the most complete list of mirror pairs of Calabi--Yau complete 
intersections in toric ambient varieties and develop the methods to solve 
the topological string and to calculate higher genus amplitudes on these 
compact Calabi--Yau spaces. These symplectic invariants are used to remove 
redundancies in examples. The construction of the B--model propagators 
leads to compatibility conditions, which constrain multi--parameter mirror 
maps. For K3 fibered Calabi--Yau spaces without reducible fibers we find 
closed formulas for all genus contributions in the fiber direction from 
the geometry of the fibration. If the heterotic dual to this geometry is 
known, the higher genus invariants can be identified with the degeneracies 
of BPS states contributing to gravitational threshold corrections and all 
genus checks on string duality in the perturbative regime are accomplished. 
We find, however, that the BPS degeneracies do not uniquely fix the 
non-perturbative completion of the heterotic string. For these geometries
we can write the topological partition function in terms of the 
Donaldson--Thomas invariants and we perform a non-trivial check of 
S--duality in topological strings. We further investigate transitions via 
collapsing $\tD_5$ del Pezzo surfaces and the occurrence of free $\mZ_2$ 
quotients that lead to a new class of heterotic duals.

\begin{quotation}\noindent

\end{quotation}

\end{titlepage}
\vfill
\eject


\newpage

\section{Introduction}

Mirror symmetry in Calabi--Yau compactifications makes the calculations
of low derivative terms in the effective action with $N=2$ and $N=1$ 
supersymmetry feasible and leads to a wealth of insights in gauge and 
string theory, many of which come from concrete examples. 
In this paper we construct  the most complete  list of Calabi--Yau mirror 
pairs. They are realized as hypersurfaces and complete intersection in 
general toric varieties.  

A motivation to go beyond the hypersurface case can already be seen locally. 
While all 2d canonical singularities are described by a singular ($ADE$) 
hypersurface, 
the 3d canonical singularities are generally  not realizable in this way.  
Simple 3d examples, such as the $\tD_5$ elliptic singularity, require a 
realization by complete intersections, here by two polynomials in 
${\mathbb C}^5$. 
Globally the complete intersection examples naturally lead to string 
backgrounds with new spectra, new interactions and  new symmetries. 

Important terms in the exact low energy effective action are given by the 
same supersymmetric family indices, calculable as topological string 
amplitudes, that are also the natural mathematical invariants to control 
the redundancy in these constructions.  In the A-model picture the expansion 
coefficients of the family indices are completely determined either by
Gromov--Witten invariants, by numbers of BPS states (Gopakumar-Vafa 
invariants) or by Donaldson--Thomas invariants. These are symplectic
invariants. Solving the topological string in the compact case is a 
challenge whose solution will determine these invariants and shed light on 
4d black hole physics~\cite{Maldacena:1999bp}, \cite{Katz:1999xq}, \cite{Ooguri:2004zv}, 
the physics of the NS five--brane~\cite{Dijkgraaf:2002ac}, non-perturbative string theory 
and low energy effective actions.

From the available techniques to calculate higher genus topological 
string amplitudes, localization and large N transitions presently fail in 
the compact case for reasons discussed in more detail at the beginning of 
Section~\ref{sec:TopString}. As direct computation of the cohomology of 
BPS moduli spaces and utilization of heterotic type II duality yields 
only information for certain classes in the K\"ahler cone, the topological 
B-model is still the most powerful tool for calculating higher genus 
amplitudes on compact Calabi--Yau spaces.
We therefore first extend the methods for the calculation of the genus zero 
and one topological string amplitudes to the case of complete intersections 
in general toric varieties. This can be viewed as a completion of the work 
of~\cite{Hosono:1993qy}, \cite{Hosono:1994ax}. An important application is the study of the 
non-perturbative completion of $N=2$ heterotic string compactifications via 
its duality to type II strings on K3 fibered Calabi--Yau spaces. It requires 
the calculation of topological string amplitudes on compact multi--parameter 
Calabi--Yau 3-folds at higher genus. This has not been done before, partly 
because the construction of the B-model propagators implies differential 
constraints on the mirror maps, which have not been solved. We solve those 
in various examples and make the first higher genus calculations for compact 
multi--parameter models. The solution to the constraint equations itself can 
shed light on the mirror maps whose integrality property is one of the most 
enigmatic features of mirror symmetry.
   
Using the Hilbert scheme for points on the K3 fiber we find closed formulas 
for higher genus amplitudes on K3 fibrations without reducible singular 
fibers in the limit of large ${\mathbb P}^1$ base, which corresponds to weak 
heterotic coupling. Unlike in the case of local limits to non-compact toric
surfaces or del Pezzos, which have no complex structure deformations, we use 
the invariance under complex structure deformations of the geometry 
to find a good model for the calculation of the BPS moduli spaces.
Even though restricted to the K3 classes in the Calabi-Yau $X$ the calculated 
invariants do contain global information of $X$.  For K3
fibrations with two parameters, the resulting formulas involve modular forms 
of subgroups of SL$(2,{\mathbb Z})$, which we obtain directly from the 
geometry. For a fixed fiber type of the normal fibration these modular forms 
for different fibrations depend on the singular fibers and can be parameterized 
in an elegant way by the Euler number of $X$. In the case where 
the heterotic dual is known they are calculated by integrals of worldsheet 
indices related to the gravitational threshold corrections and we perform all 
genus checks by comparing to our B-model results and to calculations extending 
the methods of~\cite{Katz:1999xq}. 

We also find that the partition function of the topological string has in 
general a very simple product form, which corresponds to a free field theory 
partition function with a U(1) charge.
This might give some hope that one can 
understand the indices of the non-perturbative heterotic string or the 
dependence of the topological string theory for all classes of curves in a 
Calabi--Yau by the world sheet index of some 2d conformal field theory. However,
one has to be careful with the non-perturbative completion of the heterotic 
string, because we find for well--known examples such as the $ST$ model that 
there are several dual K3 fibered Calabi--Yau spaces, which reproduce the same 
holomorphic data at weak coupling except for the triple intersection of $T$ in 
$\cF^{(g=0)}$ and the leading coefficient of $T$ in $\cF^{(g=1)}$, which have not 
been fixed on the heterotic side. These series of K3 fibrations 
are rational homotopy equivalent. They have identical perturbative BPS 
spectrum in the fiber direction, but different non-perturbative BPS 
numbers for nonzero base degrees. Turning 
on base degrees includes the non-perturbative $e^{-8 \pi^2 S}$ effects and 
one finds that the non-perturbative moduli including the heterotic dilaton are
globally different for the members of the series. 

S--duality in the topological string is conjectured to map the partition 
function of the topological A--model expressed in terms of the Gopakumar--Vafa 
invariants into a dual partition function of the B--model on the same 
Calabi--Yau manifold expressed in terms of Donaldson--Thomas invariants. 
Using our simple product form of the former partition function, restricting 
it to maps in the fiber direction, and applying S--duality, we explicitly 
compute in Section~\ref{sec:STU} a subset of the Donaldson--Thomas invariants 
of a compact Calabi--Yau manifold. We find that these invariants are indeed 
integral as expected, and hence our result can be viewed as a highly 
non-trivial check of the S--duality in the topological string. 

In order to use mirror symmetry to go from the B--model propagators to the 
invariants in the A--model, the initial data is a mirror pair of families of 
complex 3d compact complete intersections with vanishing first Chern class.
Such a pair can be given by a pair of reflexive polyhedra $(\Delta,\nabla)$ 
of dimension $\ge 4$ together with a dual pair of nef partitions of 
$(\Delta^*,\nabla^*)$. We have compiled a list with some $10^8$ new 
examples of such (2,2) vacua.
The deformation family is w.r.t. the complex 
and the symplectic structure, and it is in general not of fixed topological type. 
To fix the latter a choice of the K\"ahler cone is needed. The possible choices 
follow from the triangulations of $\Delta^*$. The enormous amount of data leads 
immediately to the question of identification of families. The Hodge numbers 
are included in the list of models. For small Picard number, many but not all 
of the models with equal Hodge numbers have diffeomorphic large radius phases 
according to the characterization of topological type by C.T.C. Wall. The 
mirrors of manifolds with diffeomorphic phases are usually in the same family 
of complex structures, but the parameterization of the complex family differs 
by coordinate changes and  different gauge choices of the $(3,0)$ form. We 
also find Calabi--Yau threefolds that are rational homotopy equivalent, but 
have different A--model amplitudes, hence different symplectic structures. 
Furthermore, we find examples of mirror pairs of diffeomorphic Calabi--Yau 
spaces that have different symplectic structures.      

We provide methods to systematically select models with desired properties 
from the list. We focus mainly on two properties: K3 fibration structures and 
free quotients. The first are of obvious interest for reasons discussed above. 
Many new Calabi--Yau spaces with non-trivial fundamental group emerge from the 
latter search. In the toric construction the abelian $\mZ_n$ group actions 
are easily obtained by specifying a sublattice $M'$ in the lattice $M$, with 
index $n=|M|/|M'|$. Such examples are phenomenologically important, as one 
can put Wilson lines to break the gauge group. They also caught attention 
because the distinction between the $K$-theory and the cohomological 
classification of D-branes on such spaces is particularly noticeable.  
   
By taking the intersection of the two searches we get a class of type II 
compactifications, which has interesting and mostly new heterotic duals.
In this class we have $\cF^{(1),{\rm het}}(S,T)= {1\over 2} S+\ldots$ and 
avoid a supergravity  argument~\cite{AL} that the $\cF^{(1),{\rm het}} R_+^2$ 
coupling has to go with $\cF^{(1),{\rm het}}(S,T)= S+\ldots$. The latter 
matches the criterion $\int_X \ch_2 S=24$ of Oguiso~\cite{oguiso} for simply 
connected K3 fibrations on the type II side. The heterotic cases seem to have 
higher level gauge groups. One special case has been studied in~\cite{FHSV}. 
In this case it was argued that the $N=2$ supergravity is of `no-scale' type, 
which implies that there are no instantons corrections to the couplings. 
Necessary is also a vanishing Euler number $\chi$, because a scale dependence 
is introduced by the four loop $\sigma$ model contribution, which is 
proportional to $\chi$. We find that all our toric examples are scale 
dependent, even the ones which have the same Hodge numbers as the model 
in~\cite{FHSV}. This is checked by explicitly calculating the instanton 
contributions to the prepotential.   
   
On the heterotic side another interesting class of compactifications with
extended supersymmetry involves local products of K3$\times T^2$. A number 
of these spaces, which have a fundamental group of rank 1, was obtained as 
Landau--Ginzburg orbifolds with $h^{11}\in\{3,5,9,13\}$ by a $\IZ_2$ or 
$\IZ_3$ acting freely on the torus~\cite{aas}. The latter two cases can be 
realized as toric complete intersections. The additional gravitini are built 
from the nonvanishing $({0,1})$--forms of these manifolds. As the possible 
automorphisms of K3 were classified by Nikulin it should be straightforward 
to construct many more examples with free $\IZ_n$ or $\IZ_n\ex\IZ_m$ actions 
on the torus, which can, however, not be realized as toric complete 
intersections. 

The organization of the paper is as follows: To set our definitions we review
in Section~\ref{sec:ToricCICY} the mirror construction of Calabi--Yau complete
intersections due to Batyrev and Borisov and describe 
the resolution of singularities. Our aim is to relate the algebraic data 
of the threefolds as directly as possible to the combinatorics of the 
polyhedra. 
We describe the criterion for fibrations as well as the construction of the 
free 
quotient examples. The lists of CICYs are available on a web page~\cite{CICY}.
We also introduce the example of the $(h^{11}=2,h^{21}=30)$ free $\mZ_2$ 
quotient which is realized as a complete intersection in codimension two. 
This is later used to exemplify the reduction of the redundancies 
via the calculation of $\cF^{(0)}$ and $\cF^{(1)}$. 
 
The geometry of the toric complete intersections is described in 
Section~\ref{sec:geometry} with special emphasis on the reduction of 
redundancies due to different ambient spaces or to different nef partitions 
of the same reflexive polytope.
We exemplify variants of 
equivalent realizations for the $(2,30)$ model. In the course of the 
discussion we explain how to get from the triangulation to the K\"ahler 
cone and to the generators of the Mori cone. These will be used later 
for the calculation of the B-model amplitudes. Further explicit examples with
small Hodge numbers are treated in Appendix~\ref{sec:other_models}. 
In Appendix~\ref{sec:smallPicard} we present the most complete list
of Calabi--Yau spaces in toric varieties with small Picard numbers. 
For codimension $r\le2$ these are probably complete.

In Section~\ref{sec:TopString} we briefly review the structure of the genus 
zero sector of the topological string. As an
example we discuss the $(2,30)$ model, whose details are relegated to 
Appendix~\ref{sec:PF}. 
The main point is the derivation of the multi parameter B-model propagators 
and the discussion of the integrability conditions. These are solved in 
Section~\ref{sec:regularK3} for many cases. 
In Section~\ref{sec:Ftbt} we give the product form 
for the holomorphic partition function of the topological string in terms
of the Gopakumar-Vafa invariants.

In Section~\ref{sec:regularK3}
we give a geometrical derivation of the general formula
for the topological string amplitudes in the fiber direction for K3 
fibrations with no reducible fibers. We apply the formula to a series
of K3 fibrations with the same fiber type and fixed self intersection of the 
fiber class $E^2=k$, but different Euler 
number. In each case we find the modular form of $\Gamma(2k)$, 
which fixes the all genus prediction in the fiber direction. We 
solve the B-model propagators and make genus 2 predictions also
for the base class. The predictions are checked against the 
direct calculation of the cohomology of the moduli space of the 
M2--brane~\cite{Katz:1999xq}. 
We then discuss the more subtle differences in the cases with the 
same generic fiber and the same Euler number. In this case we find 
rational homotopy equivalent fibrations with different GV invariants.
Finally, we discuss the mirror geometry of a diffeomorphic pair with 
$E^2=4$ fiber. It turns out that one large complex structure limit contains 
terms diverging with ${\rm Li}_3(e^{t_1-t_2})$. We then turn to models whose 
dual heterotic string is known. After a short review of the relevant heterotic
indices we make all genus checks of the heterotic type II dualities for the 
$ST$ model. We summarize the $\Gamma(2k)$ modular forms for many two parameter
fibrations with $E^2=2,4,6$ in Section~\ref{sec:K3fibrations} and 
discuss 
the subtleties
with the non-perturbative completion of the heterotic string. The section 
is 
finished 
with a discussion of the $STU$ model. Beside finding the propagators 
and making higher loop tests, we analyze the restrictions that the 
$STU$-triality symmetry imposes on the non-perturbative topological 
string amplitudes. We find that in this case the non-perturbative BPS 
numbers can also be described by quasi modular functions. In 
Appendix~\ref{sec:STUdetails} we analyze the moduli space of this model
in some detail.

Section~\ref{sec:other_app} contains an application of our 
formalism to the counting of BPS states in the $\tD_5$ del Pezzo surface. 
Collapsing of this surface leads to an elliptic $\tD_5$ singularity and 
the geometry can only be described as complete intersection. We further 
discuss the models with $\mZ_2$ torsion, i.e. free $\mZ_2$ quotients.
Particularly interesting from the point of view of the duality between the
heterotic and the type II string are free quotients which are both
elliptically and K3 fibered. Some general aspects are discussed in this
section while a list of examples is presented in Appendix~\ref{sec:FreeQ}.
Moreover, we consider a class of examples with Euler number $0$. These
have similar properties as the FHSV model but are nevertheless different,
due to instanton corrections in genus zero. We also discuss certain aspects 
of self-mirrors.

\section{Toric CICYs, fibrations and torsion}
\label{sec:ToricCICY}

In this section we first recall the construction of \CY complete intersections
in toric varieties and explain how fibrations and free quotients ({\it i.e.} 
manifolds with non-trivial fundamental group) manifest themselves in the 
combinatorics of reflexive polytopes. We will also devote a section on
the resolution of singularities.

\subsection{Complete intersections in toric varieties}
\label{sec:CICY}

The toric ambient spaces $V_\S$ are defined in terms of a fan $\S$, 
which is a collection of rational polyhedral cones $\s\in\S$, such that it 
contains all faces and intersections of its elements 
\cite{Ful,Oda,Dan,CoxRD}.
$V_\S$ is compact if the support of $\S$ covers all of the real extension 
$N_\IR$ of some $d$-dimensional lattice $N$.
A simple example is the case where $\S$ consists of the cones over the faces 
of a rational polytope $\D^*\subset N_\IR$ containing the origin 
and the 0-dimensional cone $\{0\}$. 
Let $\S(1)$ denote the set of one-dimensional cones with primitive 
generators $\rho_i$, $i=1,\dots,n$. The resulting variety $V_\S$ is smooth 
if all cones are simplicial and if all maximal cones are generated by a 
lattice basis. Singularities at codimension 4 of such an ambient space are 
irrelevant for a sufficiently generic choice of equation for a Calabi--Yau 
3-fold. Higher-dimensional singularities have to be resolved by a subdivision
of the fan, {\it i.e.} by adding to $\S(1)$ rays that are generated by 
lattice points on faces of $k\D^*$, $k \geq 1$. We will address this 
issue in more detail in Section~\ref{sec:singularities}. 

The simplest description of $V_\S$ introduces homogeneous 
coordinates $z_i$ for the $n$ generators $\rho_i$ of the rays in $\S(1)$ and
weighted projective identifications 
\BE\label{eq:weights}
        (z_1:\ldots:z_n)\sim
        (\lambda^{q_1^{(a)}}z_1:\ldots:\lambda^{q_n^{(a)}}z_n)
        \qquad a = 1,\dots,h 
\EE
where $\lambda\in\IC^*$ is a nonzero complex number, and the $n$-vectors 
$q_i^{(a)}$ are generators of the linear relations $\sum q_i^{(a)}\rho_i=0$ 
among the primitive lattice vectors $\rho_i$. To obtain a well-behaved 
quotient $V_\S=(\IC^n-Z)/(\IC^*)^{h}$ we have to exclude an exceptional set 
$Z\subset\IC^n$ that is defined in terms of the fan \cite{Cox,CoxRD}. 
There are $h=n-d$ independent $\IC^*$ identifications so that the 
dimension of $V_\S$ equals the dimension $d$ of the lattice $N$.

We have to be more precise about the quotient defining $V_\S$: Let $N'$ be 
the sublattice of $N$ that is generated by the lattice vectors in $\S(1)$. 
If $N'$ is a sublattice of $N$ with index $I>1$ then we need some additional 
identifications \cite{Cox}. Let $\S'$ be the fan obtained 
from $\S$ by relating everything to the lattice $N'$. Then $V_\S=V_{\S'}/G$ 
is a quotient of $V_{\S'}$ by a finite abelian group $G$ isomorphic to $N/N'$
that acts by multiplication with phases $\omega^{\alpha_i}$ on the homogeneous
coordinates~$z_i$. Here $\omega$ is a $|G|^{\textrm{th}}$ root of unity. 
We will denote such group actions by $(\alpha_1,\dots,\alpha_n)/|G|$, 
where $\sum_{i=1}^n \alpha_i = 0 \mod |G|$. 
To see how this comes about we note that the ring of regular functions 
on an affine coordinate patch $U_\sigma$ of $V_\S$ is spanned by the monomials 
$\prod z_i^{<m,\rho_i>}$, where $m\in \sigma\spcheck \cap M$ is a vector in 
the dual lattice $M=\mathrm{Hom}(N,\IZ)$. 
If we change from the lattice $N'$ to the finer 
lattice $N$ then we have to exclude all monomials corresponding to vectors 
$m\in M'$ that do not belong to the sublattice $M\subset M'$. Thus there 
are no more functions available to distinguish points in 
$V_{\S'}$ that live on orbits of $G$ (in turn, this can be used to define 
$G$). The quotient $V_\S=V_{\S'}/G$ is never free for a toric variety 
\cite{Dan}. If, however, a (Calabi--Yau) hypersurface or complete intersection
does not intersect the set of fixed points then we get a manifold with 
nontrivial fundamental group as will be discussed in subsection~\ref{sec:FQ}.

Batyrev showed that a generic section of the anticanonical bundle of 
$\IP_{\D^*}=V_{\S(\D^*)}$ 
defines a Calabi--Yau hypersurface if $\D$ is
reflexive, which means, by definition, that $\D$ and its dual
\BE
        \label{eq:dual}
        \D^*=\{x\in N_\IR|(x,y)\ge-1 ~\forall y\in\D\}
\EE
are both lattice polytopes. Mirror symmetry corresponds to the 
exchange of $\D$ and $\D^*$~\cite{Bat}.
The generalization of this construction to complete intersections of 
codimension $r>1$ (CICYs) was obtained by Batyrev and 
Borisov~\cite{BBnef} who introduced the notion of a nef partition. 
If $\Delta\subset M_\mR,\Delta^*\subset N_\mR$ is a dual pair of
$d$--dimensional reflexive polytopes then a partition 
$E=E_1\cup\dots\cup E_r$ of the set of vertices of $\Delta^*$ into 
disjoint subsets $E_1,\dots,E_r$ is called a nef--partition if there 
exist $r$ integral upper convex $\Sigma(\Delta^*)$--piecewise linear 
support functions $\phi_l:N_\mR\rightarrow \mR\ (l=1,\dots,r)$ such 
that
$$
\phi_l(\rho^*)=
\begin{cases}
1& \text{if $\rho^*\in E_l$,}\\
0& \text{otherwise.}
\end{cases}
$$
Each $\phi_l$ corresponds to a semi--ample Cartier divisor
\begin{equation}
  D_{0,l}=\sum_{\rho^*\in E_l} D_{\rho^*}
  \label{eq:CICY}
\end{equation}
on $\mP_{\Delta^*}$, where $D_{\rho^*}$ is the irreducible component 
of $\mP_{\Delta^*}\setminus \mT$ corresponding to the vertex 
$\rho^*\in E_l$, and $X=D_{0,1}\cap\dots\cap D_{0,r}$ defines a family 
of Calabi--Yau complete intersections of codimension $r$.
Moreover, each $\phi_l$ corresponds to a lattice polyhedron $\Delta_l$
defined as
\begin{equation}
  \label{eq:di}
  \Delta_l=\{x\in M_\mR:(x,y)\geq -\phi_l(y)\ \forall\ y\in N_\mR\}  
\end{equation}
supporting global sections of the semi--ample invertible sheaf 
${\cal O}(D_i)$. (The sum of the functions $\phi_l$ is equal to the 
support function of the anticanonical bundle of $\mP_{\Delta^*}$.) 
Since the knowledge of the decomposition $E=E_1\cup\dots\cup E_r$
is equivalent to that of the set of supporting polyhedra
$\Pi(\Delta)=\{\Delta_1,\dots,\Delta_r\}$, this data is often also called 
a nef partition. \cite{BBnef} further observed that, for a nef partition
$\Pi(\Delta)$, the polytopes 
$\nabla_l=\langle\{0\}\cup E_l\rangle\subset N_\mR$ also define a
nef partition $\Pi^\ast(\nabla)=\{\nabla_1,\dots,\nabla_r\}$ such that
the Minkowski sum $\nabla=\nabla_1+\dots+\nabla_r$ is a reflexive polytope. 
This leads to a combinatorial manifestation of mirror symmetry in terms of 
dual pairs of nef partitions of $\D^*$ and $\nabla^*$,
\begin{eqnarray}
\label{eq:polyDN}
    M_{\mR}\qquad\qquad && \qquad\qquad N_{\mR}\nonumber\\
    \D=\D_1+\ldots+\D_r && \D^*=\langle\nabla_1,\ldots,\nabla_r\rangle
        \nonumber\\[-5pt]
        &~~~~ (\D_l,\nabla_{l'})\ge-\delta_{l\,l'} ~~~~&\\[-5pt]\nonumber
        \nabla^*=\langle\D_1,\ldots,\D_r\rangle &&
                \nabla=\nabla_1+\ldots+\nabla_r
\end{eqnarray}
where the angles denote the convex hull. This yields a pair $(X,X^*)$ of 
Calabi--Yau varieties, where $X \subset \mP_{\Delta^*}$ and 
$X^* \subset \mP_{\nabla^*}$ are mirror to each other. 

Mirror symmetry identifies the quantum corrected K\"ahler moduli space of 
$X$ with the uncorrected complex structure moduli space of $X^*$. 
For details see the excellent treatise in~\cite{Cox:vi}. 
The deformations of the complex structure of $X^*$ are encoded in the 
periods $\int_{\gamma} \Omega$ and the latter can be computed from the 
equations that cut $X^*$ out of $\mP_{\nabla^*}$. For this reason, we now 
describe these equations for $X^*$ (instead of those of $X$ which is the
space of interest). Given the nef partition 
$\Delta^* = \langle \nabla_1,\dots,\nabla_r\rangle$, 
$\nabla_l = \langle E_l, \{0\} \rangle$, 
we define $a_m \in \mC$ to be the coefficients of the Laurent polynomials 
$f_1,\dots,f_r$ 
\begin{equation}
  \label{eq:CIeqs}
  f_l(t) = a_{0,l} - \sum_{m \in A_l} a_m t^m, \qquad l = 1,\dots,r
\end{equation}
where $t^m = \prod_{j=1}^d t_j^{m_j}$ and 
$t\in \mT \subset \mP_{\nabla^*}$ are the coordinates in the algebraic 
torus. The set $A_l$ is the set of all points in $\nabla_l$. 
The simultaneous vanishing of these polynomials $f_l$ then defines 
the complete intersection Calabi--Yau manifold $X^* \subset \mP_{\Delta^*}$, 
or in other words, the $\nabla_l$ are the Newton polyhedra for the $f_l$. 
Similarly, the $\Delta_l$ are the Newton polyhedra for $X$. 
In Section~\ref{sec:2,30} we will discuss the combinatorial duality
in~\eqref{eq:polyDN} in detail for an example. The computation of the periods 
starting from~\eqref{eq:CIeqs} will be exhibited in the same example in 
Appendix~\ref{sec:PF}.

In \cite{BBsth} Batyrev and Borisov were able to obtain a formula for the
generating function 
\BE                                             
        E(t,\bar t)=\sum (-1)^{p+q}~h^{pq}~t^p\,\bar t^q=\sum_{I=[x,y]}
        \frac{(-)^{\rho_x}t^{\rho_y}}{(t\bar t)^r}S(C_x,\frac{\bar t}{t})
        S( C_y^\ast,t\bar t) B_I(t^{-1},\bar t) \label{BBstrh}
\EE
of the Hodge numbers $h^{pq}$ of a toric CICY with dimension $d-r$ in terms 
of the combinatorial data of the $d+r$ dimensional Gorenstein cone 
$\Gamma(\Delta_l)$ spanned by vectors of the form $(e_l,v)$, where $e_l$ 
is a unit vector in $\IR^r$ and $v\in\Delta_l$.%
\footnote{
        The $d+r$ dimensional cone $\Gamma(\nabla_l)$ originates from the
        Cayley trick, which relates the vanishing set of the $r$ equations 
        $f_l=0$ to the complement of the hypersurfaces $1-\sum a_l f_l=0$
        \cite{DK,GKZ} with $\Delta_l$ being the Newton polytopes of $f_l$.
}
In this formula $x,y$ label faces $C_x$ of dimension $\rho_x$ of 
$\Gamma(\nabla_l)$ and $C_x^*$ denotes the dual face of the dual cone 
$\Gamma(\D_l)$. The interval $I=[x,y]$ labels all cones that are faces of 
$C_y$ containing $C_x$. The polynomials $B_I(t,\bar t)$ encode the 
combinatorics of the face lattice \cite{BBsth}. The polynomials
$S(C_x,t)=(1-t)^{\rho_x}\sum_{n\ge0}t^nl_n(C_x)$ of degree $\rho_x-1$ are
related to the numbers $l_n(C_x)$ of lattice points at degree $n$ in $C_x$ 
and hence to the Ehrhart polynomial \cite{Enum} of the Gorenstein polytope 
generating $C_x$ \cite{BBsth}.  
A Gorenstein cone $\Gamma$ is, by definition, generated by lattice points
$\rho_i$ with $\<\rho_i,n_0\>=1$ for some vector $n_0$ in the dual lattice. 
$n_0$ is unique if $\Gamma$ has maximal dimension and its duality pairing 
defines a nonnegative grading of $\Gamma$. The points of degree 1 form 
the Gorenstein polytope, whose vertices generate $\Gamma$. The points of 
higher degree will play an important role in Section~\ref{sec:singularities}.

When we define the variety $\IP_{\D^*}=V_{\S(\D^*)}$, the fan 
$\Sigma(\Delta^*)$ is not only specified by the lattice $N$, and the 
polyhedron $\Delta^* \subset N$ but also by a triangulation 
$T=T(\Delta^*)$. For simplicity, we will suppress the choice of a 
triangulation in the notation. 
However, we want to point out that in the case of complete intersections
only those triangulations are compatible with a given nef partition 
that can be lifted to a triangulation of the corresponding Gorenstein 
cone $\Gamma(\Delta^*)$~\cite{Stienstra:1998ab},
i.e. the triangulation actually is $T=T(\Gamma(\Delta^*))$. 
Note that in the case of hypersurfaces, triangulating the Gorenstein cone 
is the same as triangulating the polyhedron. Hence, the distinction does 
not play a role. In codimension $r>1$, however, there are in general 
differences due to the nef partition. An example will be discussed 
in Section~\ref{sec:phases}.

In the case of hypersurfaces, $r=1$, it is known \cite{Bat} that 
the Picard number can also be computed with the formula
\BE     h^{11}=l(\D^\ast)-1-d-\sum_{\mathrm{codim}(\Th^*)=1}l^*(\Th^*)+
                \sum_{\mathrm{codim}(\Th^*)=2}l^*(\Th^*)l^*(\Th)    
        \label{Bhodge}
\EE
where $\Th$ and $\Th^*$ are faces $\D$ and $\D^*$, respectively. $l(\Th)$ is
the number of lattice points of a face $\Th$, and $l^*(\Th)$ the number of 
its interior points. This formula has a simple interpretation (see 
also~\cite{Mavlyutov}): The principal contributions come from the toric 
divisors $D_i = \{ z_i = 0\}$ that correspond to points in $\D^*$ different 
from the origin. There are $d$ linear relations among these divisors. The 
first sum corresponds to the subtraction of interior points of facets. 
The corresponding divisors of the ambient space do not intersect a generic 
Calabi--Yau hypersurface. Lastly, the bilinear terms in the second sum can 
be understood as multiplicities of toric divisors so that their presence 
indicates that only a subspace of the cohomology ({\it i.e.} the K\"ahler 
moduli space) can be analyzed with toric methods. 

Unfortunately, the general formula (\ref{BBstrh}) does not lend itself to
a similar interpretation in any known way.%
\footnote{
        There are, however, more suggestive formulas for $h^{11}$ in 
        special cases \cite{BBnef}.
} 
As for hypersurfaces, interior
points of facets of $\D^*$ never contribute divisors to a toric Calabi--Yau,
but explicit computations of intersection rings show that for complete 
intersection it may happen that even for vertices $\rho_i$ the corresponding
divisor $D_i$ does not intersect $[X]=\bigcap_{l=1}^r D_{0,l}$, where 
corresponds to the $l^{\rm th}$ equation $f_l$ of the mirror $X^*$ of 
the Calabi--Yau variety $X$. A simple 
example is the blowup by a non-intersecting divisor of the degree (3,4) CICY 
in $\mP^5_{111112}$ with Hodge numbers (1,79) that is discussed in 
\cite{wpci}. It is very important to find a more explicit
formula for the Picard number of a complete intersection that allows for
an interpretation in terms of the multiplicities of divisors $D_i$ after
restriction to the Calabi--Yau. There is a good chance that such a formula 
exists: While intersection numbers depend on the triangulation of the fan 
we observed in many examples that these multiplicities are independent of
the triangulation and thus should depend only on the combinatorics of the
data of the polytope that enter (\ref{BBstrh}).

\subsection{Free quotients}
\label{sec:FQ}

We now come to the discussion of toric \CY spaces with non-trivial 
fundamental groups. We mainly restrict our attention to the situation where 
 they arise from free quotients coming from group actions that 
correspond to lattice quotients. We call such quotients {\it toric}.
As we noted above, a refinement of the lattice $N$ always 
entails singularities in the ambient space and no contributions to the
fundamental group \cite{Dan}. If, however, a CICY does not intersect the
singular locus of that quotient then the group acts freely on that variety 
and $\pi_1$ becomes isomorphic to $G$. This is the case if the refinement of
the lattice does not lead to additional lattice points of $\D^*$.
For a given pair of reflexive polytopes, the dual of the lattice $M'$ that
is spanned by the vertices of $\D$ is the finest lattice $N'$ with respect to 
which the polytope is reflexive. If the lattice $N_0$ that is spanned by the 
vertices of $\D^*$ is a proper sublattice of the lattice $N'$ then any 
subgroup of $N'/N_0$ corresponds to a different choice of the $N$ lattice
and hence to a different toric CY hypersurface. There is thus only a finite 
number of lattices that have to be checked to find all toric free quotients.

Some well-known examples are the free $\IZ_5$ quotient of the quintic and the
free $\IZ_3$ quotient of the CY hypersurface in $\IP^2\ex\IP^2$. 
For both cases cyclic permutation of the coordinates of the projective
spaces defines another free group action of the same order that commutes
with the toric quotient, leading to Euler numbers 8 and 18, respectively. 
These free ``double quotients'' are, however, not toric in the sense that 
the resulting manifold is not a CICY in a toric variety. Analyzing the 
complete list of 473'800'776 reflexive polytopes for CY hypersurfaces 
\cite{c4d,pwf,CYwien}
one finds 14 more examples of toric free quotients \cite{PALP}: The 
elliptically fibered $\IZ_3$ 
quotient of the degree $9$ surface in $\mP^4_{11133}$, whose group action 
on the homogeneous coordinates is given by the phases $(1,2,1,2,0)/3$,
and 13 elliptic K3 fibrations where the lattice quotient has index 2.
Among the latter there is the $\IZ_2$ quotient of $(\IP^1)^4$ with phases
$(0,\frac12)$ on each factor which admits an additional $\IZ_2$ freely 
acting on the CY hypersurface by simultaneous exchange of the coordinates of 
all $\IP^1$ factors. Models of a similar type are currently studied because 
of their promising phenomenological properties~\cite{Ovrut:2002jk},
\cite{Donagi:2003tb}.

The condition that $\D^*\cap N$ and $\D^*\cap N'$ coincide is sufficient for 
a free quotient of a CICY with dimension up to 3
because the singularities of a maximal crepant resolution are at 
codimension 4 and can be avoided by a generic choice of the defining 
equations. It is, however, not necessary:  Divisors corresponding 
to interior points of facets of $\D^*$ do not intersect the CY and hence 
do not kill the fundamental group if they are generated by a refinement
of the $N$ lattice\footnote
{       This criterion has been suggested by Victor Batyrev 
        (private communication).
}.

An analysis of the complete lists of reflexive polytopes with up to 4 
dimensions shows that the only case where the weaker condition is 
relevant for hypersurfaces is that of the torus. In that case a free $\IZ_3$
quotient in $\IP^2$, and a free $\IZ_2$ in $\mP^2_{112}$ and in $\IP^1\ex\IP^1$ 
can be realized torically in the obvious way.
Taking a product with a (toric) K3 this can be used to construct CICYs at
codimension 2 with first Betti number $b_1=2$. The resulting Hodge numbers
are $h^{11}=13$ for the $\IZ_2$ quotients and $h^{11}=9$ in case of $\IZ_3$;
the simplest examples are
$\IP^2\ex \mP^3_{1122}$ with group action $(0,1,2;0,1,0,2)/3$ for $\IZ_3$, and
$\mP^2_{112}\ex \IP^3$ with phases $(0,1,1;0,1,0,1)/2$ or 
$\IP^1\ex\IP^1\ex\IP^3$ with phases $(0,1;0,1;0,1,0,1)/2$ for $\IZ_2$.
%
%
These Hodge numbers have also been obtained for Landau--Ginzburg orbifold
models \cite{aas}, where the values $h^{11}=3, 5, 9, 13$ were found for
$h^{01}=1$ \cite{aas,CYwien}. One might expect that the models with $3$ and 
$5$ would require a larger order of the free group action. It was shown, 
however, in \cite{aas} that the 1-forms in Landau--Ginzburg models can only
arise if the model factors into $T^2\ex \mathrm{K3}$. Moreover, the twist that
reduces the Picard number must act with unit determinant on the $T^2$ and hence
must agree with the free $\IZ_3$ on the elliptic curve in $\IP^2$ or with the
free $\IZ_2$ on $\IP^2_{112}$ that we already know from the other examples.
Only the K3s can be realized in different ways. Indeed, analyzing the 
lists of \cite{aas} we found several realizations of these Hodge numbers with
K3s that correspond the generalized Calabi--Yau varieties in the sense of
\cite{Candelas:1993nd,BBgen} like the cubic in $\IP^5$, where a $\IZ_3$-twist 
with phases $(1,1,1,0,0,0)/3$ and a free action on the torus leads to 
$h^{11}=h^{12}=3$
and the degree 12 hypersurface in $\IP^5_{334446}$ where a $\IZ_2$ with
phases $(0,0,0,0,1,1)/2$ and a free action on the torus in $\IP_{112}$
leads to $h^{11}=h^{12}=5$.
For all of these models a mirror construction is available both in the 
CFT framework \cite{BH,DT} and, more geometrically, in terms of reflexive 
Gorenstein cones \cite{BBgen}.

There are many more examples of toric free quotients for CICYs with 
codimension
$r>1$, some of which will play a role later on. In that case a group may act 
freely under even weaker conditions because even divisors that correspond to 
vertices may not intersect the CY. But with the present state of the art this 
requires a case by case analysis of the intersection ring. For Calabi--Yau 
4-folds, on the other hand, the above criterion for a free quotient is no 
longer sufficient because the codimensions of the singularities in the 
ambient 
space may be too small to avoid them by an appropriate choice of the 
hypersurface equations. There are many examples where this happens.%
\footnote{
        Free quotients can be constructed easily with PALP \cite{PALP}. 
        To find all candidates among quotients of the sextic 4-fold one can
        use the following commands ({\tt zbin.aux} is an auxiliary file 
        storing polytopes on sublattices)
\\{\tt\$ cws.x -w5 6 6 -r | class.x -f -sv -po zbin.aux }
\\{\tt\$ class.x -b -pi zbin.aux | class.x -sp -f }
\\ and obtains the N-lattice polytopes in a basis where the lattice 
quotient is diagonal. 
\\{\tt\$ class.x -b -pi zbin.aux | class.x -sp -f|cws.x -N -f | poly.x -fg }
\\displays weight systems and Hodge data. This yields 6 candidates for 
free $\IZ_3$ quotients. They are, however, all singular, as the Euler numbers 
are bigger than 1/3 of those of the respective covering spaces.
}
\subsection{Resolution of terminal singularities}

\label{sec:singularities}
It is known that the points on the polyhedron $\Delta^*$ are in general not
sufficient to give a smooth ambient space~\cite{Hosono:1993qy,Hosono:1995bm}. 
In addition, one may have to take into account certain points in degree 
greater than one, {\it i.e.} points in $k\Delta^*$, $k > 1$, in order to 
resolve all singularities. In this section we 
give a general discussion of these points, in particular in the context of 
constructing smooth complete intersection Calabi--Yau spaces.

The toric ambient space is smooth if its fan is simplicial and unimodular,
{\it i.e.} if all its cones are simplicial and 
generated by (a subset of) a lattice basis.        
Suppose we have added all the lattice 
points in the polyhedron $\Delta^*$ and determined one of the possible star
triangulations of this set of points. A star triangulation is a triangulation
$T$ for which all maximal simplices contain the origin of $\Delta^*$. In other words, 
each simplex $\sigma\in T$ determines a pointed cone $C_\sigma$, {\it i.e.} 
a cone over a facet of $\Delta^*$ whose apex is the origin. 
If there is a simplex, say $\sigma$, with
$\Vol(\sigma) >1$, then the corresponding coordinate patch 
$U_\sigma=\mathrm{Spec}\,\mC[\sigma\spcheck \cap M]$ 
of the toric variety will be singular. 
Reflexivity implies that the resulting triangulation of the fan is 
unimodular if and only if the induced triangulations of the facets only
consist of
unimodular simplices. This is always the case in 3 dimensions, but for
higher-dimensional ambient spaces there can be singularities at codimension 4.
Generic complete intersection Calabi--Yau 3-fold (and even more so K3 
surfaces) are smooth because the singularities of the ambient space can 
be avoided by an appropriate choice of the parameter on 
dimensional grounds. For the evaluation of the Mori cone of the
ambient space we will, however, need to resolve all of its singularities.
Resolutions that involve points at higher degree $k>1$ are discrepant, i.e.
they contribute to the canonical class of the ambient space. This does not, 
however, spoil triviality of the canonical bundle of the Calabi--Yau complete
intersection, which does not intersect the additional exceptional divisors.

We first discuss the case of 4-dimensional polytopes, which is relevant
for Calabi--Yau hypersurfaces. For the 3-dimensional intersection of the
cone with volume $V$ with the relevant facet we can always choose a basis
with $\v_0=0$, $\v_1=e_1$, $\v_2=e_2$ and $\v_3=(x,y,V)^T$ with 
$0\le x\le y<V$.
Emptiness of this tetrahedron, i.e. absence of additional lattice points in
the convex hull, further constrains the vector $(x,y)$ and lattice 
automorphisms
make some of the allowed combinations redundant.
Clearly the number of lattice points in the semi-open parallelepiped
$\Pi_\s=\{\sum\lambda_i \v_i~|~\lambda_i\in[0,1)\subset\IR\}$ is equal 
to the volume. The $V-1$ non-zero lattice points in $\Pi_\s$
form the Hilbert basis $\mathrm{Hilb}_\s=\Pi_\s\cap N-\{0\}$ of the 
semi-group of lattice points in the cone that is generated by $\v_i$.
There is a duality $\lambda_i\to1-\lambda_i$ 
so that it is sufficient to determine all lattice points at degree up to 
half the dimension of the facet, where $\v_i$ have degree 1. 
It is natural to try to resolve the singularities by first adding the rays 
that are generated by the points in $\mathrm{Hilb}_\s$. This turned out to 
be sufficient for all our examples. There are, however, empty simplices that
span Gorenstein cones whose resolution requires additional rays (see below).

The simplest case in 4 dimensions, which are relevant for CY 
hypersurfaces, is the simplex with $x=y=1$. It obviously is 
empty for any volume. The $V-1$ interior points of $\Pi$ are
$\frac lV(\v_0+\v_3)+(1-\frac lV)(\v_1+\v_2)$ for $0 < l < V$. This leads
to the matrix
\BE                            \label{NF4d}
\left(\Msize\BM1 & 1 & 1 & 1 & ~2  &2& \ldots & 2\cr 
                0 & 1 & 0 & 1 & ~1 &1& \ldots & 1 \cr
                0 & 0 & 1 & 1 & ~1 &1& \ldots & 1 \cr
                0 & 0 & 0 & V & ~1 &2& \ldots & V-1 \cr
                \EM\right)
\EE
of column vectors,
whose first coordinate is the degree of the corresponding point in
the Gorenstein cone spanned by $\v_i$ in an appropriate basis. 
It is easily checked that the unique maximal star triangulation, 
consisting of the $4(V-1)$ simplicial cones spanned by the tetrahedra
\\1.: $\v_1$, $\v_2$, the first point at degree 2 and one of $\v_0$ or $\v_3$,
\\2.: $(\v_0,\v_1)$, $(\v_0,\v_2)$, $(\v_1,\v_3)$ or $(\v_2,\v_3)$ and two 
        adjacent points at degree 2,
\\3.: $\v_0$, $\v_3$, the last point at degree 2 and one of $\v_1$ or $\v_2$,
\\is unimodular for every $V>1$.
The first example that cannot be brought into this form shows up at volumes 5.
In an appropriate basis we find
\BE                            \label{NFvol5}
\left(\Msize\BM 1 & 1 & 1 & 1 & ~2 &2& 2 & 2\cr 
                0 & 1 & 0 & 1 & ~1 &1& 1 & 1 \cr
                0 & 0 & 1 & 2 & ~1 &1& 2 & 2 \cr
                0 & 0 & 0 & 5 & ~1 &2& 3 & 4 \cr
                \EM\right)
\EE
This is the simplest example where $\mathrm{Hilb}_\s$ is not sufficient
and a point at degree 3 is required for a unimodular triangulation. 

For complete intersections at codimension two we need a similar discussion
for 5-dimensional polytopes. 
At this point it is useful to recall the notion of a circuit 
(see e.g.~\cite{GKZ}). A circuit is a collection $Z$ of points in an affine
space such that any proper subset $Z' \subset Z$ is affinely independent but
$Z$ itself is not, {\it i.e.} the points satisfy
\begin{align}
  \label{eq:circuit}
  \sum_{\v_m \in Z} c_m \v_m &= 0 & \sum_{\v_m \in Z} c_m &= 0
\end{align}
In other words, we can obtain a circuit by adding 
one point in general position to the set of vertices of a simplex. 
It can be shown that 
the convex hull of a circuit has precisely the two triangulations 
$T_\pm = \{ \conv(Z \setminus \{\v_m\}) | \v_m \in Z_\pm\}$, 
where $Z_\pm = \{\v_m \in Z | c_m \lessgtr 0 \}$. As an easy 
consequence we note that 
the volumes $\Vol(Z \setminus \{\v_m\})$ are proportional to the
modulus $|c_m|$ of the coefficients
in the linear relation.
This gives therefore an easy way to find simplices of the triangulation of 
$\Delta^*$ whose volume is larger than one, and have to be subdivided.

The simplest case of a singularity in higher-dimensional simplices is the one 
where the volume comes from a facet, from which the remaining vertex thus
has distance 1. This effectively reduced the dimension of the
problem by one because we just need to triangulate the facet and 
joining the (unimodular) simplices of the triangulation with additional 
vertices. Clearly such singularities come in pairs and the simplex on the
other side of the facet has to be triangulated in a consistent way. As long
as the triangulation of the facet is unique (as in all our examples) this is 
automatic. 

This leaves us with a discussion of simplices whose volume is larger than
that of any of its facets.
In 5 dimensions this can only occur for volume $V>2$. We again choose a basis
of the form%
\footnote{
        This would not cover the most general case for 6-dimensional 
        Gorenstein cones because there are empty simplices in $\IR^5$ 
        with no unimodular facets \cite{HaZi00}.
}
$\v_0=0$ and $\v_i=e_i$ for $i<4$. For volume 3 there is only one case 
with points in the interior $\Pi^0$ of $\Pi$, namely $\v_4=(2,2,2,3)^T$, 
so that
$\mu=(2\v_0+\v_1+\v_2+\v_3+\v_4)/3\in\Pi^0$ has degree 2 and
$\mu'=S-\mu$ with $S=\sum\v_i$ is its dual interior point at degree 3.
Addition of the ray generated by $\mu$ splits the cone into 5 simplices,
four of which are unimodular while $\langle\mu\v_1\v_2\v_3\v_4\rangle$ has volume 2.
Since $2\mu'=\mu+\v_1+\v_2+\v_3+\v_4$ the remaining singularity is resolved
by adding the unique interior point $\mu'\in\Pi^0$ at degree 3.
We thus end up with the 9 unimodular cones generated by         
$\langle\mu0\,\widehat{\hbox{\small$1234$}}\rangle$ and 
$\langle\mu'\,\widehat{\hbox{\small$\mu1234$}}\rangle$, 
where the vertices $\v_i$ of the original simplex are represented by
their indices $i$ and the hat indicates to take all simplices that arise 
by dropping one of the respective points.

At volume 4 there must be at least one nonsimplicial facet (this is true 
for every even volume in 5 dimensions because interior points of $\Pi$ come 
in pairs at degrees 2 and 3, respectively). There is, again, only one class
of cones with an interior point $\mu$, for which we 
can choose $\v_4=(1,1,1,4)^T$ and thus
$\mu=(2\v_0+\v_1+\v_2+\v_3+3\v_4)/4\in\Pi^0$.
If we first resolve the singularity of the facet with volume 2, which does 
not contain $\v_0$, by adding its degree two point 
$\rho=(\v_1+\v_2+\v_3+\v_4)/2$ we are left with 4 simplices of 
volume 2. Since $\mu=(\rho+\v_0+\v_4)/2$ is located at the intersection of 
3 of these simplices it resolves all but the simplex 
$\langle\v_0\v_1\v_2\v_3\rho\rangle$, which has volume 2.
The remaining singularity is resolved by the interior point 
$\mu'=S-\mu=(\v_0+\v_1+\v_2+\v_3+\rho)/2\in\Pi^0$ at degree 3. 
We thus obtain the 14 unimodular cones 
$\langle\rho\,\widehat{\hbox{\small$\mu04$}}\,
        \widehat{\hbox{\small$123$}}\rangle$ and 
$\langle\mu'\,\widehat{\hbox{\small$0123\rho$}}\rangle$.

We finish our discussion with the case of volume 5.
In addition 
to the two cases where the volume comes from a facet there are two 
inequivalent situations with interior points:
\\[3pt]
For the first class of examples with 5 unimodular facets we can choose 
$\v_4=(1,2,2,5)^T$. Adding the rays generated by the two points
$\mu_1=(\v_0+\v_1+2\v_2+2\v_3+4\v_4)/5$ and 
$\mu_2=(3\v_0+3\v_1+\v_2+\v_3+2\v_4)/5$ at degree 2 we find a unique
maximal triangulation with 8 unimodular cones and the 3 singular cones
$\langle\widehat{\hbox{\small$01\mu_1$}}\,23\mu_2\rangle$ with
volume two, which are resolved by adding the rays through the points 
$\mu_1'=\2(\v_0+\v_1+\v_2+\v_3+\mu_2)$ and $\mu_2'=\2(\v_2+\v_3+\mu_1+\mu_2)$.
All together this yields the $
21$ unimodular cones
$\langle\widehat{01}\,234\mu_1\rangle$,
$\langle\widehat{\hbox{\small$01\mu_1$}}\,\widehat{23}\,4\mu_2\rangle$,
$\langle\widehat{\hbox{\small$0123\mu_2$}}\,\mu_1\mu_1'\rangle$ and 
$\langle\widehat{01}\,\widehat{\hbox{\small$23\mu_1\mu_2$}}\mu_2'\rangle$.
\\[3pt]
The last case at volume 5 is given by $\v_4=(1,1,3,5)^T$. It is slightly
more complicated because the interior points 
$\mu_1=(\v_0+\v_1+\v_2+3\v_3+4\v_4)/5$ and 
$\mu_2=(2\v_0+2\v_1+2\v_2+\v_3+3\v_4)/5$ at degree 2 now admit two different
maximal triangulations. Only the one where we first add $\mu_2$ and then
subdivide with $\mu_1=\2(\v_3+\v_4+\mu_2)$, which consists of 10 unimodular 
simplices 
and the simplex $\langle0123\mu_2\rangle$ with volume 3, is refined to a
unimodular triangulation by $\mu_1'=\frac13(\mu_2+2\v_0+2\v_1+2\v_2+\v_3)$
and $\mu_2'=\2(\mu_1'+\mu_2+\v_3)$. The resolution of the singularity
thus again requires 21 simplices, namely
$\langle\widehat{\hbox{\small$012$}}\,\widehat{\hbox{\small$34\mu_2$}}
        \,\mu_1\rangle$, $\langle0124\mu_2\rangle$,
$\langle012\,\widehat{\hbox{\small$3\mu_2$}}\,\mu_1'\rangle$ and
$\langle\widehat{\hbox{\small$012$}}\,
        \widehat{\hbox{\small$3\mu_2\mu_1'$}}\mu_2'\rangle$.

Since we are ultimately interested in complete intersection \CY spaces in 
toric varieties described by the polyhedron $\Delta^*$, we also have to 
associate these extra points with the nef partition of $\Delta^*$ 
and determine their contributions to the corresponding divisors $D_{0,l}$.
This follows from the properties of the integral piecewise linear 
functions used in the definition of the nef-partition given in 
section~\ref{sec:CICY}. Indeed, while the additional rays need not belong 
to any of the cones over $\nabla_l$, the values of the corresponding linear 
functions are integral on their generators (and obviously add up to their 
degrees). Therefore,~\eqref{eq:CICY} becomes
\begin{equation}
  D_{0,l}=\sum_{\nu^*_i\in E_l}D_i + \sum_{q} c_{\r,q,l} D_{\r,q},
  \label{eq:CICYres}
\end{equation}
where the second sum runs over the extra points $\mu,\, \mu',\, \rho,\,\dots$ 
needed for the resolution and the coefficients $c_{\r,q,l}$ are 
integers and satisfy $\sum_{l=1}^r c_{\r,q,l} = k$. For example, in the case 
$d=5$ and $r=2$, {\it i.e.} \CY threefolds, and $\Vol(\sigma) = 2$ only an 
even number of the $\nu_i^*$, $i=0,\dots,3$ in~\eqref{NF4d} can belong to 
one $E_l$, and therefore $(c_{\r,1},c_{\r,2})$ is either $(2,0)$, $(1,1)$, 
or $(0,2)$ (we dropped the index $q=1$). 

One last point concerns the toric quotients of section~\ref{sec:FQ}. If 
the variety is a quotient, the 
volumes of all the simplices will be multiples of the index $N':N_0$. 
In this case it is much simpler to work on the simply connected covering 
space so that we only need to resolve the singularities of its ambient 
toric variety.

\subsection{Fibrations}

We will now discuss some properties of fibrations. Again, we restrict
ourselves where the combinatorial data of the polytopes contain the relevant 
information. For general, also non-toric K3 surfaces and Calabi--Yau 3-folds
there exists a criterion by Oguiso for the existence of elliptic and K3
fibrations in terms of intersection numbers \cite{oguiso}. We will state it in 
Section~\ref{sec:X_A}. Like the latter, 
fibration properties thus depend on the triangulation, or in other words, 
on the choice of the phase in the extended K\"ahler moduli space. 

For toric Calabi--Yau spaces there is, however, a more direct way to 
search for fibrations that manifest themselves in the geometry of the 
polytope 
and to single out appropriate triangulations~\cite{AKMS},\cite{fft},
\cite{Hu:2000pr},\cite{pwf}. 
These fibrations descend from toric morphisms of the ambient
space \cite{Ewald,Ful}: Let $\S$ and $\S_b$ be fans in $N$ and $N_b$,
respectively, and let $\phi:N\to N_b$ be a lattice homomorphism that induces 
a map of fans $\phi:\S\to\S_b$ such that for each cone $\s\in\S$ there
is a cone $\s_b\in\S_b$ that contains the image of $\s$. Then there is a 
$T$-equivariant morphism $\tilde\phi:V_\S\to V_{\S_b}$ and the lattice $N_f$ 
for the fibers is the kernel of $\phi$ in $N$.

For our construction of a fibered Calabi--Yau variety we require the 
existence of a reflexive section $\D^*_f\subset\D^*$ of the polytope 
$\D^*\subset N_\IR$. 
The toric morphism $\phi$ is then given by the projection along the linear 
space spanned by $\D^*_f$ and $N_b$ is defined as the image of $N$ in the 
quotient space $N_\IR/\langle\D^*_f\rangle_\IR$. In order to guarantee the 
existence of the projection we choose a triangulation of $\D^*_f$ and then
extend it to a triangulation of $\D^*$.
For each such choice we can interpret the homogeneous coordinates that 
correspond to rays in $\D_f^*$ as coordinates of the fiber and the others as 
parameters of the equations and hence as moduli of the fiber space.
Reflexivity of the fiber polytope $\D_f^*$ ensures that the fiber also is a 
CICY because a nef partition of $\D^*$ automatically
induces a nef partition of $\D_f^*$. This follows immediately from the
definition by restriction of the convex piecewise linear functions defining
the partition to $\left(N_f\right)_{\IR}$.

For hypersurfaces the geometry of the resulting fibration has been worked 
out in detail in~\cite{Hu:2000pr}. The codimension $r_f$ of the fiber generically 
coincides with the codimension $r$ of the fibered space also for complete 
intersections. For $r>1$ it may happen, however, that $\S_f^1$ does not 
intersect one (or more) of the $E_l$'s, in which case the codimension 
decreases. An example of that type is the model 
\BE
    \IP{\left(\BM 2~2~2~4~1~1~0\cr0~0~0~4~1~1~2\cr\EM\right)
    \left[\BM 8\cr8\cr\EM\VL\BM 4\cr0\cr\EM 
        \right] }
                ~/~ \IZ_2: {\small\hbox{ 1 1 0 1 1 0 0}}  
\EE
with $h^{11}=3$ and $h^{12}=43$, which is a free $\IZ_2$ quotient of a
blowup of $\mP^5_{222411}$ with the 
position of the additional vertex $\v_7=-\2(4\v_4+\v_5+\v_6)$ given by the 
second linear relation, i.e. the bottom line in the parenthesis. 
(This notation will be explained in more detail in the next subsection.) 
This polytope has one nef partition with $E_1=\{\v_3,\v_4,\v_5,\v_6,\v_7\}$
and $E_2=\{\v_1,\v_2\}$. The corresponding bidegrees are separated by a
vertical line in the bracket (they are given by the sums of the gradings 
of the homogeneous coordinates that correspond to the vertices that belong 
to $E_l$). The codimension two fiber polytope $\D_f^*$ that is 
spanned by $\v_4,\ldots,\v_7$ has all of its vertices in $E_1$ so that we 
obtain a K3 fibration with the generic fiber being a degree 8 hypersurface in 
$\mP^3_{4112}$ instead of an elliptic codimension two fiber that would 
naively be expected.

In our searches for fiber spaces with certain properties we mostly
restricted attention to the generic case where $r=r_f$. We also analyzed
the intersection numbers of many spaces and found no example where a 
fibration has no toric realization, provided that the possible change 
of the codimension is taken into account. There are, however, cases 
where the fibration does not lift to a toric morphism of the ambient space.
An example is the polytope (\ref{EQdeltaB}) which will be discussed in 
subsection \ref{SECdeltaB}. Reflexivity of the $\D_f^*$ is thus sufficient 
but not necessary for a fibration structure of a complete intersection.%
\footnote{
        Note that $\D_f^*$ is the intersection of $\D^*$ with the linear fiber
        space $(N_f)_\IR$ which need not be a lattice polytope and thus
        can be larger than the convex hull of $\D^*\cap N_f$.
}

\subsection{The (2,30) example}
\label{sec:2,30}

In this subsection we will discuss a set of examples of CICYs in 
great detail. We will exhibit their non-trivial fundamental group, 
their fibrations, and their partitions.

In our analysis of the geometry and of applications of complete 
intersections it was natural to start with models with a small number $h^{11}$
of K\"ahler moduli. In this realm it is quite likely that our lists of 
toric CICYs is fairly complete, at least for codimension two. Among the
one-parameter models we found only two new Hodge numbers, namely $h^{12}=25$ 
for the free $\IZ_3$ quotient of the degree (3,3) CICY in $\IP^5$ 
and $h^{12}=37$ for the free $\IZ_2$ quotient of the degree (4,4) CICY in 
$\mP^5_{111122}$. The Picard--Fuchs equations of the respective universal 
covers were both analyzed in \cite{WPcicyKT}. 

We therefore turn to the list of 2-parameter examples, the first of which have
Hodge numbers (2,30). They will serve as our main examples in this and the 
next section.
\footnote{
        The first hypersurfaces example (2,29) is the free $\IZ_3$ quotient
        of the degree (3,3) hypersurface in $\IP^2\ex\IP^2$.
}
In the appendix we compile a brief overview of toric CICYs with small 
$h^{11}$.
%

There are three different polytopes which allow for codimension two complete 
intersections with Hodge numbers (2,30). These have eight or nine vertices 
and no additional boundary points. 
In a convenient basis the coordinates of the 
vertices of the first of these polytopes are given by the column vectors 
\BE             \D^*_{(A)}~\ni ~ \{\v_i\}=                 \label{EQdeltaA}
\left\{\Msize\BM1 & 0 & 0 & 0 & -2 & 1 & -1 & -1 \cr 
                0 & 1 & 0 & 0 & -1 & 1 & -1 & ~~0 \cr
                0 & 0 & 1 & 0 & -1 & 1 & -1 & ~~0 \cr
                0 & 0 & 0 & 1 & -1 & 0 & ~~0 & ~~0 \cr
                0 & 0 & 0 & 0 & ~~0 & 2 & -2 & ~~0 \cr\EM\right\} 
\EE
If the number of vertices is close to the simplex 
case it is most economical to describe a polytope in a coordinate independent 
way by the linear relations among the vertices. This data is sufficient if the
lattice $N$ is generated by the vertices. Otherwise it has to be supplemented 
by an abelian group action that defines the lattice. The toric variety
corresponding to the polytope $\Delta^*_{(A)}$, generalizing the notation 
$\mP^n_w$ or $\mP^n(w)$ in the simplex case, is thus 
\BE     \mP_{\D^*_{(A)}} = \IP \left(\Msize
        \BM2~1~1~1~1~0~0~0\cr0~0~0~0~0~1~1~0\cr1~0~0~0~0~0~0~1\cr\EM\right)
        ~/~ \IZ_2: \hbox{ \Msize 1 1 1 0 0 1 0 0} 
        \label{eq:P(A)}
\EE
The lines in the parenthesis indicate the linear relations among the vertices.
The first two tell us that the toric variety corresponds to a product space 
$\mP^4_{21111}\ex\IP^1$, 
while the third linear relation $\v_1+\v_8=0$ amounts to a blow-up of 
$\mP^4_{21111}$ by the last vertex $\v_8$. Finally, the group action indicates
that the lattice $N$ is not generated by the vertices alone.
It requires, as an additional generator, the lattice point 
$\2(\v_1+\v_2+\v_3+\v_6)$, {\it i.e.} the linear combination of vertices that 
corresponds to the phases of the $\IZ_2$ action on the homogeneous 
coordinates. 

The coordinates displayed in (\ref{EQdeltaA}), with the last line 
divisible by 2 for {\it all} lattice points, shows that the CICY is a free
quotient. The group action can be recovered by finding an integer linear 
combination $\v$
of the column vectors with coefficients in $\2\IZ$ whose last coordinate is 
odd (thus refining the lattice). 
The resulting generator for the $\IZ_2$ action is 
unique only up to linear combinations with the weight vectors modulo 2, 
which corresponds to a different choice 
$\v\to\v+\Delta\v$ with $\Delta\v=\2(\v_2+\v_3+\v_4+\v_5)=-\v_1$, 
$\2(\v_6+\v_7)=0$ or $\2(\v_1+\v_8)=0$.

Next, we have to look at the possible nef-partitions for $\Delta^*_{(A)}$.
It turns out that, up to symmetries,
\footnote{
        The symmetry group has order 16 and is generated by the transpositions
        $\v_2\leftrightarrow\v_3$, $\v_4\leftrightarrow\v_5$, 
        $\v_6\leftrightarrow\v_7$ and by the exchange
        $(\v_2,\v_3)\leftrightarrow(\v_4,\v_5)$.
}
there is a unique nef partition, given by $E_1=\{\v_1,\v_2,\v_4,\v_8\}$ and 
$E_2=\{\v_3,\v_5,\v_6,\v_7\}$ with the Hodge numbers (2,30). This leads to 
the partitioning $6=4+2$, $2=2+0$ and $2=2+0$ of the total degrees of the 
complete intersection $X$ into multidegrees. We will augment the previous 
notation by a bracket indicating these multidegrees and write
\begin{equation}     
        X_{(A)} = \IP \left(\Msize
        \BM2~1~1~1~1~0~0~0\cr0~0~0~0~0~1~1~0\cr1~0~0~0~0~0~0~1\cr\EM\right)
        \left[\Msize\BM4\cr2\cr2\EM\VL\BM2\cr0\cr0\EM\right]
        ~/~ \IZ_2: \hbox{ \Msize 1 1 1 0 0 1 0 0} 
        \label{eq:X(A)}
\end{equation}
In general these degrees do not specify the partition uniquely. We observed, 
however, in all examples that equal multidegrees of different partitions 
always lead to the same Hodge numbers. 
The $\IZ_2$ quotient now indicates that the lattice $M$ is replaced by the 
sublattice corresponding to monomials that are invariant under the given phase
symmetry. Since this quotient does not lead to additional lattice points in 
$\D^*_{(A)}$ the corresponding group action is free on the CICY, 
{\it i.e.} $\pi_1(X) = \mZ_2$.  

In the present example the ambient space is a product space with a $\IP^1$
factor. The CICY is, nevertheless, a nontrivial K3-fibration over $\IP^1$
because the coefficients of the equations defining the K3 fiber $X_f$ depend 
on the coordinates of the base. The K3 family containing
the generic fibers is obtained by dropping the second line and the columns 
corresponding to $\v_6$ and $\v_7$,
\begin{equation}
        X_{f} = \IP \left(\Msize
        \BM2~1~1~1~1~0\cr1~0~0~0~0~1\cr\EM\right)
        \left[\Msize\BM4\cr2\EM\VL\BM2\cr0\EM\right]
        \label{eq:Xf}
\end{equation}
Note that the $\IZ_2$ quotient does not change the fiber lattice because it 
also acts on $\IP^1$, effectively dividing the base by 2. Over the two 
fixed-points on the base we obtain, however, an Enriques fiber. (Since K3 
only admits free $\IZ_2$ quotients and since the group action on the base 
$\IP^1$ always has fixed points, a free quotient of a K3 fibration can only
have order 2.) The induced nef partition is obtained by dropping $\v_6$
and $\v_7$ from $E_2$. It does, of course, lead to the bi-degrees of the
divisors given in (\ref{eq:Xf}).

The lines in the parenthesis of our notation for toric varieties, as
in~\eqref{eq:P(A)}, \eqref{eq:X(A)}, or~\eqref{eq:Xf}, generate the
cone of non-negative linear relations among the points in $N$. We will often
call them {\it weight vectors}. 
The definition of the polytope only requires the linear relations
among the vertices (and possibly the group action defining a 
sublattice). For the discussion of fibrations and other geometrical
data it is, however, often convenient to include the linear relations
among all lattice points of $\D^*$.
When the partitioning of the total degree $d=\sum w_i$ of a weight
vector by the nef partition is specified as $(-d_1,-d_2; w_1,\ldots,w_N)$ with
$d=d_1+d_2$ we may also call them {\it charge vectors} because this is the 
data that characterizes part of the gauged linear sigma model realization of 
these geometries \cite{phases}. 
Note that for the definition of the polytope and for the 
degree data of the nef partition, it is sufficient to give the
charge vectors that correspond to the linear relations among the
vertices. A more redundant description may, nevertheless, be useful 
to make fibrations or non-free lattice quotients visible. A complete
definition of the model, on the other hand, may require a 
resolution of singularities through triangulations and the inclusion 
of additional points.
In the weighted projective case the (single) weight vector 
coincides with the generator of the Mori cone of the ambient space. In
general, however, the Mori cone will be larger than the cone that is
spanned by the charge vectors.

The other realizations of the (2,30) model are
\begin{equation}
       X_{(B)} = \IP
    \left(\Msize\BM2~1~1~1~1~0~0~0\cr2~2~1~0~1~1~1~0\cr1~0~0~0~0~0~0~1\cr\EM
        \right) \left[\Msize\BM4\cr4\cr2\EM\VL\BM2\cr4\cr0\EM\right]
        ~/~ \IZ_2: \hbox{ \Msize 1 1 1 0 0 1 0 0}
    \label{eq:X_B} 
\end{equation}
with nef partitions $E_1\cup E_2=
\{\v_1,\v_3,\v_5,\v_8\}\cup\{\v_2,\v_4,\v_6,\v_7\}$,
$\{\v_1,\v_3,\v_4,\v_7,\v_8\}\cup\{\v_2,\v_5,\v_6\}$, or
$\{\v_1,\v_2,\v_4,\v_8\}\cup\{\v_3,\v_5,\v_6,\v_7\}$
and 
\begin{equation}   
       X_{(C)} = \IP
    \left(\Msize\BM2~1~1~1~1~0~0~0~0\cr2~2~1~0~1~1~1~0~0\cr
        1~0~0~0~0~0~0~1~0\cr0~1~0~0~0~0~0~0~1\cr\EM
        \right) \left[\Msize\BM4\cr4\cr2\cr0\EM\VL\BM2\cr4\cr0\cr2\EM\right]
        ~/~ \IZ_2: \hbox{ \Msize 1 1 1 0 0 1 0 0 0} 
    \label{eq:X_C}
\end{equation}
with two possible nef partitions $E_1\cup E_2=
        \{\v_1,\v_3,\v_4,\v_7,\v_8\} \cup \{\v_2,\v_5,\v_6,\v_9\}$ or
        $\{\v_1,\v_3,\v_5,\v_8\}\cup\{\v_2,\v_4,\v_6,\v_7,\v_9\}$. 
The polytope $\Delta^*_{(B)}$ for $X_{(B)}$ has the same ``K3 fiber'' 
polytope as $\Delta^*_{(A)}$, but the two points above and below the 
fiber-hyperplane are shifted along the fiber as can be seen by the 
non-zero entries $(2,2,1,0,1)$ in the second line, below the weights of the 
fiber. The ambient space looks, at first sight, like a non-trivial
fibration over the $\IP^1$ with homogeneous coordinates $(x_6:x_7)$.
This is, however, not true because the line $\overline{\v_6\v_7}$ now 
intersects the fiber hyperplane outside the convex hull of the other lattice 
points. This line thus becomes an edge of any star triangulation of 
$\D_{(B)}^*$ so that the points in the intersection $D_6 \cap D_7$, which have
homogeneous coordinates $x_6=x_7=0$, have no image in the base $\IP^1$.

We will see in the next section that $X_{(A)}$ and $X_{(B)}$ are 
nevertheless diffeomorphic and that their Picard--Fuchs equations are related 
by a change of variables. In particular, also $X_{(B)}$ is a K3 fibration.
This is only possible if
$D_6 \cap D_7$ does not intersect the CICY (as is indeed the case).

The polytope $\Delta^*_{(C)}$ is similar to $\Delta^*_{(B)}$ except for an 
additional blowup of the fiber polytope with an exceptional divisor $D_9$ 
that, as we will see in subsection~\ref{sec:X_B,X_C}, does not intersect the 
\CY threefold. The additional point does, however, make the K3 fibration 
manifest, because $\D^*_{(C),f}$ is again reflexive.

In the next section we will discuss in more detail how different 
partitions or different polytopes leading to CICYs with the same topological 
data may be related. We conclude this section with a slight digression. We 
show how the nef-partition for $\Delta^*_{(A)}$ can be obtained from the Newton 
polytopes of the degree (4,0) and (2,2) polynomials in the double cover of the 
ambient space. The rest of this section will not be used below and can be 
skipped.

In order to arrive at the polytope (\ref{EQdeltaA}) we observe that the
ambient space of the double cover is closely related to the product space 
$\mP^4_{21111}\ex\IP^1$. We thus start with the Newton polytope $\hat\D$ of 
a degree (6,2) equation in that space. It has 10 vertices corresponding to the 
monomials
\[      x_0^3y_j^2, ~~~ x_i^6y_j^2 ~~~\hbox{ with } 1\le i\le 4, ~0\le j\le 1
\]
where $x_i$ and $y_j$ are the homogeneous coordinates in $\mP^4_{21111}$ and
$\mP^1$, respectively. The degree (4,0) and (2,2) polynomials correspond to
the Newton polytopes
\[      \hat\D_1=\<x_0^2,x_i^4\>, ~~~~~ \hat\D_2=\<x_0y_j^2,x_i^2y_j^2\>  \]
so that $\hat\D=\hat\D_1+\hat\D_2$. The $\IZ_2$ quotient acting with signs
$(- - - + + , - +)$ kills the two vertices $x_0^3y_j^2$ and generates 9
additional ones:
\begin{equation}
  x_0^3y_0y_1, x_0^2x_i^2y_j^2 ~~~\hbox{ with } 1\le i\le 4, ~0\le j\le 1 
\end{equation}
The resulting polyhedron is not reflexive, has 196 points, 17 vertices and 9
facets, but can be made reflexive by dropping the vertex $x_0^3y_0y_1$. This
yields a polyhedron $\D_{(A)}$ with 195 points and 16 vertices that
possesses a nef partition with Hodge numbers $h^{11}=2$ and $h^{21}=30$ (up to
automorphisms there is only one additional nef partition whose Hodge numbers 
are $h^{11}=4$ and $h^{21}=44$). The dual polyhedron $\D^*_{(A)}$ has 9 points
and (in an appropriate basis) the 8 vertices given in eq. (\ref{EQdeltaA}), as
can be checked using~\eqref{eq:dual}. The linear relations are 
$2\v_1+\v_2+\v_3+\v_4+\v_5=0=\v_6+\v_7$ and the facet equation corresponding 
to the last vertex $\v_8=-\v_1$ is the one that eliminates $x_0^3y_0y_1$ and 
makes $\D_{(A)}$ reflexive.

The nef partition of $X_{(A)}$ is now constructed from the Newton polyhedra 
$\hat\D_i$ as follows: In order to get $\D_{(A)}=\D_1+\D_2$ we drop the point 
$x_0y_0y_1$ (which becomes a vertex on the sublattice) from $\hat\D_2$ and 
obtain
\[      \D_1=\<x_0^2,x_i^4\>, ~~~~~ \D_2=\<x_i^2y_j^2\> \]
With $v_0=x_0^2$, $v_i=x_i^4$ and $w_{ij}=x_i^2y_j^2$ we thus find
$\D_{(A)}=\D_1+\D_2=\< v_0w_{ij}, v_iw_{ij}\>$ for the decomposition 
of the 8+8=16 vertices of $\D_{(A)}$.
Shifting $\D_1$ and $\D_2$ by the exponent vectors of $1/(x_0x_1x_3x_4)$ and
$1/(x_2y_0y_1)$ and dropping the redundant exponents of $x_4$ 
and $y_1$ we obtain the vertex-matrices
\BE \D^\s_1=      \left\{\Msize \begin{matrix}~~1 & -1 & -1 & -1 & -1 \cr
                 -1 & -1 & -1 & ~~3 & -1 \cr
                 ~~0 & ~~4 & ~~0 & ~~0 & ~~0 \cr
                 -1 & -1 & ~~3 & -1 & -1 \cr
                 ~~0 &~~0 & ~~0 & ~~0 & ~~0 \cr\end{matrix} \right\} ~~~~~~~~
   \D^\s_2=      \left\{\Msize \begin{matrix}
                 ~~0 & 0 & ~~0 &~~0 &~~0 &~~0 &~~0 &~~0 \cr
                 ~~2 & 0 & ~~0 &~~2 &~~0 &~~0 &~~0 &~~0 \cr
                  -1 & 1 &  -1 & -1 & -1 & -1 & -1 &~~1 \cr
                 ~~0 & 0 & ~~0 &~~0 &~~2 &~~0 &~~2 &~~0 \cr
                 ~~1 & 1 & ~~1 & -1 &~~1 & -1 & -1 & -1\cr\end{matrix}\right\}
\EE
with $\s_1^T=(1,1,0,1,0)$, $\s_2^T=(0,0,1,0,1)$ and $\D^\s_l\sim \D_l-\s_l$.
The shifted Newton polytopes $\D_i^\s$ can be separated by a hyperplane, 
$\langle \rho,D_1^\s\rangle\le0\le\langle \rho,D_2^\s\rangle$, with 
$\rho=(2,1,0,1,0)$. The points of $D_1^\s$ and those of $D_2^\s$ on that 
hyperplane have no common non-zero coordinates, which implies 
that $\D^\s_1\cap \D^\s_2=\{0\}$ and thus establishes the nef property. 
(Up to symmetries of $\hat\D$ there is only one other choice of the 
nonnegative integral shift vectors $\s_i$ with $\D^\s_1\cap \D^\s_2=\{0\}$, 
which leads to the same nef partition.) Converting to the basis
\[ B =  \left(\Msize \begin{matrix}1 & 0 & 0 & 0 & 0 \cr
                 0 & 1 & 0 & 0 & 0 \cr
                 0 & 0 & 1 & 0 & 0 \cr
                 0 & 0 & 0 & 1 & 0 \cr
                 1 & 1 & 1 & 0 & 2 \cr\end{matrix}\right)
\]
of the $\IZ_2$ quotient of the original lattice we find the nef partition
$\D^{(A)}_l=B^{-1}\D^\s_l$,
\[ E_1^*=\left\{\Msize\begin{matrix}~~1 & -1 & -1 & -1 & -1 \cr
                 -1 & -1 & -1 & ~~3 & -1 \cr
                 ~~0 & ~~4 & ~~0 & ~~0 & ~~0 \cr
                 -1 & -1 & ~~3 & -1 & -1 \cr
                 ~~0 & -1 & ~~1 & -1 & ~~1 \cr\end{matrix}  \right\}~~~~~~~~
   E_2^*=\left\{\Msize\begin{matrix}~~0 & 0 & ~~0 &~~0 &~~0 &~~0 &~~0 &~~0 \cr
                 ~~2 & 0 & ~~0 &~~2 &~~0 &~~0 &~~0 &~~0 \cr
                  -1 & 1 &  -1 & -1 & -1 & -1 & -1 &~~1 \cr
                 ~~0 & 0 & ~~0 &~~0 &~~2 &~~0 &~~2 &~~0 \cr
                 ~~0 & 0 & ~~1 &-1 &~~1 &~~0 & ~~0 &-1\cr\end{matrix}\right\},
\]
which is dual to the partition
\BE \label{eq:nefpartA}
   E_1= \left\{\Msize\begin{matrix}1 & 0 & 0 & -1  \cr
                 0 & 1 & 0 & ~~0 \cr
                 0 & 0 & 0 & ~~0 \cr
                 0 & 0 & 1 & ~~0 \cr
                 0 & 0 & 0 & ~~0 \cr\end{matrix}\right\} ~~~~~
   E_2= \left\{\Msize\begin{matrix}0 & -2 & 1 & -1 \cr
                 0 & -1 & 1 & -1 \cr
                 1 & -1 & 1 & -1 \cr
                 0 & -1 & 0 &~~0 \cr
                 0 &~~0 & 2 & -2 \cr
\end{matrix}\right\}
\EE
of the convex hull $\D^*_{(A)}=\<\nabla_1,\nabla_2\>=\<E_1,E_2\>$.

\section{The geometry of toric CICYs}
\label{sec:geometry}

It is well-known that the same Hodge numbers can come from different 
polyhedra and even at different codimensions, so it is important to 
identify constructions that actually give equivalent CYs. First note that 
any hypersurface or complete intersection can be reconstructed at higher 
codimension: Just multiply with an interval $[-1,1]$ and take the 
corresponding product nef partition.
\footnote{%
        In the formulas for the Hodge numbers \cite{BBsth} this leads to a 
        doubling because a quadratic equation in $\IP^1$ is solved by two 
        points.}
A less obvious redundancy is due to partitions where one of the
$\D_i$ consists of a single vertex, say $\v_1$: In that case the nef 
condition implies that the projection of $\D^*$ along $\v_1$ is reflexive.
Moreover, the CY is given by the intersection of the toric divisor $D_1$ with
the remaining divisor(s) defined by the partition of the vertices. Since 
$D_1$ can only intersect the toric divisors that correspond to points 
bounding the reflexive projection along $V_1$ we conclude that we can 
construct the same CY variety in the ambient space that is given by 
that projection of $\D^*$. We will call such a nef partition trivial.

An important but expensive calculational  tool to settle the question 
about equivalences is the classification of real six manifolds by 
C.~T.~C.~Wall~\cite{Wall:1966ab}. Specialized to Calabi--Yau manifolds 
it states that two simply connected manifolds $X$ and $X'$ are of the 
same topological type, i.e. they are diffeomorphic, if besides the 
Hodge numbers the triple intersection numbers 
$\kappa_{abc}=\int_X J_a\wedge J_b\wedge J_c$ and the linear forms 
$\int_X \ch_2\wedge J_a$ are the same, possibly up to an integer linear 
basis transformation ${\underline J}=M {\underline J}'$ 
in the K\"ahler cone of $X$ and $X'$. If the basis can only be related 
by a rational linear transformation the spaces are rational homotopy 
equivalent. Only finitely many diffeomorphic types can exist in a 
rational homotopy equivalence class.

Gromov--Witten invariants have a modest history as symplectic invariants 
in the Calabi--Yau threefold case, and in general, we indeed find that 
symplectic families with large volume limits of the same topological type 
have the same Gromov--Witten invariants, but the toric mirror may have a 
different natural parametrization of the complex structure deformation 
space, which leads to different Picard--Fuchs equations and mirror maps. 

In some cases, see Section~\ref{sec:rational_homotopy}, we find that 
rational homotopy equivalent pairs can have different Gromov--Witten 
invariants, and in Section~\ref{sec:diffeomorphic} we discuss the 
situation of diffeomorphic manifolds with different symplectic structures.
 
In this section we explain in the example of the (2,30) model introduced 
in Section~\ref{sec:2,30} how Calabi--Yau complete intersections coming 
from different polytopes and/or from different nef partitions can be 
related. Further examples are summarized in Appendix~\ref{sec:other_models}.

\subsection{Equivalence of different nef-partitions}
\label{sec:nef}

As our example for different nef partitions we have chosen one of the nine
polytopes in our list of models with Hodge numbers (2,44),
\begin{equation}     
        \IP\left(\Msize\BM 4~2~2~2~1~1~0\cr1~0~0~0~0~0~1\cr\EM\right)   
        \left[\Msize\BM8\cr2\EM\VL\BM4\cr0\EM\right]
        ~/~ \IZ_2: \hbox{ \Msize 1 1 1 0 1 0},
        \label{eq:X4a}
\end{equation}
and its double cover, which we will relate to two different hypersurface 
polytopes. The polytope $\D$ of the quotient and its dual $\D^*$ have
$\MN M:P{} V N:P^*{} V^* =\MN M:232 10 N:9 7 $ points $P$ and vertices $V$. 
The polytope for the double cover has $\MN M:461 10 N:9 7 $, so that the 
$\IZ_2$ quotient is free (the number $P^*$ of points in $\D^*$ does not 
change). This model is again a K3
fibration, but this time the fiber polytope has an additional vertex, namely
$\v_8=\2(\v_5+\v_6)$, which is a lattice point on an edge of $\D^*$.
Anticipating the structure of the nef partitions, the corresponding relation
$2\v_1+\v_2+\v_3+\v_4+\v_8=0$ shows that the fiber is again $X_f$ 
in~\eqref{eq:Xf}.
Up to permutation symmetries this polyhedron only admits two nef partitions,
$\{\v_1,\v_4,\v_5,\v_6,\v_7\}\cup\{\v_2,\v_4\}$ and
$\{\v_1,\v_3,\v_4,\v_7\}\cup\{\v_2,\v_5,\v_6\}$.  
Both lead to the same partitioning of the degrees, $12=8+4$ and $2=2+0$, as
indicated above. Note that non-vertices always belong 
to one of the $\nabla_l$'s of a partition.
\footnote{~
        The piecewise linear functions $\psi_l$ defining the nef partition
        are integral on lattice points with values $0$ or $1$ on the 
        vertices. The facets of $\D^*$ thus cannot contain lattice points 
        with other values.
}
In our example this implies that $\v_5$ and $\v_6$ always have to belong to 
the same $\nabla_l$. More cases with Hodge numbers (2,44) will be discussed
in Appendix~\ref{sec:other_models}.

Since the reflexivity constraint on $\nabla_1+\nabla_2$ is weaker on a 
sublattice all of our partitions must lift to the double cover, but additional
ones can show up. Indeed, up to automorphisms, we find a total of 
nine nef partitions with four different degrees and three different sets of 
Hodge numbers: 
\def\VRZ{\mathop{\vrule depth 1pt height 7pt width 0pt}\limits}
\BEA
        \label{eq:set1}    
        \IP\left(\Msize\BM 4~2~2~2~1~1~0\cr1~0~0~0~0~0~1\cr\EM\right)
        \left[\Msize\BM8\cr2\EM\VL\BM4\cr0\EM\right]\VRZ^{2,86}_{-168}&~~~~~~&
        {\Msize\BM \{\v_1,\v_2,\v_5,\v_6,\v_7\}\cup\{\v_3,\v_4\}
                \cr\{\v_1,\v_2,\v_3,\v_7\}\cup\{\v_4,\v_5,\v_6\}        \EM}
\\[5pt]
        \label{eq:set2}
        \IP\left(\Msize\BM 4~2~2~2~1~1~0\cr1~0~0~0~0~0~1\cr\EM\right)   
        \left[\Msize\BM8\cr1\EM\VL\BM4\cr1\EM\right]\VRZ^{2,86}_{-168}&&
        {\Msize\BM \{\v_2,\v_3,\v_4,\v_5,\v_6,\v_7\}\cup\{\v_1\}
                \cr\{\v_1,\v_2,\v_3\}\cup\{\v_4,\v_5,\v_6,\v_7\}
                \cr\{\v_1,\v_2,\v_5,\v_6\}\cup\{\v_3,\v_4,\v_7\}        \EM}
\\[5pt]
        \label{eq:set3}
        \IP\left(\Msize\BM 4~2~2~2~1~1~0\cr1~0~0~0~0~0~1\cr\EM\right)
        \left[\Msize\BM6\cr1\EM\VL\BM6\cr1\EM\right]\VRZ^{3,69}_{-132}&&
        {\Msize\BM \{\v_1,\v_5,\v_6\}\cup\{\v_2,\v_3,\v_4,\v_7\}
                \cr\{\v_1,\v_2\}\cup\{\v_3,\v_4,,\v_5,\v_6\v_7\}        \EM}
\\[5pt]
        \label{eq:set4}
        \IP\left(\Msize\BM 4~2~2~2~1~1~0\cr1~0~0~0~0~0~1\cr\EM\right)
        \left[\Msize\BM10\cr2\EM\VL\BM2\cr0\EM\right]\VRZ^{3,99}_{-192}\!\!&&
        {\Msize\BM \{\v_1,\v_3,\v_4,\v_5,\v_6,\v_7\}\cup\{\v_2\}
                \cr\{\v_1,\v_2,\v_3,\v_4,\v_7\}\cup\{\v_5,\v_6\}        \EM}
\EEA
For the double cover we thus find two trivial partitions, for which we can
construct the corresponding hypersurfaces: It is quite easy to work this
out in terms of the weight data: A projection along $\v_1$, as required
by the third partition in~\eqref{eq:set2}, just amounts to dropping that 
vertex from the linear relations. Since $\v_7=-\v_1$ the last vertex is 
projected onto the origin and we find the weighted projective space 
$\IP^4_{22211}[8]$, whose degree 8 hypersurface indeed has Hodge data (2,86).
For the trivial partition in~\eqref{eq:set4} we project along $\v_2$ and 
find the CY hypersurface 
$\IP${\small$\left(\BM\scriptsize 4~2~2~1~1~0\cr\scriptsize1~0~0~0~0~1\cr\EM
        \right)\left[\BM\scriptsize10\cr\scriptsize2\cr\EM\right]       $},
again with the expected Hodge numbers (3,99).

We also observe that the partitioning indeed fixes the Hodge numbers, 
{\it i.e.} equal charge vectors always lead to the same spectrum.
This could be expected because for weighted projective intersections the
degrees contain all information. How this result comes about in the toric
context will be seen explicitly in the examples below.
The converse is, however, not true:
The first two partitions with Euler number $\chi=-168$ are 
the ones that survive the $\IZ_2$ quotient, but there is now a different
realization of the same Hodge numbers. The difference in the charge
vectors is, however, only due to the contribution of $\v_7$ and it turns
out that the corresponding divisor $D_7$ does not intersect the CICY.
All spaces with equal Hodge numbers turn out to be topologically 
equivalent so that there do not seem to be any phase boundaries associated
with a transition among the respective partitions.

\subsection{Equivalence of different polyhedra: the (2,30) model}
\label{sec:equivalence}

We now take the main example introduced in subsection~\ref{sec:2,30} and
show that the three different models $X_{(A)}$, $X_{(B)}$, and $X_{(C)}$
are topologically equivalent. 

\subsubsection{The Mori cone, topological data and fibration structure}
\label{sec:MoriCone}

For the ambient space the K\"ahler cone is given purely combinatorial 
in terms of the secondary fan of the dual polyhedron $\Delta^*$.
One works conveniently with its dual cone called the Mori cone, 
which is generated by a certain basis of linear relations $\tilde l^{(a)}$ 
of the points in $\Delta^*$. Their entries can be 
interpreted as the intersection of the divisors $D_i$ corresponding 
to the points $\nu^*_i$ with the curves $c^{(a)}$ bounding the dual face 
of the K\"ahler cone. If the $2n$-cycles of the ambient space, whose 
positivity $\int_{C_{2n}} J^n>0$ w.r.t. to the K\"ahler form $J$ 
defines the boundary of the K\"ahler cone of the ambient space
do not descend to curves or divisors of the Calabi--Yau complete 
intersection, the K\"ahler cone of the Calabi--Yau $X$ is bigger. 
A straightforward analysis of the intersection calculation 
yields for hypersurfaces that those divisors, which correspond 
to points on codimension one faces of the dual polyhedron, do not 
intersect the Calabi--Yau $X$ as divisors and are therefore not involved 
in the linear relations spanning the Mori cone of $X$. By homology 
and cohomology pairing and Poincar\'e duality they would 
give rise to elements in $H^{1,1}$ and above is the reason for their 
absence in (\ref{Bhodge}). 
  
Even in the hypersurface case the lower dimensional toric cycles of the
ambient space can 
or cannot descend to cycles in the Calabi--Yau. This phenomenon 
was first described in~\cite{Berglund:1995gd} and requires a detailed calculation of 
the intersection ring to determine the K\"ahler cone of $X$. 
It is not surprising that this phenomenon becomes more severe 
with higher codimension of $X$ and in absence of a simple rule 
the analysis must start with the question which of the 
divisors of the ambient space become divisors of $X$. 
Details and some systematics are discussed in the present section.

Once the Mori cone generators ${\tilde l}^{(a)}$ of the ambient space are known 
we add to each of them as first entries the negatives of the intersections
$l^{(a)}_{0,m}$ of the $r$ constraints $D_{0,m}$ in (\ref{eq:CICY}), 
with the curve $c^{(a)}$, such that the general form of the Mori generators
$l^{(a)}$ for the embedded Calabi--Yau manifold becomes
\begin{equation}
l^{(a)}=( l^{(a)}_{0,1},\ldots,l^{(a)}_{0,r};l^{(a)}_1,\ldots,l^{(a)}_n),\quad {\rm for} \ \ a=1,\dots,h.
\label{eq:genmori}
\end{equation}
Note that by construction $\sum_{m=1}^r l^{(a)}_{0,m} + \sum_{i=1}^n l^{(a)}_i =0$, $\forall a$, 
which ensures the vanishing of the first Chern class of $X$. 
The knowledge of the Mori cone
is important for several reasons. In Section~\ref{sec:phases} we will use it to 
determine the phase diagram of the (2,30) model. More importantly, in Section~\ref{sec:genus-zero} 
it is needed to define the local coordinates on the complex structure moduli space of 
the mirror $X^*$ near the point of maximal unipotent monodromy. Moreover, the generators enter
the coefficients of the fundamental period in~(\ref{eq:varphi}) which is a solution of the
Picard--Fuchs equations. The latter are most easily obtained from the Mori cone as is reviewed in
Appendix~\ref{sec:PF}.

\subsubsection{The first realization of the (2,30) model}
\label{sec:X_A}

We first discuss the model $X_{(A)}$ in detail. Recall that the intersection 
ring of the ambient space is obtained as the quotient of the 
polynomial ring $\mC[D_1,\ldots,D_n]$ by the ideal generated by 
the linear relations among the points, and by the Stanley--Reisner ideal. 
The latter is obtained from the primitive collections which are collections
of vertices that do not form a cone but any proper subset forms a 
cone~\cite{Batyrev:1991ab}. The intersection ring of the \CY is then an
ideal quotient of the above ring by the ideal generated by $D_{0,l}$.

In the present example these primitive collections
can easily be seen in the geometry of the polyhedron $\D^*_{(A)}$. 
As discussed in subsection~\ref{sec:2,30}, the section in the lattice $N$ 
that corresponds to the K3 fiber is a blowup of $\IP^4_{21111}$ by the vertex 
$\v_8=-\v_1$, hence it is a double pyramid over the tetrahedron 
$\langle\v_2,\v_3,\v_4,\v_5\rangle$. Subsequently, we will identify faces 
with their index sets and simply write the indices of the points $\nu^*_i$ 
of the faces, e.g. here $\langle 2345 \rangle$. This implies the relations 
$D_1\cdot D_8=0=D_2\cdot D_3\cdot D_4\cdot D_5$ because the 
respective vertices never can belong to the same cone of any (triangulated) 
fan over the polyhedron. The complete polyhedron is a double pyramid over 
that 4-dimensional double pyramid. As $\v_6+\v_7=0$ this leads to the 
additional relation $D_6\cdot D_7=0$, which completes the generators of
the Stanley--Reisner ideal. The polyhedron is simplicial and has the 16 facets
$\langle\widehat{18}\widehat{2345}\widehat{67}\rangle$, where a hat above a 
sequence of numbers indicates to take all simplices that arise by dropping 
one of the respective vertices. The linear equivalences (up to principal 
divisors) follow from the lines of (\ref{EQdeltaA}) 
and we find altogether
\begin{align}
        \label{eq:Ilin}
        D_1\sim 2D_5&+D_8, & D_2&\sim D_3\sim D_4\sim D_5, & D_6\sim &D_7,\\
        \label{eq:ISR}
        D_1\cdot D_8&=0 & D_2&\cdot D_3\cdot D_4\cdot D_5=0, & D_6\cdot D_7&=0.
\end{align}
According to the nef partition~\eqref{eq:nefpartA} the complete intersection
$X_{(A)}$ is given by $D_{0,1}\cdot D_{0,2}$ (cf.~\eqref{eq:CICY}) with
\begin{align} 
      D_{0,1}&=D_1+D_2+D_4+D_8\sim 2D_1, &
      D_{0,2}&=D_3+D_5+D_6+D_7\sim 2D_2+2D_6,
\end{align}
so that $D_8$ does not intersect the Calabi--Yau and the K\"ahler moduli 
correspond to the volumes of, for example, $D_3$ and $D_6$.

The unique star triangulation of the simplicial polytope 
(\ref{EQdeltaA}) fixes the toric intersection
numbers and therefore the Mori cone of the ambient space $\IP_{\D^*_{(A)}}$. 
We determine this Mori cone as 
\begin{equation}
  \begin{array}{rl}
    {\hat l}^{(1)}&=(0, 0, 0, 0, 0, 1, 1,\um0)\\
    {\tilde {\hat l}}{}^{(2)}&=(0, 1, 1, 1, 1, 0, 0,-2)\\
    {\hat l}^{(3)}&=(1, 0, 0, 0, 0, 0, 0,\um1) \ . 
  \end{array}
\end{equation}
As mentioned above the divisor $D_8$ of $\IP_{\D^*_{(A)}}$ does not intersect 
the complete intersection. All other divisors descend to divisors on the 
hypersurface. The entries in the Mori vectors are the intersections of the 
curves $c^{(a)}$, which have positive volume inside the K\"ahler cone, 
with the corresponding divisor. The Mori vectors of the complete intersection
must therefore have a zero in the eighth entry. Besides ${\hat l}^{(1)}$ 
there is the unique minimal length combination 
$l^{(2)}={\tilde {\hat l}}{}^{(2)}+2 \hat
l^{(3)}=(2,1,1,1,1,0,0,0)$ with this 
property. We drop the eighth entry and add the negative value of the 
intersection of the $c^{(a)}$ with $D_{0,1}$ and $D_{0,2}$ as the first 
entries 
(these numbers correspond to the negative degrees in the charge vectors). 
Thus we get the Mori vectors for the complete intersection $X_{(A)}$
\begin{equation}
  \begin{array}{rl}
  l^{(1)}&=(-4,-2; 2,1,1,1,1,0,0)\\
  l^{(2)}&=(\um 0,-2;0,0,0,0,0,1,1)\ .
  \label{mori230A}
  \end{array}
\end{equation}
(in the present example they coincide with the charge vectors because, 
propping the non-intersecting vertex, the ambient space is a product of 
weighted projective 
spaces, but generically the Mori vectors span a larger cone).
This way of summarizing the Mori generators becomes particularly useful, when
we discuss the Picard--Fuchs system.

It is convenient to summarize all the relevant information for $X_{(A)}$ in 
the following table:
\begin{equation}
  \label{eq:X_A}
  \begin{footnotesize}
  \begin{array}{ccrrrrrrr|rrcl}
    \multicolumn{9}{c}{ }                      &c^{(1)}&c^{(2)}&&\\
    D_{0,1}&&\um1&\um0&\um0&\um0&\um0&\um0&\um0&     -4&      0&&\\
    D_{0,2}&&    0&   1&    0&   0&   0&   0&   0&     -2&     -2&&\\
    D_1    &&    1&   0&    1&   0&   0&   0&   0&      2&      0&&2H\\
    D_2    &&    1&   0&    0&   1&   0&   0&   0&      1&      0&&H\\
    D_3    &&    0&   1&    0&   0&   1&   0&   0&      1&      0&&H\\
    D_4    &&    1&   0&    0&   0&   0&   1&   0&      1&      0&&H\\
    D_5    &&    0&   1&   -2&  -1&  -1&  -1&   0&      1&      0&&H\\
    D_6    &&    0&   1&    1&   1&   1&   0&   2&      0&      1&&L\\
    D_7    &&    1&   0&   -1&  -1&  -1&   0&  -2&      0&      1&&L\\
  \end{array}  
  \end{footnotesize}
\end{equation}
On the left-hand side of the vertical line we have listed from top to bottom 
the points $\nu^*_i$ of the polyhedron $\Delta^*$, where the first two entries
refer to the nef-partition $E_l,l=1,2$, and the next five entries are their 
coordinates in $N=\mZ^5$. Together they form the coordinates of the generators
of the 7-dimensional Gorenstein cone $\Gamma(\Delta^*)$ that was defined below
(\ref{BBstrh}). To each point $\nu^*_i$ we have associated the corresponding 
divisor $D_i$. The first two rows, {\it i.e.} $D_{0,1}$ and $D_{0,2}$ 
correspond to the interior point appearing once in either partition. 
The two columns labeled by $c^{(a)}$ on the right-hand side of the vertical 
line denote the Mori generators. Its entries
are the intersection numbers of the restrictions of the divisors $D_i$ to
$X$ with the curves $c^{(a)}$. The data only refer to the \CY manifold, 
{\it i.e.} we dropped the non-intersecting divisors and the curves that 
do not descend to the complete intersection.

Let us denote the independent divisors of the complete intersection by $H$ 
and $L$. 
In our example we can choose $H= D_3 D_{0,1} D_{0,2}$ and 
$L= D_6 D_{0,1} D_{0,2}$,
as is indicated on the right in~\eqref{eq:X_A}.
The classical intersection numbers are defined as
\begin{equation}
  \label{eq:kappa}
  \kappa_{abc}=\int_X J_a \wedge J_b \wedge J_c = D_a \cap D_b \cap D_c
\end{equation}
where $J_a \in H^2(X,\mZ)$ and $D_a \in H_4(X,\mZ)$. Here, $J_1$ and $J_2$ 
are the K\"ahler forms dual to $H$ and $L$, respectively. They can be easily 
evaluated from the intersections on $\IP_{\D_{(A)}}$, {\it e.g.} 
$D_1D_2D_3D_4D_6 = \Vol \langle \v_1,\v_2,\v_3,\v_4,\v_6 \rangle = 2$,
and the relations in~\eqref{eq:Ilin} and~\eqref{eq:ISR}. The fact
that we are dealing with a free $\mZ_2$ quotient is accounted for by a
division by 4. Therefore, the intersection numbers are 
\begin{equation}
  \begin{array}{rl}
  &\kappa_{111}=\kappa_{112}=2 \\
  &\kappa_{122}=\kappa_{222}=0\ .
  \label{230cijk}
  \end{array}
\end{equation}
According to Oguiso~\cite{oguiso}, a \CY threefold admits a K3-fibration
if there exists an effective divisor $L$ such that
\begin{align}
  \label{eq:Qguiso}
    L\cdot c &\geq 0 \textrm{ for all curves } c \in H_2(X,\mZ),&
    L^2\cdot D &= 0 \textrm{ for all divisors } D \in H_4(X,\mZ).
\end{align}
Therefore, we conclude from~\eqref{230cijk} that the geometry of the \CY space
$X_{(A)}$ is a fibration with $J_1$ the K\"ahler class of the fiber and $J_2$ 
the K\"ahler class of the base. This is in agreement with the purely 
combinatorial argument in~\eqref{eq:Xf}. Normally, one expects in such cases 
the fiber to be K3 and $\int_X \ch_2 J_2=24$ \cite{oguiso,KLM,AL}. That is 
because the integral of $J_2$ over the base $\IP^1$ gives 1, the rest of the 
integral extends over the fiber and yields $\int_{\textrm{K3}}\ch_2=24$. 
Instead one has here 
\begin{equation}
  \ch_2L =\int_X \ch_2 J_2=12\qquad  \ch_2H=\int_X \ch_2 J_1=20\ ,
  \label{230c2}
\end{equation} 
which indicates that the $\IZ_2$ quotient has divided the volume of the 
$\IP^1$ by two. Indeed, we know that the model is a free $\IZ_2$ quotient 
of an ordinary K3 fibration, which might be represented as the complete 
intersection of degree $(4,0)$ 
and $(2,2)$ in $\IP^4_{21111}\times \IP^1$ with Euler number $\chi=-112$ 
and the same $l^{(a)}$ vectors as in~\eqref{mori230A}. Now recall our 
extensive discussion of the properties of $X_{(A)}$ in 
subsection~\ref{sec:2,30}. With the homogeneous coordinates 
$x_i$ for $\IP^4_{21111}$ and $y_j$ for $\IP^1$, respectively, the 
polynomials $f_1$ and $f_2$ in~(\ref{eq:CIeqs}) are
$$
\begin{array}{rl}
  f_1&=x_0^2 +x_1^4+x_2^4+x_3^4+x_4^2 + \dots\\
  f_2&=x_0^2 (y_0^2+y_0 y_1 +y_2^2) + \dots
\end{array}
$$
For any choice $(y_0:y_1)$ in $\IP^1$ the fiber is the complete intersection 
K3 $X_{f}$ in~(\ref{eq:Xf}). The $\IZ_2$ acts on the coordinates by 
$(x_0,x_1,x_2,x_3,x_4;y_0,y_1)\rightarrow (-x_0,x_1,-x_2,x_3,-x_4;y_0,-y_1)$. 
Hence the $\IP^1$ gets folded with fixed points $(1:0)$ and $(0:1)$, which 
explains the division of $\int_X \ch_2 J_a$ by $2$, while the $\IZ_2$ action 
leads to an Enriques fiber over the fixed points in the base. Note that, 
therefore, 
only $\IZ_2$ can act freely on K3 fibered Calabi--Yau manifolds.
We will come back to this observation in Section~\ref{sec:IIAhetDual}.

\subsubsection{Other realizations of the (2,30) model} 
\label{sec:X_B,X_C}

One difficulty with complete intersection realizations is a high redundancy 
in the description of a given family of Calabi--Yau spaces. For example, we 
know from subsection~\ref{sec:2,30} that there are two more polyhedra with 
three and two nef-partitions respectively, which lead to complete 
intersections with Hodge numbers $(2,30)$. In addition, these polyhedra 
admit two and five different star triangulations respectively, which 
could potentially lead to different large volume phases of the families.

We first consider the second realization of the Hodge numbers (2,30), 
$X_{(B)}$. In this case the dual polyhedron $\D^*_{(B)}$ has vertices
\label{SECdeltaB}
\BE             \D^*_{(B)}~\ni ~ \{\v_i\}=              \label{EQdeltaB}
\left\{\Msize\BM1 & 0 & 0 & 0 & -2 & 1 & -1 & -1 \cr 
                0 & 1 & 0 & 0 & -1 & 1 & -2 & ~~0 \cr
                0 & 0 & 1 & 0 & -1 & 1 & -1 & ~~0 \cr
                0 & 0 & 0 & 1 & -1 & 0 & ~~1 & ~~0 \cr
                0 & 0 & 0 & 0 & ~~0 & 2 & -2 & ~~0 \cr\EM\right\} 
\EE
and admits the three nef-partitions given below eq.~\eqref{eq:X_B}.
In order to find the triangulations we observe that the shift of $\v_7$,
as compared to its position in $\D^*_{(A)}$,
moves the intersection point of the line $\overline{\v_6\v_7}$ with the
fiber hyperplane through the facet $\langle 1345\rangle$ of $\D^*_f$ till it reaches the
linear span of $\langle 3458\rangle$ outside of that facet of the fiber polytope. This 
kills the 4 simplices $\langle\widehat{18}345\widehat{67}\rangle$ of $\D^*_{(A)}$ and 
replaces them by the 3 triangles $\langle 1\widehat{345}67\rangle$ and by the 16th 
facet $\langle 345678\rangle$. The vertices of the non-simplicial facet form a circuit, 
$\v_3+2\v_4+\v_5=\v_6+\v_7+2\v_8$, for which we introduce the short hand 
notation $\langle3_14_25_1|6_17_18_2\rangle$. It indicates the labels
$m$ of the involved vertices $\v_m$ and, as subscripts, their coefficients
$c_m$ in the linear relation~\eqref{eq:circuit}. Therefore, it can be 
triangulated in the two different ways: $\langle\widehat{345}678\rangle$ or 
$\langle 345\widehat{678}\rangle$, so that we find two different triangulations with 18 
simplices:
\begin{equation}
  \begin{array}{rl}
    T_1 &= \{ \langle\widehat{18}2\widehat{345}\widehat{67}\rangle, 
             \langle\widehat{18}\widehat{345}67\rangle \}\\
    T_2 &= \{ \langle\widehat{18}2\widehat{345}\widehat{67}\rangle, \langle1\widehat{345}67\rangle, 
    \langle345\widehat{678}\rangle \}
  \end{array}
\end{equation}
Since we deal with a free $\mZ_2$ quotient, the volume of each simplex
is divisible by two.
Because of the coefficient 2 of $\v_4$ and $\v_8$ in the circuit the 
simplices that do not contain one of these vertices, i.e. $\langle 35678\rangle\in T_1$ 
and $\langle 34567\rangle\in T_2$, have volume 4. We thus need to resolve the singularities
of the ambient space by adding points at higher degree, following the general
discussion given in Section~\ref{sec:singularities}. From that we expect another 
simplex of volume 4, sharing a facet, which has to be the same for both 
triangulations. The only possibility is the facet $\langle 13567\rangle$, which 
indeed has volume 4.
The additional point in degree two, which resolves all singularities is 
$\v_{\r} = \frac{1}{2}(\v_3 + \v_5 + \v_6 + \v_7)$. 
The corresponding triangulations are 
\begin{equation}
  \begin{array}{rl}
    \label{eq:T}
    T_1 &= \{ \langle\widehat{18}2\widehat{345}\widehat{67}\rangle, 
      \langle\widehat{18}\widehat{35}467\rangle, \langle\widehat{18}\widehat{3567\r}\rangle \}\\
    T_2 &= \{ \langle\widehat{18}2\widehat{345}\widehat{67}\rangle, \langle1467\widehat{35}\rangle, 
      \langle3458\widehat{67}\rangle, \langle\widehat{14}\widehat{3567\r}\rangle \}
  \end{array}
\end{equation}
The linear relations are 
\begin{align}
  D_6 &\sim D_7, & D_1 &\sim 2D_5+D_8+D_{\r}, & D_3 &\sim D_5 \sim D_4+D_6 \sim
  D_2-D_6-D_{\r}
\end{align}
For the first two nef-partitions in~\eqref{eq:X_B}, the face 
$\langle 3,5,6,7\rangle$ belongs to both sets of vertices, 
therefore~\eqref{eq:CICYres} becomes for the first nef-partition
\begin{align}
  D_{0,1} &= D_1+D_3+D_5+D_8+D_{\r} = 2D_1 & D_{0,2} &= D_2+D_4+D_6+D_7+D_{\r} = 2D_2
\end{align}
The second nef-partition is analogous and yields the same result. For the 
third one, however, this face lies entirely in the second set of vertices, 
so that 
\begin{align}
  D_{0,1} &= D_1 + D_2 + D_4 + D_8 = 2D_1& D_{0,2} 
        &= D_3+D_5+D_6+D_7+2D_{\r} = 2D_2
\end{align}
yields again the same result. 
We discuss the triangulations for a single nef-partition, say, 
the first one, in detail.
The Stanley--Reisner ideal, of course, always contains the generator 
$D_1\cdot D_8$ because antipodal points can never belong to the same 
simplex. The divisor $D_8$ thus never intersects the Calabi--Yau manifold 
$X_{(B)}$, whose first defining equation is a section of 
${\cal O}(D_{0,1}) = {\cal O}(2D_1)$. Furthermore, since $\v_2$ and $\v_{\r}$
never belong to the same simplex, the divisor $D_{\r}$ coming from the 
blow-up of the ambient space never intersects $X_{(B)}$ because its second
defining equation is a section of ${\cal O}(D_{0,2}) = {\cal O}(2D_2)$. 
Otherwise it depends on the triangulation, and we find, using a similar
argument, from~\eqref{eq:T} 
\begin{equation}
  \label{eq:SR}
  \begin{array}{rcl}
    \cI_{SR}(T_1) &=&\{D_1D_8,D_2D_6D_7,D_3D_4D_5,\\  
         &  &  \phantom{\{}  D_3D_5D_6D_7, D_2D_{\r}, D_4D_{\r}\}\\
    \cI_{SR}(T_2) &=&\{D_1D_8,D_2D_6D_7,D_6D_7D_8,D_1D_3D_4D_5,D_2D_3D_4D_5,\\
    &  &  \phantom{\}}  D_3D_5D_6D_7, D_2D_{\r}, D_8D_{\r}, D_1D_4D_{\r}\}
  \end{array}
\end{equation}
Note that the first lines in~\eqref{eq:SR} correspond to the Stanley--Reisner
ideal of the unresolved toric variety $\mP_{\Delta^*_{(B)}}$.
Next, we determine the Mori cone of the resolved ambient space and find
\begin{eqnarray*}
  \hat  l_{T_1} &=&
  \left(\begin{footnotesize}\begin{array}{rrrrrrrrr}
    0&   1&   0& -1&   0&  1&  1&  0&  0\\
    0&\um0&\um1&  0&\um1&  1&  1&  0& -2\\
    1&   0&   0&  0&   0&  0&  0&  1&  0\\
    0&   0&   0&  1&   0& -1& -1& -1&  1
  \end{array}\end{footnotesize}\right)\\   
  \hat  l_{T_2} &=&
  \left(\begin{footnotesize}\begin{array}{rrrrrrrrr}
    1&\um0&  0&  1&  0& -1& -1&  0&  1\\
    0&   1&  1&  1&  1&  0&  0& -2&  0\\
    0&   0& -1& -1& -1&  0&  0&  1&  1\\
    0&   0&  1&  0&  1&  1&  1&  0& -2
  \end{array}\end{footnotesize}\right)
\end{eqnarray*}
Since $D_8$ and $D_{\r}$ do not intersect the Calabi--Yau space, we consider the 
linear combinations of the vectors above for which the eighth and the ninth 
entry vanishes, and adding the intersections of $-D_{0,l}$ with 
$c^{(a)}$ as before, we get for both triangulations the two Mori generators
\begin{equation}
  \begin{array}{rl}
  \tilde l^{(1)}&=(-4  ,\um0;2,0,1,\um2,1,  -1,  -1,0,0)\\
         l^{(2)}&=(\um0,  -2;0,1,0,  -1,0,\um1,\um1,0,0)\ .
  \end{array}
\end{equation}
Now we have to check that the curves which bound the corresponding K\"ahler 
cones in $\IP_{\D^*_{(B)}}$ descend to the Calabi--Yau space $X_{(B)}$. As 
mentioned above these curves have intersection $c^{(a)}\cdot D_i=l^{(a)}_i$. 
In particular $c^{(1)}$ has negative intersection with both $D_6$ and $D_7$. 
Negative intersection numbers indicate that the curves are actually contained 
in the corresponding divisors. Since by~\eqref{eq:SR}, 
$D_6 D_7=0$ on the Calabi--Yau space, we 
conclude that $c^{(1)}$ does not descend to $X_{(B)}$. For this reason the 
Mori cone must become smaller and the K\"ahler cone becomes bigger due to the 
absence of the bounding curve in $X_{(B)}$. The minimal positive 
integer linear combination of $\tilde l^{(1)}$ and $l^{(2)}$ without two
negative entries is $l^{(1)}=\tilde l^{(1)}+l^{(2)}$
\begin{equation}
  \label{eq:mori230B}
  \begin{array}{rl}
  l^{(1)}&=(  -4,-2;2,1,1,\um1,1,0,0,0,0)\\
  l^{(2)}&=(\um0,-2;0,1,0,  -1,0,1,1,0,0)\ .
  \end{array}
\end{equation}  
We can then pick $H=D_3 D_{0,1} D_{0,2}$ and $L=D_6 D_{0,1} D_{0,2}$ and 
observe exactly the same classical intersections as in (\ref{230cijk}) and
(\ref{230c2}). According to the theorem of Wall the Calabi--Yau manifolds are 
then of the same topological type. It turns out that the world sheet 
instanton numbers on both Calabi--Yau spaces are the same. However, as we will
see in subsection~\ref{sec:PF}, the slightly different vectors $l^{(i)}$ lead 
to a different parametrization of the complex structure moduli space. 
Since these arguments used only the triangulations, which are independent of
the nef-partition, they show that the other two nef-partitions lead to the 
same topological type of the Calabi--Yau space.

The third model with Hodge numbers $(2,30)$ has the dual polyhedron
\BE             \D^*_{(C)}~\ni ~ \{\v_i\}=              \label{EQdeltaC}
\left\{\Msize\BM1 & 0 & 0 & 0 & -2 & 1 & -1 & -1  & ~~0\cr 
                0 & 1 & 0 & 0 & -1 & 1 & -2 & ~~0 & -1\cr
                0 & 0 & 1 & 0 & -1 & 1 & -1 & ~~0 & ~~0\cr
                0 & 0 & 0 & 1 & -1 & 0 & ~~1 & ~~0 & ~~0\cr
                0 & 0 & 0 & 0 & ~~0 & 2 & -2 & ~~0 & ~~0\cr\EM\right\} 
\EE
and admits the two nef-partitions given below~\eqref{eq:X_C}.
The star triangulations of $\D^*_{(C)}$ again contain the twelve simplices 
$\langle\widehat{18}2\widehat{345}\widehat{67}\rangle$ of $\D^*_{(A)}$, 
but now there are 5 additional facets, namely the two simplices 
$\langle 135\widehat{67}9\rangle$, and the facets $\langle 14\widehat{35}679\rangle$ and $\langle 3456789\rangle$. 
The circuit $\v_6+\v_7=\v_4+\v_9$ shows that the additional vertex 
$\v_9$ restores the possibility of a fibered ambient space. 
Namely, if we triangulate this circuit as $\langle 4\widehat{67}9\rangle$, we avoid the 
edge $\overline{\v_6\v_7}$. The triangulations of $\Delta^*_{(C)}$ that are 
consistent with the fibration are easily found by triangulating the 
reflexive section $\Delta^*_{(C),f}$, which is spanned by 
$\v_1,\ldots,\v_5,\v_8,\v_9$. Its single non-simplicial facet, $\langle 34589\rangle$, 
yields a circuit $\langle 3_14_15_1|8_29_1\rangle$. One triangulation, 
$\langle\widehat{345}89\rangle$, leads to a regular ambient space while the other 
contains the simplex $\langle 3459\rangle$ of volume 2. In the 4-dimensional ambient 
space of the fiber we expect a resolution of the singularity by a point in 
degree 2 in the interior of the cone. Indeed, $\v_{\r}=\2(\v_3+\v_4+\v_5+\v_9)$
is a lattice point (which actually is identical to the 
point $\v_{\r}$ 
that resolved the singularities in the previous example).
Extending these subdivisions to a triangulation of the complete polytope
we thus obtain the first two triangulations $T_2$ and $T_1$ below. $T_2$ is
regular while $T_1$ requires the subdivision of the two triangles 
$\langle 3459\widehat{67}\rangle$ through $\v_{\r}$.

\begin{figure}
\def\putlab#1)#2#3{\put#1){\makebox(0,0)[#2]{\small #3}}} 
\def\putlin#1,#2,#3,#4,#5){\put#1,#2){\line(#3,#4){#5}}}
\def\putvec#1,#2,#3,#4,#5){\put#1,#2){\vector(#3,#4){#5}}}
\def\putcx#1,#2){\put#1,#2){\circle*{1.4}}}
\begin{center}\unitlength=3pt\thicklines\begin{picture}(40,40)(-20,-20)
        \putlab(-14,14)c{\normalsize                    $\mathbf{T_1}$}
\putlin(0,0,0,1,16)\putlab(0,18)b{$\langle 8_29_1|3_14_15_1\rangle$}
        \putlab( 14,14)c{\normalsize                    $\mathbf{T_2}$}
\putlin(0,0,1,0,22)\putlab(24,0)l{$\langle 4_19_1|6_17_1\rangle$}
        \putlab(18,-8)c{\normalsize                     $\mathbf{T_3}$}
\putlin(0,0,1,-1,14)\putlab(17,-17)t{$\langle 8_29_2|3_15_16_17_1\rangle$}
        \putlab(0,-16)c{\normalsize                     $\mathbf{T_4}$}
\putlin(0,0,-1,-1,14)\putlab(-17,-17)t{$\langle 6_17_18_2|3_14_25_1\rangle$}
        \putlab(-18,-8)c{\normalsize                    $\mathbf{T_5}$}
\putlin(0,0,-1,0,24)\putlab(-25,0)r{$\langle 6_17_1|4_19_1\rangle$}
        \putcx(-20,0) \putlab(-19.5,1.5)b{$\v_8$}
        \putcx(10,0)    \putlab(10,2.5)l{$\v_3$}\putlab(10,-2.5)l{$\v_5$}
        \putcx(0,10) \putlab(-1,12)r{$\v_6$} \putlab(1,12)l{$\v_7$}
        \putcx(10,-10)  \putlab(-10,-12)l{$\v_9$}
        \putcx(-10,-10) \putlab(9.3,-12)r{$\v_4$}
\end{picture}\end{center}
\caption{\label{fig:secondaryfan} 
        Secondary fan of the facet $\langle 3456789\rangle$ of $\D^*_{(C)}$ with
        circuits relating its triangulations.}
\end{figure}

The complete set of triangulations can be found by constructing the 
secondary fan \cite{GKZ} for the facet $\langle 3456789\rangle$:
Let $A$ denote the matrix consisting of the coordinates of the respective
vertices $\v_3,\dots,\v_9$ and compute its Gale transform \cite{GKZ,BFS} 
\begin{equation}
  B=\left(\Msize\BM1&~~1&1&0&0&-2&-1\cr0&-1&0&1&1&~~0&-1\cr\EM\right),
  \label{eq:Gale}     
\end{equation}
which is the transpose of its kernel, i.e. $AB^T=0$, where we have chosen
the two circuits $\langle 3_14_15_1|8_29_1\rangle$ and
$\langle 6_17_1|4_19_1\rangle$ as generators of the kernel. 
The rays of the secondary fan are generated by the column vectors of 
$B$, which we label by the respective vertices of $\D_{(C)}^*$.
The triangulations $T_i$ can then be read off as indicated in 
figure \ref{fig:secondaryfan}: 
The vertices of the simplices $\sigma \in T_i$ are obtained as complements of 
the vertices of an appropriate number of points that span cones containing the 
phases labeled by $T_i$ (see Lemma 4.3 of \cite{BFS}). As an example consider 
$T_4$, where these complements are $\{48,39,49,59\}$, yielding the triangulation 
$\{\langle 35679\rangle,\langle\widehat{345}678\rangle\}$.
Adjacent triangulations are connected by  (bistellar flips for) 
circuits involving, on either side, vertices that can form a strictly convex cone 
with the ray that separates the corresponding phases (see Proposition 2.12
in chapter 7 of \cite{GKZ}).

A triangulation of $\langle 3456789\rangle$ induces a triangulation of the other
two non-simplicial facets $\langle 14\widehat{35}679\rangle$. 
Writing the triangulations as a union of the simplices of the
big facet $\langle 3456789\rangle$, simplices of the induced triangulations of the
circuits $\langle 14\widehat{35}679\rangle$, and simplicial facets of $\D_{(C)}^*$, 
respectively, we obtain
\BEA    T_1
                =&\{\langle 345\widehat{67}\widehat{89}\rangle\}\hspace{30pt}        
                &\cup\;\{\langle 14\widehat{35}\widehat{67}9\rangle\}        
                        \cup \{\langle\widehat{18}2\widehat{345}\widehat{67}\rangle,
                        \langle 135\widehat{67}9\rangle\}
\\      T_2
                =&\{\langle\widehat{345}\widehat{67}89\rangle\}\hspace{30pt}
                &\cup\;\{\langle 14\widehat{35}\widehat{67}9\rangle\}        
                        \cup \{\langle\widehat{18}2\widehat{345}\widehat{67}\rangle,
                        \langle 135\widehat{67}9\rangle\}
\\      T_3
                =&\hspace*{-12pt}\{\langle\widehat{35}\widehat{49}678\rangle,
                \langle 35\widehat{67}89\rangle\}
                \hspace*{-12pt}&\cup\;\{\langle 1\widehat{49}\widehat{35}67\rangle\} 
                        \cup \{\langle\widehat{18}2\widehat{345}\widehat{67}\rangle,
                        \langle 135\widehat{67}9\rangle\}
\\      T_4
                =&\{\langle\widehat{345}678\rangle,\langle 35679\rangle\}
                &\cup\;\{\langle 1\widehat{49}\widehat{35}67\rangle\}        
                \cup \{\langle\widehat{18}2\widehat{345}\widehat{67}\rangle,
                \langle 135\widehat{67}9\rangle\}
\\      T_5
                =&\{\langle 345\widehat{678}\rangle,\langle 35679\rangle\}
                &\cup\;\{\langle1\widehat{49}\widehat{35}67\rangle\}        
                        \cup \{\langle\widehat{18}2\widehat{345}\widehat{67}\rangle,
                        \langle 135\widehat{67}9\rangle\}
\EEA
$T_2$ and $T_3$ have 24 regular simplices. The triangulations
$T_1$, $T_4$ and $T_5$ have 22 simplices, two of which have volume 2. 
Inspection of the coefficients in the circuits connecting the phases 
shows that these are $\langle 3459\widehat{67}\rangle$, $\langle 3567\widehat{49}\rangle$ and 
$\langle 3567\widehat{49}\rangle$, respectively, so that the refinement induced by adding
\BE     \v_{\r}=\2(\v_3+\v_4+\v_5+\v_9)=\2(\v_3+\v_5+\v_6+\v_7)
\EE
resolves the singularities in all cases.   
Note that star triangulations are refinements of the polyhedral
subdivision induced by the cones over the facets of $\D^*$. 
Figure \ref{fig:secondaryfan} is therefore a face of the complete secondary 
fan that describes all triangulations of $\D_{(C)}^*$ (see Theorem 2.4 in 
chapter 7 of \cite{GKZ}).

We list here the data for the ambient space only for two triangulations, 
$T_2$ 
and $T_1$. For $T_2$ we do not need $\nu^*_{\r}$. Therefore, the linear 
relations are 
\begin{align}
  D_6 &\sim D_7, & D_1 &\sim 2D_5+D_8, 
  &D_2-D_6-D_9&\sim D_3\sim D_4+D_6\sim D_5
\end{align}
and the Stanley-Reisner ideal is
\begin{equation}
  I_{SR}(T_2) = \{ D_{{1}}D_{{8}},D_{{2}}D_{{9}},D_{{6}}D_{{7}},D_{{3}}D_{{4}}D_{{5}} \}
\end{equation}
The Mori generators associated to the triangulation $T_2$ are
\begin{equation}
  \hat{l}_{T_2} = 
  \left(\begin{footnotesize}\begin{array}{rrrrrrrrr} 
    0&\um1&\um0&  0&\um0&\um0&\um0&  0&  1\\
    1&   0&   0&  0&   0&   0&   0&  1&  0\\ 
    0&   0&   1&  1&   1&   0&   0& -2& -1\\ 
    0&   0&   0& -1&   0&   1&   1&  0& -1
  \end{array}\end{footnotesize}\right)
\end{equation}
The complete intersection $X_{(C)}$ for the first nef-partition below 
~\eqref{eq:X_C} is defined by
$$D_{0,1}=D_1+D_3+D_4+D_7+D_8,\qquad D_{0,2}=D_2+D_5+D_6+D_9$$
For $T_1$ the linear relations involve $\nu^*_{\r}$
\begin{align}
  D_6 &\sim D_7, & D_1&\sim 2D_5+D_8+D_{\r}, & D_2 - D_6 - D_9 - D_{\r} &\sim D_3 \sim D_4 + D_6 \sim D_5  
\end{align}
and so does the Stanley-Reisner ideal
\begin{equation}
  \label{eq:3}
  I_{SR}(T_1) = \{ D_{1}D_{8},D_{1}D_{\r},D_{2}D_{9},D_{2}D_{\r},D_{6}
  D_{7},D_{8}D_{9},D_{8}D_{\r},D_{1}D_{3}D_{4}D_{5},
  D_{2}D_{3}D_{4}D_{5},D_{3}D_{4}D_{5}D_{9} \}
\end{equation}
The Mori generators become accordingly
\begin{equation}
  \hat{l}_{T_1} = 
  \left(\begin{footnotesize}\begin{array}{rrrrrrrrrr}
    0&\um1&   1&   1&   1&\um0&\um0&  -2&   0&   0\\ 
    1&   0&   0&   0&   0&   0&   0&   0&  -1&   1\\ 
    0&   0&   1&   1&   1&   0&   0&   0&   1&  -2\\ 
    0&   0&  -1&  -1&  -1&   0&   0&   1&   0&   1\\ 
    0&   0&   0&  -1&   0&   1&   1&   0&  -1&   0
  \end{array}\end{footnotesize}\right)  
\end{equation}
The complete intersection for the first nef partition below~\eqref{eq:X_C} is 
defined by
$$D_{0,1}=D_1+D_3+D_4+D_7+D_8+D_{\r},\qquad D_{0,2}=D_2+D_5+D_6+D_9+D_{\r}$$
We find for all triangulations (and all partitions) that $D_8$ and $D_9$ do
not intersect the Calabi--Yau space $X_{(C)}$. Taking linear combinations for 
which the corresponding components of the Mori vectors vanish and going to a 
basis where curves on $X$ bound the K\"ahler cone yields again 
\begin{equation}
  \begin{array}{rl}
  l^{(1)}&=(  -4,-2;2,1,1,\um1,1,0,0,0)\\
  l^{(2)}&=(\um0,-2;0,1,0,  -1,0,1,1,0)\ .
  \label{eq:mori230C}
  \end{array}
\end{equation}  
With $H= D_3 D_{0,1} D_{0,2}$ and $L=D_6 D_{0,1} D_{0,2}$ we find the same 
intersections as (\ref{230cijk}) and (\ref{230c2}). The same conclusions arise
for the other star triangulations. 

To summarize all representations of the (2,30) model, be it different 
polyhedra, different nef-partitions, or different triangulations, are 
equivalent. Some of them exhibit, however, different parametrizations 
of the complex moduli space of the mirror.

\subsubsection{The phase diagram}
\label{sec:phases}

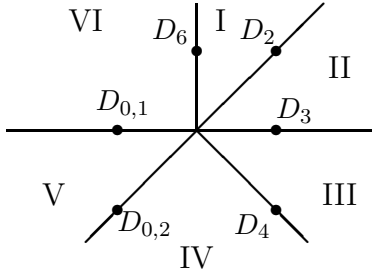
\begin{figure}
\def\putlab#1)#2#3{\put#1){\makebox(0,0)[#2]{\small #3}}} 
\def\putlin#1,#2,#3,#4,#5){\put#1,#2){\line(#3,#4){#5}}}
\def\putvec#1,#2,#3,#4,#5){\put#1,#2){\vector(#3,#4){#5}}}
\def\putcx#1,#2){\put#1,#2){\circle*{1.4}}}
\begin{center}\unitlength=3pt\thicklines\begin{picture}(40,40)(-20,-20)
        \putlab(-14,14)c{\normalsize                    $\mathrm{VI}$}
\putlin(0,0,0,1,16)
        \putlab(3,14)c{\normalsize                    $\mathrm{I}$}
\putlin(0,0,1,1,16)
        \putlab( 18,8)c{\normalsize                    $\mathrm{II}$}
\putlin(0,0,1,0,22)
        \putlab(18,-8)c{\normalsize                     $\mathrm{III}$}
\putlin(0,0,1,-1,14)
        \putlab(0,-16)c{\normalsize                     $\mathrm{IV}$}
\putlin(0,0,-1,-1,14)
        \putlab(-18,-8)c{\normalsize                    $\mathrm{V}$}
\putlin(0,0,-1,0,24)
        \putcx(-10,0) \putlab(-9.3,1.5)b{$D_{0,1}$}
        \putcx(10,0)    \putlab(10,2.5)l{$D_3$}
        \putcx(0,10) \putlab(-1,12.5)r{$D_6$} 
        \putcx(10,-10)  \putlab(9.3,-12)r{$D_4$}
        \putcx(-10,-10) \putlab(-10,-12)l{$D_{0,2}$}
        \putcx(10,10) \putlab(10,12.5)r{$D_2$}
\end{picture}\end{center}
\caption{\label{fig:phases230} 
        Secondary fan of the CICYs with $(h^{1,1},h^{2,1})=(2,30)$.}
\end{figure}

In the case of hypersurfaces it is well-known~\cite{Aspinwall:1993rj} 
that points in the interior of 
codimension one faces correspond to nonlinear automorphisms of the ambient 
space. A generic \CY hypersurface does not intersect the divisors 
corresponding to these points, and can, therefore, be neglected for most 
purposes, e.g. for computing the intersection ring and the phase diagram. In the case of complete intersections of codimension $r>1$ we have seen in the previous subsections that there are also vertices whose corresponding divisors do not intersect the generic complete intersection. We will now argue that these can be neglected a priori. We consider the example $X_{(C)}$ where $D_8$ and $D_9$ do not intersect $D_{0,1}D_{0,2}$. If we look for {\it all} the triangulations of $\langle 0_1 0_2 1 2 3 4 5 6 7 \rangle$ we find using the prescription given after~(\ref{eq:Gale})
\begin{eqnarray}
  \label{eq:triangs}
  T_{I} &=& \{ \langle0_2 1 3 4 5 6 7\rangle, \langle0_1 0_2 1 3 4 5 7\rangle, \langle0_1 0_2 1 3 4 5 6\rangle, \langle0_1 0_2 1 2 3 4 7\rangle, \langle0_1 0_2 2 3 4 5 7\rangle, \langle0_1 0_2 1 2 4 5 7\rangle,\nonumber\\
         & & \phantom{\{}\langle0_1 0_2 1 2 3 5 7\rangle, \langle0_1 0_2 1 2 3 5 6\rangle, \langle0_1 0_2 1 2 4 5 6\rangle, \langle0_1 0_2 1 2 3 4 6\rangle, \langle0_1 0_2 2 3 4 5 6\rangle\}\nonumber\\
  T_{II} &=& \{ \langle0_1 0_2 1 3 4 6 7\rangle, \langle0_1 0_2 1 2 3 4 7\rangle, \langle0_1 0_2 2 3 4 5 7\rangle, \langle0_1 0_2 1 2 4 5 7\rangle, \langle0_1 0_2 1 2 3 5 7\rangle, \langle0_1 0_2 3 4 5 6 7\rangle,\nonumber\\
        & &\phantom{\{} \langle0_1 0_2 1 4 5 6 7\rangle, \langle0_1 0_2 1 3 5 6 7\rangle, \langle0_1 0_2 1 2 3 5 6\rangle, \langle0_1 0_2 1 2 4 5 6\rangle, \langle0_1 0_2 1 2 3 4 6\rangle, \langle0_1 0_2 2 3 4 5 6\rangle \}\nonumber\\
  T_{III} &=& \{ \langle0_1 1 2 4 5 6 7\rangle, \langle0_1 1 2 3 4 6 7\rangle, \langle0_1 2 3 4 5 6 7\rangle, \langle0_1 0_2 1 2 3 6 7\rangle, \langle0_1 0_2 2 3 5 6 7\rangle, \langle0_1 0_2 1 2 5 6 7\rangle,\nonumber\\
          & & \phantom{\{}\langle0_1 0_2 1 2 3 5 7\rangle, \langle0_1 0_2 1 3 5 6 7\rangle, \langle0_1 0_2 1 2 3 5 6\rangle \} \nonumber\\
  T_{IV} &=&\{ \langle0_1 1 2 3 4 5 7\rangle, \langle0_1 1 2 3 4 5 6\rangle, \langle0_2 1 2 3 5 6 7\rangle, \langle1 2 3 4 5 6 7\rangle \}\nonumber\\
  T_{V} &=& \{ \langle0_1 1 2 3 4 5 7\rangle, \langle0_1 1 2 3 4 5 6\rangle, \langle0_2 1 2 3 4 5 6\rangle, \langle0_2 1 2 3 4 5 7\rangle, \langle0_2 1 3 4 5 6 7\rangle \}\nonumber\\
  T_{VI} &=& \{ \langle0_1 1 2 4 5 6 7\rangle, \langle0_1 1 2 3 5 6 7\rangle, \langle0_1 1 2 3 4 6 7\rangle, \langle0_1 2 3 4 5 6 7\rangle, \langle0_2 1 2 3 5 6 7\rangle \}\nonumber
\end{eqnarray}
Here $0_1$ and $0_2$ denote the two ``interior points'' corresponding to 
$D_{0,1}$ and $D_{0,2}$, respectively. One can check that if we identified 
$0_1$ with $0_2$ and triangulated the polyhedron $\Delta^*$ instead of the 
Gorenstein cone $\Gamma(\Delta^*)$, then we would find 2 more triangulations.
Looking at the columns of the Mori vectors in~(\ref{eq:mori230C}) we see, 
however, that we indeed expect only six triangulations. This confirms the 
statement that the phases are obtained by triangulating the Gorenstein cone 
instead of the polyhedron only. The corresponding phase diagram is 
drawn in figure~\ref{fig:phases230}.
We see that the K\"ahler cone that we determined to be spanned by $D_3$ and $D_6$ in the previous subsections contains two phases. This is due to the fact that for $X_{(B)}$ and $X_{(C)}$ we needed to add $\nu^*_{\r}$ in order to resolve all the singularities. This is also reflected in the different Mori cones~(\ref{mori230A}) for $X_{(A)}$ on one hand, and respectively~(\ref{eq:mori230B}), (\ref{eq:mori230C}) for $X_{(B)}$ and $X_{(C)}$ on the other hand. For $X_{(A)}$ we did not have to add a point in degree 2, so that the ray corresponding to $D_2$ points in the same direction as $D_3$.

\section{Topological strings on compact Calabi--Yau threefolds}
\label{sec:TopString}

All genus topological string amplitudes are non-trivial precisely 
on Calabi--Yau 3-folds. In the figure we give an overview over 
the various techniques to solve for these amplitudes in practice
and we will discuss their applications and limitations below.
\begin{center}
\epsfig{file=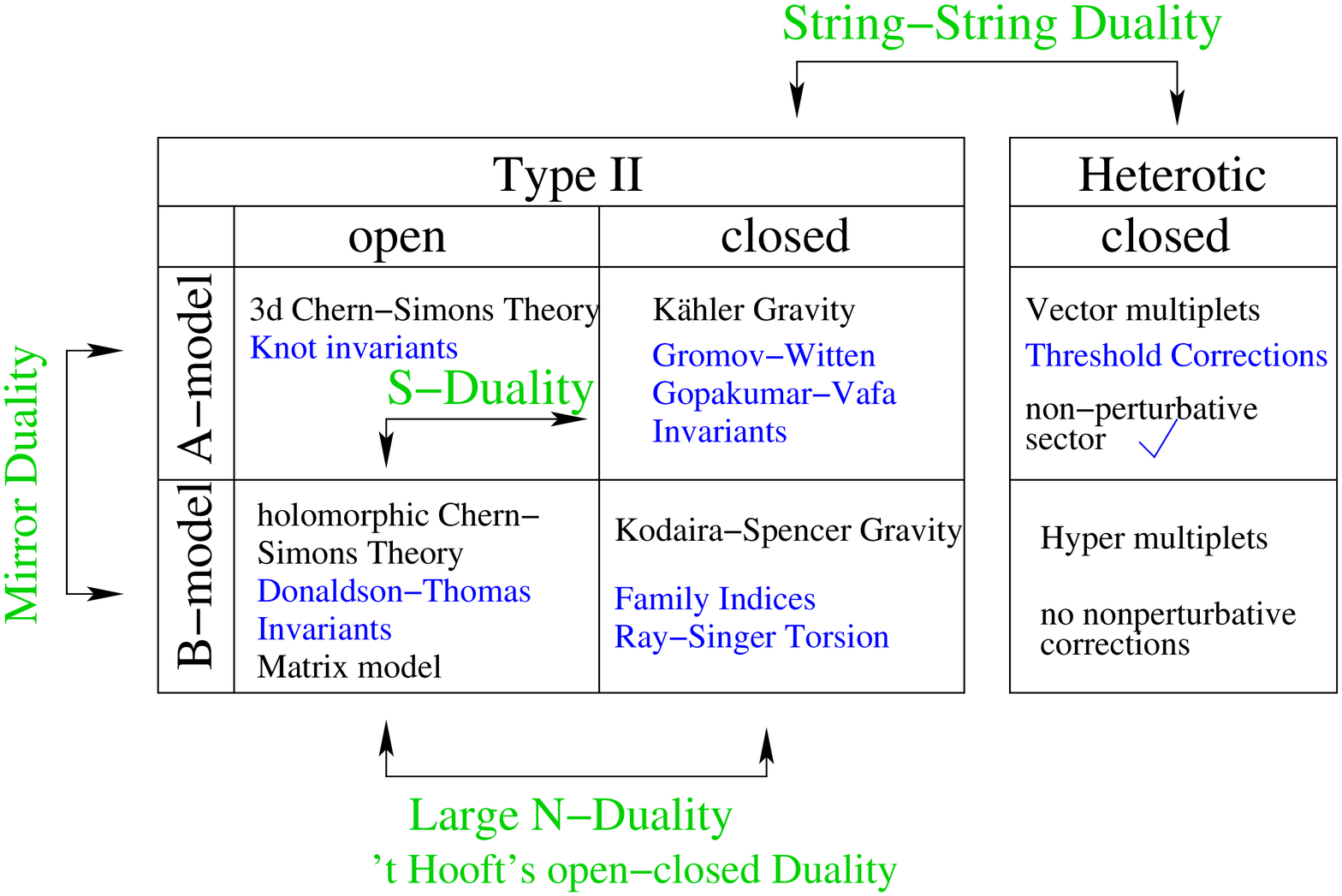, width=15cm} 
\end{center}
We want to emphasize that in this figure mirror symmetry is a map from the 
A--model on $X$ to the B--model on its mirror $X^*$, while S--duality maps 
the A--model on $X$ to the B--model on the {\em same} space. Furthermore, 
in the present work, we will use the latter to compute the 
Donaldson--Thomas invariants as an expansion of closed string amplitudes.

Easy generalizations of localization w.r.t. the torus 
action or the use of large N-transition, two
techniques that are successful for non-compact toric 
varieties, fail for the compact case. In the first case the strata 
of the moduli space of higher 
genus maps to the compact Calabi--Yau are not 
restrictions of the moduli space of maps to the ambient space. Using the 
induced torus action of the ambient space on the moduli space of maps for 
localization 
with the Atiyah-Bott fixpoint theorem will therefore not work. 
The topological vertex~\cite{Aganagic:2003db} is a tool which implements 
the large N transition idea 
systematically for all non-compact toric Calabi--Yau spaces and solves the 
open/closed topological string in these backgrounds. It relies likewise 
on the localization to equivariant maps under the torus action and does 
not generalize straightforwardly to the compact case.
The remaining tools are the B-model, heterotic type II string 
duality\cite{Kachru:1995wm} and direct calculation of the cohomology 
of D-brane moduli spaces. Unlike localizations and large N duality, 
the first two methods will give the holomorphic and the antiholomorphic 
dependence of the amplitudes. So far the second one gives only the 
amplitudes for K3 fibrations in the limit of large ${\mathbb P}^1$ base. Like 
localizations and large N techniques the last method calculates the holomorphic limit. 
But it seems that the calculational methods developed in \cite{Katz:1999xq} 
will become mathematically rigorous using 
the newly discovered relation between Gromov--Witten and Donaldson--Thomas 
invariants. For a recent comparison between Gromov--Witten, Gopakumar-Vafa and
Donaldson--Thomas invariants see also \cite{Katz:2004js}.

\subsection{Solution of the B-model on compact multi parameter Calabi--Yau spaces}

\label{sec:Bmodel}

The main physical application of mirror symmetry, manifest in 
the toric complete intersections, is the exact calculation of 
terms in the effective action of string compactification on the 
Calabi--Yau manifolds, by the variation of the Hodge structure under complex
structure deformations. These deformations are unobstructed on Calabi--Yau 
spaces and the physical interpretation of Kodairas deformation theory 
leads to the topological B-model expansion in orders $\lambda^{2 g-2}$ 
of the topological string coupling $\lambda$. 
The simplest application is to the vector multiplet moduli space $\cM$ 
in type IIB compactification on the Calabi--Yau manifold $X^*$. $\cM$ is a 
K\"ahler manifold with special geometry, which is directly identified with the
complex structure moduli space of the Calabi--Yau space $X^*$. 
In the resulting four dimensional $N=2$ 
supergravity theory the calculation yields at genus $g=0$ the exact K\"ahler 
potential for the vector moduli fields including the exact gauge couplings, 
as well as the moduli dependent BPS masses and the triple couplings, 
in lowest order in $\lambda^{2g-2}$ at genus $g=0$. The higher order terms $g>0$  
give the exact moduli dependence of the coupling of the selfdual 
part of the Riemann curvature to the selfdual part of the graviphoton 
field strength $R_{+}^2 F^{2g-2}_{+}$. If $X$ and $X^*$ are mirror pairs, mirror 
symmetry extends the application of the above calculation to the 
vector multiplet moduli space of type IIA on $X$, where the couplings are world-sheet 
instanton corrected and therefore K\"ahler structure dependent. 
(For a review and further references, see e.g.~\cite{Hori:2003ab}.)
The calculation gives only limited information about the hypermultiplet 
moduli space of both type II theories, because the latter is doubled 
by RR field background values to a quaternionic manifold, which is hard to access
by worldsheet methods. On the other hand string/string duality and 
flux compactifications extend the relevance of our analysis to vector 
multiplets of the $N=2$ heterotic string and the superpotential of 
chiral multiplets in $N=1$ compactifications.

In this chapter we will outline the techniques necessary 
to obtain the effective action for the general class of 
complete intersections described above. Calabi--Yau hypersurfaces 
in torically resolved ambient spaces have been treated in some 
generality.  The cases of complete intersections that were discussed in
the literature, however,
avoided the singularities of the ambient space apart from one class 
of examples with  $\mZ_2$ curve singularities in \cite{Hosono:1994ax}. 
One aim of the present investigation
is to overcome various technical complications 
in finding the Picard--Fuchs operators, good coordinates and 
an integral basis of solutions corresponding to the geometric periods 
for the general complete intersection Calabi--Yau in toric ambient 
spaces. This will be described in Appendix~\ref{sec:PF}. 
Our main progress however is the construction of the propagators 
of the Kodaira-Spencer Gravity for the compact multi moduli case.       

\subsection{$N = 2$ special geometry}

The B-model twist\cite{Witten:1991zz} modifies the spin operator of the $N=(2,2)$
world sheet theory by the axial $U(1)_A$ operator\footnote{This current 
is anomalous unless the target space $X^*$ is a Calabi--Yau manifold.} 
so that $Q_B={\bar Q}_+ + {\bar Q}_-$ becomes a scalar BRST operator, well 
defined for all world-sheet genus. The cohomology of local observables is 
isomorphic to $\oplus_{p,q}H^p(X^*,\wedge^q TX^*)$ and the marginal deformations,
{\it i.e.} $p=q=1$,
of the B-model are identified with the complex structure deformations\footnote{
The extended deformation space w.r.t. all operators in  
$\oplus_{p,q}H^p(X^*,\wedge^q TX^*)$ has been considered in~\cite{BK}.} 
of $X^*$.

The genus $g$ correlators with $n$ marginal B-model 
observables inserted  $C^{(g)}_{i_1,\ldots ,i_n}$ are derivable from 
generating functions $\cF^{(g)}$ as 
\begin{equation}
\begin{array}{rcl}
\ds{C^{(g)}_{i_1, \ldots, i_n}}&=&\ds{D_{i_1}\dots D_{i_n} \cF^{(g)}},\qquad {\rm with} \\ [2 mm]
\ds{D_i}&=&\ds{\partial_{t^i} - \Gamma_i- (2 - 2g)\ \partial_{t^i} K}.\\ [2 mm]
\end{array}
\end{equation}
The $D_i$ are covariant w.r.t. the metric connection on the moduli space as well as w.r.t.
K\"ahler transformations, {\it i.e.} $\Gamma^i_{jk}$ is the connection of the special K\"ahler metric on the vector 
multiplet moduli space. This metric is identified with 
the Weil--Petersson metric of the complex structure moduli space ${\cal M}$ of $X^*$. 
Its K\"ahler potential $K(t,\bar t)$ is derivable from $\cF^{(0)}$, 
the generating function of the genus zero Gromov--Witten invariants 
of $X^*$, which is called prepotential. It is a section of a holomorphic 
line bundle ${\cal L}^{2}$ over ${\cal M}$, where the power $2$ indicates the transformation, 
multiplication by $e^{2f(t)}$, under a holomorphic change of the unique 
$(3,0)$ form $\Omega\rightarrow \Omega e^{f (t)}$ of the Calabi--Yau manifold. 
More generally, the $\cF^{(g)}$ transform as a section of ${\cal L}^{2- 2g}$.

We set $h=h^{2,1}(X^*) = h^{1,1}(X)$. In inhomogeneous variables 
$t^i={X^i\over X^0}$, $i=1,\ldots, h$, with the A-periods 
$X^I=\int_{A^I}\Omega,\; I=0,\ldots,h$, one has 
\begin{equation}
\exp(-{K(t,\bar t)})= (t^i-\bar t^{\bar i})(
\partial_{t^i} \cF^{(0)}+\partial_{{\bar t}^{\bar i}} 
{\bar \cF}^{(0)})- 2 (\cF^{(0)}- {\bar \cF}^{(0)})\ .
\label{eq:kahler}
\end{equation}
Note that $\Omega\rightarrow e^{f(t)} \Omega$ induces a transformation of the
K\"ahler potential, $K\rightarrow K -f(t) -\bar f(\bar t)$, leaving the metric 
$G_{\bar \imath j}= \partial_{\bar t^i}\partial_{t^j} K$ invariant. The special 
K\"ahler property of $\cM$ manifests itself also in the special relation between
the Riemann curvature and the genus 0 correlators\footnote{
To simplify notations quantities with no superscript $^{(0)}$ 
refer also to the genus zero sector.}
\begin{equation}
R_{i{\bar k} j}^l=-{\bar \partial}_{\bar k} \Gamma^{l}_{ij}
=[D_i,\partial_{\bar k}]^l_j=G_{{\bar k} i}\delta^l_j+G_{{\bar k} j} 
\delta^l_i- C_{ijm}{\bar C}^{ml}_{\bar k}
\label{eq:special}
\end{equation}

\subsection{The genus zero sector}
\label{sec:genus-zero}

The genus zero sector is derived from the period integrals of $X^*$, 
which can be partly obtained by direct computation and more systematically by solving 
the Picard--Fuchs equations. The large complex structure 
limit in ${\cal M}$ can be defined from the K\"ahler cone of the 
original manifold $X$. 

If $\Pi=(\int_{A_i} \Omega,\int_{B_i}\Omega)=(X^i,F_i)$ is a 
symplectic\footnote{W.r.t. $\Sigma=\left(\begin{array}{cc}\phantom{-}0&\phantom{-}{\bf 1}\\ 
-{\bf 1}&\phantom{-}0\end{array}\right)$.} basis of the periods one 
has 
\begin{equation}
\begin{array}{rcl} 
e^{-K}&=& -i \int_{X^*} \Omega \wedge \bar \Omega = i\Pi^\dagger \Sigma \Pi\\ 
C^{(0)}_{ijk}&=&D_iD_jD_k \cF^{(0)}=\int_{X^*} \Omega \wedge \partial_{i}\partial_{j}\partial_{k} \Omega   \ . 
\label{eq:specialgeom}
\end{array}
\end{equation} 
The special K\"ahler equation (\ref{eq:special}) guarantees that $\cF^{(0)}$ 
can be integrated compatibly with (\ref{eq:kahler}).

Given the data for the Mori cone~\eqref{eq:genmori} and the classical intersections numbers $\kappa_{abc}$ 
we can follow~\cite{Hosono:1994ax} and write down a local expansion 
of the periods convergent near the large complex structure point, 
which is characterized by its maximal unipotent monodromy. We review 
in the following just the essentials and refer to~\cite{Hosono:1994ax} for
further details. 

A particular set of local coordinates $z_a$ on the complex structure moduli 
space on $X^*$ is defined by $z_b=\prod_{i=1}^n a_i^{l^{(b)}_i},\;b=1,\dots,h$ in terms of $a_i$, 
the coefficients in the polynomial constraints of the complete intersection 
in the torus variables~\eqref{eq:CIeqs}. A point of maximal unipotent 
monodromy is then always at $z_b=0$. Let $\varpi_{a_1,\ldots,a_s}$ be obtained 
by the Frobenius method\footnote{The holomorphic period $\varpi(z_1,\dots,z_h)$
can also be directly integrated using a residuum expression for the 
holomorphic $(3,0)$ form. An illustrative example for a complete intersection is 
given in Appendix~\ref{sec:PF}.}~\cite{Hosono:1994ax} from the coefficients of the 
holomorphic function $\varpi(\vec z,\vec \rho)$ defined as
\begin{equation}
  \begin{array}{rl} 
    \varpi(z_1,\dots,z_h,\rho_1,\dots,\rho_h)&=\ds{\sum_{\{n_a\}} c(n_1\ldots
    n_h,\rho_1\ldots \rho_h)
    \prod_{a=1}^h z_a^{n_a+\rho_a}}\\ [ 3 mm]
    c(n_1,\dots,n_h,\rho_1,\dots,\rho_h)&=\ds{ { \prod_{m=1}^r 
    \Gamma(1-\sum_{a=1}^{h} \hat l^{(a)}_m(n_a+\rho_a))\over
    \prod_{i=1}^n \Gamma(1+\sum_{a=1}^{h} 
    l^{(a)}_i (n_a+\rho_a))}}\\ [ 4 mm]
    \varpi_{a_1,\dots,a_s}(z_1,\dots,z_h)&=\left(1\over 2 \pi i\right)^s\partial_{\rho_{a_1}}
    \ldots\partial_{\rho_{a_s}}
    \varpi(z_1,\dots,z_h,\rho_1,\dots,\rho_h)|_{\{\rho_a=0\}\ . }
    \label{eq:varphi}
  \end{array}
\end{equation}     
Define also $\sigma_{a_1,\dots,a_s}=(\varpi_{a_1,\dots,a_s}
(z_1,\dots,z_h)|_{\log(z_a)=0})/\varpi(z_1,\dots,z_h,\rho_1,\dots,\rho_h)|_{\{\rho_a=0\}}$. 
At the large complex structure point
the mirror map defines natural flat coordinates on the K\"ahler moduli space 
of the original manifold $X$
\begin{equation}
t^a={X^a \over X^0}={1\over 2 \pi i}(\log(z_a) + \sigma_a), \qquad a=1,\ldots, h\ ,
\label{eq:mirrormap}
\end{equation}
where $X^0=\varpi(z_1,\dots,z_h,\rho_1,\dots,\rho_h)|_{{\underline \rho}=0}$ 
is the unique holomorphic period at $z_a=0$ and $X^a=\varpi_a$ are the 
logarithmic periods. Double and triple logarithmic solutions are given 
by 
\begin{eqnarray}
  w^{(2)}_a&=&\ds{ {1\over 2} \sum_{b,c=1}^h \kappa_{abc}  \varpi_{bc}(z_1,\dots,z_h), \ \ \ a=1,\ldots, h}.\\
  \label{eq:L_a}
  w^{(3)}&=&\ds{{1\over 6} \sum_{a,b,c=1}^h \kappa_{abc} \varpi_{abc}(z_1,\dots,z_h)} \ ,
  \label{eq:L}
\end{eqnarray}
where $\kappa_{abc}$ are the classical intersection numbers in~(\ref{eq:kappa}).
The prepotentials $F^{(0)}(X^I)$ in homogeneous or $\cF^{(0)}(t^a)$ in inhomogeneous coordinates 
can now be written as
\begin{equation} 
  \begin{array}{rcl}
  F^{(0)}&=&\ds{-{\kappa_{abc} X^a X^b X^c\over 3 !X^0} + 
  A_{ab} {X^a X^b \over 2} + c_a X^a X^0- 
  i \chi{\zeta(3)\over 2 (2 \pi)^3} (X^0)^2 + (X^0)^2 f(q)}\\ [2 mm]
  &=&(X^0)^2\cF^{(0)}=\ds{(X^0)^2\left[-{\kappa_{abc} t^a t^b t^c\over 3!} 
   +A_{ab} t^a t^b +c_a t^a - i\chi{\zeta(3)\over 2 (2 \pi)^3} + f(q)\right]}
  \end{array}
  \label{eq:prepot}
\end{equation}
where $q_a=\exp(2 \pi i t^a)$, $c_a={1\over 24}\int_X \ch_2 J_a$ and  $\chi$ is the Euler number of $X$. 
The real coefficients $A_{ab}$ are not completely fixed. They are unphysical 
in the sense that $K(t,\bar t)$ and 
$C_{abc}(q)$ do not depend on them. A key technical problem\footnote{We 
wrote an improved code for that \cite{CICY}.} in the calculation
is to invert the exponentiated mirror map (\ref{eq:mirrormap}) to obtain $z_i(\underline{t})$. 
An integral symplectic basis for the periods is given by
\begin{equation}
\Pi=X^0 \left(\begin{array}{c}
1\\ 
t^a\\ 
2 \cF^{(0)}- t^a \partial_{t^a}\cF^{(0)}\\
\partial_{t^a} \cF^{(0)}\end{array}\right)=
X^0\left(\begin{array}{c}
1\\ 
t^a\\ 
{\kappa_{abc} t^a t^b t^c\over 3!} + c_a t^a - i \chi{\zeta(3)\over (2 \pi)^3} + 2 f(q)-t^a \partial_{t^a} f(q)\\
-{\kappa_{abc}  t^b t^c\over 2}+ A_{ab}  t^b+ c_a+\partial_{t^a} f(q)\\ 
\end{array}\right)
\label{eq:periodsatinfinty}
\end{equation}
This period vector can be uniquely given in terms of~\eqref{eq:L},\eqref{eq:varphi} by adapting
the leading log behaviour. The $A_{ab}$ are further restricted by the requirement
that the Peccei-Quinn symmetries $t^a\rightarrow t^a+1 $ act as integral 
${\rm Sp}(2 h^{11}+2,\mZ)$ transformations on $\Pi$. Note that $\cF^{(0)}$ can be read off from 
the periods and since $t^a$ are flat coordinates, we have
\begin{equation}
  C_{abc}(q)= \partial_{t^a} \partial_{t^b}\partial_{t^c} \cF^{(0)}=\kappa_{abc}+
\sum_{d_a,d_b,d_c\geq 0} n_d^{(0)}d_a d_b d_c \frac{q^d}{1-q^d},
  \label{eq:rational_inst}
\end{equation}
where the sum counts the contribution of the genus zero worldsheet instantons. We defined 
$q^d=\prod_a e^{-2\pi i d_a t^a}$ where the tuple $(d_1,\dots,d_h)$ specifies a 
class $Q$ in $H^2(X,{\mathbb Z})$. 

An alternative way of obtaining $C_{abc}(q)$ is to first calculate
$Y_{ijk}:=C_{ijk}(z)$ from the last equation in (\ref{eq:specialgeom}) and
the Picard--Fuchs operators~(\ref{eq:DL}). This leads to linear differential 
equations, which determine $Y_{ijk}$ up to a common constant, 
see again~\cite{Hosono:1994ax} for
details.  $C_{abc}(q)$ follows from  $Y_{ijk}(z)$ using the inverse mirror 
map (\ref{eq:mirrormap}) $z=z(t)$ in a special gauge w.r.t. ${\cal L}^2$ as 
\begin{equation}
C_{abc}(q)=\frac{1}{X_0^2}\frac{\partial z_i}{\partial t^a}
\frac{\partial z_j}{\partial t^b} \frac{\partial z_k}{\partial t^c}
  Y_{ijk}(z(q))\ .
\end{equation}
Knowledge of the $Y_{ijk}(z)$ will be essential for determining the B-model 
propagators in the next subsection.

\subsection{Multi parameter B-model propagators}
\label{sec:propagators}

The key to the solution of the B-model are the holomorphic anomaly 
equations~\cite{Bershadsky:1993cx}. The derivatives of $\cF^{(g)}$ or 
$C^{(g)}_{i_1,\dots,i_n}$ w.r.t. the antiholomorphic coordinates lead to 
an integral over an exact form over the moduli space ${\cal M}_{g,n}$ of the
world sheet curve $\Sigma_g$, which receives contributions only from the 
boundary of ${\cal M}_{g,n}$. In particular, $C_{abc}$ is holomorphic because 
the sphere is rigid and three points can be fixed by ${\rm SL}(2,\mC)$, 
and therefore 
the boundary of the moduli space is empty. In the other cases the boundary 
contributions leads to recursive differential equations, the so called 
holomorphic anomaly equations\cite{Bershadsky:1993cx}, which read for $g>1$   
\begin{equation}
{\bar \partial}_{\bar k} \cF^{(g)}={1\over 2}{\bar C}_{\bar k}^{ij}
\left(D_i D_j \cF^{(g-1)}+ \sum_{r=1}^{g-1}D_i \cF^{(r)}  D_j\cF^{(g-r)}\right)
\label{eq:anomaly}  
\end{equation} 
and for $g=1$ \cite{Bershadsky:1993ta}
\begin{equation}
{\bar \partial}_{\bar k} \partial_m \cF^{(1)}={1\over 2}
{\bar C}_{\bar k}^{ij} C_{mij} + \left( {\chi\over 24} -1\right )
G_{\bar k,m}
\label{eq:anomalyg1}\ .
\end{equation} 
Both terms in (\ref{eq:anomaly}) have an obvious interpretation as coming from 
boundary components  of ${\cal M}_{g,n}$; the first from pinching a handle and 
the second from pinching the connection between two irreducible components of 
genus $r$ and $g-r$. In (\ref{eq:anomalyg1}) the first and second term 
correspond, respectively, to also pinching a handle and to a contact term 
which arises when the locations of the two operators needed to fix the conformal 
symmetry coincide.

Here ${\bar C}_{\bar k}^{ij}=e^{2K} G^{i\bar \imath} G^{j\bar \jmath} 
{\bar C}_{\bar \imath \bar \jmath \bar k}$ is related to the antiholomorphic 
coupling ${\bar C}_{\bar i \bar j  \bar k}$, which is symmetric in its 
indices and fulfils an integrability condition,
known as the WDVV equations, 
$D_{\bar \imath} {\bar C}_{\bar \jmath \bar k  \bar l}= 
D_{\bar \jmath} {\bar C}_{\bar \imath \bar k \bar l}$. It can therefore be 
derived from a section $S$ of ${\cal L}^{-2}$ as 
\begin{equation} 
{\bar C}_{\bar \imath \bar \jmath  \bar k}=e^{-2 K} D_{\bar \imath} 
D_{\bar \jmath} D_{\bar k} S\ .
\label{eq:S}
\end{equation}  
One of the main prerequisites to solve (\ref{eq:anomaly}) recursively is the 
construction of $S$. It is convenient to define as intermediate steps of the 
above integration $S_{\bar \jmath}$, $S^i_{\bar \jmath}$, $S^i$ and $S^{ij}$ by
\begin{equation}
\begin{array}{rcl}
S^i&=&G^ {i{\bar \jmath}}S_{\bar \jmath}= G^{i{\bar \jmath}} {\bar \partial}_{\bar \jmath}S \\
S^{ik}&=&G^{k \bar k} S_{\bar k}^i= G^{k \bar k} {\bar \partial}_{\bar k} S^i\\
{\bar C}_{\bar k}^{ij}&=& \partial_{\bar k} S^{ij}\ .
\end{array}
\label{eq:integrationsteps}
\end{equation}
These are used in rewriting the right hand side of 
(\ref{eq:anomaly}) in successive steps as $\partial_{\bar k}[\ldots]$. 
I.e. one gets in the first step $\partial_{\bar k} {1\over 2}S^{ij} (\ldots)-
{1\over 2} S^{ij}\partial_{\bar k}(\ldots)$ on the right hand side.  
Using the commutator~(\ref{eq:special}) from special geometry 
and~(\ref{eq:anomaly}), (\ref{eq:anomalyg1}) in the 
second term one gets again terms in which antiholomorphic derivatives of 
$S,S^i,S^{ij}$ can be substituted using (\ref{eq:integrationsteps}). 
Repeating the partial integration and the application of 
$[\partial_{\bar k},D_i]$ and (\ref{eq:anomaly}), (\ref{eq:anomalyg1})  
one gets after $g$ steps the desired total derivative 
$\partial_{\bar k}[\ldots ]$. In the final expression we set $K^{\phi\phi}:=-S$, 
$K^{i,\phi}:=-S^{i}$ and $K^{ij}:=-S^{ij}$ which can be interpreted as 
propagators of a Feynman expansion, where $i$ labels moduli fields and 
$\phi$ the dilaton. 
Using the formula for the Ricci tensor in K\"ahler geometry 
$R_{i\bar \imath}=-{1\over 2} \partial_i \partial_{\bar \imath} \log \det(G)$ 
and (\ref{eq:special}) one can integrate the anomaly (\ref{eq:anomalyg1}) to
\footnote{The normalization of $\cF^{(1)}$ differs from the normalization
of $\cF^{(1),{\rm het}}$ by $\frac{16\pi^2}{24}$~\cite{AL}. This will be taken
up in the discussion of free quotients in Section~\ref{sec:IIAhetDual}}
\begin{equation}
\cF^{(1)}= {1\over 2} \log\left( \det(G^{-1}) \exp\left( {K } 
\left(3 + h^{11}-{\chi\over 12}\right)\right) |f_1|\right)\ .
\label{eq:f1a}
\end{equation}
Here $f_1=\prod_i \Delta_i^{r_i}$ is a function on the moduli space ${\cal M}$, 
which encodes the behaviour of $\cF^{(1)}$ at the various components $\Delta_i$ 
of the discriminant loci. We can alternatively use the propagator $S^{ij}$ to 
write
\begin{equation}
\partial_k \cF^{(1)}= {1\over 2} S^{ij} C_{ijk} + 
({\chi\over 24}-1) \partial_k (K + \log f_1).
\label{eq:f1b}
\end{equation} 
Here no partial integration was required since $C_{ijk}$ is holomorphic. For 
the same reason one can rewrite (\ref{eq:special}) as 
$\bar \partial_{\bar k}[S^{ij}C_{jkl}]= \bar \partial_{\bar k}
[\delta^i_l\partial_k K + \delta^i_k \partial_l K+\Gamma^{i}_{kl}]$. At the 
expense of introducing a holomorphic ambiguity $A^p_{ij}$ the 
$\bar \partial_{\bar k}$ derivative can be removed and the equation is solved
for $S^{ij}$
\begin{equation}
S^{ij}=(C_{k}^{-1})^{jl}\left((\delta^i_k\partial_l+\delta^i_l \partial_k) K
+ \Gamma_{kl}^i+A^i_{kl}\right)=: 
(C_{k}^{-1})^{jl}(Q_{kl}^i+A^i_{kl}),\ \ k=1,\ldots,h
\end{equation}
In this last step we must assume that the matrix $(C_k)_{ij}= C_{ijk}$ is 
invertible, see below. The left hand side $S^{ij}$ should be a unique and 
globally defined section of ${\rm Sym}^2 T^*{\cal M}_{g,n}\times {\cal L}^{-2}$. 
This has three important consequences: First $A^i_{kl}$ is not a tensor but 
has to compensate for the non covariant transformation of $Q^i_{kl}$, i.e. 
\begin{equation}
A^i_{kl}=\delta^i_k \partial_l \log(f)+ \delta^i_l \partial_k \log(f)
-v_{l,a} \partial_k v^{i,a}+T^i_{kl}\ ,
\label{eq:Acov}
\end{equation}  
where $f\in {\cal L}$, $v_{i,a}$ are $h$ sections of $T{\cal M}_{g,n}$ and
$T^i_{kl}$ does transform as a tensor. Secondly we must satisfy the 
compatibility   
\begin{equation}
(C_{\kappa}^{-1})^{il}(Q_{\kappa l}^i+A^i_{\kappa l})=(C_{\mu}^{-1})^{il}(Q_{\mu l}^i+A^i_{\mu l}) \qquad  
\forall \mu\neq \kappa,\ \ \mu,\kappa=1,\ldots, h
\label{eq:comp1}
\end{equation}
and finally ensure the symmetry  
\begin{equation}
(C_{\kappa}^{-1})^{jl}(Q_{\kappa l}^i+A^i_{\kappa l})=(C_{\kappa}^{-1})^{il}(Q_{\kappa l}^j+A^j_{\kappa l}) \qquad  
\forall i\neq j, \ \ i,j = 1,\ldots, h \ .
\label{eq:comp2}
\end{equation}

We solve the equations (\ref{eq:comp1}), (\ref{eq:comp2}) by first finding 
$A^i_{jk}$ algebraically. It turned out that one could always make the solution 
covariant by finding $v_{i,a}$ and $f$ and that $T^i_{jk}$ could be set to zero. 
From $\partial_{\bar \imath} S^j=G_{i\bar \imath} S^{ij}$ and 
$S^{ij}C_{jkl}=Q^i_{kl}+A^i_{kl}$ we get 
\begin{equation}
S^i C_{ikl}=(\partial_k K \partial_l K- \partial_k \partial_l K +
A_{kl}^i \partial_i K)+A_{kl}=:Q_{kl}+A_{kl}\ .
\end{equation}
If $A^i_{kl}$ transforms as in (\ref{eq:Acov}), which we always could achieve, 
then $A_{ij}$ transforms as ${\rm Sym}^2 T{\cal M}_{g,n}$. We get again 
constraints on the $A_{jk}$ from the uniqueness of 
$S^i=(C_{\kappa}^{-1})^{ij}(Q_{\kappa j}+A_{\kappa j})$, 
$\kappa=1,\ldots,h$, namely
\begin{equation}
(C_{\kappa}^{-1})^{ij}(Q_{\kappa j}+A_{\kappa j})=(C_{\mu}^{-1})^{ij}(Q_{\mu j}+A_{\mu j}), 
\qquad \mu\neq \kappa,\ \ \mu,\kappa=1,\ldots, h
\label{eq:comp3}
\end{equation}
In this step we found the necessity of non zero rational $A_{ij}$ in some 
cases. The last propagator $S$ involves no further algebraic consistency 
conditions and a special solution is given by~\cite{Bershadsky:1993cx}
\begin{equation}
S={1\over 2 h^{11}}[(h^{11}+1) S^i- D_j S^{ij}- S^{ij} S^{kl} C_{jkl}] \partial_{i}(K+\log(|f|)/2)+
  {1\over h^{11}}(D_i S^i+ S^i S^{jk} C_{ijk})\ .
\end{equation}

\subsection{Compatibility constraints on the multi parameter mirror maps}

So far (\ref{eq:comp1}), (\ref{eq:comp2}) and (\ref{eq:comp3})  
have not been solved for global 
multi parameter Calabi--Yau manifolds. The one parameter case is of course 
trivial in this respect. What we have found is that the conditions 
(\ref{eq:comp1}) to (\ref{eq:comp3}) are in general non-trivial and can 
nevertheless be solved with rational functions 
$A^i_{kl}$, $A_{kl}$ over the complex 
structure moduli space ${\cal M}$. In the holomorphic limit 
$\bar{t}^\imath \to 0$, one has
\begin{equation} 
\begin{array}{rcl}
G_{\bar pj}&=&\delta_{\bar p p} {\partial t_j\over \partial z_p},\qquad  
G^{j\bar p}=\delta^{\bar p p} {\partial z_p\over \partial t_j}, \\ 
\Gamma^i_{kj}&=&-{\partial z^i\over \partial t_p} {\partial \over 
\partial z_k} 
{\partial t_p\over \partial z_j}= {\partial t_p\over \partial z_j} 
{\partial \over \partial z_k} {\partial z^i\over \partial t_p}, \\ 
K&=&\log(X^0)\ .
\end{array}
\end{equation}
The equations (\ref{eq:comp1}) to (\ref{eq:comp3}) and the fact that we can find
rational $A^i_{kl}$, $A_{ij}$ in $z$ do therefore constitute non-trivial relations for the 
derivatives of the mirror maps and derivatives of $X^0$. This generalizes in 
an interesting way observations that the mirror maps of local multi parameter 
Calabi--Yau manifolds fulfil rational relations $Q(z_1,\dots,z_h;t^1,\dots,t^h)=0$. 
For example, in the case of the local Calabi--Yau 
${\cal O}(-2,-2)\rightarrow \mP^1\times \mP^1$ it was found in~\cite{Aganagic:2002wv} 
that the relation $t_1 z_2 = z_1 t_2$ holds and with $X^0=1$ in the local case, 
one can use the rational relation to find a 
solution of (\ref{eq:comp1}), (\ref{eq:comp2}). In the local case 
(\ref{eq:comp3}) is always trivial as was shown in~\cite{Klemm:1999gm}.
Here we find that~(\ref{eq:comp3}) is in general non-trivial, 
while rational expressions in $z$ for $A_{ij}$ could always be found.
 
The number of constraints does depend furthermore on the fibration structure 
of $X$. 
We found e.g. that for elliptically fibered Calabi--Yau manifolds such as 
$\mP_{1,1,1,6,9}[18]$ and  $\mP_{1,1,2,8,12}[24]$ the $(C_k)_{ij}$ are not 
invertible over the index $k$
that corresponds to the class of the elliptic fiber,
hence there are $h-1$ constraints less in (\ref{eq:comp1}).

\subsection{The solutions to the holomorphic anomaly equations} 

Solutions to the holomorphic anomaly equations~\eqref{eq:anomaly} have 
been constructed in~\cite{Bershadsky:1993cx} for genus two and 
in~\cite{Katz:1999xq} up to genus four for the compact and genus seven
for the non-compact case. As an example we quote from~\cite{Bershadsky:1993cx}
the solution of~\eqref{eq:anomaly} for $g=2$, which we will compute in the 
examples below:
\begin{equation}
  \label{eq:F2an}
  \begin{array}{rcl}
    \cF^{(2)}(t,\bar t) &=& \ds{\frac{1}{2}S^{ij}C^{(1)}_{ij} + 
        \frac{1}{2} C^{(1)}_iS^{ij}C^{(1)}_j - \frac{1}{8}S^{ij}S^{kl}C_{ijkl}}\\ [ 3mm]
    & & \ds{-\frac{1}{2}S^{ij}C_{ijk}S^{kl}C^{(1)}_l 
        + \frac{\chi}{24} S^i C^{(1)}_i} \\ [ 3mm]
    & & \ds{+\frac{1}{8}S^{ij}C_{ijk}S^{kl}C_{lmn}S^{mn} 
        + \frac{1}{12}S^{il}S^{jm}S^{kn}C_{ijk}C_{lmn}}\\ [ 3mm]
    & & \ds{-\frac{\chi}{48}S^iC_{ijk}S^{jk} 
        + \frac{\chi}{24}\left(\frac{\chi}{24}-1\right)S + f_2(t)}
  \end{array}
\end{equation}
where $f_2(t)$ is the holomorphic ambiguity coming from the integration
of~(\ref{eq:anomaly}). We should stress that the solutions for 
$\cF^{(g)}$ in~\cite{Bershadsky:1993cx},\cite{Katz:1999xq} together with
the K\"ahler potential and the Weil--Petersson metric (\ref{eq:special}) 
are sufficient to obtain the full $(t,\bar t)$ dependence of $\cF^{(g)}$. 
However the determination of $f_g(t)$ is the main difficulty in finding 
an explicit solution for a given $X$. The general form of $f_g(t), g>0$ 
is expected to be, written in terms of the complex structure moduli 
$z=z(t)$,
\begin{equation}
  \label{eq:f_g}
  f_g(z) = \sum_{i=1}^D \sum_{k=0}^{2g-2} \frac{p^{(k)}_i(z)}{{\Delta_i}^k}
\end{equation}
where $D$ is the number of components $\Delta_i$ of the discriminant, and 
$p^{(k)}_i(z)$ are polynomials of degree $k$. At present, it is not known 
how to
determine the $p^{(k)}_i(z)$ a priori. Their structure is presumably 
related to
the compactification of the complex structure moduli space ${\cal M}$, i.e. 
it is encoded in the boundary of ${\cal M}$. It would be interesting to
make this relation precise. For our purpose, we will make an ansatz and try
to constrain this ansatz by using all the knowledge about the $\cF^{(g)}$ 
that is already available. We will address this point in detail in 
Section~\ref{sec:Ansatz}.

\section{Topological free energy and partition function}
\label{sec:Ftbt}

The only currently known way to derive the free energy of the 
topological A-model on a Calabi--Yau manifold $X$
\begin{equation}
\cF(\lambda,t,\bar t)=\sum_{\lambda=0}^\infty \lambda^{2g-2} \cF^{(g)}(t,\bar t)
\label{eq:fullfreeenergy}
\end{equation}
is to use mirror symmetry and the solutions to the holomorphic 
anomaly equation described in the last section. The topological 
partition function
\begin{equation}
Z(\lambda,t,\bar t)=e^{\cF(\lambda,t,\bar t)}
\end{equation}
has a very elegant interpretation essentially as a wave function $\Psi$ in an 
auxiliary Hilbert space, which arises by geometric quantization of 
$H^3(X,{\mathbb Z})$~\cite{Witten:1993ed}. 
The holomorphic anomaly equations are interpreted as granting the independence
of $\Psi$ on the choice of the symplectic structure on $H^3(X,{\mathbb Z})$ 
relative e.g. to the symplectic structure induced by a given complex 
structure on $X$. To specify the state $\Psi$, which corresponds to
the topological string, a huge amount of holomorphic boundary data are needed,
namely the holomorphic $f^{(g)}$. To make progress here, we discuss what 
is known
in various holomorphic limits of $\cF^{\rm hol}(\lambda,t)=\lim_{{\bar t}
\rightarrow \infty} \cF(\lambda,t,\bar t)$.

\subsection{Holomorphic free energy and partition function}
\label{sec:Fhol}

In this limit $\cF^{\rm hol}(\lambda,t)$ has a 
conjectured expansion in terms of BPS states associated to M2 branes wrapped
around a holomorphic curve $C$ and the M-theory circle. The charge $Q$ of the 
BPS state is labelled by the homology class of the curve, i.e. $Q\in H_2(X,\mZ)$. 
The genus of the curve is related to the SU$(2)_L$ spin~\cite{Gopakumar:1998ii},
\cite{Gopakumar:1998jq}. 

More precisely, the authors of~\cite{Gopakumar:1998jq} consider an M-theory 
compactification on $X$ to five dimensions. The space time BPS states fall into 
representations of the little group of the 5d Lorentz group 
$L={\rm SO}(4)\simeq {\rm SU}(2)_L\times {\rm SU}(2)_R$. 
The low energy interpretation of the free energy $F$ in 4d relates it to the 5d 
BPS spectrum through a Schwinger one loop calculation of the 4d $R_+^2 F_+^{2g-2}$ 
effective terms. A similar one loop calculation corrects the effective 
gauge coupling ${1\over g^2(G,p^2)}$ through threshold effects~\cite{Kaplunovsky:1995jw}. 
Note that these 4d calculations are sensitive to the off shell 
quantum numbers, i.e. to  ${\rm SU}(2)_L\times {\rm SU}(2)_R$. 
Only BPS particles annihilated by the supercharges in the $({\bf 0},{\bf {1\over 2}})$
representation contribute to the loop. They couple to the anti-selfdual graviphoton 
field strength $F_+$ and the anti-selfdual curvature $R_+$ only via their left spin 
eigenvalues of their representation under $L$. The right representation content enters 
solely via its multiplicity and a sign $(-1)^{2 j^3_R}$, in particular any
contribution of long multiplets is projected out by these signs. 
To summarize,
the dependence of $\cF$ on  the BPS spectrum is via a supersymmetric index 
\begin{equation}
I(\alpha,\beta)={\rm Tr}_{\cal H} (-1)^{2 j^3_L+2 j^3_R} e^{-\alpha j^3_L- \beta H}\ 
\end{equation} 
and all information entering $\cF$ is carried by the following combination
\begin{equation} 
\sum_{j^3_L,j_R^3} (-1)^{2 j_R^3}(2 j_R^3+1) N^Q_{j^3_R,J^3_L}[{\bf j_L}]=
\sum_{g=0}^\infty n^{(g)}_Q I_g\ . 
\label{eq:rightsum}
\end{equation}
of the multiplicities of the BPS states $N^Q_{j^3_R,J^3_L}$. 
The last basis change of the left spin from $[{\bf j_L}]$ to 
\begin{equation} 
I_g=\left[\left({\bf 1\over 2}\right)_L +2 ({\bf 0})_L\right]^{\otimes g}
\label{torusspin}
\end{equation}
relates the left spin to the genus $g$ of $C$ and defines the integer 
Gopakumar-Vafa invariants $n^{(g)}_Q$ associated to a holomorphic 
curve $C$ of genus
$g$ in the class $Q=[C]\in H^2(X,{\mathbb Z})$. In contrast to
the $n^{(g)}_Q$, the $N^Q_{j^3_R,j_L^3}$ are no symplectic invariants. 
They change when lines of marginal stability in 
the complex moduli space are crossed\footnote{Notice that the successful 
microscopic interpretation of the 5d  black hole entropy requires deformation invariance and 
relies on the index-like quantity and not on the total number of BPS states.}~\cite{Katz:1999xq}. 
The $N^Q_{j^3_R,j^3_L}$ are interpreted as the dimension of ${\rm SU}(2)_L\times {\rm SU}(2)_R$ representations 
w.r.t. a natural action of this group on the cohomology of the moduli space of the M2 brane. 
This moduli space is given by the moduli of flat U(1) connections on $C$ and the moduli of 
the curve. 
A model for this space is the Jacobian fibration over the moduli space of the curve $C$ in $Q$.
The expansion of $\cF^{\rm hol}$ in terms of these BPS state sums is obtained by 
performing the Schwinger loop calculation~\cite{Antoniadis:1995zn},\cite{Gopakumar:1998jq} as 
\begin{equation}
\begin{array}{rcl}
\label{gova}
\cF^{\rm hol}(\lambda,t)&=&\ds{\sum_{g=0}^\infty \lambda^{2 g-2} \cF^{(g)}}(t)\\
&=&\ds{{c(t)\over \lambda^2}+l(t)+\sum_{g=0}^\infty \sum_{Q\in H_2(X,\mZ)}\sum_{m=1}^\infty n_{Q}^{(g)} {1\over m}
\left(2 \sin {m \lambda \over 2}\right)^{2 g-2}  q^{Q m}}\\
&=&\ds{{c(t)\over \lambda^2}+l(t)+\sum_{g=0}^\infty \sum_{Q\in H_2(X,\mZ)}\sum_{m=1}^\infty n_{Q}^{(g)} 
(-1)^{g-1}{[m]^{(2 g-2)} \over m}  q^{Q m}},
\end{array}
\end{equation}
with
\begin{align*}
  q^Q &= e^{i\sum_{i=1}^{h^{1,1}}t_i \int_Q J_i}, & [x]&:=q_\lambda^{x\over 2}-q_\lambda^{-{x\over 2}},& q_\lambda&=e^{i\lambda}. 
\end{align*}
The cubic term $c(t)$ in the K\"ahler parameters $t_i$ is the classical part of the 
prepotential ${\cal F}^{(0)}$ given in (\ref{eq:prepot}) without the constant term, 
and $l(t)=\sum_{i=1}^h{t_i\over 24}\int_X \ch_2 J_i$ is the classical part
of ${\cal F}^{(1)}$. 
Using the expansion
$${1\over m}{1\over \left(2 \sin {m \lambda\over 2}\right)^2}=\sum_{g=0}\lambda^{2 g-2} 
(-1)^{g+1} {B_{2g}\over 2 g (2 g-2)!} m^{2g-3}$$
and a $\zeta(x)=\sum_{m=1}^\infty {1\over m^x}$ regularization of the sum over
$m$ with $\zeta(-n)=-{B_{n+1}\over n+1}$, we see that for $g\ge 2$ the $Q=0$ constant
map terms from localization~\cite{faberpandharipande} 
\begin{equation}
\langle 1\rangle^X_{g,0}=(-1)^g{\chi\over 2} \int_{{\cal M}_g}c^3_{g-1}=
(-1)^g\frac{\chi}{2} {|B_{2 g} B_{2g-2}| \over 2 g (2g-2)(2g-2)!}
\label{eq:rahul}
\end{equation}
are reproduced if we set $n^{(0)}_0=-{\chi \over 2}$. This choice also 
reproduces the constant term proportional to $\zeta(3)$ in ${\cal F}^{(0)}$.
In ${\cal F}^{(1)}$ there is a $\zeta(1)$ term which requires an 
additional regularization. 

In terms of the invariants $n_Q^{(g)}$ the partition function 
$Z^{\rm hol}=\exp(\cF^{\rm hol})$ has the following product 
form\footnote{Here we dropped the $\exp({c(t)\over \lambda^2}+l(t))$ 
factor of the classical terms  at genus $0,1$.}
\begin{equation}
Z^{\rm hol}_{\rm GV}(X,q_{\lambda},q)=\prod_{Q}\left[
\left(\prod_{r=1}^\infty (1-q_\lambda^r q^Q)^{r n_Q^{(0)}}\right)
\prod_{g=1}^\infty \prod_{l=0}^{2g-2}(1-q_\lambda^{ g-l-1} q^Q)^{(-1)^{g+r} 
\left(2 g-2\atop l\right) n_Q^{(g)}}\right]\ . 
\label{zhol}
\end{equation}
This product form resembles the Hilbert scheme of symmetric products 
written in terms of partition sums over
free fermionic and bosonic fields with an integer $U(1)$ charge as well
as the closely related product form for the elliptic genus of 
symmetric products. 
As it has already been pointed out in~\cite{Gopakumar:1998ii}, it is also 
reminiscent of the Borcherds product form of automorphic forms of $O(2,n,\mZ)$, 
see \cite{Borcherds} and \cite{kontsevich1} for a review. Here the idea is 
that integrality of the $n^{(g)}_Q$ is related to the fact that they are 
Fourier coefficients of other (quasi)automorphic forms, see also~\cite{kawai2}.

\subsection{Donaldson--Thomas expansion}
\label{sec:DT}

Another way to obtain BPS states is by wrapping D--branes on supersymmetric 
cycles in $X$. In particular, we can wrap Euclidean 6-branes on $X$ itself 
and Euclidean 2-branes on a curve $C \subset X$, possibly bound to some 0-branes.
At the level of RR charges such a configuration can be cast into a short 
exact sequence of the form
\begin{equation}
  \label{eq:Idealsheaf}
  0 \longrightarrow \cI \longrightarrow \cO_X \longrightarrow \cO_Z \longrightarrow 0
\end{equation}
where $\cI$ is the ideal sheaf describing this configuration and $Z$ is the 
subscheme of $X$ consisting of the curve $C$ and the points at which the
0-branes are supported. Counting BPS states therefore leads to the study of
the moduli space $I_m(X,Q)$ of such ideal sheaves $\cI$, which has two discrete
invariants: the class $Q=[Z] \in H_2(X,\mZ)$ and, roughly speaking, the number 
of 0-branes $m = \chi(\cO_Z)$. Due to the Calabi--Yau condition the virtual
dimension of $I_m(X,Q)$ is zero, and the number of BPS states with these 
charges is therefore obtained by counting the points in $I_m(X,Q)$. It is,
however, not quite as simple as that because as is well--known from 
Gromov--Witten theory, these configurations can appear in families, and one
has to work with the virtual fundamental class. Putting this important subtlety 
aside, this number is called the Donaldson--Thomas invariant 
${\tilde n}^{(m)}_{Q}$~\cite{Donaldson:1998ab}, \cite{Thomas:1998uj}. These
invariants are expected to be integral as they count BPS states.

Since both invariants, Gopakumar--Vafa and Donaldson--Thomas, keep track of the
number of BPS states, they should be related. The relation is in fact a 
consequence of the S--duality in topological 
strings~\cite{Nekrasov:2004js}, and takes the following form. 
The factor in~(\ref{zhol}) coming from the constant maps gives the McMahon 
function $M(q_\lambda)=\prod_{n\ge 0} {1\over (1-q_\lambda^n)^n}$ to the 
power ${\chi\over 2}$. This function appears also in Donaldson--Thomas 
theory~\cite{DTGW}, calculable on local toric Calabi--Yau spaces e.g. with 
the vertex~\cite{Aganagic:2003db}. However, in Donaldson--Thomas theory the 
power of the McMahon function is\footnote{We would like to thank Jim Bryan for
discussion on this point. The formula (\ref{zhol}) has been independently 
derived by him, S.~Katz and us.} $\chi$. Note also that if (\ref{gova}) holds 
then $\cF$ or $Z$ restricted to this class is always a finite degree rational 
function in $q_\lambda$ symmetric in $q_\lambda\rightarrow {1\over q_\lambda}$, 
since the genus is finite in a given class $Q$. Thanks to this observation
one can read from the comparison of the expansion of $Z^{\rm hol}$ in terms of 
Donaldson--Thomas invariants ${\tilde n}^{(m)}_{Q} \in {\mathbb Z}$ 
\begin{equation}
Z^{\rm hol}_{\rm DT}(X,q_{\lambda},q)=\sum_{Q,m\in {\mathbb Z}} {\tilde n}^{(m)}_{Q} q_\lambda^m q^Q
\label{eq:DTdef}
\end{equation}
with the expansion in terms of Gopakumar--Vafa invariants~\cite{DTGW}
\begin{equation}
Z^{\rm hol}_{\rm GV}(X,q_{\lambda},q) M(q_\lambda)^{\frac{\chi(X)}{2}}=Z^{\rm hol}_{\rm DT}(X,-q_{\lambda},q)\  
\label{eq:GVDT}
\end{equation}
the precise relation between ${\tilde n}^{(m)}_{Q}$ and $n_Q^{(g)}$. Eq. (\ref{gova}) 
and (\ref{zhol}) then relate the two types of invariants to the Gromov--Witten 
invariants $r_Q^{(g)}\in {\mathbb Q}$ as in 
$${\cal F}^{\rm hol}_{\rm GW}(\lambda,q)=\sum_{g=0}^\infty \lambda^{2g-2} \sum_{Q} r_Q^{(g)} q^Q\ .$$

\subsection{Constraints on the Ansatz}
\label{sec:Ansatz}

If we expand~(\ref{gova}) we find for the genus 2 contribution
\begin{equation}
  \label{eq:F2gv}
  \cF^{(2)}(t) = \frac{\chi}{5760} 
  + \sum_{Q} \left(\frac{1}{240} n_Q^{(0)} + n_Q^{(2)} \right) {\rm Li}_{-1}(q^Q)
\end{equation}
We can compare this expression with the $\bar{t}\to 0$ limit of~(\ref{eq:F2an}).
However, in order to find the genus 2 instanton numbers $n_Q^{(2)}$ we need in 
general additional information. We have seen in the previous section that the 
system of equations determining 
$\cF^{(2)}$ is overdetermined, and we have solved it using an ansatz. 
Furthermore, we made an ansatz for the holomorphic ambiguity $f_2(t)$ 
in~(\ref{eq:f_g}). We therefore need additional consistency checks in order to 
fix all the ambiguities. We can obtain them from the following six sources. 
\begin{itemize}
  \item[-] For low degrees $Q$, some of the $n_Q^{(g)}$ can be computed using 
  classical algebraic geometry, in particular Castelnuovo theory, as is 
  explained in~\cite{Katz:1999xq}.
  \item[-] Next, the authors of~\cite{Ghoshal:1995wm} have given an argument for 
  the expected behaviour of the $\cF^{(g)}$ near the conifold locus in $\cM$,
  described by the vanishing of $\Delta_{con}$. It is given by the 
  asymptotic expansion of the $c=1$ string at the selfdual radius
  \begin{equation}
    \label{eq:c=1}
    \cF(\mu) = \frac{1}{2}\mu^2\log\mu-\frac{1}{12}\log\mu+\sum_{g=2}^\infty
    \frac{B_{2g}}{(2g-2)2g}\mu^{2-2g} + \dots
  \end{equation}
  The leading coefficient in $f_2(z)$ in~(\ref{eq:f_g}), 
  $a_2=p^{(2g-2)}_{con}(0)|_{g=2}$ 
  is $a_2=s {1\over 240}$. The factor
  ${1\over 240}=B_{2g}/(2 g-2)(2 g)|_{g=2}$ comes
  from~(\ref{eq:c=1}) while the scaling factor
  $s={\prod_{i_1}^r d_i \over \prod_{i=1}^k w_i}$ 
  of the natural coordinate in 
  terms of the degrees $d_i$ of the constraints of $X$ and the weights $w_i$ 
  of the ambient weighted projective space $\mP^4_w$ was already observed 
  in~\cite{Katz:1999xq}. In the case of general toric ambient spaces, we 
  expect 
  $s$ to be given in terms of the generators of the Mori cone~(\ref{eq:genmori}). 
  \item[-] Furthermore, at other singularities in the moduli space, a 
  Calabi--Yau manifold $X$ can degenerate to another Calabi--Yau manifold $X'$, 
  having less moduli in general. This degeneration is typically described via a
  birational map $f:X \dashrightarrow X'$. The instanton numbers in this case 
  are related by 
  \begin{equation}
    \label{eq:birat}
    n_{Q'}^{(g)}(X') = \sum_{Q:f(Q)=Q'} n_Q^{(g)}(X)    
  \end{equation}
  An example is the curve of $\mZ_2$ singularities discussed in detail 
  in~\cite{Candelas:1993dm}.
  \item[-] Fourth, if $X$ admits an elliptic fibration, we can take the local 
  limit of a large elliptic fiber, and obtain a non-compact \CY manifold $Y$.
  Therefore, we can use the results for local Calabi--Yau manifolds 
  from~\cite{Klemm:1999gm}, \cite{Katz:1999xq}, \cite{Aganagic:2002qg} 
  and~\cite{Aganagic:2002wv} to check that for $Q \in H_2(Y) \subset H_2(X)$ 
  we have
  \begin{equation}
    \label{eq:local}
    n_Q^{(g)}(X) = n_Q^{(g)}(Y)
  \end{equation}
  \item[-] On the other hand, if $X$ admits a K3 fibration $\cF^{\rm hol}$ has 
  been evaluated using the heterotic-type II duality in the limit where the 
  base of the fibration is large by~\cite{Antoniadis:1995zn} 
  and~\cite{Marino:1998pg}. We will extend their argument in several 
  directions, and discuss it in detail in Section~\ref{sec:regularK3}.
  \item[-] Finally, closely related to the last result, as well as to the 
  birational maps above, is the fact that one can in principle use the 
  information contained in the monodromies about the singularities in the 
  complex structure moduli space ${\cal M}$. For example, if $Q$ and $Q'$ 
  are related by a Weyl reflection of the form
  $Q' = Q + \langle Q, D \rangle [C]$ for some divisor $D$ and a rational 
  curve $C$ such that $D\cdot C < 0$ then~\cite{Candelas:1993dm},
  \cite{Klemm:1996kv}, \cite{Katz:1996ht}
  \begin{equation}
    \label{eq:monodr}
    n^{(g)}_Q = n^{(g)}_{Q'}
  \end{equation} 
\end{itemize}
We will apply these additional consistency checks, where applicable, in the 
examples in the next sections.

\section{Higher genus calculations for regular K3 fibrations}
\label{sec:regularK3}

In this section we solve the higher genus amplitudes for K3 fibrations
without reducible fibers, which we call regular. For simplicity we also
assume that there is no monodromy on the algebraic cycles in the fiber, even 
though we expect that the general form of the formulas for ${\cal F}$ below
will not be affected by lifting this restriction.  

\subsection{${\rm K3} \times T^2$}

For $C$ a curve in the class $[C]$ in the K3 with $C^2=2 g-2$ a formula for
the topological free energy was given in~\cite{Katz:1999xq}. It is based on
a specific model of the moduli space of M2 branes, which leads 
to the Hilbert scheme of points on K3  
\begin{equation}
{\cal H}^{\rm hol}(\lambda,t) = \ds{\left(1\over 2 \sin({\lambda\over 2})\right)^2
\prod_{n\ge 1}{1\over (1-e^{i\lambda} q^n)^2 (1-q^n)^{20}(1-e^{-i\lambda}q^n)^2}}\ .
\label{eq:hilbk3t2}
\end{equation}
As we will review below, this formula is basically a reorganization of the 
formula of~\cite{Yau:1995mv} keeping track of the 
${\rm su}(2)_L\times {\rm su}(2)_R$ quantum numbers. 
Closely related situations on $T^4\times S^1$ were discussed 
in~\cite{Maldacena:1999bp} and a very similar logic for the counting of the 
BPS states in the $Q = [B]+ g[F]$ class, in an elliptic del Pezzo surface with
base $B$ and fiber $F$, appeared in~\cite{Hosono:1999qc}. In the latter case 
there is no complex structure deformation and one could in principle
calculate $N^{Q}_{j^3_L,j^3_R}$. A generalization of~(\ref{eq:hilbk3t2}) for 
classes wrapping the $T^2$ base involving the elliptic genus of 
${\rm Hilb}^n({\rm K3})$ was also given in~\cite{Katz:1999xq}. 

As it stands (\ref{eq:hilbk3t2}) is not very useful, because in 
the $N=4$ supersymmetry preserving geometry $T^2\times {\rm K3}$ the index 
calculating ${\cal F}$ vanishes, 
since the factor 
from the torus involving $\left({\bf 1\over 2}\right)_R+2 ({\bf 0})_R$ 
vanishes under the sum $\sum_R (-1)^{2 j^3_R}(2 j_R^3+1)$ over the right 
SU$(2)_R$ representations. 
There are three possibilities to make it applicable to $N=2$ geometries. 
\begin{itemize}
  \item[-] Use the deformation invariance to make predictions for classes in
  the K3 fiber of regular K3 fibrations. This will be done in the next section. 
  \item[-] Consider a local model ${\cal O}(K)\rightarrow S$ where $S$ is an
  elliptic del Pezzo surface as in~\cite{Hosono:1999qc}.
  \item[-] Closely related to the first one, one can mod out the 
  ${\mathbb Z}_2$ which converts ${\rm K3}\times T^2$ to a special K3
  fibration over ${\mathbb P}^1$ with 4 Enriques fibers, a Calabi--Yau
  manifold which has ${\rm SU}(2)\times {\mathbb Z}_2$ holonomy. Such a
  manifold has been constructed in~\cite{FHSV}. A first step for this model
  will be taken in Section~\ref{sec:Enriques}.
\end{itemize}

\subsection{Regular K3 fibrations}

Extending the argument in~\cite{Katz:1999xq} for $T^2\times {\rm K3}$ we 
will argue in this section that for regular K3 fibrations the higher genus 
invariants in the fiber classes are given by
\begin{equation}
  \cF^{\rm hol}_{\rm K3}(\lambda,t) = \ds{{\Theta(q)\over q} 
  \left(1\over 2 \sin({\lambda\over 2})\right)^2
  \prod_{n\ge 1}{1\over (1-e^{i\lambda} q^n)^2 (1-q^n)^{20}
  (1-e^{-i\lambda}q^n)^2}}\ ,
  \label{eq:hilb}
\end{equation}
where the subscript K3 indicates that only maps into fiber classes are 
counted. $\Theta(q)$ is determined from the lattice embedding of 
\begin{equation} 
  i:{\rm Pic}({\rm K3}) \hookrightarrow H^2(X,\mZ)\ .
  \label{eq:mappiclattice}
\end{equation} 
As such it depends on global properties of the fibration and in particular 
not only on ${\rm Pic}({\rm K3})$. (\ref{eq:hilb}) is clearly inspired by 
the results of heterotic-type II 
duality~\cite{Harvey:1995fq},~\cite{Marino:1998pg}, which are reviewed below. 
However we can construct $\Theta(q)$ purely from the geometric data of the 
fibration and in particular in cases where the heterotic dual is not known. 

As explained above the BPS states are counted 
by complex structure deformation invariant indices related to the cohomology 
of the M2 brane moduli space, i.e. the deformation space
of the genus $g$ curves in the 
class $[C]$ of the K3 together with the flat U(1) bundle moduli. 
Global issues of the fibration affect only the $\Theta(q)$ part and can be 
ignored for the sake of determining the BPS moduli space of a genus $g$ 
curve in a generic class $[C]$  in the K3 fiber with $C^2=2g-2$. 
In particular, one can ignore the global embedding into $X$ and vary the 
complex 
structure of the K3 fiber so that it admits an elliptic fibration with base
$B$ and fiber $F$~\cite{Beauville:1999ab}. Further by a diffeomorphism on the 
fiber, $[C]$ can be brought into the class $[B]+g [F]$. In fact, this argument
applies even to fibers which have Picard number one, however, not all classes 
$[B]+g [F]$ can be related to classes in the image of 
$i:{\rm Pic}({\rm K3})\hookrightarrow H^2(X,\mZ)$.  
A degenerate configuration of the curve which has $g$ fiber 
components over $g$ points on the $\mP^1$ section is a sufficient model to determine the 
relevant cohomology of the moduli space. The U(1) bundle modulus over each 
fiber $T^2$ corresponds to a choice of a point in the dual $T^2$ fiber. 
Hence the unresolved moduli space of a genus $g$ curve is given by
the choice of $n=g$ points on the dual K3$=Y$, irrespectively of their ordering, i.e. by 
${\rm Sym}^n(X)=X^n/{\rm Sym}_n$, where ${\rm Sym}_n$ is the symmetric group. 
It is convenient to define the formal sum of vector spaces such as 
$V=H^*(Y)$ by
$S_q V=\bigoplus_{n\ge 0} q^n {\rm Sym}^n(V)$~\cite{dijkgraafreview}.  
The relevant model of the moduli space $\cM_n$ is given by the orbifold 
resolution of ${\rm Sym}^n(Y)$, i.e  $\cM_n={\rm Hilb}^n(Y)$. 
In general, for a surface $X$ with Euler number 
$\chi(Y) = \sum_{i}(-1)^{i} b_{i}(Y)$ 
one can write a generating function for $\chi(\cM_n)$  by using the partition 
function of ${\chi(Y)}$ free fields 
$\chi_{orb}(S_p Y)=\sum_{n} q^n {\rm sdim}H^*(\cM_n)=\prod_k {1\over ( 1 - q^k)}^{\chi(Y)}$. 
One can get more detailed information about the individual cohomology groups. 
In particular~\cite{goettschesoergel} obtain for the Poincar\'e polynomial 
$P(Y,y)=\sum_{0\le k \le d} (-1)^k y^k b_k(Y)$, $d={\rm dim} Y$, of the orbifold resolution
\begin{equation}
  P_{orb}(S_q Y,y)=\prod_{n>0\atop 0\le k \le d}
  \left(1-y^{k-{\rm dim}_\mC(X)} q^n\right)^{-(-1)^k b_k}.
  \label{eq:goettsche}
\end{equation}
The principle to obtain this finer information is to assign to the bosonic and 
fermionic free field oscillators the ${\rm su}(2)$ eigenvalues of the natural 
Lefshetz action on the cohomology of $Y$ as their zero modes. For the K3 one 
has only bosonic operators and the cohomology decomposes into 
$21({\bf 0}) + {\bf 1}$ under this  ${\rm su}(2)$ action. 
 
The eigenvalues of the cohomology of $\cM_g$ under the  
${\rm su}(2)_L\times {\rm su}(2)_R$ Lefshetz action~\cite{Gopakumar:1998jq} 
on the cohomology of $\cM_n$ can similarly be recovered by assigning to 
each of the oscillators of the bosons ($b_{\rm odd}=0$) the representations 
of the unique splitting of the K3 cohomology, i.e. the ones of their zero 
modes,   
$$[\alpha^i_n]=20({\bf 0},{\bf 0})+\left({\bf {1\over 2}},{\bf {1\over 2}}\right)\ .$$
Summing now over the right ${\rm su}(2)_R$ eigenvalues with 
$(-1)^{2j^3_R}(2 j^3_R+1)$ yields formula (\ref{eq:hilb}). 
Note that the knowledge of the genus zero invariants on the Calabi--Yau manifold in 
the fiber direction alone is in general not sufficient to determine $\Theta(q)$. 
We need in addition the information of the embedding of the Picard lattice of 
$Y$ into the Picard lattice of $X$ and global properties of $X$, in particular the 
Euler number. Let us assume for simplicity that a heterotic dual does exist. The 
Narain lattice is then
\begin{equation}
\Gamma_{2,r}=H\oplus \Lambda\ , 
\label{eq:narainlattice}
\end{equation}
with $H$ being the
even unimodular lattice of signature $(1,1)$, $\Lambda$ a rational 
lattice of signature $(1,r)$.
$\Gamma_{2,r}$ and can be identified in the stringy K3 cohomology lattice 
$\Gamma_{4,20}$. In particular, $\Lambda$ gets identified with the Picard lattice of the 
K3 fiber. (\ref{eq:hilb}) counts the index of BPS states. To relate it to 
$n_Q^{(g)}$ of~(\ref{gova}), (\ref{zhol}) one has to change the basis of the 
left sum according to~(\ref{eq:rightsum}). Furthermore, to actually write 
down the free energy of the topological string for the classes in
the image of $i$ one must replace powers of $\lambda$, $q$ by 
\begin{equation}
\lambda^{2g -2}q^l\rightarrow {1\over (2 \pi i)^{3-2 g}}
\sum_{\alpha^2/2=l}{\rm Li}_{3-2g}(e^{2 \pi i (\alpha \cdot t)}) \ ,
\label{eq:latticespacing}
\end{equation}
where $\alpha$ is an element of ${\rm Pic }({\rm K3})\cong \Lambda$ and $t$ 
stands for the associated complex volumes. We can determine $\Theta(q)$ from 
the spacing of the exponents in $\Theta(q)$ determined 
by~(\ref{eq:latticespacing}) and the genus $0$ results.

It is noticeable that this  formula makes sense even for $\alpha^2=0$. 
The $q^0$ coefficient of $\Theta/\eta^{24}$ is always the Euler number of $X$. 
In the heterotic string this is related to the one loop contribution to the 
gravitational coupling. Focusing on genus zero first, and using 
${\rm Li}_3(1)=\zeta(3)$ the replacement gives 
${\zeta(3) \chi(X)\over (2 \pi i)^2}$. This is up to a factor $-1/2$ the 
constant map contribution to ${\cal F}^0$ from the famous 4-loop $\sigma$ 
model calculation. The factor can eventually be explained as follows. 
For the moduli space of the curves in K3 the total moduli space contains an 
additional factor from the $\mP^1$ for which the sum over
$(-1)^{2 J_R^3}(2J_R^3+1)$ 
is $-2$. For this reason the modular forms in~(\ref{eq:thetaSTU}) etc. below
contain a factor of 2. However, this factor is not appropriate for the 
constant maps as their moduli space does not factorize. The same 
regularization 
which lead from (\ref{gova}) to (\ref{zhol}) will reproduce the constant map 
term for the higher genus ${\cal F}^{(g)}$ for $g>1$. It has been noticed 
in~\cite{Harvey:1995fq} that the correct normalization comes out naturally  
from the integral ${\tilde I}_{2,2}$.  

In order to see the effect of topologically different types of fibrations 
with the same K3 fiber we determine the higher genus numbers for the three 
fibrations
\begin{align*} 
  A&=\mP^4_{1,1,2,2,2}[8], & 
  B&=\mP\left(\Msize
        \BM 1 & 1 & 1 &  1 & 2 & 0 & 0 \cr  
            0 & 0 & 0 &  0 & 0 & 1 & 1 \cr \EM\right) 
        \left[\Msize\BM4\cr0\EM\VL\BM2\cr2\EM\right], &
  C&=\mP\left(\Msize
        \BM 1 & 1 & 1 &  1 & 1 & 0 & 0 \cr  
            0 & 0 & 0 &  0 & 0 & 1 & 1 \cr \EM\right) 
        \left[\Msize\BM4\cr 1\EM\VL\BM1\cr1\EM\right]
\end{align*}
$A$ is a hypersurface of degree 8 in the weighted projective space 
$\mP_{1,1,2,2,2}$ and $B$ a complete intersection of degree $(4,0)$ and $(2,2)$ 
in $\mP^4_{1,1,1,1,2}\times \mP^1$ with Hodge numbers $(h^{11},h^{21}) = (2,58)$.
Both fibres $\mP^3_{1,1,1,1}[4]$  and $\mP^4_{1,1,1,1,2}[2,4]$ have a one 
dimensional Picard lattice with self intersection $E^2=4$ and the same modular
properties of the mirror map, but $\Theta(q)$ will be different. 
The fibration $C$, with $(h^{11},h^{21}) = (2,86)$, has the same generic
fibers and the same type of degenerate fibers as the first one. 
Nevertheless, the fibration turns out to be topologically different.
We next discuss these fibrations as well as a closely related fibration that 
was treated in~\cite{Ruan:1996ab} in detail.

\subsection{The $\mP^4_{1,1,2,2,2}[8]$ model}
\label{sec:11222}

This is a K3 fibration whose heterotic dual has not yet been 
precisely identified. 
Its topological data are $\chi=-168$, 
$h^{1,1}=2$, $h^{2,1}=86$, ${\cal R}= 8 t_1^3+ 4 t_1^2 t_2$,
$\int_X \ch_2 J_1=56$ and $\int_X \ch_2 J_2=24$.
The K\"ahler class $t_1=T$ is identified with the unique class 
of the K3 fiber and $t_2=4\pi i S$ is the size of the base, which,
in leading order,
would be proportional to the heterotic dilaton $S$. The generic manifold 
$X$ is defined as the vanishing locus of $p=x_1^8+x_2^8+x_3^4+x_4^4+x_5^4+
\sum_{\sum w_i d_i=8} a_{\underline d} {\underline x}^{\underline d}$ in  
$\mP^4_{1,1,2,2,2}$, where $a_{\underline d}$ are the complex structure 
moduli, 
$d_1,d_2\le 6$, $d_3,d_4,d_5\le 2$ and $w_i$ are the weights of the ambient 
space. $X^*$ can be described by the same equation in 
$\mP^4_{1,1,2,2,2}/{\mathbb Z}_4^3$ where the only invariant perturbations  
are $a x_1x_2x_3x_4x_5$ and $b x_1^4 x_2^4$. 
Clearly, the Euler number of $X$ is determined by the singular fibres. 
$X$ admits a contraction along a K3 to a $\mP^1$, which is defined by the 
first 
two coordinates. The dependence of a smooth K3 fiber on the $\mP^1$ 
coordinate $\mu$ is given by
$$(1+\mu^8)z^4+x_3^4 + x_4^4+x_5^4=0$$
which is a quartic in $\mP^3$. This quartic degenerates for $\mu^8=-1$ 
to a complex cone over the genus 3 curve $\mP^2[4]$ with Euler number $-4+1$.
The Euler number of $X$ hence is $\chi(X)=2(24) -8(24) + 8(-3)=-168$.

The complex structure moduli of the mirror $X^*$ with $\log(x=b/a^4) \sim t_1$ and 
$\log(y=1/b^2)\sim t_2$ at the large complex structure limit are suitably rescaled 
so that the Yukawa couplings read~\cite{Hosono:1993qy}
\begin{align}
  \label{eq:Yukawa11222}
  Y_{111} &= {4 \over x^3 \Delta_{con}}, &  
  Y_{112} &= {2 (1 - x)\over x^2 y \Delta_{con}}\notag\\
  Y_{122} &= { 1 - 2 x\over x y \Delta_{con} \Delta_{s}}, &
  Y_{222} &=  {1 - y + x (1 + 3y) \over 2y^2 \Delta_{con} \Delta_{s}}
\end{align}
where the discriminants are $\Delta_{con}= 1 - 2 x + x^2 (1- y)$ and 
$\Delta_{s}=1-y$.  In contrast to elliptic toric fibrations 
$Y_{1jk}$ and $Y_{2jk}$ are invertible at a generic point in the moduli 
space away from $\Delta_{con}=0, \Delta_{s}=0, x=0$ and $y=0$.   

The first task is to fulfil the algebraic compatibility 
conditions~(\ref{eq:comp1}) and~(\ref{eq:comp2}) for the ambiguity 
$A^i_{kl}$ inherent in the two ways of constructing the topological 
propagators $S^{ij}$. With 
\begin{align}
  \label{eq:A3_11222}
  A^1_{1,1}&= -{1\over x}, &A^1_{1,2}&= -{1\over 4 y}, & A^1_{2,2}&=0
  \notag\\
  A^2_{1,1}&=0 &A^2_{1,2}&={1\over 2 x}, & A^2_{2,2}&=-{1\over y}
\end{align}
we find the leading behaviour  
\begin{equation}
\begin{array}{rcl} 
S^{11}&=&-{1\over 16}-44 q_1 q_2 +O(q^3),\\ 
S^{12}&=&{1\over 8}-22 q_1+214 q_1^2 +22 q_1 q_2+O(q^3)\\
S^{22}&=& -{1 \over 4}-1272 q_1^2+O(q^3)\ . 
\end{array}
\end{equation} 
We checked (\ref{eq:f1b}) and found that 
$$\partial_{t_k} \cF^{(1)} ={1\over 2} C_{k,j,i} S^{ij} + \partial_{t_k}\left({1-\chi \over 24} 
\log(\omega)-{1\over 12} \log(\Delta_{con})- {5\over 12} \log(\Delta_{s})- {31\over 12}\log(x)- {7\over 8}\log(y)\right)$$
in accordance with the expected behaviour of $\cF^{(g)}$ near the conifold 
coming from the 1-loop $\beta$-function of one 
hypermultiplet~\cite{Strominger:1995cz},~\cite{Vafa:1995ta}. 
The $-{5\over 12}\log(\Delta_{s})$ is the same behaviour 
as in the conventional calculation of $\cF^{(1)}$~\cite{Candelas:1993dm}
\footnote{The difference by a factor of 2 w.r.t.~\cite{Candelas:1993dm} originates 
in the different normalization of $\cF^{(1)}$ between~\cite{Bershadsky:1993ta} 
and~\cite{Bershadsky:1993cx}.}. The total leading coefficient is 
$\cF^{(1)} =- {1\over 12} \log(\Delta_{s})+ O(\Delta_{s})$ 
coming from the 1-loop $\beta$-function of $3$ massless hypermultiplets and 
a non-contributing  massless $N=4$-like spectrum of an SU(2) gauge 
symmetry enhancement~\cite{Klemm:1996kv}. This implies that the 
$S^{ij}$ terms contribute ${1\over 4} \log(\Delta_{s})$.
The $S^{i}$ can also be derived in two ways.
We find that with the choice (\ref{eq:A3_11222})
\begin{align}
  \label{eq:A2_11222}
  A_{1,1}&=0, & A_{2,1}&=0, & A_{2,2}&=0
\end{align}
is a possible solution to the constraints (\ref{eq:comp3}). The leading 
behaviour of $S^{i},S$ is then
\begin{equation}
\begin{array}{rcl} 
S^{1}&=& \frac{3}{2} q_1 - 33 q_1^2 + \frac{9}{2} q_1 q_2 +  O(q^3),\\ [2 mm]
S^{2}&=& 3 q_1 + 186 q_1^2 - 3 q_1 q_2 +  O(q^3)\\ [2 mm]
S    &=& -18 q_1^2+ 792 q_1^3 -180 q_1 q_2 + O(q^4)\ . 
\end{array}
\end{equation} 
We found the following ansatz for the holomorphic ambiguity $f_2$
\begin{equation}
\label{eq:ansatz11222} 
f_2(x,y)=\left({107\over 9216}+ {193\over 1280} y - 2 x(1 - 4 y)\right){1\over \Delta_{s}}+
      \frac{(831 - 199936x)(1-2^8x)}{46080\Delta_{con}}+\frac{(1-2^8x)^3}{240\Delta_{con}^2} 
\end{equation}  
in~(\ref{eq:f_g}), which satisfies the integrability constraints as well as the 
vanishing of some low degree invariants as mentioned in Section~\ref{sec:Ansatz}.
With this ansatz we get integral genus 2 invariants. For comparison we list
the genus 0 and 1 invariants first.

\noindent $g=0$
\vskip 5 mm {\vbox{\footnotesize{
$$
\vbox{\offinterlineskip\tabskip=0pt \halign{\strut \vrule#& &\hfil
~$#$ &\hfil ~$#$ &\hfil ~$#$ &\hfil ~$#$ &\hfil ~$#$ &\hfil ~$#$
&\hfil ~$#$ &\hfil ~$#$ 
&\vrule#\cr \noalign{\hrule} 
&d_1 &d_2=0 &1           &2         &3          &4       &5       &6      & \cr
\noalign{\hrule}
&0&& 4& & & & & &\cr
&1&640& 640& & & & & &\cr
&2& 10032& 72224& 10032& 0& 0&0& 0&\cr
&3& 288384& 7539200& 7539200& 288384& 0& 0& 0& \cr  
&4&10979984& 757561520&2346819520& 757561520& 10979984& 0& 0 &\cr 
&5&495269504& 74132328704& 520834042880& 520834042880& 74132328704& 495269504& 0& \cr
\noalign{\hrule}}\hrule}$$}}}

\noindent $g=1$
\vskip 5 mm {\vbox{\footnotesize{
$$
\vbox{\offinterlineskip\tabskip=0pt \halign{\strut \vrule#& &\hfil
~$#$ &\hfil ~$#$ &\hfil ~$#$ &\hfil ~$#$ &\hfil ~$#$ &\hfil ~$#$
&\hfil ~$#$ &\hfil ~$#$ 
&\vrule#\cr \noalign{\hrule} 
&d_1 &d_2=0 &1           &2         &3          &4       &5       &6      & \cr
\noalign{\hrule}
&2& &  &   &   &     &        &       &\cr 
&3&-1280 & 2560  & 2560  &-1280   & & &  &\cr 
&4&-317864   &1047280  &  15948240  & 1047280  & -317864  &   &     &\cr 
&5&-36571904      &   224877056&  12229001216   & 12229001216&224877056&-36571904&& \cr
&6&-3478899872&36389051520&4954131766464&13714937870784&4954131766464&36389051520
&-3478899872&\cr
\noalign{\hrule}}\hrule}$$}}}

\noindent $g=2$
\vskip 5 mm {\vbox{\footnotesize{
$$
\vbox{\offinterlineskip\tabskip=0pt \halign{\strut \vrule#& &\hfil
~$#$ &\hfil ~$#$ &\hfil ~$#$ &\hfil ~$#$ &\hfil ~$#$ &\hfil ~$#$
&\hfil ~$#$ &\hfil ~$#$ 
&\vrule#\cr \noalign{\hrule} 
&d_1 &d_2=0 &1           &2         &3          &4       &5       &6      & \cr
\noalign{\hrule}
&2& &  &   &   &     &        &       &\cr 
&3& &  &   &   & & &  &\cr 
&4&472 &  -1232&  848  & -1232  & 472  &   &     &\cr 
&5&875392      &         -2540032 &  9699584   &   9699584          &     -2540032    & 875392     &           & \cr
&6&220466160   &  -1005368448     &21816516384 &  132874256992      & 21816516384     &-1005368448 & 220466160  &\cr
\noalign{\hrule}}\hrule}$$}}}

It should be stressed that~\eqref{eq:ansatz11222} is only the simplest ansatz in 
the sense that it matches all known properties of the $n^{(g)}_{i,j}$, as 
discussed in Section~\ref{sec:Ansatz}. In particular, we find that in the strong 
coupling limit~\cite{Klemm:1996kv} this model has a non-conifold 
transition at $\Delta_s=0$ to the complete intersection $X'=\mP^5[2,4]$ 
with the topological data $\chi=-176$, $h^{1,1}=1$. This transition 
occurs for $x_2=q_2=1$ 
and~(\ref{eq:birat}) yields 
$$n^{(g)}_d=\sum_{i=0}^{2d+1} n^{(g)}_{d,i}$$
for the invariants with one modulus, 
which is consistent with the $n_{d}^{(2)}(X')$ in the following table
\vskip 5 mm
{\vbox{\footnotesize{
$$
\vbox{\offinterlineskip\tabskip=0pt \halign{\strut \vrule#& &\hfil
~$#$~&\hfil ~$#$ &\hfil ~$#$ &\hfil ~$#$ &\hfil ~$#$ &\hfil ~$#$~&~ \vrule#\cr \noalign{\hrule} 
& d=1 & 2        &3                 &4                      &5                     &6                          &  \cr
\noalign{\hrule}
&   0& 0    &  0        &         -672    &     16069888    & 174937485184       & \cr 
\noalign{\hrule}}\hrule}$$}}} 

As we mentioned we can reconstruct $\Theta(q)$ from the 
lattice embedding and $g=0$ results. We use the fact that $E^2=4$ and since 
the one dimensional Picard lattice is generated by $\delta$ with 
$\delta^2={1\over 4}$, the embedding of the Picard lattice of the K3 into 
$H^2(X,\mZ)$ in~(\ref{eq:latticespacing}) is
\begin{equation}
\lambda^{2g-2}q^l\rightarrow {1\over (2 \pi i)^{3-2 g}}\sum_{n^2/8=l}{\rm Li}_{3-2g}
(e^{2 \pi i\ n T}) \ 
\label{eq:latticespacingST4} ,
\end{equation}  
which yields
\begin{equation}
{\Theta\over \eta^{24}}=-\frac{2}{q} + 168 + 640\,q^{\frac{1}{8}} + 
  10032\,{\sqrt{q}} + 158436\,q + 
  288384\,q^{\frac{9}{8}} + 1521216 \,q^{\frac{3}{2}} + 10979984\,q^2 + \ldots
\label{eq:etax8}
\end{equation}
Now we can realize $\Theta$ as a modular form of $\Gamma^0(8)$ 
with weight $\frac{21}{2}$ in terms of
\begin{align}
  \label{eq:UV}
  U&=\theta_3(\tfrac{\tau}{4})\ , & V&=\theta_4(\tfrac{\tau}{4})
\end{align}
with $\theta_3(\tau)=\sum_{n\in\mZ}q^{\frac{n^2}{2}}$ 
and $\theta_4(\tau)=\sum_{n\in\mZ}(-1)^nq^{\frac{n^2}{2}}$,
\begin{eqnarray}
  \label{eq:ThetaA}
  \Theta &=& 2^{-21}\left(3U^{21} - 81U^{19}V^2 - 627U^{18}V^3 - 14436U^{17}V^4 - 
       20007U^{16}V^5 - 169092U^{15}V^6 \right.\nonumber\\  
    && -120636U^{14}V^7 - 621558U^{13}V^8 - 292796U^{12}V^9 - 
      1038366U^{11}V^{10} - 346122U^{10}V^{11} \nonumber\\ 
    && -878388U^9V^{12} -207186U^8V^{13} - 361908U^7V^{14} 
       - 56364U^6V^{15} - 60021U^5V^{16}\nonumber\\
    && \left.- 4812U^4V^{17} - 1881U^3V^{18} - 27U^2V^{19} + V^{21}\right) .
\end{eqnarray}
It is quite remarkable that we reconstruct the threshold correction of a not 
yet identified heterotic string from the higher genus calculation on the 
K3 fibration.

In fact, compatibility with~(\ref{eq:hilb}) gives us consistency checks on the genus 
2 invariants in the first column and allows to predict all higher genus numbers 
$n^{(g)}_{d,0}$ such as
\vskip 5 mm {\vbox{\small{
$$
\vbox{\offinterlineskip\tabskip=0pt \halign{\strut \vrule#& 
&\hfil~$#$ 
&\hfil~$#$ 
&\hfil~$#$ 
&\hfil~$#$ 
&\hfil~$#$
&\hfil ~$#$
&\hfil ~$#$
&\hfil ~$#$
&\hfil ~$#$
&\vrule#\cr \noalign{\hrule} 
&g&d=1&2         &3          &4       &5       &6       &7 & 8& \cr
\noalign{\hrule}
&0& 640&10032&288384&10979984&495269504&24945542832&1357991852672&78313183960464&\cr 
&1&0&0&-1280&-317864&-36571904&-3478901152&-306675842560&-26077193068200& \cr 
&2&0&0&0&472&875392&220466160&36004989440&4824769895208&\cr 
&3&0&0&0&8&-2560&-6385824&-2538455296&-599694313488&\cr 
&4&0&0&0&0&0&50160&101090432&51094399168&\cr 
&5&0&0&0&0&0&0&-1775104&-2848329800&\cr 
&6& 0&0&0&0&0&0&4480&92179128&\cr 
&7& 0&0&0&0&0&0&0&-1286576&\cr
&8& 0&0&0&0&0&0&0&1192&\cr 
&9& 0&0&0&0&0&0&0&20&\cr  
\noalign{\hrule}}\hrule}$$}}}

Some predictions of this table can be checked with the
methods\footnote{We would like to thank Sheldon Katz for many explanations 
regarding the geometry of the instanton moduli spaces.} described 
in~\cite{Katz:1999xq} as follows. We denote the divisors dual to $t_1,t_2$
by $H={\cal O}(2)$ and $L={\cal O}(1)$, i.e. $H^3=8$ and $H^2 L=4$ are the
non vanishing intersection numbers. The authors of~\cite{Katz:1999xq} use a
Jacobian fibration over the universal curve as model for the moduli 
space of M2 branes and construct the mixed sums of dimensions 
of the cohomology groups of this moduli space $n^{(g)}_Q$ with the
Abel-Jacobi map. It follows from~\cite{Katz:1999xq} that 
the $n_Q^{(g-\delta)}$ 
of a curve $C_g$ of genus $g$ with $\delta$ nodes can be computed 
by $n_Q^{(g-\delta)}=(-1)^{{\rm dim}({\cal M}_\delta)}e({\cal M}_\delta)$. 
In particular, if $C_g$ is a smooth curve of genus $g$ in a class in 
$H^2(X,\mZ)$, i.e. $\delta=0$, one obtains  
\begin{equation}
  n_Q^{(g)}=(-1)^{{\rm dim}({\cal M}_0)}e({\cal M}_0) ,
  \label{eq:ngsmoothcurve}
\end{equation} 
where ${\cal M}_0$ is the moduli space of the smooth curve $C_g$. 

We therefore first look at smooth curves of higher genus in the K3 fiber. 
If such a curve is a complete intersection $C=(k H) L$, its canonical class is
$K_C=kH +L$ and its genus is given by adjunction as
\begin{equation} 
{K_C C\over 2}+1=g\ .
\label{eq:adjunction}
\end{equation}
Such a curve hence has genus $g=2k^2+1$, degree $0$ w.r.t. 
$L$ and degree $4k$ w.r.t. $H$. Its moduli space is given by a fibration of 
${\mathbb P} H^0({\cal O}(2k H))$ over ${\mathbb P}^1$. The dimensions 
$p(n)=\dim {\mathbb P} H^0({\cal O}(n H))$ can be described for a general 
hypersurface in a weighted projective space by
\begin{equation}
  \label{eq:dimH0}
  \sum_{n=0}^\infty p(n) q^n=(1-q)\prod_{i=1}^5 \frac{1}{1-q^{w_i}},
\end{equation}
which for $H\subset X=\mP^4_{1,1,2,2,2}[8]$ becomes 
\begin{equation}
\sum_{n=0}^\infty p(n) q^n={1\over (1-q) (1-q^2)^3}
\label{eq:dimH011222}
\end{equation}
and  we conclude that 
\begin{equation}
n^{(2k^2+1)}_{4k ,0}=(-1)^{p(2k)} 2 p(2k)
\end{equation} 
in perfect agreement with the table above. By a similar method one
can count the $n_Q^{(g)}$ of smooth curves with degree 1 w.r.t.
$L$ and gets $n^{(k^2+1)}_{2k,1}=(-1)^{p(2k)-1} 4 (p(2k)-1)$.

Curves with one node in the class $(4k,0)$ have genus $g=2 k^2$. The formula
for the Euler number of moduli space with one node is~\cite{Katz:1999xq} 
\begin{equation}
e({\cal M}_1)=e({\cal C}) + (2g-2) e({\cal M}_0),
\label{eq:eonenode}
\end{equation}
where $\cal C$ is the universal curve $\pi:{\cal C}\rightarrow {\cal M}_0$,
which in our case has $e({\cal C})=-168 (p(2k)-1) g$. Combining this
with~(\ref{eq:eonenode}) 
and using the fact that we already calculated $e({\cal M}_0)$ we obtain
\begin{equation}
n^{(2k^2)}_{4k ,0}= (-1)^{p(2k)-1} (8 k^2 p (2 k) +168 (1-p(2k)) \ ,  
\end{equation} 
again in perfect agreement with the table.

\subsection{The fibration B}
\label{sec:fibrationB}

Comparing the fibrations $A$ and $B$ which have the same Picard lattice is very 
instructive. The model $B$ has $h^{11}=2$ and $h^{21}=58$. Its intersection data are 
${\cal R}=4J_1^3+ 4 J_1^2 J_2$ and $\int_{X} \ch_2 J_1=40$, $\int_{X} \ch_2 J_2=24$.

Again we can realize $\Theta$ as a modular form of $\Gamma^0(8)$ 
with weight $\frac{21}{2}$ in terms of $U=\theta_3({\tau\over 4})$ 
and $V=\theta_4({\tau\over 4})$, which has a quite similar, but different 
form than in the previous case
\begin{eqnarray}
  \label{eq:ThetaB}
\Theta &=& 2^{-21}\left(3U^{21} - U^{20}V - 60U^{19}V^2 - 580U^{18}V^3 - 
    14223U^{17}V^4 - 20027U^{16}V^5 - 169536U^{15}V^6 \right.\nonumber\\  
    && -120800U^{14}V^7 - 621858U^{13}V^8 - 292746U^{12}V^9 - 1037592U^{11}V^{10} - 
    345864U^{10}V^{11} \nonumber\\ 
    && -878238U^9V^{12} - 207286U^8V^{13} - 362400U^7V^{14} - 56512U^6V^{15} - 
    60081U^5V^{16}\nonumber\\
    && \left.- 4741U^4V^{17} -1740U^3V^{18} - 20U^2V^{19} - 3UV^{20} + V^{21} \right) ,
\end{eqnarray}
which yields 
\begin{equation}
{\Theta\over \eta^{24}}=\frac{-2}{q} + \frac{4}{{\sqrt{q}}} + 112 + 768\,q^{\frac{1}{8}} + 9376\,{\sqrt{q}} + 154820\,q + 293632\,q^{\frac{9}{8}} + 
  1505904\,q^{\frac{3}{2}} + {O(q^2)}
\end{equation}
 
From this we get the following higher genus numbers $n^{(g)}_{d,0}$ in the
fiber direction
\vskip 5 mm {\vbox{\small{
$$
\vbox{\offinterlineskip\tabskip=0pt \halign{\strut \vrule#& 
&\hfil~$#$ 
&\hfil~$#$ 
&\hfil~$#$ 
&\hfil~$#$ 
&\hfil~$#$
&\hfil ~$#$
&\hfil ~$#$
&\hfil ~$#$
&\hfil ~$#$
&\vrule#\cr \noalign{\hrule} 
&g&d=1&2         &3          &4       &5       &6       &7 & 8& \cr
\noalign{\hrule}
&0& 768&9376&293632&10924128&495966976&24935882144&1358136021760&78310908911200&\cr 
&1& 0&-8&-1536&-310296&-36755968&-3474708168&-306768929792&-26075148138264 &\cr 
&2& 0&0&0&304&893184&219647904&36034776064&4823807971056 &\cr 
&3&0&0&0&8&-3072&-6305376&-2544193024&-599398380992&\cr 
&4&0&0&0&0&0&47072&101752576&51032102080 &\cr 
&5&0&0&0&0&0&-24&-1815552&-2839525176 &\cr 
&6&0&0&0&0&0&0&5376&91402192 &\cr 
&7&0&0&0&0&0&0&0&-1250592 &\cr
&8&0&0&0&0&0&0&0&688 &\cr 
&9&0&0&0&0&0&0&0&20&\cr  
\noalign{\hrule}}\hrule}$$}}}

As the Picard lattice of the fiber is identical to the model discussed
in the last section, i.e. $H^2 L=4$, we also get 
\begin{equation}
n^{(2 k^2+1)}_{4k ,0}=(-1)^{{p}(2k)} 2 {p}(2k)\ ,
\end{equation} 
but since  $e({\cal C})=-112  $ we now get 
\begin{equation}
n^{(2k^2)}_{4k ,0}= (-1)^{p(2k)-1} (8 k^2 p (2 k) +112 (1-p(2k))  
\ .  
\end{equation}

\subsection{The fibration C}
\label{sec:rational_homotopy}

The fibration $C$ with intersection ${\cal R}=5 J_1^3+4 J_1^2 J_2$ 
and $\int_{X} \ch_2 J_1=50$,$\int_{X} \ch_2 J_2=24$ is well known, as it
has a transition to the quintic hypersurface in ${\mathbb P}^4$, see 
\cite{Candelas:1989js}.  This is reflected by the fact that the GV 
invariants in the rows of the tables below sum up to the GV invariants 
of the quintic.

\noindent $g=0$
\vskip 5 mm {\vbox{\small{
$$
\vbox{\offinterlineskip\tabskip=0pt \halign{\strut \vrule#& 
&\hfil ~$#$ &\hfil ~$#$ &\hfil ~$#$ &\hfil ~$#$ &\hfil ~$#$ &\hfil ~$#$ &\hfil ~$#$ &\hfil ~$#$ 
&\hfil ~$#$ &\hfil ~$#$ &\hfil ~$#$ &\vrule#\cr \noalign{\hrule} 
&d_1 &d_2=0 &1           &2         &3          &4       &5       &6      & 7 & 8 & 9 & \cr
\noalign{\hrule}
&0& &16&&&&&&&&& \cr 
&1& 640&2144&120&-32&3&&&&&&\cr 
&2& 10032&231888&356368&14608&-4920&1680&-480&80&-6&&\cr 
&3& 288384&23953120&144785584&144051072&5273880&-1505472&512136&-209856&75300&
-21600& \cr
\noalign{\hrule}}\hrule}$$}}}

\noindent $g=1$
\vskip 5 mm {\vbox{\small{
$$
\vbox{\offinterlineskip\tabskip=0pt \halign{\strut \vrule#& 
&\hfil ~$#$ &\hfil ~$#$ &\hfil ~$#$ &\hfil ~$#$ &\hfil ~$#$ &\hfil ~$#$ &\hfil ~$#$ &\hfil ~$#$ 
&\hfil ~$#$ &\hfil ~$#$ &\hfil ~$#$ &\hfil ~$#$ &\hfil ~$#$ &\hfil ~$#$ 
&\vrule#\cr \noalign{\hrule} 
&d_1 &d_2=0 &1           &2         &3          &4       &5       &6      & 7 & 8 & 9 & 10 & 11 & 12 & \cr
\noalign{\hrule}  
&0& &&&&&&&&&&&&&\cr 
&1& -1280&10240&356368&243328&-15708&34320&-32032&21840&-10920&3920&-960&144&-10& \cr
\noalign{\hrule}}\hrule}$$}}} 

The model admits a projection to $\mP^2$, which is visible in the 
instanton numbers as the $n^{(g)}_{k+2, 4 k}$ numbers agree with
the local ${\cal O}(-3)\rightarrow {\mathbb P}^2$ numbers. 

What is physically intriguing is that the BPS numbers $n^{(g)}_{k,0}$ 
and hence the $\Theta(q)$ coincide with the one of the fibration $A$, 
while the non-perturbative ones, i.e. $n^{(g)}_{k,m}$ for $m>0$ 
do not coincide.  This is geometrically  
expected as the generic fiber is the same and the degenerations of 
the fibres are the same in both cases. However, it shows 
that the degeneracy of BPS states as calculated by the gauge and 
gravitational threshold corrections, i.e. the indices of the perturbative 
$N=2$ heterotic string associated to the vector multiplets, do not 
suffice to fix a unique non-perturbative completion. 
It should be stressed that globally the non-perturbative
completions are very different. This might be surprising as the heterotic 
dilaton does not couple to the hypermultiplets, i.e. potential differences 
(which we have not controlled) in this sector between the two models 
should not affect the completion. 
However, the moduli space of the fibration $A$ has been analyzed in 
sufficient detail~\cite{Candelas:1993dm} to exclude e.g. a transition to the 
quintic in ${\mathbb P}^4$, which the fibration $C$ admits. On the other hand, some 
local features of the perturbative region, in particular, the occurrence of 
the Seiberg-Witten theory~\cite{Kachru:1995fv} at the blow up of the point 
$T=i,S\rightarrow \infty$ does not differ for the different fibrations. 
The same  conclusion will  be reached in section~\ref{sec:STandP11226} 
for a model where a heterotic dual is known. 

Mathematically it is noticeable that the manifolds 
$A$ and $C$ are rational homotopy  equivalent by the criterion of C.T.C. 
Wall~\cite{Wall:1966ab}.
Calling the K\"ahler classes of $C$ $J_1$ and $J_2$ 
as above and the ones of $A$ $\tilde J_1$ and $\tilde J_2$ the transformation
$\tilde J_1=J_1+{1\over 4} J_2$, $\tilde J_2=J_2$ identifies the topological 
data, i.e. these models  are rational homotopy equivalent. The example shows
that Gromov--Witten invariants can be different on rational homotopy 
equivalent manifolds.

\subsection{Diffeomorphic CY with $\mP^3[4]$ fiber}
\label{sec:diffeomorphic}

In this section we discuss two further K3 fibrations with fiber $\mP^3[4]$.
They are defined as the anticanonical divisors in 
$\mP({\cal O}^4)\rightarrow \mP^1$ and 
$\mP({\cal O}(-1)\oplus {\cal O}^2\oplus {\cal O}(1))\rightarrow \mP^1$ and
were discussed in detail in~\cite{Gross:1997ab} and~\cite{Ruan:1996ab} 
for the distinction of families of symplectic structures on diffeomorphic
manifolds. Thomas~\cite{Thomas:1998uj} used them as examples for the 
application of holomorphic Casson invariants, now known as 
Donaldson--Thomas invariants. To investigate the GV invariants and the 
mirror picture of these models we note that the polyhedra are realized 
as follows
\begin{equation*}
  \footnotesize
  \begin{array}{ccrrrrr|rrcl}
    \multicolumn{7}{c}{ }    &l^{(1)}&l^{(2)}&&\\
    D_0    &&     1&     0&   0&   0&   0&     -4&     -2&&\\
    D_1    &&     1&     1&   0&   0&   0&      1&      0&&\\
    D_2    &&     1&     0&   1&   0&   0&      1&      0&&\\
    D_3    &&     1&     0&   0&   1&   0&      1&      0&&\\
    D_4    &&     1&    -1&  -1&  -1&   0&      1&      0&&\\
    D_5    &&     1&     0&   0&   0&   1&      0&      1&&\\
    D_6    &&     1&     0&   0&   0&  -1&      0&      1&&
  \end{array} \qquad\quad
  \begin{array}{ccrrrrr|rrrl}
    \multicolumn{7}{c}{ }    &{l}^{(1)}&{\tilde l}^{(1)}&l^{(2)}&\\
    D_0    &&     1&     0&   0&   0&   0&     -4& -2&   -2&\\
    D_1    &&     1&     1&   0&   0&   0&      1&  1&    0&\\
    D_2    &&     1&     0&   1&   0&   0&      1&  0&    1&\\
    D_3    &&     1&     0&   0&   1&   0&      1&  2&   -1&\\
    D_4    &&     1&    -1&  -1&  -1&   0&      1&  1&    0&\\
    D_5    &&     1&     0&  -1&   0&   1&      0& -1&    1&\\
    D_6    &&     1&     0&   0&   1&  -1&      0& -1&    1&
  \end{array}   
\end{equation*}
We indicated here, in addition to the points of the polyhedra in the first
four columns, the generators $l^{(i)}$ of the Mori cone. 
Both polyhedra are 
reflexive and define mirror pairs according to Batyrevs 
construction. Both manifolds admit K3 fibrations which have 
Euler number $\chi=-168$ and $h^{1,1}=2$, $h^{2,1}=86$. The 
first case has intersection ring ${\cal R}=2 t_1^3+ 4 t_1^2 t_2$ and 
$\int \ch_2 J_1=44$, as well as $\int \ch_2 J_2=24$. We can 
readily calculate the genus zero Gopakumar-Vafa invariants 
$n^{(0)}_{i,j}$ as discussed in the previous sections.
\vskip 5 mm {\vbox{\small{
$$
\vbox{\offinterlineskip\tabskip=0pt \halign{\strut \vrule#& 
&\hfil ~$#$ &\hfil ~$#$ &\hfil ~$#$ &\hfil ~$#$ &\hfil ~$#$&\hfil ~$#$   
&\vrule#\cr \noalign{\hrule} 
&d_1 &d_2=0 &1           &2         &3          &4& \cr
\noalign{\hrule}
&0&   &64&&&&\cr 
&1& 640&6912&14400&6912&640&\cr 
&2& 10032&722784&8271360&3134400&48098560& \cr
\noalign{\hrule}}\hrule}$$}}} 

The second case is slightly more involved to analyze. In 
the basis of the K\"ahler cone dual to ${\tilde l}^{(1)}$ 
and ${l}^{(2)}$ the intersections are equal to the ones of 
the first case. Therefore the two manifolds are 
diffeomorphic. In this phase of the generalized K\"ahler cone 
a curve in the ambient space is contracted and it is not immediate 
to calculate the GV invariants by mirror symmetry using the formulas 
in the previous section in this case. In order to do so we 
use the basis of a larger Mori cone 
${\tilde l}^{(1)}=l^{(1)}-l^{(2)}$ and  ${\tilde l}^{(2)}=l^{(2)}$. 
This results in a smaller K\"ahler cone in which the edge dual 
to ${\tilde l}^{(1)}$ is bounded by the curve mentioned above, 
which now descends to the  hypersurface. 
In this basis the topological data  are ${\cal R}=3 
{\tilde t}_1^3+ 6 {\tilde t}_1^2 {\tilde t}_2 +10  {\tilde t}_1 
{\tilde t}_2^2 +14 {\tilde t}_2^3 $ and 
$\int \ch_2 {\tilde J}_1=42$, as well as $\int \ch_2 {\tilde J}_2=68$.
It is noteworthy that this phase of the generalized K\"ahler 
cone is topologically inequivalent to the previous one and in 
particular it does not admit a K3 fibration. Now it is easy to 
calculate the genus zero GV invariants ${\tilde n}^{(0)}_{ij}$ in 
the non K3 fibration phase
\vskip 5 mm {\vbox{\small{
$$
\vbox{\offinterlineskip\tabskip=0pt \halign{\strut \vrule#& 
&\hfil ~$#$ &\hfil ~$#$ &\hfil ~$#$ &\hfil ~$#$ &\hfil ~$#$&\hfil ~$#$   
&\vrule#\cr \noalign{\hrule} 
&d_1 &d_2=0 &1           &2         &3          &4& \cr
\noalign{\hrule}
&0&   &20&-2&&&\cr 
&1& 1&640&6474&1152&-1023&\cr 
&2&  &   &10032&733560&4458084& \cr
\noalign{\hrule}}\hrule}$$}}} 
\noindent
and use  $n^{(0)}_{i,j-i}={\tilde n}^{(0)}_{i,j}$ to relate 
them to the K3 fibration phase. The result is that the analytic
continuation of ${\rm Li}_3$
\begin{equation} 
  \label{eq:Li3}
  {\rm Li}_{3}(e^x)={\rm Li}_3 (e^{-x})-{x^3\over 6}-
  {\pi i\over 2} x^2+{\pi^2 x\over 3}
\end{equation}
applied to the GV invariant $\tilde n^{(0)}_{1,0}$ contributes 
${\rm Li}_3(q_2/q_1)$ and a term $-\frac{(2\pi i)^3}{12} \big(2(t_1-t_2)^3+3(t_1-t_2)^2
+t_1-t_2\big)$ that connects precisely 
the classical intersection data in ${\cal F}^{(0)}$ in 
the two phases. Similarly, the contribution from the analytic
continuation of ${\rm Li}_1$
\begin{equation} 
  \label{eq:Li1}
  {\rm Li}_{1}(e^x)={\rm Li}_1 (e^{-x})+x
\end{equation}
connects the $\int_X \ch_2 J_i$ terms in ${\cal F}^{(1)}$. The
term  ${\rm Li}_3(q_2/q_1)$ vanishes in the polarization  
$B+N F$ with $N \gg 0$ chosen
in~\cite{Thomas:1998uj}. The genus zero GV 
invariants in the K3 fibration phase are
\vskip 5 mm {\vbox{\small{
$$
\vbox{\offinterlineskip\tabskip=0pt \halign{\strut \vrule#& 
&\hfil ~$#$ &\hfil ~$#$ &\hfil ~$#$ &\hfil ~$#$ &\hfil ~$#$&\hfil ~$#$   
&\vrule#\cr \noalign{\hrule} 
&d_1 &d_2=0 &1           &2         &3          &4& \cr
\noalign{\hrule}
&0&0   &20&-2&&&\cr 
&1&640&6474&1152&-1023&1536&\cr 
&2&10032&733560&4458084&1080696&-762128& \cr
\noalign{\hrule}}\hrule}$$}}} 
They are clearly different from those on the diffeomorphic K3
fibration discussed first, which means that the two 
cases have disjoint families of symplectic 
structures. As in the cases $A$ and $C$ these cases have the
same weak coupling limit when the base of the $\mP^1$ goes to infinity. 
Our construction shows also that it is 
possible to put two different complex structures on 
mirrors of one topological type of K3 fibrations. 
Given the polyhedra  it is possible to 
decide whether these mirrors have topologically equivalent phases, 
but we have not done this.

Our conclusion, however, is that the BPS numbers on the 
heterotic side fix the topology up to rational equivalence,
but even on topologically equivalent manifolds they do not necessarily fix
the symplectic family. The difference between the heterotic data is 
now even more subtle than in the comparison between the fibrations $A$ and  
$C$. While the higher genus data agree, even the classical terms of 
$\cF^{(1)}$, we see that there is an unconventional term ${\rm Li}_3(q_2/q_1)$ 
in the genus zero prepotential. This term gets exponentially suppressed  
in the weak coupling limit $q_2\sim\exp(-\frac{1}{g^2})$. One may already 
think of it as a non-perturbative effect, whose suppression 
can, however, be counterbalanced by tuning the geometrical modulus to the
limit $T\rightarrow i \infty$ in which $q_1\rightarrow 0$. The visibility 
of non-perturbative effects even at weak coupling at special points of the
geometric moduli space is just as for small instantons.

\subsection{Heterotic duals and threshold corrections}
\label{sec:heterotic}
 
In this section we focus on the situation of Calabi--Yau manifolds $X$ that 
admit a K3 fibration. For such an $X$ with a known heterotic-type II dual, 
$\cF^{\rm hol}$ has been evaluated in the limit where the base of 
the fibration is large. More specifically, for the $STU$ model
(see Section~\ref{sec:STU} below)~\cite{Harvey:1995fq}, \cite{Marino:1998pg} 
have obtained results which can be summarized in the holomorphic limit as 
follows
\begin{equation}
\cF^{\rm hol}_{\rm K3}(\lambda,t)= - {\Theta(q) \over 4 \pi^2 \eta^{24}}  
  \sum_{g=0}^\infty S_{g} (G_2,{1 \over 2} G_4,\ldots,{1\over g} G_{2g}) \left({\lambda\over 2 \pi} \right)^{2g-2}\ .
\label{eq:mooremarino}
\end{equation}
Here $G_k(\tau)= 2 \zeta(k) E_k(\tau)$\vphantom{x}\footnote{Note that 
$\zeta(2 n)= -{1\over (2 n)!} (-1)^n 2^{2 n-1} \pi^{2 m} B_{2 n}$.} and the Schur 
polynomials are obtained by $\sum_{k=0}^\infty S_k(x_1,\ldots,x_k)z^k:=$ $\exp \sum_{k=1}^\infty x_k z^k$.
The Eisenstein series are defined as 
\begin{equation}
  \label{eq:Eisenstein}
  E_k(\tau)= {1\over 2}\sum_{n,m\in \mZ\atop (n,m)=1} {1\over (m \tau +n)^k}=
  1 - {2 k\over B_k} \sum_{n=1}^\infty \sigma_{k-1}(n) q^n , 
\end{equation}
where $q=e^{2 \pi i \tau}$, $B_k$ are the Bernoulli numbers defined through 
$\sum_{k=0}^\infty B_{k} {x^k\over k!}:= (e^x-1)^{-1}$ and 
$\sigma_k(n):= \sum_{d|n} d^k$ is the $\sigma$ divisor function. $\Theta(q)$ is related 
to an automorphic form of the classical duality group  
${\rm SO}(2,h-1,\mZ)$ by the Borcherds lifting. The connection to the 
classes $T, U_i$ in the fiber of the K3 is defined by an analog 
of~(\ref{eq:latticespacing}).

The expression~(\ref{eq:mooremarino}) was obtained from a one loop string 
calculation in the heterotic theory. The information at genus 0 and 1 is 
contained in the well known index calculation for the one loop threshold 
dependence of the gauge coupling~\cite{Harvey:1995fq},\cite{Forger:1997tu}
\footnote{Differences $\Delta(G_1) - \Delta(G_2)$ were first computed 
in~\cite{Dixon:1990pc}, but for these the important non-holomorphic term
proportional to $({\rm Im}\tau)^{-1}$ in the integrand drops out.}
\begin{equation}
  \Delta(G)=-i\int_{\cF} {{\rm d}^2\tau \over {\rm Im} \tau}{1\over \eta^2} {\rm Tr}_{\cal H} F_L (-1)^{F_L} 
  q^{H_L} q^{H_R}\left[Q^2-{k\over 8 \pi {\rm Im }\tau}\right],
  \label{treshhold}
\end{equation} 
where $k$ is the level and $Q$ is the operator of the gauge group in 
${\cal H}$. For the $STU$ model the integrand is given by
$-{i\over 12}\sum_{p \in \Gamma_{2,2}} q^{{1\over 2} P_L^2} {\bar q}^{{1\over 2} P_R^2}\left( 
{E_4 E_6\over \eta^{24}}(E_2-{3\over \pi {\rm Im}{\tau}})-{ E_6^2 \over \eta^{24}}\right)$.
Only the first terms in the second bracket are relevant for the 
calculation\footnote{They constitute up to the constant 264 the integrand of 
$\tilde I_{2,2}$ evaluated in \cite{Harvey:1995fq}.} of
$\cF^{(0)}$ and $\cF^{(1)}$. The $\hat E_2=(E_2-{3\over \pi {\rm Im}{\tau}})$ 
term has a beautiful generalization to all higher genus $\cF^{(g)}$ found 
in~\cite{Antoniadis:1995zn}, see also~\cite{Serone:1996bk}, to include the gravitational 
corrections to the gauge coupling. Using the identity 
\begin{equation}
{\ 2 \pi \eta^3(\tau)z\over \theta_1(z,\tau)}=-\exp\left[ \sum_{k=1}^\infty 
{\zeta(2 k)\over k} E_{2 k}(\tau) z^{2 k}\right],
\label{eq:thetaidentity}
\end{equation}
this generalization can be written as 
\begin{equation} 
\left( 2 \pi i {\tilde \lambda}\eta^3\over \theta_1(\lambda,\tau)\right)^2
e^{-{\pi {\tilde \lambda}^2\over {\rm Im}\tau}}=-\sum_{k=0}^\infty 
{\tilde \lambda}^{2k} S_{2k} \left(\hat G_2,{1\over 2} G_4, \ldots ,{1\over k} G_{2k}\right),
\label{eq:narainint}
\end{equation}
where $\tilde{\lambda} = \tau_2 \lambda f(T, U_i)$ with $f$ given in~\cite{Antoniadis:1995zn}.   
The integration over the fundamental region was performed 
in~\cite{Marino:1998pg} 
using techniques of~\cite{Dixon:1990pc},~\cite{Harvey:1995fq} 
and~\cite{Borcherds}. 
Evaluation of the integral leads to the full $(t,\bar t)$ 
dependence of $\cF(\lambda,t,\bar t)$, but in the holomorphic limit there 
is a 
simplification which will help us to make contact with the geometry.
With the Jacobi triple product identity 
$(q=\exp( 2 \pi i \tau),y=\exp(2 \pi i z))$
\begin{equation}
\theta_1(z,\tau)=i(y^{-{1\over 2}} - y^{1\over 2}) q^{1\over 8} 
\prod_{n=1}(1-q^n)(1-q^n y)(1-q^n y^{-1})
\end{equation}
and the definition $\eta=q^{1\over 24}\prod_{n=1}^\infty (1-q^n)$ 
the holomorphic limit~(\ref{eq:mooremarino}) is rewritten in the form~(\ref{eq:hilb}). 
Note the important difference that~(\ref{eq:narainint}) 
is not holomorphic but decomposes into 
$(2k,0)$ ${\rm SL}(2,\mZ)$ forms, which is 
essential to perform the integrals, while~(\ref{eq:mooremarino}) contains 
only 
quasimodular forms but is holomorphic. Also $\tilde \lambda$ is shifted 
w.r.t. $\lambda$,
which introduces a non-holomorphic space time moduli dependence.

\subsection{The $ST$ model and dual type II K3 fibrations}
\label{sec:STandP11226}

Many aspects of the $ST$ heterotic model on K3$\times T^2$ 
have been discussed in the context of the $N=2$ heterotic-type 
II string duality. It has two vectors multiplets, the complexified size of 
the $T^2$ and the dilaton $S$. Because of the Hodge numbers the type II dual
has been identified with the hypersurface $X=\mP^4_{1,1,2,2,6}[12]$ 
with $\chi=-252$, $h^{1,1}=2$ and $h^{2,1}=128$~\cite{Kachru:1995wm}, which 
is a K3 fibration whose generic fiber has Picard number one and self 
intersection $E^2=2$. The significance of K3 fibrations in the context 
of heterotic-type II duality was first pointed out in~\cite{KLM}.
Later a supergravity argument was given in~\cite{AL}. The intersection ring of  
$X=\mP^4_{1,1,2,2,6}[12]$ is ${\cal R}= 4 t_1^3+ 2 t_1^2 t_2$, and the 
heterotic modulus $T$ is identified with the unique class of the K3 fiber 
$t_1=T$, while the complexified size of the base $t_2$ is identified with the 
heterotic dilaton $t_2=4 \pi i S$. Furthermore, $\int_X \ch_2 J_1=52$ 
and $\int_X \ch_2 J_2=24$. 

The propagators and the higher genus amplitudes have very similar structure 
as for the model $\mP^4_{1,1,2,2,2}[8]$ discussed in Section~\ref{sec:11222}. 
In particular the Yukawa couplings are, after a suitable 
rescaling~\cite{Hosono:1993qy}, identical to those in~(\ref{eq:Yukawa11222}). 
The algebraic compatibility conditions~(\ref{eq:comp1}) and~(\ref{eq:comp2}) 
for the ambiguity $A^i_{kl}$ can therefore be satisfied with~(\ref{eq:A3_11222}).
The leading behaviour of the propagators $S^{ij}$ becomes 
\begin{equation}
  \begin{array}{rcl} 
  S^{11}&=&-{1\over 8}-552 q_1 q_2 +O(q^3),\\ 
  S^{12}&=&{1\over 4}-276 q_1^2+42732 q_1^2 +276 q_1 q_2+O(q^3)\\
  S^{22}&=& -{1 \over 2}-77040 q_1^2+O(q^3)\ . 
  \end{array}
\end{equation} 
We checked (\ref{eq:f1b}) and found that 
$$\partial_{t_k} \cF^{(1)} ={1\over 2} C_{k,j,i} S^{ij} + \partial_{t_k}\left({1-\chi \over 24} 
\log(\omega)-{1\over 12} \log(\Delta_{con})- {1\over 3} \log(\Delta_{s})- {29\over 12}\log(x)- {7\over 8}\log(y)\right)$$
again in accordance with the expected behaviour of $\cF^{(g)}$ near the conifold. 
The $-{1\over 3}\log(\Delta_{s})$ is the same behaviour as in 
the conventional calculation of $\cF^{(1)}$. The total leading coefficient is 
$\cF^{(1)} =- {1\over 12} \log(\Delta_{s})+ O(\Delta_{s})$ 
coming from the 1-loop $\beta$-function of $2$ massless hypermultiplets and 
a non-contributing  massless $N=4$-like spectrum of an SU(2) gauge 
symmetry enhancement. This implies that the $S^{ij}$ terms contribute 
${1\over 4} \log(\Delta_{s})$. The ambiguity $A_{ij}$ for the $S^{i}$ can also be
solved by~(\ref{eq:A2_11222}). The leading behaviour of the $S^{i}$ and $S$ is 
then
\begin{equation}
  \begin{array}{rcl} 
  S^{1}&=& 15 q_1 - 3330 q_1^2 + 45 q_1 q_2 +  O(q^3),\\ [2 mm]
  S^{2}&=& 30 q_1 + 5940 q_1^2 - 30 q_1 q_2 +  O(q^3)\\ [2 mm]
  S&=& -900 q_1^2+ 399600q_1^3 -9000 q_1 q_2 + O(q^4)\ . 
  \end{array}
\end{equation} 
The ansatz for the holomorphic ambiguity $f_2$ in~(\ref{eq:f_g}) satisfying all known
constraints of Section~\ref{sec:Ansatz} is
\begin{eqnarray}
\label{eq:ansatz11226} 
f_2(x,y)&=&\left({2051\over 103680}+ {965\over 5184} y - {615\over 4}x(1-4y)\right){1\over \Delta_{s}}
      \nonumber\\ [2 mm]
&&    +\frac{-2+6757x - 5643648 x^2}{60\Delta_{con}}+\frac{(1-1728x)^3}{120\Delta^2_{con}}
\end{eqnarray} 
Observe that $p^{(3)}_{con}$ has a very simple form, similar to the one in~(\ref{eq:ansatz11222}).
This leads to the following predictions for the invariants $n^{(2)}_{i,j}$ for curves of genus 2.
We list for comparison the invariants $n^{(g)}_{i,j}$ for curves of genus $g=0,1,2$.

\noindent $g=0$
\vskip 5 mm {\vbox{\footnotesize{
$$
\vbox{\offinterlineskip\tabskip=0pt \halign{\strut \vrule#& &\hfil
~$#$ &\hfil ~$#$ &\hfil ~$#$ &\hfil ~$#$ &\hfil ~$#$ &\hfil ~$#$
&\hfil ~$#$ &\hfil ~$#$
&\vrule#\cr \noalign{\hrule} 
&d_1 &d_2=0 &1           &2         &3          &4       &5       &      & \cr
\noalign{\hrule}
&0&  &2& & & & &  &  \cr 
&1& 2496&2496& & & & & & \cr 
&2& 223752&1941264&223752&&&&&\cr 
&3& 38637504&1327392512&1327392512&38637504&&&&\cr
&4& 9100224984&861202986072&2859010142112&861202986072&9100224984&&& \cr 
&5& 2557481027520&540194037151104&4247105405354496&4247105405354496&
540194037151104&2557481027520&&\cr 
\noalign{\hrule}}\hrule}$$}}} 

\noindent $g=1$
\vskip 5 mm {\vbox{\footnotesize{
$$
\vbox{\offinterlineskip\tabskip=0pt \halign{\strut \vrule#& &\hfil
~$#$ &\hfil ~$#$ &\hfil ~$#$ &\hfil ~$#$ &\hfil ~$#$ &\hfil ~$#$
&\hfil ~$#$ &\hfil ~$#$
&\vrule#\cr \noalign{\hrule} 
&d_1 &d_2=0 &1           &2              &3           &4       &     5    & & \cr
\noalign{\hrule}
&0&           &          &               &             &       &                 &&    \cr 
&1&           &          &               &             &       &                 &&\cr 
&2& -492      &      480 & -492          &             &       &                 &&\cr 
&3&-1465984   &   2080000&2080000        &-1465984&    &         &                 &\cr  
&4&-1042943028&3453856440&74453838480&3453856440&-1042943028&    &                 &  \cr 
&5&-595277880960&3900245149440&313232037949440&313232037949440&3900245149440&-595277880960&&
\cr
\noalign{\hrule}}\hrule}$$}}} 

\noindent $g=2$
\vskip 5 mm {\vbox{\footnotesize{
$$
\vbox{\offinterlineskip\tabskip=0pt \halign{\strut \vrule#& &\hfil
~$#$ &\hfil ~$#$ &\hfil ~$#$ &\hfil ~$#$ &\hfil ~$#$ &\hfil ~$#$
&\hfil ~$#$ &\hfil ~$#$
&\vrule#\cr \noalign{\hrule} 
&d_1 &d_2=0 &1           &2         &3          &4       &5       &      & \cr
\noalign{\hrule}
&0   &    &              &          &           &        &        &       &\cr 
&1&  &    &              &           &        &        &       &\cr 
&2&-6& 8      &   -6&   &     &        &       &\cr 
&3& 7488&  0  & 0 &  7488& & &  &\cr 
&4& 50181180 &  -73048296 &    32635544  & -73048296  & 50181180 &   &     &\cr 
&5&72485905344& -194629721856 &  2083061531520   &  2083061531520       &  -194629721856         &  72485905344   &   &
\cr 
\noalign{\hrule}}\hrule}$$}}} 

In the strong coupling limit~\cite{Klemm:1996kv} this model has a non-conifold 
transition at $\Delta_s=0$ to the complete intersection $X'=\mP^5_{1,1,1,1,1,3}[2,6]$ 
with the topological data $\chi=-256$, $h^{1,1}=1$ whose invariants $n^{(2)}_d(X')$
are given by
\vskip 5 mm
{\vbox{\footnotesize{
$$
\vbox{\offinterlineskip\tabskip=0pt \halign{\strut \vrule#& &\hfil
~$#$~&\hfil ~$#$ &\hfil ~$#$ &\hfil ~$#$ &\hfil ~$#$ &\hfil ~$#$~&~ \vrule#\cr \noalign{\hrule} 
& d=1 & 2        &3                 &4                      &5                     &6                          &  \cr
\noalign{\hrule}
& 0 & -4 & 14976 & -13098688 & 3921835430016 & 128614837503143532 & \cr
\noalign{\hrule}}\hrule}$$}}} 
We therefore see that the $n^{(2)}_{i,j}$ and $n^{(2)}_d$ are in agreement with 
the constraint~(\ref{eq:birat}).

Moreover, we find a good consistency check in the weak coupling limit of the 
heterotic string dual to all orders in $n=d_1$ for $d_2=0$. 
Combining the results of~\cite{Antoniadis:1995zn},~\cite{Kawai:1996te} 
and~\cite{Marino:1998pg} we conclude that~(\ref{eq:hilb}) extends the 
analysis of~\cite{Kawai:1996te} to all genus with
\begin{equation}
  \label{eq:Kawai}
{\Theta(q)\over \eta^{24}}= - {2 \theta E_4 F_6\over \eta^{24}}
={-2 \over q}+252+2496 q^{1\over 4}+223752 q+\ldots,
\end{equation}
where 
\begin{align}
  \theta(q) &=\sum_{n\in \mZ} q^{n^2\over 4}, & F_2(q) &= \sum_{n \in
    \mZ_{>0}, {\rm odd}} \sigma_1(n) q^{n\over 4}
\end{align}
generate the ring of modular forms for the congruence subgroup $\Gamma^0(4)$ 
and $F_6=E_6 - 2F_2(\theta^4- 2F_2)(\theta^4-16 F_2)$. The self intersection of the
unique divisor class in the generic fiber is $E^2=2$ and the one dimensional 
Picard lattice is generated by $\delta$ with $\delta^2={1\over 2}$. Hence 
we get for the embedding of the Picard lattice of the K3 into $H^2(X,\mZ)$ 
\begin{equation}
  \lambda^{2g-2}q^l\rightarrow {1\over (2 \pi i)^{3-2 g}}\sum_{n^2/4=l}{\rm Li}_{3-2g}
  (e^{2 \pi i\ n T}) \ .
  \label{eq:latticespacingST}
\end{equation}  
For $g=0$ this was already checked in~\cite{Kawai:1996te}. For $g=1$ the 
subtraction scheme used in~\cite{Bershadsky:1993ta},\cite{Kawai:1996te} 
involves 
the M\"obius function for the covering of the torus to itself and differs 
from~(\ref{eq:hilb}),(\ref{eq:latticespacingST}).  
Our generalization to all genus is perfectly consistent with the above B-model
one and two loop results and the lowest $n$ non-vanishing genus $g$ invariants
can be checked to all $g$ using~\cite{Katz:1999xq} as we will do below.
The predictions for the $n^{(g)}_{d,0}$, $g\ge 0$, are as follows
\vskip 5 mm {\vbox{\small{
$$
\vbox{\offinterlineskip\tabskip=0pt \halign{\strut \vrule#& 
&\hfil~$#$ 
&\hfil~$#$ 
&\hfil~$#$ 
&\hfil~$#$ 
&\hfil~$#$
&\hfil ~$#$
&\hfil ~$#$
&\hfil ~$#$
&\hfil ~$#$
&\vrule#\cr \noalign{\hrule} 
&g&d=1&2         &3          &4       &5       &6       &7 & 8& \cr
\noalign{\hrule}
&0&2496&223752&38637504&9100224984&2557481027520&805628041231176&\ldots&\ldots&\cr 
&1&0&-492&-1465984&-1042943520&-595277880960&-316194812546140&\ldots&\ldots& \cr 
&2&0&-6&7488&50181180&72485905344&70378651228338&\ldots&\ldots& \cr 
&3&0&0&0&-902328&-5359699200&-10869145571844&\ldots&\ldots& \cr 
&4&0&0&0&1164&228623232&1208179411278&\ldots&\ldots& \cr 
&5&0&0&0&12&-4527744&-94913775180&\ldots&\ldots& \cr 
&6&0&0&0&0&17472&4964693862&\ldots&\ldots & \cr 
&7&0&0&0&0&0&-152682820&\ldots&\ldots&\cr
&8&0&0&0&0&0&2051118&\ldots&\ldots&\cr 
&9&0&0&0&0&0&-2124&\ldots&\ldots&\cr
&10&0&0&0&0&0&-22&605915136&\ldots& \cr
&11&0&0&0&0&0&0&-9419904&\ldots& \cr
&12&0&0&0&0&0&0&32448&\ldots& \cr
&13&0&0&0&0&0&0&0&\ldots& \cr
&14&0&0&0&0&0&0&0&\ldots&  \cr
&15&0&0&0&0&0&0&0&\ldots&\cr
&16&0&0&0&0&0&0&0&3132& \cr
&17&0&0&0&0&0&0&0&36& \cr
\noalign{\hrule}}\hrule}$$}}} 
Repeating the discussion at the end of the subsection~\ref{sec:11222} 
with $H^3=4$ and $H^2 L=2$ we obtain  
\begin{equation}
n^{(k^2+1)}_{2k ,0}=(-1)^{{\tilde p}(2k)} 2 {\tilde p}(2k)\ ,
\end{equation} 
where we define ${\tilde p}$ using~(\ref{eq:dimH0}) as
\begin{equation}
  \sum_{n=0}^\infty {\tilde p}(n) q^n={1\over (1-q) (1-q^2)^2 (1-q^6)}\ .
  \label{eq:dimH011226}
\end{equation}
Furthermore, with $e({\cal C})=-252 ({\tilde p}(2k)-1) g$ we get for the $n_Q^{(g)}$ 
with one node
\begin{equation}
  n^{(k^2)}_{2k ,0}= (-1)^{{\tilde p}(2k)-1} (4 k^2 {\tilde p} (2 k) +252 (1-{\tilde p}(2k)) \ ,  
\end{equation} 
again in perfect agreement with the table.
In a way very similar to the first calculation one can also
calculate the Euler number of the moduli space of smooth
curves in the class $(2 k,1)$ with the result
\begin{equation}
  n^{(k^2+1)}_{2k ,1}=(-1)^{{\tilde p}(2k)-1} 4 ({\tilde p}(2k) -1) \ .  
\end{equation} 

What we have seen above is a very detailed perturbative check of 
heterotic-type II duality based on the BPS degeneracies for certain
degrees to all genus. However, just as in 
the case of the fibration  discussed in section~\ref{sec:11222}, 
we find three fibrations with the same Hodge numbers as 
listed in table~\ref{tab:K3} below. From the dimension of the vector- and 
hypermultiplet moduli space they could be equally well type II duals 
of the ST model. The difference between these cases is in the classical 
intersection ring $\cF^{(0)}= -\eta t_1^3/6 - t_1^2 t_2$ and in the 
classical terms 
in $\cF^{(1)}\sim \frac{36+ 4 \eta}{24} t_1 +t_2$. We parameterized these by 
$\eta=4,2,0$, so that $\eta=4$ is the case presented above. The complete 
intersection
with $\eta=2$ has Picard--Fuchs operators
\begin{equation}
\begin{array}{rl}
{\cal L}_1&=\theta_1^2(\theta_1 - \theta_2)-
8 (6 \theta_1 + 2 \theta_2-5) (6 \theta_1 + 2 \theta_2-3) (6 \theta_1 + 2 \theta_2-1) z_1 \\
{\cal L}_2&= \theta_2^2 - 2(\theta_1 - \theta_2+1)(6\theta_1 + 2\ \theta_2-1)z_2  \\
\end{array}
\end{equation}
from which we obtain, with suitably rescaled variables $a=\frac{z_1}{1728}$ 
and  $b=\frac{z_2}{16}$, the B-model triple couplings
\begin{align}
  \label{eq:Yukawab252}
  Y_{111} &= {2(3(1 + 3b)^2 + a(3 + 13b + 16b^2)) \over 3 a^3\Delta_{con}}, & 
  Y_{112} &= {2((1 + 3b)^2 - a(1 + 9b + 16b^2)) \over a^2 b \Delta_{con}}
        \notag\\
  Y_{122} &= { 2(3 + 9b + 3a(5 + 16b)) \over a b \Delta_{con}}, &
  Y_{222} &= {18(1 - a + b - 16ab)\over b ^2 \Delta_{con}}
\end{align}
where the conifold discriminant is
$$\Delta_{con}=(1-a)^2-b(6 + 9b - 2a(21 + 96b + 128b^2))\ . $$
The fact that the fiber has $E^2=2$ guarantees that 
$\frac{1}{z_1}\bigr|_{z_2=0}=J(t_1)$, and together with the discriminant
of the form $\Delta_{con}=(1-1728 z_1 )^2-O(z_2)$ we get a clear hint
of a SU(2) gauge enhancement for the value $t_1:=T=i$, i.e. the
gauge symmetry enhancement expected for the torus modulus of the
heterotic torus in the weak coupling. Using the Picard--Fuchs 
operators transformed to the variables $x=(1-a)=\epsilon \tilde u$ 
and $y=\frac{a^2 b}{3 (1-a)^2}=\frac{1}{\tilde u^2}$ we have established
that we get the embedding of the Seiberg-Witten non-perturbative completion 
of the rigid  $N=2$ SU(2) gauge symmetry at this point. 
The Seiberg-Witten periods $a(u)$ and $a_D(u)$ with 
$\frac{1}{\tilde u^2}= \frac{\Lambda^4}{u^2}$ occur in the double
scaling limit of the Calabi-Yau periods like in the work of 
\cite{Kachru:1995fv}, 
which considers the fibration $n=4$. The Calabi-Yau periods for the various 
cases $\eta=4,2,0$ differ in the $\epsilon$ corrections near the SW point.

The Seiberg-Witten enhancement, even though non-perturbative in the rigid 
theory, is an effect that is local near the weak coupling divisor $z_2=0$ 
of the moduli space. When we venture deeper into the non-perturbative regime 
of the full heterotic string we see a completely different picture. 
The easiest way 
to detect the differences is to look at the GV invariants in 
arbitrary classes. Here we see the different non-perturbative completions
of the three theories.  

\noindent $g=0$
\vskip 5 mm {\vbox{\footnotesize{
$$
\vbox{\offinterlineskip\tabskip=0pt \halign{\strut \vrule#& &\hfil
~$#$ &\hfil ~$#$ &\hfil ~$#$ &\hfil ~$#$ &\hfil ~$#$ &\hfil ~$#$
&\hfil ~$#$ &\hfil ~$#$
&\vrule#\cr \noalign{\hrule} 
&d_1 &d_2=0 &1           &2         &3          &4       &5       &      & \cr
\noalign{\hrule}
&0& &24&-2&&& && \cr 
&1& 2496&24000&4544&-4096&6144&-8192&& \cr 
&2& 223752&17244192&92555600&22945824&-12818168&24969216&&\cr 
&3& 38637504&11552340480&239152764672&911674812096&283693717248&-89080677888&& \cr
&4& 9100224984&7397501182992&387426434941992&4444162191527440&,13387851546648736&\ldots & & \cr  \noalign{\hrule}}\hrule}$$}}} 

\noindent $g=1$
\vskip 5 mm {\vbox{\footnotesize{
$$
\vbox{\offinterlineskip\tabskip=0pt \halign{\strut \vrule#& &\hfil
~$#$ &\hfil ~$#$ &\hfil ~$#$ &\hfil ~$#$ &\hfil ~$#$ &\hfil ~$#$
&\hfil ~$#$ &\hfil ~$#$
&\vrule#\cr \noalign{\hrule} 
&d_1 &d_2=0 &1           &2              &3           &4       &     5    & & \cr
\noalign{\hrule}
&0&           &          &               &             &       &                 &&    \cr 
&1&           &          &               &             &       &                 &&\cr 
&2& -492      &   5760   &  43080        &   -108928  &638420 &   -2703360      &&\cr 
&3&-1465984   &  23531520&3651733760     &14152031872  &2833959168&5870161920 && \cr  
&4&-1042943028&35586706080&21017735917528&346074753826240&1064755476016840&364053943057856&&\cr 
\noalign{\hrule}}\hrule}$$}}} 
We have also checked with the B-model calculation that the 
$n^{(2)}_{d_1,0}$ genus 2 BPS numbers for the $\eta=2,0$ cases 
are the same as in the $\eta=4$ case for a choice of the ambiguity.
Since the discriminant is of higher order in $b$, it becomes 
harder to fix the ambiguity to uniquely determine the genus 2 
numbers for general $(d_1,d_2)$ classes in the $\eta=2,0$ cases. 
The agreement for all genus amplitudes for the perturbative
classes $(d_1,0)$ follows from formulae (\ref{eq:hilb}), 
(\ref{eq:chie2=2}), (\ref{eq:e2=2}). 

The complex structure data of the $\eta=0$ case are
\begin{equation}
\begin{array}{rl}
{\cal L}_1&=\theta_1^3-
8 (6 \theta_1 + 4 \theta_2-5) (6 \theta_1 + 4 \theta_2-3) (6 \theta_1 + 4 \theta_2-1) z_1 \\
{\cal L}_2&= \theta_2^2 - 4(6\theta_1 + 4\theta_2-1)(6\theta_1 + 4\ \theta_2-3)z_2  \\
\end{array}
\end{equation}
and with 
$a=\frac{z_1}{1728}$, $b=\frac{z_2}{64}$ the B-model triple couplings are
\begin{align}
  \label{eq:Yukawac252}
  Y_{111} &= {4(3+b) \over 3 a^2\Delta_{con}}, &  
  Y_{112} &= {2((1 - b)^2 - a(1 + b)) \over a^2 b \Delta_{con}}\notag\\
  Y_{122} &= { 3(2+ a- 2b) \over a b \Delta_{con}}, &
  Y_{222} &= {9(1 - a + 3 b)\over 2 b ^2 \Delta_{con}}
\end{align}
where the conifold discriminant is
$$\Delta_{con}=(1-a)^2-b(3(1+2 a) -b (3-b))\ . $$
Establishing the rigid SU(2) Seiberg-Witten embedding is  
very similar as in the case above and the non-perturbative BPS
numbers are

\noindent $g=0$
\vskip 5 mm {\vbox{\footnotesize{
$$
\vbox{\offinterlineskip\tabskip=0pt \halign{\strut \vrule#& &\hfil
~$#$ &\hfil ~$#$ &\hfil ~$#$ &\hfil ~$#$ &\hfil ~$#$ &\hfil ~$#$
&\hfil ~$#$ 
&\vrule#\cr \noalign{\hrule} 
&d_1 &d_2=0 &1           &2         &3          &4          &    \cr
\noalign{\hrule}
&0&  & 288&252&288&252&& \cr 
&1& 2496& 216576&6391296&104994816&1209337344&& \cr 
&2& 223752&152031744&19638646848&1180450842624&43199009739072&&\cr 
&3& 38637504&100021045248&34832157566976&4962537351009792&401057938191181824&& \cr
&4& 9100224984& 63330228232704&47042083144050624&13025847457256417280&\ldots & & \cr  \noalign{\hrule}}\hrule}$$}}} 

\noindent $g=1$
\vskip 5 mm {\vbox{\footnotesize{
$$
\vbox{\offinterlineskip\tabskip=0pt \halign{\strut \vrule#& &\hfil
~$#$ &\hfil ~$#$ &\hfil ~$#$ &\hfil ~$#$ &\hfil ~$#$ &\hfil ~$#$
&\hfil ~$#$ 
&\vrule#\cr \noalign{\hrule} 
&d_1 &d_2=0 &1           &2              &3           &4       &     & \cr
\noalign{\hrule}
&0&           &          &  3            &             &       &  &                   \cr 
&1&           &          &-4992&-433152&-12797568     &        &\cr 
&2& -492      & 69120&104982288&11531535360&582562937232 &&\cr 
&3&-1465984   & 265236480&1257004076544  &351343656784896& 42652659615054336& & \cr  
&4&-1042943028&360976775040&4290798245087472&2287148665427369472&506382335153204333472&&\cr 
\noalign{\hrule}}\hrule}$$}}}

To summarize, these three models have the same degeneracies of
BPS states in the perturbative limit and these are the data on which 
the impressive checks of the duality \cite{Kachru:1995wm}
\cite{Kaplunovsky:1995tm}\cite{KLM} rely. The differences in 
the classical terms $\cF^{(0)}$ and $\cF^{(1)}$ are suppressed in 
the perturbative limit $g\rightarrow 0$, where we use the identification 
$t_2=4 \pi i S=4 \pi i \left(\frac{1}{g^2}- i\frac{\theta}{8 \pi^2}\right)$, 
and for the $ST$ model they have not been determined on the 
heterotic string side. Above we also see that the BPS 
degeneracies alone do not suffice to fix the non-perturbative 
completion. There is an additional data necessary on the heterotic 
side, which corresponds to picking one of the three (and possibly more) 
fibrations on the type II side. This discrete data might be 
a subtle choice in the construction of the gauge bundle before 
higgsing it, or a flux. It determines the non-perturbative 
completion of the heterotic string in a decisive way.

\subsection{Further 2--parameter K3 fibrations}
\label{sec:K3fibrations}

In this section we give an overview over other 2--parameter K3 fibrations. Using
an extension of PALP~\cite{PALP} to include the nef-partitions of complete 
intersections discussed in Section~\ref{sec:CICY}~\cite{Riegler:2004ab} we have
searched for K3 fibered complete intersection Calabi--Yau manifolds in 
codimension 2. The number of five-dimensional reflexive polyhedra with a 
nef-partition is enormous. Hence, we have restricted ourselves to polyhedra with
at most ten points. Nevertheless, we have come up with a large number of
examples as Table~\ref{tab:K3} shows.
\begin{table}           \vspace*{-3.8mm}                
  $\footnotesize
  \begin{array}{|c|cccc|ccc|c|}
  \hline
  X & \chi & H^3 & E^2  & \ch_2 H & n^{(0)}_{0,1}  & n^{(0)}_{0,2} & n^{(g)}_{i,k}   &k\\[1mm]
  \hline 
  \IP\left(\tiny\BM 3&3&2&2&1&1&0  \cr  
        3&1&0&0&1&1&2 \cr  \EM\right) 
        \left[\Msize\BM 6\cr6  \EM\VL\BM 6\cr2   \EM\right]
        & -84  & 1            & 2            & 22       & 16       & 2       & 
        n^{(g)}_{i,k} = n^{(g)}_{i,4i-k},\; 0\leq k \leq 2i     &1\\[3mm]
  \IP\left(\tiny\BM  3&1&1&1&1&1&0 \cr  
        3&1&1&1&0&0&2 \cr  \EM\right) 
        \left[\Msize\BM 6\cr6  \EM\VL\BM 2\cr2   \EM\right]
        & -140 & 2            & 2            & 32       & 8        & 0       & 
        n^{(g)}_{i,k} = n^{(g)}_{i,2i-k},\; 0\leq k \leq i      &2\\[3mm]
  \IP\left(\tiny\BM  3&2&1&1&1&0&0 \cr  
        1&0&0&0&1&1&1 \cr  \EM\right) 
        \left[\Msize\BM 6\cr2  \EM\VL\BM 2\cr2   \EM\right]
        & -140 & 0            & 2            & 24       & 96       & 148     & 
        n_{0,i} \not = 0, i > 2               &2\\[3mm]
  \IP\left(\tiny\BM  3&2&1&1&1&1&0 \cr  
        3&2&0&0&1&1&1 \cr  \EM\right) 
        \left[\Msize\BM 6\cr6  \EM\VL\BM 3\cr2   \EM\right]
        & -196 & 3            & 2            & 42       & 4        & 0       
        & n^{(g)}_{i,2i} = n^{(g)}_i(\textrm{local E}_8)           &3\\[3mm]
  \IP\left(\tiny\BM  5&2&2&1&1&1&0 \cr  
        3&0&0&1&1&1&2 \cr  \EM\right) 
        \left[\Msize\BM 10\cr6  \EM\VL\BM 2\cr2   \EM\right]
        & -196 & 1            & 2            & 34       & 48       & -2      
        & n^{(g)}_{i,k} = n^{(g)}_{i,4i-k},\; 0\leq k \leq 2i     &3\\[3mm]
  \mP^4_{1,1,2,2,6}[12]
        & -252 & 4            & 2            & 52       & 2        & 0       
        & n^{(g)}_{i,k} = n^{(g)}_{i,i-k}, \; 0\leq k \leq [i/2]  &4\\[3mm]
  \IP\left(\tiny\BM  4&2&1&1&1&1&0 \cr  
        3&1&0&0&1&1&1 \cr  \EM\right) 
        \left[\Msize\BM 8\cr6  \EM\VL\BM 2\cr1   \EM\right]
        & -252 & 2            & 2            & 44       & 24       & -2      &
                                             &4\\[3mm]
  \IP\left(\tiny\BM  3&3&1&1&1&0&0 \cr  
        0&0&0&2&2&1&1 \cr  \EM\right) 
        \left[\Msize\BM 6\cr4  \EM\VL\BM 3\cr2   \EM\right]
        & -252 & 0            & 2            & 36       & 288      & 252     & 
        n_{0,i} \not = 0, i > 2               &4\\[3mm]
  \hline
  \mP\left({\tiny\BM 2&1&1&1&1&0&0\cr
        0&1&1&1&3&2&2\cr\EM}\right) 
        \left[\Msize\BM   4\cr4  \EM\VL\BM   2\cr6   \EM\right]
        & -84  & 2            & 4            & 32       & 32       & 4       
        & n^{(g)}_{i,k} = n^{(g)}_{i,4i-k},\; 0\leq k \leq 2i     &1\\[3mm]
  \mP\left(\Msize\BM 2 & 1 & 1 &  1 & 1 & 0 & 0 \cr  
            0 & 0 & 0 &  0 & 0 & 1 & 1 \cr \EM\right) 
        \left[\Msize\BM4\cr0\EM\VL\BM2\cr2\EM\right]
        & -112 & 4            & 4            & 40       & 16       & 0       
        & n^{(g)}_{i,k} = n^{(g)}_{i,2i-k},\; 0\leq k \leq i      &2\\[3mm]
  \IP\left(\tiny\BM  2&1&1&1&1&0&0 \cr  
        0&0&0&0&1&1&1 \cr  \EM\right) 
        \left[\Msize\BM 4\cr1  \EM\VL\BM 2\cr2   \EM\right]
        & -112 & 1            & 4            & 34       & 64       & 24      
        & n^{(g)}_{i,k} = n^{(g)}_{i,8i-k},\; 0\leq k \leq 4i     &2\\[3mm]
  \IP\left(\tiny\BM  1&1&1&1&1&1&0 \cr  
        0&0&2&1&1&1&1 \cr  \EM\right) 
        \left[\Msize\BM 4\cr4  \EM\VL\BM 2\cr2   \EM\right]
        & -140 & 6            & 4            & 48       & 8        & 0       
        & n^{(g)}_{i,2i} = n^{(g)}_i(\textrm{local E}_7)                    &3\\[3mm]
  \IP\left(\tiny\BM  2&1&1&1&1&0&0 \cr  
        0&0&0&0&0&1&1 \cr  \EM\right) 
        \left[\Msize\BM 4\cr1  \EM\VL\BM 2\cr1   \EM\right]
        & -140 & 3            & 4            & 42       & 32       & 0       
        & n^{(g)}_{i,4i} = n^{(g)}_i(\textrm{local }\mP^1\times\mP^1)      &3\\[3mm]
  \IP\left(\tiny\BM  2&1&1&1&1&0&0 \cr  
        0&0&0&1&1&1&1 \cr  \EM\right) 
        \left[\Msize\BM 4\cr2  \EM\VL\BM 2\cr2   \EM\right]
        & -140 & 0            & 4            & 36       & 128      & 144     & 
        n_{0,i} \not = 0, i > 2               &3\\[3mm]
  \mP^4_{1,1,2,2,2}[8]
        & -168 & 8            & 4            & 56       & 4        & 0       
        & n^{(g)}_{i,k} = n^{(g)}_{i,i-k}, \; 0\leq k \leq [i/2]  &4\\[3mm]
  \mP\left(\Msize \BM 1 & 1 & 1 &  1 & 1 & 0 & 0 \cr  
            0 & 0 & 0 &  0 & 0 & 1 & 1 \cr \EM\right) 
        \left[\Msize\BM4\cr 1\EM\VL\BM1\cr1\EM\right]
        & -168 & 5            & 4            & 50       & 16       & 0       
        & n^{(g)}_{i,4i} = n^{(g)}_i(\textrm{local }\mP^2)      &4\\[3mm]
  \IP\left(\tiny\BM  2&1&1&1&1&0&0 \cr  
        0&0&0&0&0&1&1 \cr  \EM\right) 
        \left[\Msize\BM 4\cr2  \EM\VL\BM 2\cr0   \EM\right]
        & -168 & 2            & 4            & 44       & 64       & 0       
        & n^{(g)}_{i,k} = n^{(g)}_{i,4i-k},\; 0\leq k \leq 2i     &4\\[3mm]
  \IP\left(\tiny\BM  3&1&1&1&1&1&0 \cr  
        2&0&0&1&1&1&1 \cr  \EM\right) 
        \left[\Msize\BM 6\cr4  \EM\VL\BM 2\cr2   \EM\right]
        & -196 & 4            & 4            & 52       & 32       & -2      &
                                             &5\\[3mm]
  \hline
  \IP\left(\tiny\BM  1&1&1&1&1&0&0 \cr  
        0&0&0&0&0&1&1 \cr  \EM\right) 
        \left[\Msize\BM 3\cr0  \EM\VL\BM 2\cr2   \EM\right]
        & -108 & 6            & 6            & 48       & 24       & 0       
        & n^{(g)}_{i,k} = n^{(g)}_{i,2i-k},\; 0\leq k \leq i      &\\[3mm]
  \IP\left(\tiny\BM  1&1&1&1&1&0&0 \cr  
        0&0&0&0&1&1&1 \cr  \EM\right) 
        \left[\Msize\BM 3\cr1  \EM\VL\BM 2\cr2   \EM\right]
        & -116 & 2            & 6            & 44       & 72       & 18      
        & n^{(g)}_{i,k} = n^{(g)}_{i,6i-k},\; 0\leq k \leq 3i     &\\[3mm]
  \IP\left(\tiny\BM  1&1&1&1&1&1&0 \cr  
        0&0&1&1&1&1&1 \cr  \EM\right) 
        \left[\Msize\BM 3\cr3  \EM\VL\BM 3\cr2   \EM\right]
        & -120 & 9            & 6            & 54       & 12       & 0       
        & n^{(g)}_{i,2i} = n^{(g)}_i(\textrm{local E}_6)                    &\\[3mm]
  \IP\left(\tiny\BM  1&1&1&1&1&0&0 \cr  
        0&0&0&0&0&1&1 \cr  \EM\right) 
        \left[\Msize\BM 3\cr1  \EM\VL\BM 2\cr1   \EM\right]
        & -128 & 5            & 6            & 50       & 36       & 0       
        & n^{(g)}_{i,3i} = n^{(g)}_i(\textrm{local D}_5)                    &\\[3mm]
  \mP^5_{1,1,2,2,2,2}[4,6]
        & -132 & 12           & 6            & 60       & 6        & 0       
        & n^{(g)}_{i,k} = n^{(g)}_{i,i-k}, \; 0\leq k \leq [i/2]  &\\[3mm]
  \IP\left(\tiny\BM  1&1&1&1&1&1&0 \cr  
        0&0&1&1&1&1&1 \cr  \EM\right) 
        \left[\Msize\BM 4\cr3  \EM\VL\BM 2\cr2   \EM\right]
        & -140 & 8            & 6            & 56       & 18       & 0       
        & n^{(g)}_{i,3i} = n^{(g)}_i(\textrm{local }\mP^1\times\mP^1)      &\\[3mm]
  \IP\left(\tiny\BM  1&1&1&1&1&0&0 \cr  
        0&0&0&0&0&1&1 \cr  \EM\right) 
        \left[\Msize\BM 3\cr2  \EM\VL\BM 2\cr0   \EM\right]
        & -148 & 4            & 6            & 52       & 54       & 0       
        & n^{(g)}_{i,k} = n^{(g)}_{i,3i-k},\; 0\leq k \leq [3i/2] &\\[3mm]
  \hline
  \end{array}
  $
  \caption{\label{tab:K3} An overview of K3 fibered Calabi--Yau threefolds 
                with $h^{1,1}=2$}               
\end{table}
In the first column, we have indicated the Calabi--Yau threefold $X$ in terms of weight vectors. We give only one of the typically many realizations of $X$ as a complete intersection with the same topological data. From the weight vectors it is also easy to read off the reflexive section yielding the K3 fiber. The possible K3 fibers together with the self intersection number $E^2$ of the Picard lattice are given by
\begin{equation*}
  \begin{array}{|c|c|}
    \hline
    \mP^3_{1,1,1,3}[6] & 2\\
    \mP^4_{1,1,1,2,3}[2,6] & 2\\
    \mP^4_{1,1,1,3,3}[3,6] & 2\\
    \hline
    \mP^3_{1,1,1,1}[4] & 4\\
    \mP^4_{1,1,1,1,2}[2,4] & 4\\
    \mP^4_{1,1,1,1,3}[3,4] & 4\\
    \hline
    \mP^4_{1,1,1,1,1}[2,3] & 6\\
    \hline    
  \end{array}
\end{equation*} 
In the three cases of weighted projective spaces it is implicitly understood that the singularities coming from the weights with common factor are resolved.
The next three columns in Table~\ref{tab:K3} contain the complete information about the intersection ring and the second Chern class of $X$. The topological numbers not indicated are fixed by Oguiso's criterion~(\ref{eq:Qguiso}). More information about the spaces $X$ is contained in the Gopakumar-Vafa invariants $n^{(g)}_{i,j}$. For most examples, the $n^{(0)}_{0,i}$ are non-vanishing only for $i=1, 2$. These values are explicitly given in the table, as well as some special properties of the general $n^{(g)}_{i,j}$ which will be explained below.

We have determined the modular form $\Theta(q)$ in~(\ref{eq:hilb}) for all the examples with $E^2=2$ and $E^2=4$. The result depends on the value of $k$ in the last column of Table~\ref{tab:K3}. For the examples with $E^2=2$, $k$ is related to the Euler number as
\begin{equation}
  \label{eq:chie2=2}
  \chi = -28(1+2k)
\end{equation}
and we find for $k=1,\dots,4$ 
\begin{equation}
  \label{eq:e2=2}
  \Theta^{(k)}(q) = 2^{-8} W E_4 F_6^{(k)}  
\end{equation}
with
\begin{equation}
  F_6^{(k)}(q) = 4\,{W}^{12} - (76 + 7k)\,{W}^{8}{X}^{4}
-(180 - 6k)\,{W}^{4}{X}^{8} - (4-k)\,{X}^{12}
\end{equation}
and $W = \theta_3(\frac{\tau}{2})$, $X=\theta_4(\frac{\tau}{2})$, and $E_4(\tau)$ 
is as in~(\ref{eq:Eisenstein}). Higher genus amplitudes for $k=4$ have been discussed in 
Section~\ref{sec:STandP11226} in detail, while some topological 
aspects of the $k=1$ 
example are given in Appendix~\ref{sec:other_models}. $F_6^{(4)}(q)$ 
coincides with the expression in~(\ref{eq:Kawai}) found in~\cite{Kawai:1996te}.
 
For the examples with $E^2=4$, $k$ is related to the Euler number as
\begin{equation}
  \label{eq:chie2=4}
  \chi = -4(4+7k)
\end{equation}
and we find for $k=1,\dots,5$
\begin{eqnarray}
  \label{eq:e2=4}
  \Theta^{(k)}(q) &=& 2^{-22}\left(6U^{21}-(4-k)U^{20}V-(78+21k)U^{19}V^2-(1066+47k)U^{18}V^3\right.\nonumber\\
  &&-(28020+213k)U^{17}V^4-(40094-20k)U^{16}V^5-(339960-444k)U^{15}V^6\nonumber\\
  &&-(241928-164k)U^{14}V^7-(1244316-300k)U^{13}V^8-(585392+50k)U^{12}V^9\nonumber\\
  &&-(2073636+774k)U^{11}V^{10}-(691212+258k)U^{10}V^{11}-(1756176+150k)U^9V^{12}\nonumber\\
  &&-(414772-100k)U^8V^{13}-(725784-492k)U^7V^{14}-(113320-148k)U^6V^{15}\nonumber\\
  &&-(120282-60k)U^5V^{16}-(9340+71k)U^4V^{17}-(3198+141k)U^3V^{18}\nonumber\\
  &&\left.-(26+7k)U^2V^{19}-(12-3k)UV^{20}+2V^{21}\right)
\end{eqnarray}
with $U=\theta_3(\frac{\tau}{4})$ and $V=\theta_4(\frac{\tau}{4})$ as 
in~(\ref{eq:UV}). The cases with $k=2$ and $k=4$ have been discussed in detail 
in Sections~\ref{sec:fibrationB} and~\ref{sec:11222} respectively, 
cf.~(\ref{eq:ThetaB}) and~(\ref{eq:ThetaA}). Again, some topological aspects 
of the $k=1$ example are given in Appendix~\ref{sec:other_models}.
We emphasize that unlike the $E^2=2$ case~(\ref{eq:e2=2}), it is not possible to 
factor the $E_4$ Eisenstein series. Since this Eisenstein series is nothing but
the theta function $\Theta_{\tE_8}(q,0)$ for the $\tE_8$ lattice (see also 
Section~\ref{sec:BPScount}) this hints at the possibility that an $\tE_8$
symmetry which was realized in the $E^2=2$ case, is broken in the $E^2=4$ case.

Observe that in all the examples
\begin{equation}
  \label{eq:Euler}
  \textrm{Coeff}_{q^0}\frac{2\Theta^{(k)}}{\eta^{24}} = \chi
\end{equation}
as was already explained in~\cite{Antoniadis:1995zn}.

It turns out that there is a much simpler basis for the space of modular forms than $\theta_3$ and $\theta_4$. In this basis more properties of these modular forms can be read off. This will be fully explored in a future publication~\cite{wipI} where we also intend to include the expressions for $\Theta$ for the examples with $E^2=6$.

The two-parameter K3 fibrations $X$ with the same modular form $\Theta(q)$ and hence the same Euler number can be distinguished by their different topology, i.e. by the numbers $H^3$ and $\ch_2H$. This is also reflected in the instanton numbers as we will discuss now in detail. In the case that the Calabi--Yau threefold $X$ admits a contraction in which the fibers of a ruled surface get blown down, there is a $\mZ_2$ involution on $H_2(X,\mZ)$~\cite{Candelas:1993dm}. This implies a relation for the $n^{(g)}_Q$ of the type~(\ref{eq:monodr}). 
This explains the entries in Table~\ref{tab:K3} which are of the form $n^{(g)}_{i,k} = n^{(g)}_{i,Mi-k}, \; 0\leq k \leq [Mi/2], i>0$ and $n_{i,k} = 0$ for $i>0$ and $k > Mi$. The constant $M$ can be determined from the geometry of $X$. Alternatively, if $X$ admits a contraction to a local Calabi--Yau whose compact part is a del Pezzo surface $S$ or a $\mP^2$, there are again restrictions on the $n^{(g)}_Q$, given by~(\ref{eq:local}). In this situation, they read $n^{(g)}_{i,M'i} = n^{(g)}_i(S)$ for $i>0$ and $n^{(g)}_{i,k} = 0$ for $i>0$ and $k>M'i$. $M'$ is another constant which is also encoded in the geometry of $X$. The invariants $n^{(0)}_i(S)$ for low degrees and for all $S$ which appear in toric Calabi--Yau complete intersections are listed below~\cite{Klemm:1996hh},~\cite{Katz:1999xq}.
\begin{equation*}
  \begin{array}{|l|rrrrrrr|}
  \hline
              S                   & d &   1 &     2 &      3 &          4 &           5 &              6 \\
  \hline
  \textrm{local }\mP^1\times\mP^1 &   &  -4 &    -4 &    -12 &        -48 &        -240 &          -1356 \\
  \textrm{local }\mP^2            &   &   3 &    -6 &     27 &       -192 &        1695 &         -17064 \\
  \textrm{local D}_5              &   &  16 &   -20 &     48 &       -192 &         960 &          -5436 \\
  \textrm{local E}_6              &   &  27 &   -54 &    243 &      -1728 &       15255 &        -153576 \\
  \textrm{local E}_7              &   &  56 &  -272 &   3240 &     -58432 &     1303840 &      -33255216 \\  
  \textrm{local E}_8              &   & 252 & -9252 & 848628 & -114265008 & 18958064400 & -3589587111852 \\
  \hline
  \end{array}
\end{equation*}
Finally, there are a few other cases where there seem to be no restrictions 
on the instanton numbers. This is in particular the case for the examples 
with $H^3=0$. As for the $n^{(0)}_{0,1}$, we observe that they are in 
general multiples of $E^2$. For those $X$ which admit a contraction of a 
ruled surface or to a del Pezzo surface, $n^{(0)}_{0,1}$ is also a multiple 
of the constants $M$ and $M'$, respectively. 

We conclude with a remark on the completeness of the table. We have seen 
that the Euler numbers for the 2-parameter K3-fibrations with $E^2=2,\ 4$ 
take only particular values,~(\ref{eq:chie2=2}) and~(\ref{eq:chie2=4}). 
Looking at the table, we see that for a fixed Euler number there is a 
maximal value of $H^3$. This value decreases by 2 for $E^2=2$ and by 3 
for $E^2=4$. This is related to the rational homotopy equivalence of these 
models as follows. If we change $H$ by a rational multiple of the class $L$ 
of the fiber, $H\rightarrow H'=H-\frac{1}{n}L$, then the requirement that 
$H'{}^3$ and $\int_X \ch_2 H'$ be integers tells us that $n\mid 24$ and 
$n\mid 3E^2$. However, not all values of $n$ that are allowed by these 
conditions actually appear. The lower bound is given by the fact that
$H^3$ must be non-negative. The maximal value of $H^3$ is $k$ for $E^2=2$
and $2k$ for $E^2=4$, so it seems to be directly related to the Euler number.
All these bounds taken together suggest that it is possible to 
classify such Calabi--Yau families, at least up to rational homotopy. For 
example, these observations suggest the existence of a 2-parameter K3 
fibration with $\chi = -196$, $E^2=4$, $H^3=10$ and $\ch_2 H = 64$. 
Moreover, the same argument show that the list of the families with $E^6$
is far from being complete.

\subsection{The $STU$ model and its type II duals }
\label{sec:STU}

In this section we will study the Gopakumar-Vafa invariants for various type 
IIA duals $X$ of the $STU$ model. We begin with 
$X=\IP\left({\footnotesize\BM 6~4~1~0~1~0\cr6~4~0~1~0~1\cr\EM}\right)
\left[\footnotesize\BM12\cr12\EM\right]$, 
which has a K3 fibration structure
${\rm K3}\rightarrow X\rightarrow \mP_{(3)}^1$. The K3 fiber 
itself has an elliptic fibration $E \rightarrow {\rm K3} \rightarrow \mP_{(2)}^1$ compatible with an
elliptic fibration structure of $X$, $E \rightarrow X \rightarrow
\mF_0$ where the base is the Hirzebruch surface $\mF_0 = \mP^1_{(2)}
\times \mP^1_{(3)}$.  The fibrations have no reducible fibers and hence
the model has 3 K\"ahler classes: The class of the elliptic fiber $E$
and the classes of $\mP^1_{(2)}$, $\mP^1_{(3)}$, which we identify in the
$STU$ model as follows
\begin{equation}
\label{eq:classes}
t_1=U, \qquad t_2=T-U, \qquad  {\rm and} \qquad t_3=S, 
\end{equation}
respectively. 
With $\chi =-480$ and nonzero classical intersections 
$\kappa_{111}=8$, $\kappa_{112}=2$, 
$\kappa_{113}=2$, $\kappa_{123}=1$ and  $\int_X \ch_2 J_1=92$, 
$\int_X \ch_2 J_2=\int_X \ch_2 J_3=24$, the expansion of the prepotential is 
\begin{equation}
\cF^{(0)}=-{4\over 3} t_1^3-t_1^2 t_2- t_1^2 t_3 - t_1 t_2 t_3 + 
{23\over 6} t_1 + t_2 +  t_3- {480\over 2(2 \pi i)^3}\zeta(3)+O(q_i) \ .  
\end{equation}

For the $STU$ model we have 
\begin{equation} 
\Theta(q)=2 E_4 E_6
\label{eq:thetaSTU}
\end{equation} 
in~(\ref{eq:hilb})  
and to restore the $T,U$ dependence the $q^l$ powers in the expansion 
of the automorphic form  multiplying $\lambda^{2g-2}$ have to be replaced by 
\begin{equation}
\label{eq:latticespacingSTU}
q^l\rightarrow {1\over (2 \pi i)^{3- 2g}}\sum_{m n=l}{\rm Li}_{3-2g}
(e^{2 \pi i (m T+ n U)}) \ .
\end{equation}
Heterotic/Type II duality predicts that~(\ref{eq:hilb}) counts 
the invariants $n_Q^{(g)}$ for curves in the classes of 
a two-dimensional sublattice of $H^2(X,\mZ)$ generated by $\mP^1_{(2)}$ and $E$,
corresponding to the lattice $H^2({\rm K3},\mZ)$ of the K3 fiber. This 
identifies the $t_3\rightarrow -\infty$ and $q_3\rightarrow 0$ limit as 
the weak coupling limit, but otherwise the most useful geometric 
identification of the heterotic dilaton is an open question.

\subsubsection{Propagators and compatibility}
\label{sec:STUprops}

In Appendix~\ref{sec:monodr-stu} we have analyzed in part the complex structure
moduli space of the mirror of $X$, in particular the discriminants and the 
monodromies. We also computed the Yukawa couplings for this model which are
ingredients in the computation of the propagators. From these results and
from (\ref{eq:yukef0}), it is easily checked that ${\rm det}[(Y_1)_{ij}]=0$, 
while ${\rm det}[(Y_2)_{ij}]\neq 0\neq{\rm det}[(Y_3)_{ij}]$. Hence we get less 
compatibility relations of the type~(\ref{eq:comp1}), (\ref{eq:comp2}, 
(\ref{eq:comp3}). They can be fulfilled by choosing for the non-vanishing 
$A^i_{jk}$ and $A_{ij}$
\begin{align}
  A^1_{12}&=-\tfrac{1}{4 b},& 
  A^1_{13}&=-\tfrac{1}{4 c},& 
  A^2_{22}&=-\tfrac{1}{b},&
  A_{12}  &=-\tfrac{5}{144 b},\cr
  A_{13}  &=-\tfrac{5}{144 c},&
  A^3_{33}&=-\tfrac{1}{c},&
  A^2_{23}&=-\tfrac{1}{4 c},&
  A^3_{23}&=-\tfrac{1}{4 b},\cr
  A_{22}  &=-\tfrac{5}{72 a b},&
  A_{33}  &=-\tfrac{5}{72 a b},&
  A_{23}  &=\tfrac{5}{144 a}\left(\tfrac{1}{b}+\tfrac{1}{c}\right) & &
\end{align}
Note the $2 \leftrightarrow 3$ symmetry.  
The propagators then become in leading order
\begin{align}
S_{11}&=q_2 q_2 +O(q^3),& 
S_{12}&=-\tfrac{1}{4}-60 q_1 q_2-2 q_2q_3+O(q^3)  \cr
S_{22}&=\tfrac{1}{2}+240 q_1 q_3+ 4 q_2 q_3+O(q^3),&
S_{23}&=\tfrac{1}{2}+60 q_1+13320 q_1^2+ 60 q_1(q_2+q_3)+ 4 q_2 q_3+O(q^3) ,\cr
S_1&=90 q_1(q_2+q_3)+O(q^3),&
S_2&=30 q_1-150 q_1 q_2-12690 q_1^2+O(q^3),\cr
&&S&=-\tfrac{5}{2}q_1(1+(q_2+q_3)+1560 q_1^2)+O(q^4),  
\end{align}
where the remaining 
propagators follow from the  $2 \leftrightarrow 3$ symmetry.
We have checked with these propagators that we can satisfy the constraints from the 
local ${\mathbb P}^1\times {\mathbb P}^1$ and the predictions from the heterotic 
string from Mari\~no and Moore~\cite{Marino:1998pg} and get integer GV invariants. 
However, the easiest ansatz~(\ref{eq:f_g}) for the ambiguity $f_2$ has still to be 
parameterized by $6$ unknown integer GV invariants.

\subsubsection{Gopakumar--Vafa versus Donaldson--Thomas invariants}
Using the scheme~(\ref{gova}) we can construct the reduced series 
whose coefficients of $q^r$ count  e.g. $n^{(g)}_{(r,1,0)}$, or more generally curves
with $[C]^2=2 r -2$. Because of the adjunction formula $[C]^2+ [C][K]=2 g-2$ and 
the triviality of the canonical bundle, smooth curves in this class have genus $g=r$.  

\vskip 5 mm {\vbox{\footnotesize{
$$
\vbox{\offinterlineskip\tabskip=0pt \halign{\strut \vrule#& 
&\hfil~$#$ 
&\hfil~$#$ 
&\hfil~$#$ 
&\hfil~$#$ 
&\hfil~$#$
&\hfil ~$#$
&\hfil ~$#$
&\hfil ~$#$
&\hfil ~$#$
&\hfil ~$#$
&\vrule#\cr \noalign{\hrule} 
&g&r=0&1&2         &3          &4       &5       &6       &7 & 8& \cr
\noalign{\hrule}
&0&-2&480&282888&17058560&477516780&8606976768&115311621680&1242058447872&11292809553810&\cr
&1&0&4&-948&-568640&-35818260&-1059654720&-20219488840&-286327464192&-3251739174540&\cr
&2&0&0&-6&1408&856254&55723296&1718262980&34256077056&506823427338&\cr
&3&0&0&0&8&-1860&-1145712&-76777780&-2455936800&-50899848132&\cr
&4&0&0&0&0&-10&2304&1436990&98987232&3276127128&\cr
&5&0&0&0&0&0&12&-2740&-1730064&-122357100&\cr
&6&0&0&0&0&0&0&-14&3168&2024910&\cr 
&7&0&0&0&0&0&0&0&16&-3588&\cr
&8&0&0&0&0&0&0&0&0&-18&\cr
\noalign{\hrule}}\hrule}$$}}}      
The genus two predictions appeared already in~\cite{Marino:1998pg}, however 
with the opposite overall sign. Using the ideas of~\cite{Gopakumar:1998jq},~\cite{Katz:1999xq} 
we can now make a simple quantitative check of string duality to all genus in the 
type IIA description. 

The moduli space of the smooth curve $c_g$ in the class $Q=(g,1,0)$ is given 
by the product the base $\mP^1_3$ with the space parameterizing the locations 
of the $g$ points, where $c_g$ intersects $\mP^1_2$, the section of the 
elliptic K3 fibration. The moduli space of such a curve  
is ${\cal M}_0=\mP^1\times \mP^g$ and no curves of genus $g$ exist in the class $(r,1,0)$, 
if $r<g$. Hence we get
$$n^{(g)}_{(g,1,0)}=(-1)^{(g+1)} 2 (g+1),\qquad\qquad  n^{(g)}_{(r,1,0)}=0, \quad r<g $$
in agreement with the heterotic string prediction.
             
Curves with one node in the class $(g+1,1,0)$ have genus $g$. The formula
for the Euler number of moduli space with one node is~\cite{Katz:1999xq} 
$e({\cal M}_1)=e({\cal C}) + (2g-2) e({\cal M}_0)$, 
with $e({\cal M}_0)=2(g+1)$ as calculated above. ${\cal C}$ is the universal curve over 
the moduli space. In our case ${\cal C}$ fibres over the Calabi--Yau space 
$X$. Every $p\in X$ determines a fiber and the section. The choices of the
$g$ remaining points in the section gives a $\mP^{g-1}$ and so 
$e({\cal C})=-480 g$, hence 
$n^{(g-1)}_{(g,1,0)}=(-1)^{(g+1)} ( 2 (g+1) (2 g-2) - 480 g)$
or 
$$n^{(g)}_{(g+1,1,0)}=(-1)^{(g+1)} ( 4 (g+2) g - 480 g-480)\ , $$
again in perfect agreement with the heterotic prediction. For curves 
of genus $g$ in the class $(g+2,1,0)$, i.e. with two nodes, there are 
reducible components in ${\cal C}^{(2)}$. Unfortunately, this makes the 
application of~\cite{Katz:1999xq} more complicated. In particular, 
the genus $g$ BPS numbers $n^{(g)}_{(g+2,1,0)}$  for curves in the 
class $(g+2,1,0)$ do not lie on a degree $3$ polynomial in $g$.
Because of the vanishing of the $n^{(g)}_Q$ for fixed $Q$ and for 
$g>g_{max}$ we can apply (\ref{eq:GVDT}) and reorganize the partition 
function in terms of the Donaldson--Thomas invariants 
${\tilde n}^{(m)}_{Q=(r,1,0)}$. The first few 
${\tilde n}^{(m)}_{(r,1,0)}\in {\mathbb Z}$ are listed below  
\vskip 5 mm {\vbox{\footnotesize{
$$
\vbox{\offinterlineskip\tabskip=0pt \halign{\strut \vrule#& &\hfil
~$#$ &\hfil ~$#$ &\hfil ~$#$ &\hfil ~$#$ &\hfil ~$#$ &\hfil ~$#$
&\vrule#\cr \noalign{\hrule} 
&m &r=0 &1           &2              &3        & & \cr
\noalign{\hrule}
&-3&           &                     &                     &  -114                        &&\cr 
&-2&           &                     &             29      &   -93120                       &&\cr 
&-1&           &    -6               &            21120    &     -23125422                     &&\cr 
&0&     1      &     -3840           &          4959252    &      -148113736                    &&\cr  
&1&   480      &     -861918         &         222239296   &       828968673654                   &&\cr 
&2&   114000   &     -81368016       &       -96779956415  &      135202938212616                    &&\cr
&3& 17857600   &     2384015304      &    -21060955624896  &       4245487378821074                  &&\cr
\noalign{\hrule}}\hrule}$$}}} 
It is remarkable that the ${\tilde n}^{(m)}_{(r,1,0)}$ actually turn out to
be integers, and this result therefore constitutes a non-trivial check of 
the S--duality in topological strings~\cite{Nekrasov:2004js}.

\subsubsection{Symmetries at large complex structure} 

Here, we discuss the symmetries of the $STU$ model. Exchanging the two 
$\mP^1$'s corresponds to an $S\leftrightarrow T-U$ duality on the heterotic 
side, subject to the question of the best identification of the dilaton $S$. By 
construction this symmetry is realized on all amplitudes on $X$, in particular 
on $\cF^{(g)}$. In the heterotic theory there is a 
${\rm PSL(2,\mZ)}\times {\rm PSL}(2,\mZ)$ action on the complex and K\"ahler 
structure moduli $U, T$ of the torus. Furthermore, there is mirror symmetry 
$U\rightarrow T$, charge conjugation $(T,U)\rightarrow (-\bar T,-\bar U)$ and 
parity reversal $T\rightarrow -\bar T$.
The $U\rightarrow T$ symmetry is realized in the weak coupling limit as
\begin{equation}
\begin{array}{rcl}
t_1 &\rightarrow &  t_1+t_2 \\
t_2 &\rightarrow & -t_2\
\label{eq:utexchange}
\end{array}
\end{equation}
and is reflected on the BPS numbers as constraints 
$n^{(g)}_{i,j,0}=n^{(g)}_{i,i-j,0}$ and $n^{(g)}_{i,j,0}=0$ for $j\geq i$ and 
$i>0$. However, the realization among these instanton numbers is not perfect 
as there is an unpaired number $n^{(g=0)}_{0,1,0}=-2$ corresponding to a 
$-2 {\rm Li}_3(q_2)$ contribution to $\cF^{(0)}$, which has an interesting 
interpretation. 
From the analytic continuation of ${\rm Li}_3(x)$ in~(\ref{eq:Li3}) one has 
\begin{equation} 
  {\rm Li}_{3}(q_2)={\rm Li}_3 (q_2^{-1})-\frac{2\pi i}{12}
  \left(2t_2^3+3 t_2^2 + t_2\right)
\end{equation}
In the conifold flop case a similar change of the classical terms leads to
the (weak or birational) topology change. Here we have the prediction of a 
symmetry in the weak coupling limit and a linear transformation on the $t_i$ 
should be sufficient to leave $\cF^{(0)}$ in the limit invariant. 
We find that in addition to (\ref{eq:utexchange}) 
we have to transform the dilaton $t_3\sim S$ as follows 
\begin{equation}
\label{eq:dilatonshift}
t_3 \rightarrow   t_3 - 2 t_2 \ . 
\end{equation}
Now the changes in the cubic as well 
as the linear terms in $t_i$ in $\cF^{(0)}$ 
cancel. There is a shift in the quadratic term $\cF(t (t'))=
\cF(t')+{1\over 2}(t')^2$, which however is physically irrelevant 
as the Weyl-Petersson metric $G_{i\bar \jmath}$ and the $C_{ijk}$ are 
unchanged.
There are no unpaired instanton numbers at higher genus. 
However, notice that the bubbling contribution at genus 1 from the genus 0 
number $-2 {1\over 12}{\rm Li}_1(q_2)$  together with~(\ref{eq:Li1}) 
cancels precisely the change in the leading terms 
$\cF^{(1)}\sim {t_i\over 24} \int_X \ch_2 J_i$. The constant map contribution 
to $\cF^{(g)}$, $g>1$, is just proportional to the Euler number (\ref{eq:rahul})
and, consistently, the $-2 {\rm Li}_{3-2g}(q_2)$ bubbling contribution at 
$\cF^{(g)}$ is invariant under $q_2\rightarrow 1/q_2$.  

We could identify $S=t_1+t_3$, $T=t_1+t_2$
and $U=t_1$ without changing the heterotic weak coupling limit. In this
case the exchange symmetry between the two $\mP^1$, $t_2\leftrightarrow t_3$, 
corresponds to an $S \leftrightarrow T$ exchange symmetry but mirror symmetry 
$T \leftrightarrow U$ is still accompanied by a shift in the dilaton. 
It was conjectured in~\cite{Duff:1996rs} that there is a symmetry 
acting as an $S_3$ permutation on $STU$ parameter. However, it is precisely 
the shift 
of the dilaton (\ref{eq:dilatonshift}) that modifies this symmetry  
for any choice of flat coordinates, as the group $\Gamma$ generated 
by the two $\mZ_2$ elements 
\begin{align}
  W &=
\left(\begin{array}{ccc}
1&0&0\cr
0&0&1\cr
0&1&0 \end{array}
\right) &
  M &=
\left(\begin{array}{ccc}
1&1&0\cr
0&-1&0\cr
0&-2&1 \end{array}
\right)  
\label{eq:WM}\end{align}
contains with $$T=WM=\left(\begin{array}{ccc}
1&0&1\cr
0&0&-1\cr
0&1&-2 \end{array}
\right)$$ an element of infinite order $$T^n=
\left(\begin{array}{ccc}
1&(-)^n\left[n\over 2\right]&-(-)^n\left[n+1\over 2\right]\cr
0&(-)^n(1-n)&(-)^n n\cr
0&-(-)^n n &(-)^n (n+1) \end{array}
\right).$$ 
The existence of a shift symmetry in the non-perturbative
heterotic string, which is not related to the usual Peccei-Quinn
symmetry on the vector multiplet fields is remarkable and leads 
to useful restrictions on  the symplectic invariants or the 
BPS spectrum.  

\subsubsection{Symmetries on the symplectic invariants}

The afore-mentioned symmetries preserve the large volume limit. 
They identify e.g. the symplectic invariants in certain classes. 
We have already discussed them in the strict weak coupling 
limit. To analyze the non-perturbative heterotic corrections we have 
to understand some properties of asymptotic expansions of the GV 
invariants. We illustrate the point by comparing the GV invariants at 
genus zero for $d_3=0$  and $d_3=2$.
\vskip 5 mm {\vbox{\small{
$$
\vbox{\offinterlineskip\tabskip=0pt \halign{\strut \vrule#& &\hfil
~$#$ &\hfil ~$#$ &\hfil ~$#$ &\hfil ~$#$ &\hfil ~$#$ &\hfil ~$#$
&\hfil ~$#$ 
&\vrule#\cr \noalign{\hrule} 
&d_1 &d_2=0 &1           &2         &3          &4       &5       & \cr
\noalign{\hrule}
&0&    & -2          &              &               &           &        &\cr 
&1&480 &480          &              &               &           &        &\cr 
&2&480 & 282888      & 480          &               &           &        &\cr 
&3&480& 17058560     & 17058560     &  480          &           &        &\cr 
&4&480& 477516780    & 8606976768   &  477516780    & 480       &        &\cr 
&5&480& 8606976768   &1242058447872 & 1242058447872 & 8606976768& 480    &\cr
\noalign{\hrule}}\hrule}$$}}} 
For $d_3=0$ the Gopakumar-Vafa invariants fulfill $n^{(g)}_{i,j,0}=n^{(g)}_{i,i-j,0}$  
and vanish for $j>i$ except for $n^{(0)}_{0,1,0}=-2$ with the discussed 
change of  the classical terms. Note that the numbers 
depend only  on the combination $ d_2(d_1-d_2)$, which is manifest from 
the formulae (\ref{eq:classes}, (\ref{eq:latticespacingSTU}).

For $d_3=2$ we get the following GV invariants
\vskip 5 mm {\vbox{\small{
$$
\vbox{\offinterlineskip\tabskip=0pt \halign{\strut \vrule#& &\hfil
~$#$ &\hfil ~$#$ &\hfil ~$#$ &\hfil ~$#$ &\hfil ~$#$ &\hfil ~$#$
&\hfil ~$#$ 
&\vrule#\cr \noalign{\hrule} 
&d_1 &d_2=0 &1           &2         &3          &4       &5       &\cr
\noalign{\hrule}
&0&          & -6          & -32          &  -110         &  -288     & -644   & \cr 
&1&          & 2400        & 16800        &  64800        &184800     &436800  & \cr 
&2&480       & -452160     & -4093920     &-18224640      &-56992800  &-143674560 & \cr 
&3&17058560  & 103374720  &  789875200    &3955306880     &13696352640&  -37214805760& \cr 
&4&8606976768&-16013531460  &-115113738240 &-6921322409900 & -2765391112320&-8288446754100 & \cr 
\noalign{\hrule}}\hrule}$$}}}  
The symmetry $M$, which should occur for fixed $d_3=s$, is 
not manifest in the GV invariants for $d_3\neq 0$ as it was for $d_3=0$. 
However, this does not mean that it is not present. The symmetry on the indices 
of the GV invariants $n^{(g)}_{i,j,k}=n^{(g)}_{i,i-j-2k,k}$ with 
$j=0\ldots \infty$ requires a resummation of the infinite series to negative 
powers of $q_2$. The key is to choose instead of ${\rm Li}_3(\underline q)$ an
expansion scheme of the GV invariants that is manifestly symmetric under $M$. 
We therefore expand $\cF^{(0)}$ as
\begin{equation}
\cF^{(0)}=\sum_{u,s=0}^\infty q_1^{u} q_3^{s} f_{u,s}(q_2)
\label{eq:fus}
\end{equation}
where the mirror symmetry on the heterotic $T^2$ requires 
\begin{equation}
f_{u,s}(1/q)=q^{2s-u} f_{u,s}(q)\ .
\label{eq:ms}
\end{equation}
We make an ansatz for $f_{u,s}(q)={P_{u,s}(q)\over  (1-q)^{4s -2}}$, $s>0$, 
so that (\ref{eq:ms}) imposes that $P_{u,s}(q)=\sum_{k=n}^d c_{u,r,s} 
q^r$ is a finite polynomial with symmetric coefficients 
$c_{u,r,s}=c_{u,d+n-r,s}$ for $k=n\ldots d$ in $q$. The required degree 
is $d=u+2s-2 -n$. Indeed, if we organize the GV invariants according 
to~(\ref{eq:fus}) we determine the $c_{u,r,s=d_3=2}$ as
\vskip 5 mm {\vbox{\small{
$$
\vbox{\offinterlineskip\tabskip=0pt \halign{\strut \vrule#& &\hfil
~$#$ &\hfil ~$#$ &\hfil ~$#$ &\hfil ~$#$ &\hfil ~$#$ &\hfil ~$#$
&\hfil ~$#$ &\hfil ~$#$
&\vrule#\cr \noalign{\hrule} 
&u &r=0 &1           &2         &3          &4       &5       &6      & \cr
\noalign{\hrule}
&0&-1          & -18          &  -1         &       &   &    & & \cr 
&1&0           & 2400        &  2400        &     &  & & & \cr 
&2& 540     & -455400     &-13726800      &-455400  &540 & & & \cr 
&3&17058560  & 103360  &    425505280   &425505280  &103360& 17058560&&\cr 
&4&8607012129 & -67655604234  &110072604195&-413793025380 &110072604195 
&-67655604234 & 8607012129 & \cr 
\noalign{\hrule}}\hrule}$$}}} 
This establishes $M$ as symmetry in higher order corrections in the heterotic
dilaton, i.e. non-perturbative contributions to the heterotic string.

Below we  list $c_{1,r,s}$. As in the $c_{u,r,s=fixed}$ table the
definition (\ref{eq:fus}) resums an infinite number of GV invariants
to a finite number of $c_{u,r,s}$
$$
\vbox{\offinterlineskip\tabskip=0pt \halign{\strut \vrule#& &\hfil
~$#$ &\hfil ~$#$ &\hfil ~$#$ &\hfil ~$#$ &\hfil ~$#$ &\hfil ~$#$
&\hfil ~$#$ &\hfil ~$#$
&\vrule#\cr \noalign{\hrule} 
&s &r=0 &1           &2         &3          &4       &5       &6      & \cr
\noalign{\hrule}
&0& 480  & 480          &           &       &   &    & & \cr 
&1& 480  & 480         &              &     &  & & & \cr 
&2&      & 2400     &2400      &         &    & & & \cr 
&3&      & 3360     &31200     & 31200   & 3360 & & &\cr 
&4&      & 4320     &124320    & 576480 & 576480 & 124320 & 4320 & \cr 
\noalign{\hrule}}\hrule}$$ 
   
Let us give a simple example how the triality symmetry (\ref{eq:WM}) can actually be 
used to fix the Gromov--Witten invariants in some classes completely. 
We consider a non-compact Calabi--Yau ${\cal O}(-2,-2)\rightarrow \mP^1\times \mP^1$ 
which emerges in the limit of large elliptic fiber $t_1\rightarrow \infty$.
This model can be solved by various methods, e.g. local mirror symmetry, 
localization or the Chern-Simons dual. We can solve the genus zero sector 
much more efficiently from the $STU$ triality. From (\ref{eq:ms}) and the wall crossing
behaviour the ansatz for $\cF^{(0)}$ has to be 
$$\cF^{(0)}(q_2;q_3)=C_{clas.}-2 {\rm Li}_3(q_3)+ \sum_{n=1}^\infty {q_2^s\over s
^3} 
{P_{0,s}(q_3)\over (1-q_3)^{4 s-2}}\ . $$       
Here $P_{0,s}=\sum_{r=0}^{2s-2} c_{s,r} q_3^r$ is a polynomial whose 
coefficients fulfill $c_{s,r}=c_{s,d-r}$. Moreover, the symmetry $W$ just 
states that there must be a symmetric ansatz with the same $c_{s,r}$ but 
$q_2\leftrightarrow q_3$ exchanged,  
i.e. $\cF^{(0)}(q_2;q_3)=\cF^{(0)}(q_3;q_2)$. 
It is easy to see that this requirement leads to a linear system of equations 
for the $c_{s,r}$, which determines them completely. E.g.    
$$
\vbox{\offinterlineskip\tabskip=0pt \halign{\strut \vrule#& &\hfil
~$#$ &\hfil ~$#$ &\hfil ~$#$ &\hfil ~$#$ &\hfil ~$#$ &\hfil ~$#$
&\hfil ~$#$ 
&\vrule#\cr \noalign{\hrule} 
&s &r=0 &1           &2         &3          &4       &5       & \cr
\noalign{\hrule}
&0&     &          &           &       &   &    & \cr 
&1& -2  &          &          &     &  & &  \cr 
&2& -2  & -36      &          &         &    & &  \cr 
&3& -2  & -196     & -900     &     &   & & \cr 
&4& -2  & -612     & -9702    & -26376 &  &  &   \cr
&5& -2  & -1464    & -53806    &  -412868  &-852120  &  &   \cr
&6& -2  & -2980    & -210630    & -3321208  & -16716584 & -29303208 &   \cr 
\noalign{\hrule}}\hrule}$$ 
It is easy to write recursion relations and sum them up to closed formulas. 
We find e.g. that 
$c_{r,1}=2\,\left( 1 - r \right) \,\left( 2 - 2\,r - 2\,r^2 - r^3 \right)$ 
and more generally that the degree of $c_{r,m}$ in $r$ is $ 4 m$.

\subsubsection{Quasi modular properties in the non-perturbative regime}

The $\cF^{(g)}$ of the $STU$ model admit not only in strict weak coupling
limit but also for finite $S$ an expansion in terms of quasi modular
forms. We organize the potentials as $\cF^{(g)}=\sum_{m,n} 
\cF^{(g)}_{m,n}(q_U=:q) q_{U-T}^m q^n_S$. Then we find that 
$$\cF^{(g)}_{m,n}=\left({q\over \eta^{24}}\right)^{n+m} 
P_{g+6 (m+n)-1}(E_2,E_4,E_6) =:Y_{m+n} P_{g+6(n+m)-1}
(E_2,E_4,E_6)\ .$$
Here $P_d$ is a weighted homogeneous polynomial of the indicated degree, 
which can be 
calculated recursively using the modular transformations on the periods.   
$E_2$ appears only in the form $X=E_2 E_4 E_6$ and we have, for the first 
few $\cF^{(g)}_{m,n}=\cF^{(g)}_{n,m}$, 
\begin{equation}
\begin{array}{rcl}
\cF^{(0)}_{1,0} &=& \ds{ -2  Y_1 E_4 E_6 }\\[3pt]
\cF^{(0)}_{1,1} &=& \ds{ - Y_2 E_4 E_6  \left({67\over 36} E_4^3+{65 \over 36} E_6^2 + {1\over 3} X\right)}\\[15pt]
\cF^{(0)}_{1,2} &=& \ds{- Y_3 E_4 E_6 \left({7751\over 6912} E_4^6+ {3863\over 1152} E_4^3 E_6^2+ {5551\over 6912} E_6^4 +
{319\over 864} X E_4^3+{281\over 864} E_6^2 X + {X^2\over 36}\right)}\\[15pt]
\cF_{1,3}^{(0)} &=& \ds{- Y_4E_4 E_6  \biggl( 
{1519183\over 26873856} E_4^9+
{30434765 \over 895752} E_4^6 E_6^2+ 
{23643511\over 895752} E_4^3 E_6^4 +
{7040645 \over 26873856} E_6^6+ } \\[15pt]
 & & \ds{
\frac{189871}{746496} E_4^6 X+ 
\frac{245249}{373248} E_4^3 E_6^2 X +
\frac{117967}{746496} E_6^4 X+
\frac{185}{5184} E_4^3 X^2+
  \frac{151}{5184} E_6^2X^2 + 
\frac{X^3}{648} \biggr)}\\[15pt]
\cF^{(0)}_{2,2} &=& \ds{- Y_4E_4 E_6  \biggr( 
\frac{2619935}{1327104} E_4^9+ 
\frac{51044081}{3981312} E_4^6 E_6^2 + 
\frac{45565703}{3981312} E_4^3 E_6^4 + 
\frac{5007275}{3981312} E_6^6 +}\\[15pt]
& & \ds{
\frac{260449}{248832} E_4^6 X + 
\frac{348887}{124416} E_4^3 E_6^2 X + 
\frac{201265}{248832} E_6^4 X + 
\frac{151}{864} E_4^3 X^2 + 
\frac{137}{864} E_6^2X^2 + 
\frac{X^3}{108}}\biggr)\\
\end{array}\end{equation}

For the genus one case we obtain 
\begin{equation}
\begin{array}{rcl}
\cF^{(1)}_{1,0} &=& \ds{ - {1\over 6} Y_1 X }\\[12pt]
\cF^{(1)}_{1,1} &=& \ds{  Y_2 \left(
\frac{E_4^3 E_6^2}{72} - 
\frac{67 E_4^3 X}{432} - 
\frac{65 E_6^2 X}{432} - 
\frac{X^2}{24}\right)}\\[15pt]
\cF^{(1)}_{1,2} &=& \ds{ Y_3 \biggl(
\frac{85 E_4^6 E_6^2}{3456} + 
\frac{25 E_4^3 E_6^4}{1152} - 
\frac{3991 E_4^6 X}{41472} - }\\[15pt] & &\ds{
\frac{5963 E_4^3 E_6^2 X}{20736} - 
\frac{2803 E_6^4 X}{41472} - 
\frac{7 E_4^3 X^2}{144} - 
\frac{E_6^2 X^2}{24} - 
\frac{X^3}{216}\biggr)}\\[15pt]
\cF^{(1)}_{1,3} &=& \ds{ Y_4 \biggl(
\frac{146191 E_4^9 E_6^2}{5971968} + 
\frac{169577 E_4^6 E_6^4}{2985984} + 
\frac{99871 E_4^3 E_6^6}{5971968} - 
\frac{5463727 E_4^9 X}{107495424} -}\\[15pt] & &\ds{
\frac{10819933 E_4^6 E_6^2 X}{35831808} - 
\frac{2775229 E_4^3 E_6^4 X}{11943936} - 
\frac{2424725 E_6^6 X}{107495424} - 
\frac{637631 E_4^6 X^2}{17915904} -} \\[15pt] & &\ds{
\frac{794569 E_4^3 E_6^2 X^2}{8957952} - 
\frac{375599 E_6^4 X^2}{17915904} - 
\frac{791 E_4^3 X^3}{124416} - 
\frac{625 E_6^2 X^3}{124416} - 
\frac{5 X^4}{15552} \biggr)}\\[15pt] 
\end{array}\end{equation}
\begin{equation*}
\begin{array}{rcl}
\cF^{(1)}_{2,2} &=& \ds{ Y_4 \biggr(  
\frac{54847 E_4^{12}}{995328} + 
\frac{1542107 E_4^9 E_6^2}{995328} + 
\frac{10956793 E_4^6 E_6^4}{2985984} + 
\frac{3811975 E_4^3 E_6^6}{2985984} +} \\[15pt] & &\ds{
\frac{38969 E_6^8}{1492992} - 
\frac{1309 E_4^9 X}{1327104} + 
\frac{22031 E_4^6 E_6^2 X}{1327104} + 
\frac{67243 E_4^3 E_6^4 X}{3981312} -} \\[15pt] & &\ds{
\frac{385 E_6^6 X}{3981312} - 
\frac{132911 E_4^6 X^2}{1492992} - 
\frac{176713 E_4^3 E_6^2 X^2}{746496} - 
\frac{101183 E_6^4 X^2}{1492992} -} \\[15pt] & &
\ds{\frac{193 E_4^3 X^3}{7776} - 
\frac{173 E_6^2 X^3}{7776} - 
\frac{X^4}{576}\biggr)}
\end{array}\end{equation*}
Another remarkable fact is that in the function $\cF^{(0)}_1(U,T)$ in 
$\cF^{(0)}={\rm class.}+\sum_{k} \cF^{(0)}_k(U,T) q_S^k$ has also a product type
expansion very similar as in (\ref{eq:hilb}). Following  the approach of 
\cite{Henningson:1996jf} the authors of \cite{Berglund:1997eb}
derive an expression for $\cF_1(U,T)$ which can be written in the following
form
\begin{equation}
\cF_1(U,T)=2
\frac{E_4(U) E_6(U) E_4(T) E_6(T)q_U^2 q_T}{(1-q_T/q_U)^2
(\prod_{k>0}(1-q_U^k)^{24} (1-q_T^k)^{24})\prod_{m,n>0}(1-q_U^n q_T^m)^{d(n m)} } \ ,
\label{prodstu}
\end{equation}
where $\sum_{n=-1} d(n) q^n={E_4^3\over \eta^{24}}-q {{\rm d}\over {\rm d} q} 
{E_4 E_6\over \eta^{24}}$.
The methods of \cite{Henningson:1996jf} are not so closely related
to the BPS expansion. One reason is that \cite{Henningson:1996jf}
work with the modular invariant dilaton $S_{inv}$, which is not a 
flat coordinate. In this coordinate one does not expect an integer BPS
expansion. Secondly they search for a modular expression for $\cF^{(0)}_k(U,T)$ 
and not for the BPS degeneracies. However, for $\cF^{(0)}_1(U,T)$ both objections
are irrelevant, and this might be the reason that $\cF^{(0)}_1(U,T)$ can be
written in the form (\ref{prodstu}).  

\subsection{K3 fibration with Enriques fibers}
\label{sec:Enriques}

The intersection form on the middle cohomology of K3 is
$$[E_8(-1)\oplus H] \oplus [E_8(-1) \oplus H] \oplus H,$$   
where $E_8(-1)$ is the lattice defined by the negative Cartan matrix 
of $E_8$, and $H$ is the even selfdual lattice of signature 
$(1,1)$ defined by $H=\Big(\footnotesize\begin{matrix}0&1\\1&0\end{matrix}\Big)$. 
We describe the bilinear form of this even and selfdual lattice by the matrix 
$M^{IJ}_{\rm K3}$, $I,\,J=1,\dots,22$. We denote by $\Sigma^I$ a basis of 
2-cycles of the K3 satisfying $\Sigma^I\cdot \Sigma^J=M^{IJ}_{\rm K3}$, and by 
$\eta_I$ a dual basis of 2-forms satisfying $\int_{\Sigma^I} \eta_J=\delta^I_J$.
The Enriques involution $I_E$ exchanges the cohomology elements corresponding 
to the first two factors and acts by a $(-1)$ on the last factor. 
The invariant and anti-invariant lattices are 
$L_+=(E_8(-2)\oplus H(2))$ and $L_-=(E_8(-2)\oplus H(2)\oplus H)$.
The dual type II geometry $X$ of the FHSV model~\cite{FHSV} is constructed as a 
free quotient $X=({\rm K3}\times T^2)/{\mathbb Z}_2$ by simultaneously modding 
out the Enriques involution $I_E$ on the K3 and a ${\mathbb Z}_2$ action that 
inverts the $T^2$ coordinate $z$. This model has four singular fibres, 
which are Enriques surfaces and the Euler number is accordingly 
$\chi(M)=(0- 4\cdot 24)/2+ 4 \cdot 12=0$. The $(2,0)$-form $\eta$ of the 
Enriques fiber, as well as $\bar \eta$ and further 2-forms 
$\tilde \eta_i=\eta_i-\eta_{i+10}$, $i=1,\ldots,10$, are in $L_-$. 
The cohomology of $X$ is hence constructed as follows: The $(3,0)$ form is 
$\Omega= \dd z \wedge \eta$. Ten $(2,1)$-forms are 
$\Omega_i=\frac{1}{2}\dd z \wedge \tilde \eta_i $, $i=1,\ldots,10$ and one $(2,1)$-form is 
$\Omega_{11}=\dd {\bar z} \wedge  \eta $. 
The $11$ $(1,2)$-forms and the one $(0,3)$-form are constructed analogously. 
It is likewise straightforward to determine the invariant homology in terms of 
the one of the covering space. With $A$ and $B$ the standard basis of 1-cycles on 
$T^2$ we choose  the following invariant cycles 
\begin{equation}
\begin{array}{rclrclrcl}
A^i&=&A\times(\Sigma^i-\Sigma^{i+10}),&  A^{11}&=&B\times \Sigma^{21},& A^{0}&=&A\times \Sigma^{21}\\
B_i&=&B\times M_{ik}(\Sigma^k-\Sigma^{k+10}),&  B_{11}&=&A\times \Sigma^{22}, &B_{0}&=&B\times \Sigma^{22}\ .
\end{array}
\end{equation}
This choice is different from the one in~\cite{Dijkgraaf:2002ac}, because we 
want to go from the homogeneous coordinates $X^i=\int_{A^i}\Omega$ to the 
inhomogeneous coordinates $t_i=X^i/X^{0}$, $i=1,\ldots, 11$ so that 
$\tau=X^{11}/X^{0}$ is the modulus of the base and proportional to the 
heterotic dilation $S$. The intersection matrix is 
$A^i\cap B_j=2 \delta^i_{\phantom{i}j}$, $A^{11}\cap B_{11}= A^{0}\cap B_{0}=1$. 
The expansion of the $(3,0)$-form in terms of $\alpha_I,\,\beta^I$, the 
cohomology basis dual to $A^I$ and $B_I$, is 
$\Omega=X^I\alpha_I - F_I \beta^I$, where 
$X^I= \int_{A^I} \Omega$  and  $F_I = \int_{B_I} \Omega$ are the period 
integrals. From Griffith transversality 
$\Omega_I=\partial_I\Omega=k_I \Omega+\chi_I\in H^{3,0}\oplus H^{2,1}$, 
i.e. $\int_M \Omega\wedge \partial_I \Omega=0$  
($\int_M \Omega\wedge \partial_I \partial_J \Omega=0$)
with $\partial_I=\frac{\partial}{\partial X^I}$ it follows that 
$F_I=\partial_I F$ with $F={1\over 2} X^I F_I$ a homogeneous function of degree 
$2$ in $X^I$. Other direct consequences are 
$\int_{B_I} \Omega_J=\partial_I \partial_J F$ and 
$\Omega=Z^I\Omega_I$. With the invariant bases of homology and cohomology above, 
and the choice of $X^0$, we can write down the prepotential for $X$ in 
inhomogeneous coordinates
\begin{equation} 
{\cal F}=-{1 \over 2} t^s M_{ij} t^i t^j-2 (t^s)^2+t^s,
\end{equation}
where $M_{ij}$ is the matrix of the lattice $E_8(-2)\oplus H(2)$, $t^s$ is the 
modulus of the base, identified with the heterotic dilation, 
and $t_i$, $i=1,\ldots,10$ are the moduli of the classes in
the fiber. This prepotential is compatible with the heterotic weak coupling limit 
and is indeed not corrected by genus zero instantons as argued in~\cite{FHSV} 
from the self mirror property of the FHSV model and the fact that the 
hypermultiplet moduli space metric is uncorrected by 
world-sheet instantons and is given by the K\"ahler potential\footnote{This simplifications 
makes it also much easier to find the attractor points for this model \cite{Moore:1998pn}.}
\begin{equation} 
K=-\log(t^s-{\bar t}^s) -\log  M_{ij} (t^i+{\bar t}^i) (t^j+{\bar t}^j)
\end{equation}
Going through the analysis in section \ref{sec:propagators} reveals that the 
solution to the holomorphic anomaly equation does not contain any $q_s,\,q_i$ terms. 
However, this does not mean that all world-sheet instanton contributions 
vanish. This situation is analogous to the one for the two torus. In this case 
the holomorphic anomaly equation states  
$\partial_{\tau}{\bar \partial}_{\bar \tau} \cF^{(1)}(\tau,\bar \tau)=-\frac{1}{2(\tau-{\bar \tau})^2}$, 
where the last term is the Poincar\'e metric on the upper halfplane. 
This is easily solved to yield $\cF^{(1)}(\tau,\bar \tau)=-\log(\tau_2 |f(\tau)|)$. 
Modular invariance now requires $f$ to be a modular form of weight 1 and its 
behaviour at $\tau=i \infty$ fixes it to be $f=\eta(\tau)^2$. 

In the more complicated case at hand Borcherds constructs in~\cite{Borcherds:1996ab} 
a modular form of weight 4 for $O_{L_-}(\mathbb Z)^+$ on $L\otimes {\mathbb R}+i C$, where 
$L=E_8(-2)\oplus H \subset L_-$ and $C$ is an open positive cone in  
$L\otimes {\mathbb R}$. This modular form is given by
\begin{equation}
\Phi(y)=e^{ 2\pi i (\rho,y)}\prod_{r\in \Pi^+} (1- e^{2 \pi i (r,y)})^{(-1)^{(r,\rho'-\rho)} c((r,r)/2)}\ ,
\label{eq:borcherdsI}
\end{equation}  
where $y\in L \otimes {\mathbb C}$ can be thought of as a parametrization 
of the period domain of the Enriques surface 
$D^0=\left(O\setminus (\bigcup_d H_d) \right)/O_{L_-}(\mathbb Z)$, and
$O=\{\omega\in {\mathbb P}(L_-\otimes \mC)|(\omega,\omega)=0,(\omega,\bar \omega)>0\}$ 
is the period domain of the anti-invariant periods. For all $d$ with 
$(d,d)=-2$ one excludes orthogonal sets $H_d\in O$ and $r$ runs over the 
positive roots $\Pi^+$ of the fake monster Lie 
superalgebra which consists of nonzero vectors in $L_+$ of the form $(v,n,m)$, 
where $v\in L$ with norm $(r,r)=v^2+ 2 nm\ge -2$ and either $m>0$ or $m=0$ and
$n>0$. Finally, $\rho=(0,0,1)$ and $\rho'=(0,1,0)$.  
Most remarkably, the $c(n)$ are themselves Fourier coefficients of a modular
form for ${\rm SL}(2,\mathbb{Z})$
\begin{equation}
  \sum_{n} c(n) q^n = \frac{ \eta(2 \tau)^8}{\eta(\tau)^8 \eta(4 \tau)^8}=\frac{1}{q}+8+36 q+128 q^2+ 402 q^3+\ldots
  \label{eq:borcherdsII}
\end{equation}
The modular transformation properties and the discriminant were sufficient for 
Harvey and Moore to conclude that the gravitational threshold corrections to 
the heterotic string are \cite{Harvey:1996ts}
\begin{equation}
  {\cal F}^{(1)}=\log ||\frac{1} {\eta(2 t^s)^{24}\Phi(y)}||^2\ ,
  \label{eq:harveymoore}
\end{equation}
where $t^s$ is the heterotic dilaton.
The double norm in this formula has to be read as the inclusion of the metric 
and K\"ahler potential dependent terms as in (\ref{eq:f1a}). These terms 
get no $q_i,q_s$ corrections and give the 
classical contributions to ${\cal F}^{(1)}$ in the holomorphic limit. Moreover, 
due to (\ref{gova}), ${\cal F}^{(1)}$ has the expansion 
${\cal F}^{(1)}=\sum_{Q} n^{(1)}_{Q} \log(1-q^Q)$, where $q^Q$ can be identified
with $e^{2 \pi i (r,y)}$ using the identification of the mirror coordinates on K3 
surfaces~\cite{Dolgachev:1996xw}, \cite{Aspinwall:1996mn}. Due to the product form 
of (\ref{eq:borcherdsI}) and taking the holomorphic limit of (\ref{eq:harveymoore}) 
the GV invariants become the coefficients of the modular 
form~(\ref{eq:borcherdsII}), which explains their integrality. The physical reason 
requiring the product form (\ref{eq:borcherdsI}) seems to be the same as for 
the product form of $\eta(\tau)$ in the case of the two torus.    

We expect that the calculation of the M2--brane moduli space using the Hilbert
scheme of points on the fiber, which lead to (\ref{eq:hilb}) applies with some modifications 
to the case at hand. This should be checked by calculating the index on the heterotic side 
using the methods of \cite{Harvey:1995fq}\cite{Borcherds}\cite{Marino:1998pg} and
taking the limit \cite{wipII}. It is tempting to make a prediction taking into account 
that in the expansion of (\ref{eq:hilb}) only the even degrees in $q$ contribute 
to the BPS counting on $X$. We further use that the genus zero GV invariants 
are zero and have just argued that the genus one GV invariants are given by the 
expansion of (\ref{eq:borcherdsII}). With this information we can calculate 
$\Theta$ and find
$$\Theta={1\over \eta(8\tau)^8 \eta(4 \tau)^4}=q^{-3} + \frac{4}{q} + 14\,q + 40\,q^3 + 113\,q^5 +\ldots .$$
This expansion has only odd powers and hence does not give rise to genus zero 
invariants. Note moreover that the absence of the constant term is compatible 
with $\chi(X)=0$. The GV invariants at genus one are given by the even powers 
in (\ref{eq:hilb}) and reproduce (\ref{eq:borcherdsII}). In addition, we can read 
off the GV invariants at all genus. 
\vskip 5 mm {\vbox{\small{
$$
\vbox{\offinterlineskip\tabskip=0pt \halign{\strut \vrule#& 
&\hfil~$#$ 
&\hfil~$#$ 
&\hfil~$#$ 
&\hfil~$#$ 
&\hfil~$#$
&\hfil ~$#$
&\hfil ~$#$
&\hfil ~$#$
&\hfil ~$#$
&\hfil ~$#$
&\vrule#\cr \noalign{\hrule} 
&g&\frac{[C]^2}{4}=0&1&2         &3          &4       &5       &6       &7 &8& \cr
\noalign{\hrule}
&0&0&0& 0& 0& 0& 0& 0& 0& 0&\cr 
&1&1&8& 36& 128& 402& 1152& 3064& 7680& 18351&\cr 
&2&0&-8& -80& -448& -1888& -6720& -21344& -62208& -169216&\cr 
&3&0&2& 68& 688& 4280& 20176& 79480& 275904& 870464&\cr 
&4&0&0&-24& -544& -5488& -36512& -187088& -800896& -3005312&\cr
&5&0&0&3& 228& 4206& 42084& 293274& 1602704& 7349936&\cr 
&6&0&0&0& -48& -1952& -31736& -314848& -2270016& -13057760&\cr
&7&0&0&0& 4& 536& 15786& 235304& 2316272& 17140232&\cr 
&8&  0& 0&0& 0& -80& -5120& -123136& -1721984& -16834384&\cr 
&9&  0& 0& 0& 0& 5& 1040& 44848& 936968& 12475573&\cr 
&10& 0& 0&0& 0& 0& -120& -11120& -371952& -7002272&\cr
&11& 0& 0&0& 0& 0& 6& 1788& 106316& 2971880&\cr 
&12& 0& 0&0& 0& 0& 0& -168& -21280& -945904&\cr 
&13& 0& 0&0& 0& 0& 0& 7& 2828& 221898&\cr 
\noalign{\hrule}}\hrule}$$}}} 
As we argued above the anomaly equation will only carry information about 
the holomorphic-antiholomorphic mixing, while the holomorphic information
has to be fixed essentially from the modular properties. If the 
monodromies on $t^s$ and $t^i$ do not mix, the factorisation in  
(\ref{eq:harveymoore}) should hold at higher genus and the model  
would be solved. This will be investigated further in~\cite{wipII}.

\section{Other applications to dualities}
\label{sec:other_app} 

In this section we give two more applications of the techniques and results presented in the previous sections. First, we count BPS states in a type II compactification on a complete intersection \CY manifold which admits an elliptic fibration of $\tD_5$ type. Second, we extend heterotic type II duality to free toric $\mZ_2$ quotients, and we show how to construct examples of pairs of \CY manifolds with non-trivial fundamental group whose reflexive polyhedra are equal, but the corresponding \CY spaces are not self-mirror. In both cases, we illustrate that it is now possible to construct very non-trivial complete intersections satisfying a given set of desired properties.

\subsection{Counting BPS states on del Pezzo surfaces}
\label{sec:BPScount}

We start with a short review of the work~\cite{Klemm:1996hh} which studied the BPS states of exceptional noncritical strings. These noncritical strings appear in a transition which can be seen in different dual pictures as follows. In the heterotic string we consider $\tE_8$ instantons of zero size. If we compactify the theory on a circle, we can turn on Wilson lines so as to break $\tE_8$ to $\tE_d$. $\tE_d$ instantons are then believed to be identical to $\tE_8$ instantons deformed by these Wilson lines. In the context of M-theory, the configuration dual to an $\tE_8$ instanton of finite size is an M5-brane 
approaching the 9-brane at the end of the world. After the instanton has shrunk to zero size, there is a second phase where the 5-brane moves away from the 9-brane world-volume. An M2-brane stretched between the M5-brane and the 9-brane lives as a string in the common six-dimensional space-time. Since the tension for this string is proportional to the distance between the M5-brane and the 9-brane, we get a tensionless string at the transition point between the two phases. 

By the duality to F-theory, the six-dimensional theory is obtained by compactifying on a (elliptically and K3-fibered) \CY threefold $X$. The transition from having an $\tE_d$ instanton of finite size to the M5-brane departing from the 9-brane is locally the same as the transition in the K\"ahler moduli space of $X$ obtained by a blow-up of the base of the fibration $\mP^2 \to \mF_1$. In type IIB language, the tensionless string can be interpreted as a D3-brane wrapping the exceptional curve $e$ which appears after the blow-up~\cite{Seiberg:1996vs}. By compactifying the theory on a circle to five dimensions it turns out, however, that the relevant transition is not where the 2-cycle $e$ has shrunk to zero size but where an entire 4-cycle vanishes. This realizes one of the possibilities for singularities in the K\"ahler moduli space~\cite{Wilson:1992ab}. 

A shrinking 4-cycle yields an isolated canonical singularity with a crepant blow-up (for details see e.g.~\cite{Morrison:1996pp}). The 4-cycle in this case is a (generalized) del Pezzo surface $S=S_d$ of degree $d$. These are either a $\mP^1 \times \mP^1$ or the blow-up $\tBl_{p_1,\dots,p_d} \mP^2$ of $\mP^2$ at $d \leq 8$ general points. The rank of the cohomology lattice is $\rank H^{1,1}(S) = d+1$. If $S$ is embedded into a \CY manifold $X$, however, then the map $H^{1,1}(X) \to H^{1,1}(S)$ has rank $k$, $1 \leq k \leq d+1$. The sublattice of divisors which are orthogonal to the canonical divisor has the important property that it is isomorphic to the root lattice of $\tE_d$. 
  \begin{equation}
      \Lambda_d = \{ x \in H^{1,1}(S,\mZ) | x\cdot \ch_1(S) = 0 \} \cong R_{\tE_d}
  \end{equation}
  This induces a natural action of the Weyl group on $H^{1,1}(S)$. Curves of a given degree therefore fall into representations of the Weyl group of $\tE_d$. After the 4-cycle is shrunk, the resulting singularity is still characterized by the degree $d$. The possible singularity types can be realized as follows \cite{Saito:1974ab}: For $ 6 \leq d \leq 8$ they can be represented as hypersurfaces in $\mC^4$ and for $ d  = 5 $ as a complete intersection in $\mC^5$. For $ d \leq 4 $ they cannot be represented as a transverse intersection of hypersurfaces.

Finally, the number of BPS states of the exceptional non-critical string in six-dimensions is given by the number of BPS states coming from a single D3-brane wrapping the curve $e$, in other words the number of such curves. By a mirror symmetry computation, this number is given by certain world-sheet instanton numbers. Due to the properties of the homology lattice of the vanishing 4-cycle, the generating functional for the number of BPS states in the $\tE_8$ case is a modular form associated to the root lattice of $\tE_8$.
\begin{equation}
  \label{eq:Z_E8}
  Z_{\tE_8} = \frac{\Theta_{\tE_8}}{\Delta^{\frac{1}{2}}} = \frac{1}{2} \sum_{\alpha=2,3,4} \frac{\theta^8_a(\tau)}{q^{-\frac{1}{2}}\eta(\tau)^{12}} = 1 + 252q + 5130q^2 + 54760 q^3 + \dots
\end{equation}
In the $\tE_d$ case, turning on the Wilson line, the $\tE_8$ BPS states split into states which form representations of $\tE_d \times \textrm{U}(1)^{8-d}$.

As shown for the cases $6 \leq d \leq 8$ in~\cite{Klemm:1996hh}, one possibility to realize \CY manifolds admitting such singularities is to take an elliptic fibration of $\tE_d$ type over a Hirzebruch surface $\mF_1$. The reason for using this choice is that we can take the elliptic fiber of $S_d$ to be that of the \CY threefold. Here we are going to complete their analysis by providing the toric description of the $\tD_5$ case which, as just mentioned, necessarily requires a complete intersection. 

We will first determine the Hodge numbers of this complete intersection. Comparing with the $\tE_6$ case in~\cite{Klemm:1996hh} we will need one more K\"ahler class to account for the complete intersection of the fiber, hence $h^{1,1} = 6$. The requirement that the elliptic fiber be of $\tD_5$ type yields for the Euler number $\chi(X) = -2c_{\tD_5}\ch_1(\mF_1)^2 = -128$ (see~\cite{Klemm:1996ts}), and hence $h^{2,1} = 70$. 

The simplest way of constructing an elliptic fibration of $\tD_5$ type over an $\mF_1$ is to take the polyhedron $\Delta_{\mF_1}^* = \langle (-1,-1), (1,0), (0,-1), (0,1) \rangle$ of $\mF_1$ and the polyhedron of $\Delta_{\mP^3}^* = \langle (-1,-1,-1), (1,0,0), (0,1,0), (0,0,1)$ of $\mP^3$ (the ambient space of the $\tD_5$ type elliptic curve), and combine them to $\{(\Delta_{\mF_1}^*,-1,-1,-1),(0,0,\Delta_{\mP^3}^*)\}$. It can be checked (using PALP~\cite{PALP}) that this polyhedron has the required Hodge numbers. However, since it has eight vertices, five of which correspond to $(\mC^*)^5$ rescalings of the ambient toric variety, we see that only three of the six K\"ahler moduli will be realized torically. In order to get six toric K\"ahler moduli we need to deform the above polyhedron by adding three more vertices. The deformation which has the desired properties is given in the following table, in which we have collected most of the toric data in the same way as in~\eqref{eq:X_A}:
  \begin{equation}
    \begin{footnotesize}
    \begin{array}{lcrrrrrrr|rrrrrrcl}
      \multicolumn{9}{c}{ }                &   f_1&\sigma_1&c_1&  c_2&   c_3&   c_4&&\\
      D_{0,1}&&1&\um0&0&0&0&0&0&0&0&-2&0&0&0&&\\
      D_{0,2}&&0&1&0&0&0&0&0&0&0&0&-2&0&0&&\\
      D_1&&0&1&-1&-1&-1&-1&-1&0&1&0&0&0&0&&L_1\\
      D_2&&0&1&1&0&-1&-1&-1&0&1&0&0&0&0&&L_1\\
      D_3&&0&1&0&-1&-1&-1&-1&1&-1&0&0&0&0&&S_1\\
      D_4&&0&1&0&1&-1&-1&-1&1&0&0&0&0&0&&S_2\\
      D_5&&0&1&0&0&-1&-1&-1&-2&-1&1&0&0&0&&F_1\\
      D_6&&0&1&0&0&-1&1&-1&0&0&-1&1&0&0&&F_2\\
      D_7&&0&1&0&0&1&1&-1&0&0&0&-1&1&0&&F_3\\
      D_8&&0&1&0&0&1&-1&1&0&0&0&0&1&-1&&F_4\\
      D_9&&0&1&0&0&1&0&0&0&0&0&2&-2&1&&D_9\\
      D_{10}&&1&0&0&0&0&1&0&0&0&2&0&0&-1&&D_{10}\\
      D_{11}&&1&0&0&0&0&0&1&0&0&0&0&0&1&&L_2
    \end{array}
    \end{footnotesize}
\end{equation}
The Mori generators on the right-hand side of the vertical line correspond to the triangulation that is relevant for the counting of BPS states, in other words to the phase which contains the 4-cycle $S_1$ (we will comment on other triangulations below). On the top, we have labeled these by generators by $f,\dots,c_4$ which will become clear below when we discuss the geometry of this \CY manifold. In the same discussion we will relabel the divisors according to their geometric interpretation as indicated on the far right of the table. The generators of $\Delta^*_{\mF_1}$ are $\nu^*_1,\dots,\nu^*_4$, those of $\mP^3$ are $\nu^*_5,\nu^*_9,\dots,\nu^*_{11}$, and the deformation from the simple polyhedron we started with is given by the vertices $\nu^*_6$, $\nu^*_7$ and $\nu^*_8$.

Implicitly encoded in the above toric data is also the Stanley--Reisner ideal
\begin{equation}
\cI_{SR} = [D_{{1}}D_{{2}},D_{{3}}D_{{4}},D_{{5}}D_{{10}},D_{{6}}D_{{8}},D_{{6}}D
_{{9}},D_{{7}}D_{{8}},D_{{7}}D_{{11}},D_{{9}}D_{{11}}]
\end{equation}
from which we can determine the intersection numbers of $X$. For this purpose we need the basis of divisors dual to the basis of Mori generators
\begin{align}
J_{{1}} &= D_1 + D_3, & J_{{3}} &= 3D_1 + 2D_3 + D_5, & J_{{5}} &= 3D_1 + 2D_3 + D_5 + D_6 + D_7,\notag\\
J_{{2}} &= D_1, & J_{{4}} &= 3D_1 + 2D_3 + D_5 + D_6, & J_{{6}} &= 3D_1 + 2D_3 + D_5 + D_6 + D_7 - D_8, 
\end{align}
and the fact that the \CY space $X$ is defined by $4J_3J_4$. 
The non-zero intersection numbers $\kappa_{i,j,k} = J_i\cdot J_j \cdot J_k$ are
\begin{align}
\kappa_{{1,1,3}}
&=1
&
\kappa_{{1,1,4}}
&=2
&
\kappa_{{1,1,5}}
&=3
&
\kappa_{{1,1,6}}
&=2
&
\kappa_{{1,2,3}}
&=1
&
\kappa_{{1,2,4}}
&=2
\notag \\
\kappa_{{1,2,5}}
&=3
&
\kappa_{{1,2,6}}
&=2
&
\kappa_{{1,3,3}}
&=3
&
\kappa_{{1,3,4}}
&=6
&
\kappa_{{1,3,5}}
&=9
&
\kappa_{{1,3,6}}
&=6
\notag \\
\kappa_{{1,4,4}}
&=6
&
\kappa_{{1,4,5}}
&=9
&
\kappa_{{1,4,6}}
&=6
&
\kappa_{{1,5,5}}
&=9
&
\kappa_{{1,5,6}}
&=6
&
\kappa_{{2,3,3}}
&=2
\notag \\
\kappa_{{2,3,4}}
&=4
&
\kappa_{{2,3,5}}
&=6
&
\kappa_{{2,3,6}}
&=4
&
\kappa_{{2,4,4}}
&=4
&
\kappa_{{2,4,5}}
&=6
&
\kappa_{{2,4,6}}
&=4
\notag \\
\kappa_{{2,5,5}}
&=6
&
\kappa_{{2,5,6}}
&=4
&
\kappa_{{3,3,3}}
&=8
&
\kappa_{{3,3,4}}
&=16
&
\kappa_{{3,3,5}}
&=24
&
\kappa_{{3,3,6}}
&=16
\notag \\
\kappa_{{3,4,4}}
&=16
&
\kappa_{{3,4,5}}
&=24
&
\kappa_{{3,4,6}}
&=16
&
\kappa_{{3,5,5}}
&=24
&
\kappa_{{3,5,6}}
&=16
&
\kappa_{{4,4,4}}
&=16
\notag \\
\kappa_{{4,4,5}}
&=24
&
\kappa_{{4,4,6}}
&=16
&
\kappa_{{4,5,5}}
&=24
&
\kappa_{{4,5,6}}
&=16
&
\kappa_{{5,5,5}}
&=24
&
\kappa_{{5,5,6}}
&=16
\notag \\
\end{align}
Finally, the linear forms are
\begin{align}
\ch_2 \cdot J_1 &=36,
&
\ch_2 \cdot J_2 &=24,
&
\ch_2 \cdot J_3 &=92,
&
\ch_2 \cdot J_4 &=88,
&
\ch_2 \cdot J_5 &=84,
\notag \\
\ch_2 \cdot J_6 &=24,
&
\end{align}
There are two more phases whose Mori generators are:
\begin{equation}
  l^{(1)}, l^{(2)}, l^{(3)}, l^{(4)}, -l^{(6)}, l^{(5)}+l^{(6)}, l^{(1)} +  
l^{(2)} + l^{(5)} + 2 l^{(6)}
\end{equation}
and
\begin{equation}
  l^{(1)}, l^{(3)}, l^{(4)}, l^{(5)}, -l^{(5)} - l^{(6)}, l^{(2)} - 2l^{(6)},
 l^{(1)} +  l^{(2)} + l^{(5)} + 2 l^{(6)}
\end{equation}
Note that both of these Mori cones are not simplicial. The first phase is obtained by blowing down the curve $D_8 \cap D_{10} = 32c_4$ and resolving the resulting singularity by blowing up the curve $D_9 \cap D_{11}$. If we then blow down the curve $D_9 \cap D_{10}$ and resolve the resulting singularity by blowing up the curve $D_7 \cap D_{11}$, we get the last phase. All of these curves lie in the \CY manifold, hence these phases descend to phases of the \CY.

With all these data at hand, we can describe the geometrical interpretation 
of these divisors and curves. The divisors $F_i$, $i=1,\,2,\, 3,\, 4$ are all 
Hirzebruch surfaces $\mF_1$. This can be seen as follows: We claim that 
$f_1 = F_1\cdot L_1$ is the class of a fiber of $F_1=\mF_1$, and $\sigma_1 = 
F_1 \cdot S_1$ is the class of the section of $F_1$ with self-intersection 
number $-1$. This is consistent with the intersection numbers $f_1\cdot f_1 =
L \cdot L|_{F_1} = L^2F_1 = 0$, $f_1\cdot \sigma_1 = L_1 \cdot S_1|_{F_1} = 
L_1\cdot S_1\cdot F_1 = 1$, and $\sigma_1\cdot \sigma_1 = S_1 \cdot S_1|_{F_1} =
{S_1}^2F_1 = -1$. Furthermore, the canonical class of $F_1$ is $K_{F_1} = 
F_1\cdot F_1 = -2\sigma_1 - 3 f_1$ which is precisely the canonical class of 
a Hirzebruch surface $\mF_1$. 

We then set $\sigma_2 = F_2 \cdot S_1 = \sigma_1 + c_1$, $f_2 = F_2 \cdot L_1 =
 f_1 + 2c_1$, $\sigma_3 = F_3 \cdot S_1 = \sigma_1 + c_1 + c_2$, $f_3 = 
F_3 \cdot L_1 = f_1 + 2c_1 + 2c_2$, $\sigma_4 = F_4 \cdot S_1 = 
\sigma_1 + c_1 + c_2 + c_3 + 2 c_4$ and $f_4 = F_4\cdot L_1 = 
f_1+2c_1+2c_2+2c_3+4c_4$. We can check that $f_i \cdot f_i = 0$, 
$\sigma_i \cdot f_i = 1$, $\sigma_i \cdot \sigma_i = -1$, and $K_{F_i} = 
-2\sigma_i - 3f_i$. Finally, from $\ch_2\cdot F_i = -4$ we find that 
$\chi(F_i) = 4$ and $\chi(\cO_{F_i}) = 1$, hence showing that the $F_i$ are 
$\mF_1$'s for $i=1,\,2,\,3$. $F_4$ is slightly different since 
$\ch_2 F_4 = 60$ implies that $\chi({\cal O}_{F_4}) = 1$, but $\chi(F_4)=36$. 

The class of the elliptic fiber is $h=c_1+2c_2+3c_3+2c_4$, and since in 
particular the intersection number $h\cdot F_1 =1$, we can identify $F_1$ 
as the base of the elliptic fibration.

Next, we consider the divisor $S_1$. It is a rational elliptic surface, also 
known as del Pezzo surface $\tdP_9$. First, from ${S_1}^3 = 0$ and 
$\ch_2\cdot S_1 = 12$ we see that $\chi(S_1) = 12$ and 
$\chi(\cO_{S_1}) = 1$. From the previous paragraph we already know that e.g. 
${S_1}^2F_1 \not = 0$, so $S_1$ is an elliptic surface fibered over 
$\sigma\cong\mP^1$. More precisely, since $S_1\cdot \sigma < 0$, the curve 
$\sigma$ must lie inside $S_1$. Furthermore, the negative of elliptic fiber 
$h$ of $X$ intersects $\sigma$ as $(-h)\cdot \sigma|_{S_1} = 1$. On the other 
hand, the canonical class of $S_1$ is $K_{S_1} = S_1\cdot S_1 = -h$, and 
satisfies ${K_{S_1}}^2 = 0$, as it should for a $\tdP_9$. We conclude that 
the elliptic fiber of $S_1$ lies in the class $-h$. In addition, we see that 
the classes $\sigma_i = F_i \cdot S_1$ also belong to the cohomology of $S_1$. 

The divisor $S_2$ is a properly elliptic surface. Its canonical class is 
$K_{S_2} = S_2\cdot S_2 = h = -K_{S_1}$, and therefore, obviously 
${K_{S_2}}^2 = 0$. In other words, the class of the elliptic fiber of $S_2$ 
is $h$. The base of the fibration is $f_1 \cong \mP^1$ because 
$S_2\cdot f_1 < 0$ tells us that $f_1$ lies inside $S_2$, and 
$h\cdot f_1|_{S_2} = 1$ tells us that $h$ intersects $f_1$ once.

Finally, since $L_i^2\cdot D = 0$ for any divisor $D$, the divisors $L_1$ and 
$L_2$ are the fibers of two different K3 fibrations.

In order to translate to the notation used in~\cite{Klemm:1996hh}, we set
\begin{align}
  J_1 &= F, & J_2 &= D, & J_3 &= E_1, & J_4 &= E_2, & J_5 &= W_1, & J_6 &= W_2.
\end{align}
where we have two classes $E_1$, $E_2$ instead of only $E$.

Let $d_c$ be the degree of the curve $c$ in $X$. In the following tables we list all the non-vanishing instanton numbers in~\eqref{eq:rational_inst} $n_{d_f,\dots,d_{c_4}}$ with $d_f=0$ and $d_{\sigma_1}=1$ up to order $d=\sum d_i = 21$. In table~\ref{tab:(0,1,1)} we have listed the invariants for $(d_f,d_{\sigma_1},d_{c_1}) = (0,1,1)$. The three tables correspond to $d_{c_2} = 1,2,3$, respectively. In addition, we also find that $n_{0,1,1,0,0,0} = 1$ and $n_{0,1,1,4,6,4} = 1$. We see that these two invariants together with the three sums of the last columns of table~\ref{tab:(0,1,1)} precisely agree with the invariants in table 4 of~\cite{Klemm:1996hh}.
\begin{table}[htbp]
  \centering
  $
  \footnotesize
  \begin{array}{ccc}
  \begin{array}{|c|rrr|r|}
     \hline
     &0&1&2&\sum\\
     \hline
     0&1&&&1\\
     1&10&16&1&27\\
     2&1&16&10&27\\
     3&&&1&1\\
     \hline
  \end{array}
  &
  \begin{array}{|c|rrrrr|r|}
     \hline
     &0&1&2&3&4&\sum\\
     \hline
     1&&&&&&\\
     2&1&16&10&&&27\\
     3&&16&52&16&&84\\
     4&&&10&16&1&27\\
     \hline
  \end{array}
  &
  \begin{array}{|c|rrr|r|}
     \hline
     &2&3&4&\sum\\
     \hline
     3&1&&&1\\
     4&10&16&1&27\\
     5&1&16&10&27\\
     6&&&1&1\\
     \hline
  \end{array}
  \end{array}
  $
  \caption{\label{tab:(0,1,1)}{ Decomposition of $\tE_6$ reps into the $\tD_5$ reps forming the degree $(d_{c_3},d_{c_4})$ invariants for $d_{c_2} = 1,\,2,\, 3$, respectively.}}
\end{table}

Increasing $d_{c_1}$ by one we obtain table~\ref{tab:(0,1,2)} containing the invariants for $(d_f,d_{\sigma_1},d_{c_1}) = (0,1,2)$. Again, the sums in the last columns provide a complete agreement with the invariants of the $\tE_6$ case in table 5 of~\cite{Klemm:1996hh}.

\begin{table}[htbp]
  \centering
  $
  \footnotesize
  \begin{array}{ccc}
  \begin{array}{|r|rrrrr|r|}
     \hline
     &0&1&2&3&4&\sum\\
     \hline
     1&&&&&&\\
     2&1&16&10&&&27\\
     3&&16&52&16&&84\\
     4&&&10&16&1&27\\
     5&&&&&&\\
     \hline
  \end{array}
  &
  \begin{array}{|r|rrrrr|r|}
     \hline
     &1&2&3&4&5&\sum\\
     \hline
     2&&&&&&\\
     3&16&52&16&&&84\\
     4&16&200&272&52&&540\\
     5&&52&272&200&16&540\\
     6&&&16&52&16&84\\
     \hline
  \end{array}
  &
  \begin{array}{|r|rrrrr|r|}
     \hline
     &2&3&4&5&6&\sum\\
     \hline
     4&10&16&1&&&27\\
     5&52&272&200&16&&540\\
     6&10&272&660&272&10&1224\\
     7&&16&200&272&52&540\\
     8&&&1&16&10&27\cr
     \hline
  \end{array}
  \end{array}
  $
  \caption{\label{tab:(0,1,2)}{ Decomposition of $\tE_6$ reps into the $\tD_5$ reps forming the degree $(d_{c_3},d_{c_4})$ invariants 
 for $d_{c_2} = 2,\,3,\, 4$, respectively.}}
\end{table}

We can also look at doubly wound states, {\it i.e.} states with $d_{\sigma_1} = 2$. Table~\ref{tab:(0,1,2)} contains the invariants for $(d_f,d_{\sigma_1},d_{c_1}) = (0,2,2)$. The vertical sums in the last columns fully agree with the first three invariants of the $\tE_7$ case in table 3 of~\cite{Klemm:1996hh}.

\begin{table}[htbp]
  \centering
  $
  \footnotesize
  \begin{array}{cc}
  \begin{array}{|r|rrrrr|r|}
     \hline
     &0&1&2&3&4&\sum\\
     \hline
     1&&&&&&\\
     2&-2&-32&-20&&&-54\\
     3&&-32&-100&-32&&-164\\
     4&&&-20&-32&-2&-54\\
     5&&&&&&\\
     \hline
  \end{array}
  &
  \begin{array}{|r|rrrrr|r|}
     \hline
     &1&2&3&4&5&\sum\\
     \hline
     2&&&&&&\\
     3&-32&-100&-32&&&-164\\
     4&-32&-360&-480&-100&&-972\\
     5&&-100&-480&-360&-32&-972\\
     6&&&-32&-100&-32&-164\\
     \hline
  \end{array}
  \\
  \\
  \multicolumn{2}{c}{
  \begin{array}{|r|rrrrr|r|}
     \hline
     &2&3&4&5&6&\sum\\
     \hline
     4&-20&-32&-2&&&-54\\
     5&-32&-360&-480&-100&&-972\\
     6&-20&-480&-1112&-480&-20&-2112\\
     7&&-32&-360&-480&-100&-972\\
     8&&&-2&-32&-20&-54\cr
     \hline
  \end{array}}
  \end{array}
  $
  \caption{\label{tab:(0,2,2)}{ Decomposition of $\tE_6$ reps into the $\tD_5$ reps forming the degree $(d_{c_3},d_{c_4})$ invariants 
 for $d_{c_2} = 2,\,3,\, 4$, respectively.}}
\end{table}

Similar to the $\tE_8$ case in~(\ref{eq:Z_E8}), we can use the theta function for the $\tD_5$ lattice, $\Theta_{\tD_5}(q) = \frac{1}{2}\left(\theta_3(q)^5+\theta_4(q)^5\right)$, to write down a generating function for some of the numbers of BPS states in the $\tD_5$ case
\begin{equation}
  \label{eq:Z_D5}
  Z_{\tD_5} = \frac{1}{2} \sum_{\alpha=3,4} \frac{\theta_a(q)^5}{q^{-\frac{1}{2}}\eta(q)^{12}} = 1 + 52q + 660q^2 + 5440 q^3 + \dots  
\end{equation}
If we compare with the instanton expansion, we find $n_{0,1,0,0,0,0}=1$, $n_{0,1,1,2,3,2}=52$ (see Table~\ref{tab:(0,1,1)}), and $n_{0,1,2,4,6,4}=660$ (see Table~\ref{tab:(0,1,2)}). In general, one would therefore expect that $Z_{\tD_5} = \sum_k n_{0,1,k,2k,3k,2k} q^k$. We are unable to check the coefficient of the $q^3$ term since it appears at total degree 25 but we only computed up to the total degree 21. However, going back to the discussion of curves and divisors of $X$, we see that the curve $(0,0,1,2,3,2) = c_1 + 2c_2 + 3c_3 + 2c_4$ has a direct geometric interpretation: It is the anticanonical divisor $-K_{S_1}$ of the del Pezzo surface $S_1$. This is the same curve as $(0,0,1)$ in the $\tE_8$ case in~\cite{Klemm:1996hh}. The curves contributing to $Z_{\tE_8}$, or $Z_{\tD_5}$, are always of the form $\sigma-k K_{S_1}$, $k=0,1,2,\dots$, where $\sigma$ is the exceptional section of the Hirzebruch surface $F_1=\mF_1$. Therefore, we have in general
\begin{equation}
  \label{eq:Z_G}
  Z_G = \frac{\Theta_G(q)}{q^{-\frac{1}{2}}\eta(q)^{12}} = \sum_k n(\sigma-k K_{S_1}) q^k
\end{equation}
For the instanton numbers with other degrees a similar relation should hold, but with $\Theta_G(q)$ replaced by the full theta function of the root lattice of $G$, $\Theta_G(q,z)$, cf. e.g.~\cite{Hosono:2002xj}.

\subsection{Type II/heterotic duality of free quotients}
\label{sec:IIAhetDual}

In this section we first present free quotients that are both elliptically 
and K3 fibered. We also show that there are also hypersurfaces with that 
property. In Appendix \ref{sec:FreeQ} we give a list of the Hodge numbers of 
such free quotients, both in codimension one and two.

We computed $\cF^{(1)}$ in Appendix~\ref{sec:PFsub} for the (2,30) model. 
This model was shown to be a free $\mZ_2$ quotient in Section~\ref{sec:2,30}.
The result for $\cF^{(1)}$ is that
\begin{equation}
  \label{eq:F1}
  \cF^{(1)} \sim \frac{t_2}{2} + \dots
\end{equation}
has a different normalization than~(\ref{eq:f1a}). This holds for any free 
$\mZ_2$ quotient of a K3 fibration, and is confirmed by the integrality of
the instanton expansion. With the identification 
of the complexified volume of the base $t_2=4\pi i S$ with the dilaton $S$ 
of the dual heterotic string the prepotential of the Type II string on the 
above fibration is immediately compatible with the one of a perturbative 
$N=2$ heterotic string~\cite{KLM}. However, the comparison of $\cF^{(1)}$ 
is not consistent with the analysis of~\cite{AL} of type II and heterotic 
string duality in 4d, because the argument in~\cite{AL} leads to the 
requirement that $\int_X \ch_2 J_2=24$.

Heterotic--type II duality requires K3 fibered Calabi--Yau threefolds on the
type II side. The base of such fibrations is a $\mP^1$, with homogeneous
coordinates $(z_1:z_2)$. The algebraic $\mZ_n$ action on the base can be 
chosen to be of the form $(z_1,z_2) \mapsto (e^{\frac{2 \pi i}{n}}z_1,z_2)$. 
Over the fixed points the $\mZ_n$ action on the K3 must be free otherwise 
the action on the full Calabi--Yau will not be free. The only free actions 
on K3 surfaces are $\mZ_2$ involutions which restricts the possible freely 
acting groups to $\mZ_2$. Hence, the action on the base is 
$(z_1,z_2) \mapsto (-z_1,z_2)$. A rather general method to find the heterotic 
dual is the adiabatic argument of~\cite{Vafa:1995gm}. The double covers of 
our toric free quotients are very similar to a dual of the $STU$ model, 
$\mP^4_{1,1,2,8,12}[24]$, and everything goes through with the exception that 
at the end of the argument we get a lattice which is different from 
$\Gamma^{2,18}$. When we apply the fiberwise duality, then the 
monodromy-invariant part of the lattice should be the same, up to the 
$\mZ_2$ quotient. We would expect a splitting of the heterotic fiber $T^4$ 
into an invariant $T^2$ and a variable $T^2$, although the splitting is more 
complicated than in the case of the $STU$ model. The variable $T^2$ fibered 
over the $\mP^1$ will give the K3 factor on the heterotic side. Now, note 
that for the quotient $\mZ_2$ acts on the base $\mP^1$ by $-1$ on the affine
coordinate $\frac{z_1}{z_2}$. Therefore, the K3 factor gets replaced by a 
fibration of a $T^2$ over ``half'' a $\mP^1$. 

Finally, we need to know the effect of this quotient on $\cF^{(1),{\rm het}}$. 
It is well--known that there is a term in the 10d low energy effective action 
of the heterotic string of the form
\begin{equation}
  \label{eq:LEEA10}
  \int \dd^{10}{x} \sqrt{-g} e^{-\phi_{10}} \left(A (R^2 + {\rm tr} F^2)^2 + B\zeta(3)t_8 t_8 R^4\right)
\end{equation}
which comes from string tree-level scattering amplitudes~\cite{Gross:1986mw}. 
One term in this expression is of the form $(R^2)^2$. Dimensional reduction 
on the K3 factor of K3$\times T^2$ down to six dimensions yields a $N=(1,0)$ 
theory with a term in the action of the form
\begin{equation}
  \label{eq:LEEA6}
  \chi_{\rm K3} \Vol({\rm K3}) \int \dd^{6}{x} \sqrt{-g} e^{-\phi_{10}} R^2
\end{equation}
Further reduction on the $T^2$ with radii $R_1,\ R_2$ down to four dimensions 
yields a term of the form
\begin{equation}
  \label{eq:LEEA4}
  24 \Vol({\rm K3}) R_1 R_2 \int \dd^{4}{x} \sqrt{-g} e^{-\phi_{10}} R^2  
\end{equation}
Defining the dilaton as ${\rm Re} S = \Vol({\rm K3}) R_1 R_2  e^{-\phi_{10}}$ 
this term becomes
\begin{equation}
  \label{eq:LEEA4S}
  24 \int \dd^{4}{x} S R^2
\end{equation}
Since this describes a coupling to $R^2$ it contributes to $\cF^{(1),{\rm het}}$.
In the case of the free $\mZ_2$ quotient, we argued above that the K3 factor 
gets replaced by ``half'' a K3. Therefore, we would expect that the contribution 
from the $R^2$ term in the reduction to six dimensions is reduced by a factor 
of 2. This would then yield a term
\begin{equation}
  \label{eq:LEEA4Q}
  12 \int \dd^{4}{x} S R^2
\end{equation}
in four dimensions. This is in agreement with the computation from the 
type II side~(\ref{eq:F1}).

Next, we give an example for each of the two types of elliptic fibers which 
occur for the free $\IZ_2$ quotients that we discussed in 
Section~\ref{sec:FQ}. Both of these examples belong to the class of free 
quotients that are both elliptically and K3 fibered. The fiber of the first 
example is of $\tE_7$ type, i.e. a hypersurface $\mP^2(1,1,2)[4]$. 
The simplest model has Hodge numbers $(3,43)$ and is realized by the polyhedron
\begin{equation}
  \begin{footnotesize}
  \begin{array}{ccrrrrr|rrrcl}
    \multicolumn{7}{c}{ }              &c^{(1)}&c^{(2)}&c^{(3)}&&\\
    D_0    &&    1&   0&    0&   0&   0&     -4&      0&      0&&\\
    D_1    &&    1&   1&    0&   0&   0&      2&      0&      0&&\\
    D_2    &&    1&   0&    1&   0&   0&      1&     -2&      0&&\\
    D_3    &&    1&  -1&    0&   0&   0&      0&      0&      0&&\\
    D_4    &&    1&  -2&   -1&   0&   0&      1&      0&     -2&&\\
    D_5    &&    1&   0&    0&   1&   0&      0&      1&      0&&\\
    D_6    &&    1&   0&    2&  -1&   0&      0&      1&      0&&\\
    D_7    &&    1&   1&    1&   1&   2&      0&      0&      1&&\\
    D_8    &&    1&  -5&   -3&  -1&  -2&      0&      0&      1&&\\
  \end{array}
  \end{footnotesize}  
\end{equation}
where the divisor corresponding to the third point does not intersect 
the hypersurface. The non-zero intersections numbers are
\begin{align}
  \kappa_{111}&=4, & \kappa_{112}&=2, & \kappa_{113}&=2, & \kappa_{123}&=1\\
  \ch_2 \cdot J_1&=28, & \ch_2 \cdot J_2&=12, & \ch_2 \cdot J_3&=12\notag
\end{align}
The second type of elliptic fiber is a degree (2,2) hypersurface in 
$\IP^1\times \IP^1$. The simplest model has 
Hodge numbers $(4,36)$ and is realized by the polyhedron
\begin{equation}
  \begin{footnotesize}
  \begin{array}{ccrrrrr|rrrrcl}
    \multicolumn{7}{c}{ }              &c^{(1)}&c^{(2)}&c^{(3)}&c^{(4)}&&\\
    D_0    &&    1&   0&    0&   0&   0&     -2&     -2&      0&      0&&\\
    D_1    &&    1&   1&    0&   0&   0&      1&      0&      0&      0&&\\
    D_2    &&    1&   0&    1&   0&   0&      0&      0&      1&      0&&\\
    D_3    &&    1&  -1&    0&   0&   0&      1&      0&      1&      0&&\\
    D_4    &&    1&  -2&   -1&   0&   0&      0&      0&     -2&      0&&\\
    D_5    &&    1&   0&    0&   1&   0&      0&      1&      0&      0&&\\
    D_6    &&    1&   0&    0&  -1&   0&      0&      1&      0&     -2&&\\
    D_7    &&    1&   1&    1&   1&   2&      0&      0&      0&      1&&\\
    D_8    &&    1&  -1&   -1&  -3&  -2&      0&      0&      0&      1&&\\
  \end{array}
  \end{footnotesize}  
\end{equation}
This is a fibration of $\tD_5$ type over a Hirzebruch surface $\mF_1$. 
The non-zero intersections numbers are
\begin{align}
  \kappa_{112}&=4, & \kappa_{113}&=2, & \kappa_{122}&=4, & \kappa_{123}&=2\notag\\
  \kappa_{124}&=2, & \kappa_{134}&=1, & \kappa_{224}&=2, & \kappa_{234}&=1\\
  \ch_2 \cdot J_1&=24, & \ch_2 \cdot J_2&=14, & \ch_2 \cdot J_3&=12 & \ch_2 \cdot J_4&=12\notag
\end{align}
A further example stems from a close cousin of the $STU$ models discussed 
in Section~\ref{sec:STU}. It is obtained by replacing the $\tE_8$ type fiber 
by a $\tE_7$ type fiber, which can be realized by the hypersurface 
$\mP^4_{1,1,2,4,8}[16]$. This space has several candidates for a heterotic dual, 
given e.g. in~\cite{Kachru:1995wm}, \cite{Aldazabal:1996du}. In principle, 
one could use them to explicitly carry out the $\mZ_2$ quotient as indicated 
above, and study the heterotic dual. There are also many realizations of such 
free $\mZ_2$ quotients as complete intersections, see Appendix~\ref{sec:FreeQ}.
There are also free quotients of K3 fibrations which are not elliptically 
fibered e.g. the (2,30) model and the space $X_4$ in 
Appendix~\ref{sec:other_models}.

In the remainder of this section we want to compare our toric construction of
free quotients with another construction which does not refer to a toric
ambient space. As we reviewed in Section~\ref{sec:Enriques} Ferrara, Harvey, 
Strominger, and Vafa constructed in~\cite{FHSV} a \CY manifold as follows:
\begin{equation}
  X = \frac{{\rm K3} \times T^2}{(I_E,-1)}
\end{equation}
where $I_E$ is the Enriques involution. $X$ has 
$(h^{1,1},h^{1,2}) = (11,11)$ and fundamental group $\pi_1(X) = \mZ_2$. Its 
double cover $\widetilde{X}$ has $(h^{1,1},h^{1,2}) = (21,21)$. $X$ is mirror 
to itself up to discrete torsion. Due to peculiar properties of the vector 
and hyper multiplet moduli spaces at tree level, this model has no instanton 
corrections. This is a local statement and seems not to be affected by the
discrete torsion which is a global property.

With the techniques developed in the previous sections and with the package
for analyzing lattice polytopes~\cite{PALP}, it is possible to construct 
complete intersection \CY spaces in toric varieties with very similar properties. 
One can find polyhedra $\Delta_{\rm K3}$ and $\Delta_{T^2}$ and build from 
them a polyhedron $\Delta_Y$ such that
    \begin{equation}
      Y = \frac{{\rm K3} \times T^2}{\sigma}
    \end{equation}
where $\sigma$ is a free involution.  $Y$ has $(h^{1,1},h^{1,2}) = (11,11)$ and 
fundamental group $\pi_1(Y) = \mZ_2$. Its double cover $\widetilde{Y}$, 
however, has $(h^{1,1},h^{1,2}) = (19,19)$. We find two pairs of polyhedra 
$(\Delta_Y,\nabla_Y)$ in five dimensions such that $Y$ and $Y^*$ are both 
quotients by such free involutions. However, as can be seen from their double 
covers, or from their intersection rings below, or from the fact that they 
admit instanton corrections, that
they are not self-mirror in the sense that the mirror map is non-trivial.
The reason for our failure to construct a toric realization of the FHSV model
$X$ is that all nef-partitions of the free quotient models do not respect the 
factorization ${\rm K3} \times T^2$ of the polytopes for the toric ambient 
spaces. 

On the other hand, we do find free quotients of ${\rm K3} \times T^2$ 
polytopes whose nef-partitions respect the factorization, but all of 
these manifolds have $h^{0,1}(X) = 1$, hence lead to extended supersymmetry 
in 4d. Their fundamental groups and their Hodge data were already discussed 
in Section~\ref{sec:FQ}.

In the following we will give a detailed discussion of some realizations of 
the \CY spaces $Y$ mentioned above. In codimension 2 there are 58 
free quotient polyhedra admitting nef-partitions with Hodge numbers 
$(h^{1,1},h^{1,2}) = (11,11)$. For only 4 of those their mirror is also a 
free quotient and thus contained in this list. We show that two of them are 
self-mirror in the sense that the mirror construction by Batyrev and Borisov 
reviewed in Section~\ref{sec:CICY} yields the same nef partition for both $Y$ 
and $Y^*$ in the same polyhedron, i.e. in terms of~(\ref{eq:polyDN}) we have 
$\Delta = \nabla$ and $\Delta_l = \nabla_l$, $l=1,2$. We explicitly present 
these two polyhedra and some of their properties in the following subsections. 
The remaining two are ordinary mirror pairs with the additional property that
both $Y$ and $Y^*$ are free $\mZ_2$ quotients. All the other spaces have 
mirrors which are not free quotients. This is the generic situation. As a
final remark we want to point out that in contrast to the FHSV model, our two
models are self-mirror without turning on discrete torsion. It would be 
interesting to understand whether it is possible to turn on discrete torsion, 
and how this affects our picture. Unfortunately, the role of discrete torsion 
in the mirror symmetry of complete intersections in toric varieties is barely 
understood, see e.g.~\cite{Aspinwall:1994uj}.

\subsubsection{Model $\tE_7$}

We consider the product of an elliptic curve of $\tE_7$ type, {\it i.e.} a degree 4 hypersurface $E_1$ in $\mP^2_{112}$, with a K3 surface given as a hypersurface $Y$ in $\mP_{\Delta_Y^*}$ where the polyhedron $\Delta_Y^*$ is
\begin{equation}
  \label{eq:K3surface1}
  \begin{array}{c}
  \footnotesize
  \left(\begin{array}{rrrrrrrrrrr}
    -1&  \um0&  \um0&  \um1&     1&    -1&     2&  \um1&  \um1&     0&  \um0\\
     1&     1&     0&     0&    -1&     0&    -1&     0&     0&     0&     0\\
    -1&     0&     1&     2&    -1&    -1&     0&     0&     1&    -1&     0
  \end{array}\right)
  \end{array}
\end{equation}
The first seven columns denote the vertices of $\Delta_Y^*$. The Picard number of $Y$ is $\rho(Y) = 6$. This K3 surface admits two different elliptic fibrations. The one we are interested in is as follows. The first, third and fifth vertex span the polyhedron for $\mP^2_{1,1,2}$ in which we take a degree 4 hypersurface $E_2$. Finally, we take a free quotient by a $\mZ_2$ action on the product which yields the following Gorenstein cone
\begin{equation}
  \label{eq:model45}
  \begin{array}{c}
  \footnotesize
  \left(\begin{array}{rrrrrrrrrrrrrrrr}
     1&  \um0&    0&    0&    1&    0&  \um1&  \um1&  \um0&  \um0&    1&    0&    1&    0&    0&    0\\
     0&     1&    1&    1&    0&    1&     0&     0&     1&     1&    0&    1&    0&    1&    1&    1\\
     0&     0&    0&    0&   -2&    0&     0&     2&     0&     0&    0&    0&    0&    0&    0&    0\\
     0&     0&   -1&   -1&   -2&   -2&     1&     0&     0&     0&    1&    1&   -1&   -1&    0&   -1\\
     0&     0&   -2&   -1&    0&   -3&     0&     0&     0&     1&    1&    2&    0&   -1&    0&   -1\\
     0&     0&   -1&   -1&   -1&   -2&     0&     1&     0&     0&    1&    1&    0&   -1&    0&   -1\\
     0&     0&   -1&    2&    0&    0&     0&     0&     1&     0&   -1&   -1&    0&    0&   -1&    1
  \end{array}\right)
  \end{array}
\end{equation}
The \CY space we are interested in is given as a complete intersection in this ambient space. The first two lines determine the partition for the two defining equations of the complete intersection. The resulting \CY manifold has Hodge numbers $h^{1,1}=h^{2,1}=11$. We denote the points of $\Delta^*$ by $\nu^*_{01}$, $\nu^*_{02}$, $\nu^*_1$, $\nu^*_2, \dots,\nu^*_{14}$ and correspondingly the coordinates $x_{01},x_{02},x_1,x_2,\dots,x_{14}$ and the divisors $D_i=\{x_i=0\}$. In these terms, the $\mZ_2$ acts by $-1$ on the coordinates $x_1$, $x_3$, $x_5$ and $x_7$. $x_3$, $x_6$ and $x_5$ form the coordinates of the $\mP^2_{112}$ in which the torus $E_1$ lives. $x_1$, $x_{10}$ and $x_7$ are the coordinates of the $\mP^2_{112}$ in which the elliptic fiber $E_2$ of the K3 surface lives. Note that the $\mZ_2$ has the same action on both of the $\mP^2_{1,1,2}$.

The mirror polyhedron $\nabla^*$ is by~(\ref{eq:polyDN})
\begin{equation}
  \label{eq:model45dual}
  \begin{array}{c}
  \footnotesize
  \left(\begin{array}{rrrrrrrrrrrrrrrr}
     1&  \um0&    1&    1&    0&    0&    1&    0&    0&    1&    0&    0&    1&    0&    0&    0\\
     0&     1&    0&    0&    1&    1&    0&    1&    1&    0&    1&    1&    0&    1&    1&    1\\
     0&     0&   -1&    2&    1&    0&    0&   -1&    2&   -1&    0&    1&    1&    1&    1&    0\\
     0&     0&    0&    0&   -1&    1&    0&   -1&   -1&    0&   -1&    1&    0&   -1&    0&   -1\\
     0&     0&    1&    3&    0&    0&   -1&    0&    1&   -1&    1&    1&    1&    1&    1&    0\\
     0&     0&    2&   -4&    1&   -1&    0&    1&   -1&    2&   -1&   -3&   -2&   -1&   -2&    1\\
     0&     0&    1&   -1&    0&    0&   -1&    0&    0&    0&    0&    0&   -1&    0&    0&    0\\
  \end{array}\right)
  \end{array}
\end{equation}
If we denote the points by $\nu_{01}$, $\nu_{02}$, $\nu_1$, $\nu_2, \dots, \nu_{14}$, one can check that this polyhedron is the same as $\Delta^*$, and has the same partition as in~\eqref{eq:model45} if we make the following identifications on the vertices $\nu_1,\dots,\nu_{10}$ and $\nu^*_1,\dots,\nu^*_{10}$
\begin{align}
  \notag \nu^*_5 &\leftrightarrow \nu_1 & \nu^*_9 &\leftrightarrow \nu_8 & \nu^*_7 &\leftrightarrow \nu_4 & \nu^*_2 &\leftrightarrow \nu_{10} & \nu^*_1 &\leftrightarrow \nu_6\\
  \nu^*_{10} &\leftrightarrow \nu_3 & \nu^*_6 &\leftrightarrow \nu_5 & \nu^*_3 &\leftrightarrow \nu_2 & \nu^*_4 &\leftrightarrow \nu_9 & \nu^*_8 &\leftrightarrow \nu_7
\end{align}
The polyhedron $\Delta^*$ in~\eqref{eq:model45} admits 24 star-triangulations which yield twelve simplicial and twelve non-simplicial Mori cones. The twelve triangulations corresponding to the simplicial Mori cones each have nine generators, seven of which descend to the complete intersection. These triangulations form seven phases. If we combine the linear forms for the Mori basis $J_i$ in to a vector $(\ch_2\cdot J_1,\dots,\ch_2\cdot J_7)$ then we find
\begin{equation}
  \label{eq:model45c2}
  \begin{array}{@{(}c@{,\; }c@{,\; }c@{,\; }c@{,\; }c@{,\; }c@{,\; }c@{)}r}
    0&12&24&24&24&36&48&\textrm{ three times}\\
    0&12&12&24&24&24&24&\textrm{ twice}\\
    0&12&12&24&24&24&24&\textrm{ once}\\
    0&12&12&12&24&24&36&\textrm{ twice}\\
    0&12&12&24&24&24&48&\textrm{ once}\\
    0&12&12&24&24&24&48&\textrm{ twice}\\
    0&12&12&12&24&24&24&\textrm{ once}
  \end{array}
\end{equation}
If such a vector appears in more than one line, then the corresponding intersection numbers are different.

\subsubsection{Model $\tD_5$}

We consider the product of an elliptic curve of $\tD_5$ type, {\it i.e.} a degree $(2,2)$ hypersurface $E_1$ in $\mP^1\times\mP^1$, with with a K3 surface given as a hypersurface $Y$ in $\mP_{\overline{\Delta}_Y^*}$ where the polyhedron $\overline{\Delta}_Y^*$ is
\begin{equation}
  \label{eq:K3surface2}
  \begin{array}{c}
  \footnotesize
  \left(\begin{array}{rrrrrrrrrrr}
     0&  \um0&  \um1&  \um1&    -1&     0&     0&     1&  \um1&  \um1&  \um0\\
     1&     0&     2&     1&    -1&     0&    -1&     1&     0&     1&     0\\
     0&     1&     0&     1&     0&    -1&     0&    -1&     0&     0&     0
  \end{array}\right)
  \end{array}
\end{equation}
The first nine columns denote the vertices of $\overline{\Delta}_Y^*$. The Picard number of $Y$ is $\rho(Y) = 6$. This K3 surface admits two different elliptic fibrations. The fiber of the first fibration is given by a hypersurface in the del Pezzo surface $\tdP_2$ spanned the vertices 1, 3, 5, 7, and 9. The fiber of the second fibration is given by a hypersurface in another del Pezzo surface $\tdP_2$ spanned by the vertices 2, 4, 5, 6, and 8. We choose the first fibration and call the fiber $E_2$. Finally, we take a free quotient by a $\mZ_2$ action on the product which yields the following Gorenstein cone
\begin{equation}
  \label{eq:model58}
  \begin{array}{c}
  \footnotesize
  \left(\begin{array}{rrrrrrrrrrrrrrrr}
     1&  \um0&     1&  \um0&  \um0&  \um1&    1&  \um0&  \um0&  \um1&    1&  \um0&    0&    0&    0&  \um0\\
     0&     1&     0&     1&     1&     0&    0&     1&     1&     0&    0&     1&    1&    1&    1&     1\\
     0&     0&     0&     2&     2&     0&    0&     0&     0&     0&    0&     0&    0&   -2&   -2&     0\\
     0&     0&    -1&     1&     0&     1&    0&     2&     0&     0&   -1&     1&   -1&    1&    0&     1\\
     0&     0&     0&     0&     0&     1&   -1&     0&     0&     1&   -1&     0&    0&    0&    0&     0\\
     0&     0&     0&     1&     1&     1&    0&     0&     0&     0&   -1&     0&    0&   -1&   -1&     0\\
     0&     0&    -1&     1&     0&     0&    0&     1&     1&     0&    0&     0&    0&    1&    0&     1
  \end{array}\right)
  \end{array}
\end{equation}
The \CY space we are interested in is given as a complete intersection in this ambient space. The first two lines determine the partition for the two defining equations of the complete intersection. The resulting \CY manifold has Hodge numbers $h^{1,1}=h^{2,1}=11$. We denote the points of $\Delta^*$ by $\nu^*_{01}$, $\nu^*_{02}$, $\nu^*_1$, $\nu^*_2, \dots,\nu^*_{14}$ and correspondingly the coordinates $x_{01},x_{02},x_1,x_2,\dots,x_{14}$ and the divisors $D_i=\{x_i=0\}$. In these terms, the $\mZ_2$ acts by $-1$ on the coordinates $x_3$, $x_4$, $x_5$, and $x_{11}$. $x_4$, $x_5$, $x_8$, and $x_9$ form the coordinates of the $\mP^1\times\mP^1$ in which the torus $E_1$ lives. $x_2$, $x_6$, $x_7$, $x_{10}$, and $x_{12}$ are the coordinates of the $\tdP_2$ in which the elliptic fiber $E_2$ of the K3 surface lives.

The mirror polyhedron $\nabla^*$ is by~(\ref{eq:polyDN})
\begin{equation}
  \label{eq:model58dual}
  \begin{array}{c}
  \footnotesize
  \left(\begin{array}{rrrrrrrrrrrrrrrr}
     1&  \um0&   0&   1&   0&   1&   0&  \um1&   1&   1&   0&  \um1&   1&   1&   0&\um  1\\
     0&     1&   1&   0&   1&   0&   1&     0&   0&   0&   1&     0&   0&   0&   1&     0\\
     0&     0&   0&  -1&  -1&   1&   0&     0&  -1&   0&   1&     0&   0&   1&   0&     0\\
     0&     0&   1&   0&  -1&   0&  -1&     0&   0&   0&   1&     0&   0&   0&   0&     0\\
     0&     0&   0&  -1&   0&   1&   0&     1&  -1&  -1&   0&     1&  -1&   1&   0&     0\\
     0&     0&  -1&   2&   1&  -2&   1&     0&   2&   0&  -1&     0&   0&  -2&   0&     0\\
     0&     0&  -1&   1&   1&   1&   1&     1&   0&   1&  -1&     0&   0&   0&  -1&     1
  \end{array}\right)
  \end{array}
\end{equation}
If we denote the points by $\nu_{01}$, $\nu_{02}$, $\nu_1$, $\nu_2, \dots, \nu_{14}$, one can check that this polyhedron is the same as $\Delta^*$, and has the same partition as in~\eqref{eq:model58} if we make the following identifications on the vertices $\nu_1,\dots,\nu_{13}$ and $\nu^*_1,\dots,\nu^*_{13}$
\begin{align}
  \notag \nu^*_1 &\leftrightarrow \nu_{13} & \nu^*_{11} &\leftrightarrow \nu_7 & \nu^*_{13} &\leftrightarrow \nu_{10} & \nu^*_9 &\leftrightarrow \nu_9\\
  \notag \nu^*_5 &\leftrightarrow \nu_5 & \nu^*_8 &\leftrightarrow \nu_1 & \nu^*_4 &\leftrightarrow \nu_3 & \nu^*_3 &\leftrightarrow \nu_{11}\\
  \notag \nu^*_{10} &\leftrightarrow \nu_{12} & \nu^*_2 &\leftrightarrow \nu_8 & \nu^*_6 &\leftrightarrow \nu_4 & \nu^*_7 &\leftrightarrow \nu_2\\
  \nu^*_{12} &\leftrightarrow \nu_6 & & & & & &
\end{align}
The polyhedron $\overline{\Delta}^*$ in~\eqref{eq:model58} admits 16 star-triangulations which yield eight simplicial and eight non-simplicial Mori cones. The eight triangulations corresponding to the simplicial Mori cones each have nine generators, eight of which descend to the complete intersection. There they form a unique phase. If we combine the linear forms for the Mori basis $J_i$ in to a vector $(\ch_2\cdot J_1,\dots,\ch_2\cdot J_8)$ then we find
\begin{equation}
  \label{eq:model58c2}
  \begin{array}{@{(}c@{,\; }c@{,\; }c@{,\; }c@{,\; }c@{,\; }c@{,\; }c@{,\; }c@{)}}
  0&12&12&12&12&12&24&36
  \end{array}
\end{equation}

\section{Conclusions and Outlook}
\label{sec:conclusions}

We have analyzed a number of aspects of string dualities for Calabi--Yau
complete intersections in toric varieties. Toric geometry, in spite of
its limitations, makes it easy to construct and analyze large numbers
of examples and we found, compared to the hypersurface case, surprisingly
rich geometries, like new types of K3 fibrations and free quotients with
small Hodge numbers, as well as interesting new physical phenomena.
We discussed the large redundancy of the construction due to degeneracies
in polytopes and nef partitions and it turned out, for example, that the
geometry only depends on the degrees of the partitions. 

We developed efficient computational tools that combine and extend existing 
software and demonstrated the feasibility of a systematic analysis of a 
large number of examples, including the computation of the mirror map and 
of symplectic invariants for multi-parameter models. Our examples were 
found either by scanning our list of some $2\cdot10^8$ reflexive
polyhedra in 5d for manifolds with special properties, like in the 
discussion of 2-parameter K3 fibrations in section \ref{sec:regularK3}, or 
by targeted engineering of polytopes with certain characteristics, like 
elliptic fibrations of $\tD_5$ type or self-mirror free quotients that were 
constructed in section \ref{sec:other_app}.

It turned out, however, that given the large number of polyhedra, together
with the large number of nef partitions, our methods still are not very 
efficient. Improvements in efficiency can be achieved if the redundancy in
the combinatorial information can be reduced, i.e. if the following 
problems can be solved. We need a better formula for the generating 
function of the Hodge numbers, both for computational and conceptual reasons.
We found e.g. empirical evidence that the non-intersecting divisors are
independent of the triangulation, hence this information should be already
contained in the combinatorics of the Gorenstein cones. As we mentioned
above one should also find an argument that the geometry only depends on
the degrees of the nef partition. 

An obvious question is about the completeness of the list of 
Calabi--Yau families that we present here. An interesting estimate 
is given by the one parameter models. One can 
classify them to some extend by classifying 
fourth order linear differential operators with given number 
of regular singular points, a monodromy group in 
${\rm Sp}(4,\mathbb{Z})$ and integral instanton expansion. 
For three regular singular points the result is that there 
are 14 such differential equations~\cite{Chuck}. 
Thirteen can be identified directly with the Picard--Fuchs equations of 
mirrors of smooth hypersurfaces or complete intersections in 
toric varieties with Picard number one, that appear in our list. 
The last case can be obtained as one parameter restriction 
of the mirror of the $STU$ model\footnote{The first restriction leads to the 
two parameter deformation space of the mirror of a $(2,12)$ complete 
intersection in a blow-up of ${\mathbb P}^4_{1,1,1,1,4,6}$ as discussed in 
appendix \ref{sec:restrict}. We like to thank C. Doran for an e-mail 
correspondence regarding this.}. If the restriction to three singular points is dropped, 
one gets more one parameter models, though no classification has been completed. 
For some of them it is clear by construction that they come from restrictions 
of multi moduli hypersurfaces or complete intersections in toric varieties~\cite{duco}, 
but it is an interesting question whether all of them can be reached in this way. 

The approach via the classification of the monodromy problem is 
particularly general because it makes no assumptions on the geometrical realization 
of the deformation space. Since the generalized complex structure families 
are also special K\"ahler\footnote{We are grateful to N. Hitchin for pointing 
that out to us.}, as expected from the $N=2$ supergravity point of view, 
one might wonder whether those deformation spaces are also realized by or 
closely related to the models we represent. Unfortunately, there is no 
non-trivial example of those generalized deformation spaces known 
that one could check.      

The table~\ref{tab:K3} of 2-parameter K3 fibrations in section
\ref{sec:K3fibrations} gives a hint on the 
completeness of our list of Calabi--Yau families. Even though this list is
far from being complete, a classification of 2-parameter K3 fibrations in 
toric ambient spaces seems feasible due to the observations made from that 
table. The K3 fibers must necessarily have 1 parameter. Apart from those 
mentioned in Section~\ref{sec:K3fibrations} there seem to be only a few more 
in higher codimension, e.g. $\mP^5[2,2,2]$. Furthermore, it is experimentally 
observed that there seems to be an upper bound on the Euler number of toric 
Calabi--Yau threefolds and
for elliptic fibrations this is even proven~\cite{Gross:1993ab} in general. 
The tools developed for analyzing lattice polytopes~\cite{PALP} can be used 
to construct further polytopes with the desired properties, although their 
efficiency has to be improved for this purpose. At any rate, more entries in
the list are expected to give more clues for tackling a classification.

By providing the propagators of Kodaira-Spencer gravity we are 
now able to compute the higher genus Gopakumar--Vafa invariants in all classes and study 
the truly non-perturbative BPS states of the heterotic string. Beyond 
genus two this was possible in practice only up the knowledge of a 
rational function of the mirror maps. These mirror maps $z_i({\underline t})$ 
are the natural invariants of the full non-perturbative duality group $G$ in 
${\rm Sp} (2 h^{1,1}+2,\mathbb{Z})$, like the $J$-invariant is for 
${\rm SL}(2,\mathbb{Z})$, and like the $J$-invariant they have 
an integral expansion. We found very non-trivial relations between
the multi-parameter mirror maps, but in order to fix the rational 
function one needs  additional physical information about the leading order behaviour 
of the $\cF^{(g)}$ at the boundaries in the non-perturbative regime 
of the heterotic moduli space, which goes beyond the knowledge of 
the non-perturbative duality symmetries. In some non-compact limits these information 
is available using large N transitions and the resulting Chern-Simons or 
matrix model decriptions of the topological string \cite{Ghoshal:1995wm,Aganagic:2002wv,
Aganagic:2003db}. This information was used, but has 
in general not fixed the rational functions completely. In the perturbative limit 
the modular properties and simple models for the D-brane moduli spaces allowed 
us to fix the all genus information at this boundary component of the moduli 
space completely. This information is  known to be intimately tied to root multiplicities 
of generalized Kac-Moody superalgebras and the corresponding vertex 
algebras~\cite{Harvey:1995fq}. The staggering integrality properties of the 
non-perturbative BPS expansions, confirmed here to higher genus, nourishes the speculations 
that generalized structures of this type will enable us to understand the non-perturtbative 
heterotic string. However we also saw that these extensions are not uniquely fixed by the 
perturbative BPS expansion and the quest how to fix the algebraic structure of the
non-perturbative completion is open. In this context it was intriguing to see in the $STU$ model that quasi-modular forms 
appear in the higher $F^{(g)}$, i.e. in higher orders in the type II string 
coupling, as well as in higher orders of non-perturbative heterotic effects, i.e. in 
$\exp(2 \pi i S)$.

\section*{Acknowledgement}

It is a pleasure to thank V. Batyrev, J. Bryan, R. Dijkgraaf, G. Moore, Y. Ruan,
S. Stieberger, S. Theisen, D. van Straten, E. Verlinde, D. Zagier, E. Zaslow for discussions 
and in particular S. Katz for verifying various higher genus invariants. 
A. K. thanks the AEI in Golm, the ESI in Vienna and the Aspen Center for physics 
for hospitality. This work is supported in part by the FWF projects P15553 and 
P15584.

\newpage        \appendix

\section{Periods, Picard--Fuchs equations, and instanton numbers 
of the (2,30) model}
\label{sec:PF}

\subsection{The fundamental period for a toric complete intersection Calabi--Yau space}

{
\long\def\@ifundef#1#2#3{\expandafter\ifx\csname #1\endcsname\relax
  #2\else #3\fi}
\@ifundef{nc}{\newcommand{\nc}{\newcommand}}{\renewcommand{\nc}{\newcommand}}
\@ifundef{rnc}{\newcommand{\rnc}{\renewcommand}}{\renewcommand{\rnc}{\renewcommand}}
\@ifundef{l}{\nc{\l}[1]{$l_{#1}$}}{\rnc{\l}[1]{$l_{#1}$}}
\@ifundef{a}{\nc{\a}[1]{a_{#1}}}{\rnc{\a}[1]{a_{#1}}}
\@ifundef{b}{\nc{\b}[1]{b_{#1}}}{\rnc{\b}[1]{b_{#1}}}
\@ifundef{z}{\nc{\z}[1]{z_{#1}}}{\rnc{\z}[1]{z_{#1}}}
\@ifundef{m}{\nc{\m}[1]{m_{#1}}}{\rnc{\m}[1]{m_{#1}}}
\@ifundef{n}{\nc{\n}[1]{n_{#1}}}{\rnc{\n}[1]{n_{#1}}}
\@ifundef{N}{\nc{\N}[1]{\nabla_{#1}}}{\rnc{\N}[1]{\nabla_{#1}}}
\@ifundef{d}{\nc{\d}[1]{$\Delta_{#1}$}}{\rnc{\d}[1]{$\Delta_{#1}$}}
\@ifundef{v}{\nc{\v}[1]{\nu^\ast_{#1}}}{\rnc{\v}[1]{\nu^\ast_{#1}}}
\@ifundef{D}{\nc{\D}[1]{D_{#1}}}{\rnc{\D}[1]{D_{#1}}}
\@ifundef{E}{\nc{\E}[1]{E_{#1}}}{\rnc{\E}[1]{E_{#1}}}
\@ifundef{t}{\nc{\t}[1]{t_{#1}}}{\rnc{\t}[1]{t_{#1}}}
\@ifundef{nn}{\nc{\nn}{\nonumber}}{\rnc{\nn}{\nonumber}}
\@ifundef{Res}{\nc{\Res}{\operatorname{Res}}}{\rnc{\Res}{\operatorname{Res}}}
\@ifundef{dt}{\nc{\dt}{\frac{d\t1d\t2d\t3d\t4d\t5}{\t1\t2\t3\t4\t5}}}
             {\rnc{\dt}{\frac{d\t1d\t2d\t3d\t4d\t5}{\t1\t2\t3\t4\t5}}}
\@ifundef{zpif}{\nc{\zpif}{\frac{1}{(2\pi
      i)^5}}}{\rnc{\zpif}{\frac{1}{(2\pi i)^5}}}
\@ifundef{bra}{\nc{\bra}[0]{\langle}}{\rnc{\bra}[0]{\langle}}
\@ifundef{ket}{\nc{\ket}[0]{\rangle}}{\rnc{\ket}[0]\rangle{}}
\@ifundef{IP}{\nc{\IP}[1]{\bra #1 \ket}}{\rnc{\IP}[1]{\bra #1 \ket}}
\def\<#1\>{\bra #1 \ket}
\@ifundef{B}{\nc{\B}[1]{|#1|}}{\rnc{\B}[1]{|#1|}}
\@ifundef{P}{\nc{\P}[0]{\varpi}}{\rnc{\P}[0]{\varpi}}

We start our discussion with the construction of the fundamental
period $\varpi$ of a toric complete intersection Calabi--Yau manifold.
This is a straightforward generalization of the construction for
hypersurfaces as is explained in detail e.g. in~\cite{Cox:vi}.  
Given the cycle $\gamma \subset \mT$ defined by $|t_1| = \dots = |t_d|=1$ 
and the Laurent polynomials $f_l(t)$ in~(\ref{eq:CIeqs}),
this period integral is
\begin{equation}
  \label{eq:pairingCICY}
  \varpi = \int_{\gamma'} \Omega =\frac{1}{(2\pi i)^d}\int_\gamma
   \left(\prod_{l=1}^{r}\frac{a_{0,l}}{f_l}\right)
  \frac{dt_1}{t_1}\wedge\dots\wedge\frac{dt_d}{t_d},
\end{equation}
where $\gamma'$ is the restriction of $\gamma$ to $X^*$.
The set $A_l$ of lattice points of $\nabla_l$ can be reduced to those 
points which actually correspond to the complete intersection. Recall
that in the hypersurface case one can exclude all
interior points of facets of $\Delta$. In the present situation, we
have seen that there are even divisors corresponding to vertices that 
do not intersect $X^*$, so we can exclude those vertices as well. 
They can be found after performing a careful analysis of the 
intersection ring. Using the expansions
\begin{equation}
  \frac{a_{0,l}}{f_l}=\frac{1}{1-\sum_{m\in A_l} a_m(-a_{0,l})^{-1}t^m}
  =\sum_{K_l=0}^\infty\left(\sum_{m\in A_l} a_m(-a_{0,l})^{-1}t^{m}
   \right)^{K_l}
\end{equation}
and further expand the expressions $(\dots)^{K_l}$ as
\begin{equation}
  -l_{0,l}:= K_l =\sum_{m\in A_l} l_m.
\end{equation}
where the powers $l_m$ of the monomials $t^m$ are partitions of $K_l$.
It is clear that the integral in~(\ref{eq:pairingCICY}) gets
a non--zero contribution if and only if the vectors
\begin{align}
  \label{eq:relCICY}
  \begin{aligned} 
  l=(l_{0,1},\dots,l_{0,r};l_1,\dots,l_s) && \mathrm{with} &&
  s+r=\sum_{l=1}^{r}\#A_l &&\mathrm{and}&&l_i \in \mZ_\geq0
  \end{aligned}
\end{align}
are relations of $\Gamma(\nabla_l) \cap \overline{N}$, where 
$\Gamma(\nabla_l) \subset \overline{N}_{\mR}$ is the Gorenstein 
cone of the nef partition $\Pi(\nabla)=\{\nabla_1,\dots,\nabla_l\}$ 
($\Gamma(\nabla_l)\cap \overline{N}$ is constructed by the lattice 
points of $e_l\times A_l \in \overline{N} = \mZ^r \oplus N$ 
$(l=1,\dots,r)$, cf. also Section~\ref{sec:CICY}). 
Here is another instance of the fact mentioned in Section~\ref{sec:CICY} 
that we actually need a triangulation $T=T(\Gamma(\Delta^*))$ of the 
Gorenstein cone to determine a basis for the relations $l$.
Thus we get
\begin{equation}
  \label{eq:sumpairingCICY}
  \varpi_0 =\sum_{l_1,\dots,l_s}
  \frac{(-l_{0,1})!\dots (-l_{0,r})!}{l_1!\dots l_s!}
  (-a_{0,1})^{l_{0,1}}\dots (-a_{0,r})^{l_{0,r}}a_1^{l_1}\dots a_s^{l_s},
\end{equation}
We choose the basis~\eqref{eq:genmori} for the Mori generators and 
introduce torus invariant coordinates:
\begin{equation}
  z_b=\prod_{l=1}^{r}{a_{0,l}}^{l^{(b)}_{0,l}}
  \prod_{i=1}^{s}{a_{i}}^{l^{(b)}_{i}}.
\end{equation}
Writing each relation $l$ as $\sum_{a=1}^{h} n_a l^{(a)}$, we end up with
\begin{equation}
  \label{eq:finaly0CICY}
   \varpi=\sum_{n_1,\dots,n_h}{}\left[
   \frac{\prod_{l=1}^{r}\left(-\sum_{a=1}^{h}n_a l^{(a)}_{0,l}\right)!}
   {\prod_{j=1}^{s}\left(\sum_{a=1}^{h}n_al^{(a)}_{j}\right)!}
   \prod_{a=1}^h
   \left((-1)^{\sum_{l=1}^{r}l^{(a)}_{0,l}}z_a\right)^{n_a}\right].
\end{equation}
Note that we have to restrict the sum over the integers $n_a$ to those
linear combinations of the relations that are of the form
(\ref{eq:relCICY}). With the abbreviation $z_i {d\over dz_i}=:\theta_i$ the set of
Picard--Fuchs differential operators is generally given by 
\begin{equation}
  \mathcal L_a = \prod_{l_i^{(a)} > 0} \prod_{j=0}^{l_i^{(a)} - 1} 
  \left(\sum_{b=1}^h l_i^{(b)} \theta_b -j\right) 
  - \prod_{l_i^{(a)} < 0} \prod_{j=0}^{-l_i^{(a)} - 1} 
  \left(\sum_{b=1}^h l_i^{(b)} \theta_b -j\right)z_a
  \label{eq:DL}
\end{equation}
Once we have determined the fundamental period (\ref{eq:finaly0CICY})
and the Picard--Fuchs operators~(\ref{eq:DL}), we can determine the Yukawa 
couplings~(\ref{eq:specialgeom}) and the mirror map~(\ref{eq:mirrormap}) 
in the same way as for toric hypersurfaces. In the next subsection, 
we explicitly demonstrate this considering as example the (2,30) model 
introduced and discussed in Sections~\ref{sec:2,30} and~\ref{sec:equivalence}.

\subsection{Periods, Picard--Fuchs equations, and instanton numbers 
of the (2,30) model}
\label{sec:PFsub}

Let us first calculate explicitly the fundamental period $\varpi(z_1,z_2)$ 
of the mirror of $X_{(A)}$, the first realization (\ref{EQdeltaA}) of 
the (2,30) model with nef partition
$\E1=\{\v1,\v3,\v4,\v8\},\ \E2=\{\v2,\v5,\v6,\v7\}$. The polytopes
$\N1=\IP{\{0\},\E1}$ and $\N2=\IP{\{0\},\E2}$ defined before 
(\ref{eq:polyDN}) determine the mirror as a complete intersection in terms
of the Newton polynomials as in~(\ref{eq:CIeqs})  
\begin{eqnarray}
f_1(t) &=& 1 - \sum\limits_{i:\v{i}\in E_1}\a{i} t^{\v{i}} =
1 - \a1\t1-\a2\t3-\a3 \t4-\frac{\a4}{\t1} = 1-p_1(t)\nn\\
f_2(t) &=& 1 - \sum\limits_{i:\v{i}\in E_2}\b{i} t^{\v{i}} =
1 -
\b1\t2-\frac{\b2}{\t1^2\t2\t3\t4}-\b3\t1\t2\t3\t5^2-
\frac{\b4}{\t1\t2\t3\t5^2}= 1-p_2(t)\nn.
\end{eqnarray}
The relevant points of the $\N{l}$ are 
$\E{l}\cup\{0\}\ (l=1,2)$. The origin $\{0\}$ contributes as monomial 
$\a0t^0=\a0$ ($\b0t^0=\b0$) which gives $1$ after rescaling the Newton
polynomials. Plugging the basis of relations (\ref{mori230A}) in
(\ref{eq:finaly0CICY}) yields for the fundamental period
\begin{equation}
  \label{eq:period2_30}
   \P(\z1,\z2) = {\sum\limits_{\n1,\n2\geq 0}
       \frac{(4\n1)!(2(\n1+\n2))!}{(\n1!)^4(2\n1)!(\n2!)^2}
       \z1^{\n1}\z2^{\n2}}.
\end{equation}}
with
\begin{align}
  z_1&=\frac{a_1^2a_2a_3a_4a_5}{a_{0,1}^4a_{0,2}^2},&
  z_2&=\frac{a_6a_7}{a_{0,2}^2}.
\end{align}
{\def\L#1.#2{l^{(#1)}_{#2}} \def\H#1.#2{\hat{l}^{(#1)}_{#2}}
In this example, the Picard--Fuchs operators~(\ref{eq:DL}) factorize into
$-2\theta_1(2\theta_1-1)(\theta_1+\theta_2)\mathcal{D}_1=
\mathcal{L}_1-2\theta_1(2\theta_1-1)
\theta_1^2\mathcal{L}_2$ and $\mathcal{D}_2=\mathcal{L}_2$ so that we end up
with the two reduced differential operators
\begin{eqnarray}
  \mathcal{D}_1 &=&
  \theta_1^2(\theta_2-\theta_1)+(2\theta_1+2\theta_2-1)(8(4\theta_1-1)
  \label{pf230I}
  (4\theta_1-3)z_1-2\theta_1^2z_2)\nonumber\\
  \mathcal{D}_2 &=& \theta_2^2-2(\theta_1+\theta_2)(2\theta_1+2\theta_2-1)z_2
\end{eqnarray}
annihilating $\varpi(z_1,z_2)$.}
The solutions can be easily characterized by the topological triple 
couplings~(\ref{230cijk}) and the Mori generators~(\ref{mori230A}) of 
$X_{(A)}$ as discussed in Section~\ref{sec:genus-zero}. The first solution 
is of course $X^0=\varpi$ given in~\eqref{eq:period2_30}. The next two 
solutions are
$$
\begin{array}{rl}
  X^1=\varpi_1(z_1,z_2)&=\frac{\varpi}{2\pi i} (\log(z_1) + \sigma_1)\\
  X^2=\varpi_2(z_1,z_2)&=\frac{\varpi}{2\pi i} (\log(z_2) + \sigma_2)\\
\end{array}
$$
For the present example, the integral expansion of the instanton 
contribution~(\ref{eq:rational_inst}) with respect to $(d_1,d_2)$ yields: 
\vskip 5 mm {\vbox{\footnotesize{
\[
\vbox{\offinterlineskip\tabskip=0pt \halign{\strut \vrule#& &\hfil
~$#$ &\hfil ~$#$ &\hfil ~$#$ &\hfil ~$#$ &\hfil ~$#$ &\hfil ~$#$
&\hfil ~$#$ &\hfil ~$#$
&\vrule#\cr \noalign{\hrule} 
&d_1 &d_2=0 &1           &2         &3          &4       &5       &6      & \cr
\noalign{\hrule}
&0 &    &            8&         &           &        &        &       &\cr 
&1& 384 &1088         &      384&           &        &        &       &\cr 
&2&4688 &117088       &   247680&     117088&    4688&        &       &\cr 
&3&146816&    12092928& 84309504&  148640576&84309504&12092928& 146816&\cr 
&4& 5462064& 1205851824         &   20072874752        &  86051357872         &
  135328662848      &   86051357872     & 20072874752      &\cr 
\noalign{\hrule}}\hrule}\]}}}
The discriminant of the Picard--Fuchs system has two components
\begin{align}
  \Delta_1&=(1-a)^2- b(2+2 a-b),& \Delta_2&=1-b\ , \label{dis230II}
\end{align}
where we rescaled $a=256 z_1$ and $b=4 z_2$. We have also calculated the 
triple intersection couplings and found
\begin{align}
    Y_{111}&=2\,\frac{1+a-b}{a^3 \Delta_1 } &
    Y_{112}&=2\,\frac{1-a+b}{a^2 b \Delta_1}\notag\\
    Y_{122}&=2\,\frac{3-a+b}{ab\Delta_1\Delta_2} & 
    Y_{222}&=2\,\frac{1-a+b(6+b-a)}{b^2 \Delta_1  \Delta_2^2}
  \label{yuk230I}
\end{align}
We calculate $\cF^{(1)}$~\cite{Bershadsky:1993ta},~\cite{Berglund:1995gd} with the 
boundary conditions at infinity $\cF^{(1)}\sim {20\over 24} t_1+ {12\over 24} t_2+O(q)$
and $\cF^{(1)} \sim {1\over 12} \log(\Delta_1)$, $\cF^{(1)}\sim -{21\over 12}\log(\Delta_2)$ at
the corresponding discriminants.  The integrality of the Gromov--Witten 
invariants for this choice confirms the different normalization of the 
classical terms, found in subsection~\ref{sec:X_A}, in $\cF^{(1)}$ with respect 
to the K3 fibration case. 
\vskip 5 mm 
{\vbox{\footnotesize{
$$\hss
\vbox{\offinterlineskip\tabskip=0pt \halign{\strut \vrule#& &\hfil
~$#$ &\hfil ~$#$ &\hfil ~$#$ &\hfil ~$#$ &\hfil ~$#$ &\hfil ~$#$
&\hfil ~$#$ &\hfil ~$#$
&\vrule#\cr \noalign{\hrule} 
&d_1 &d_2=0 &1           &2         &3          &4       &5       &6      & \cr \noalign{\hrule}
&0 &    &             &         &           &        &        &       &\cr 
&1& -132 &-704 &    -132&           &        &        &       &\cr 
&2& -1714 & -186112      &  -459736  &   -186112&           -1714  &  &     &\cr 
&3&-166656&  -36828672   &   -335463120 &    -634764672         &   -335463120       & -36828672&   -166656     &\cr 
&4&-9733904 &   -5892361760   &      -132167284928    &  -647978879168    &  -1060524299520  &-647978879168 &  -132167284928 & \cr 
\noalign{\hrule}}\hrule}\hss$$}}}

As we mentioned before the large complex structure variables defined by the 
polyhedron can differ even for topologically equivalent families. In particular, the 
Picard--Fuchs equations for the variables $(\tilde z_1,\tilde z_2)$ 
defined by (\ref{eq:mori230C}) are formally different from (\ref{pf230I}), namely
\begin{equation}
\begin{array}{rl}
\tilde{\cal D}_1&=\tilde \theta_1^2(\tilde \theta_1-\tilde \theta_2)- 8 (4 \tilde 
\theta_1 - 3) (4 \tilde \theta_1 - 1)
(2 \tilde \theta_1+2\tilde \theta_2-1)\tilde z_1\\
\tilde{\cal D}_2&=\tilde \theta_2^2 - 2 (\tilde \theta_1 - 
\tilde\theta_2+1)(2\tilde\theta_1 + 2\tilde\theta_2-1) \tilde z_2\ .
\label{pf230II}
\end{array}
\end{equation}
The triple intersections are 
\begin{align}
    Y_{111}&=2\,{1+\tilde a\over \tilde a^3 \tilde \Delta_1}, &  
    Y_{112}&=2\,{1-\tilde a\over {\tilde a}^2 \tilde b \tilde\Delta_1}\notag\\
    Y_{122}&=2\,{1+\tilde a \over \tilde a \tilde b  \tilde \Delta_1\tilde \Delta_2}, &
    Y_{222}&=2\,{1-\tilde a\over {\tilde b}^2 \tilde \Delta_1 \tilde\Delta_2}
    \label{yuk230II}
\end{align}
with 
$$\tilde \Delta_1= (1-\tilde a)^2- 4 \tilde a \tilde b,\qquad \tilde\Delta_2=1+\tilde b$$
As the A-models are topologically equivalent we should find a rational
transformation of variables preserving the large complex structure limit 
$(z_1=0,z_2=0)\mapsto ({\tilde z}_1=0,{\tilde z}_2=0)$ and identifying 
(\ref{pf230I}) with (\ref{pf230II}). To find this transformation we
make the following ansatz:
\begin{align}
  \frac{\Delta_1(a,b)}{P(a,b)}=\tilde \Delta_1(\tilde a,\tilde b)
  &&\mathrm{and} &&
  \frac{\Delta_2(a,b)}{P(a,b)}=\tilde \Delta_2(\tilde a,\tilde b),
\end{align}
where $P(a,b)$ is some rational function and 
$\tilde a = 256\tilde z_1$, $\tilde b = 4 \tilde z_2$. 
Solving for $\tilde a$ and $\tilde b$ gives:
\begin{eqnarray}\label{eq:trans230}
 \tilde b &=& \frac{-1-P+b}{P},\nonumber\\
 \tilde a &=& \frac{-4-2P+4b\pm 2\sqrt{
  4(-1+b)^2-P(-3+a^2+2b+b^2-2a(1+b))}}{2P}.\nonumber 
\end{eqnarray}
This transformation is rational if and only if we can get rid of the
square root. Setting
\begin{equation}
  P=-(b-1)^2
\end{equation}
completes the square in the expression under the square root and we
find (for the plus sign in (\ref{eq:trans230})):
\begin{equation}
  {\tilde a}={a\over {1-b}},\qquad {\tilde b}={b\over {1-b}}\ , 
  \label{variableident}
\end{equation}
It is straightforward to check that~(\ref{yuk230II}) transform by 
(\ref{variableident}) into (\ref{yuk230I}). Here we used 
the transformation property that $Y_{x_i,x_j,x_k}$ transforms in ${\rm Sym}[ 
(T^*M)^3] \otimes {\cal L}^2$, i.e. $Y_{x_i,x_j,x_k}= {l_x\over l_z} \sum_{a,b,c} 
{\partial z_a \over \partial x_i} {\partial z_b\over\partial x_j}  
{\partial z_c\over\partial x_k} Y_{z_a,z_b,z_c}$, where $l_x,l_z$ correspond to 
choices of sections in  ${\cal L}^2$. In our case we have 
${\tilde l\over l}={1 \over 1-b}$.          

This phenomenon is not special to complete intersections, it can also occur for 
hypersurfaces. We list here a few examples of topologically equivalent 
realizations of well-known hypersurfaces, some of which have made their 
appearance in Section~\ref{sec:regularK3}.
\begin{align*}
        \mP^4_{1,2,2,3,4}[12] &\cong
        \IP\left(\Msize\BM 2~1~1~1~1~0\cr1~2~1~1~0~1\cr\EM\right)   
        \left[\Msize\BM6\cr6\EM\right] & (h^{1,1},h^{2,1}) &= (2,74)\\ 
        \mP^4_{1,1,2,2,2}[8] &\cong
        \IP\left(\Msize\BM 1~1~1~0~1~0\cr1~1~0~1~0~1\cr\EM\right)   
        \left[\Msize\BM4\cr4\EM\right] & (h^{1,1},h^{2,1}) &= (2,86)\\ 
        \mP^4_{1,1,2,2,6}[12] &\cong
        \IP\left(\Msize\BM 3~1~1~0~1~0\cr3~1~0~1~0~1\cr\EM\right)   
        \left[\Msize\BM6\cr6\EM\right]  & (h^{1,1},h^{2,1}) &= (2,128)\\
        \mP^4_{1,1,2,8,12}[24] &\cong
        \IP\left(\Msize\BM 6~4~1~0~1~0\cr6~4~0~1~0~1\cr\EM\right)   
        \left[\Msize\BM12\cr12\EM\right] & (h^{1,1},h^{2,1}) &= (3,243)
\end{align*}
In each example, the Mori cones of the two \CY threefolds are different.
In the first example, the corresponding polyhedra arise from different blowups 
of the same simplex associated to $\mP^4_{1,1,1,1,2}$. In the other examples,
the reflexive section $\mP^3_{w'}$ of the K3 fiber is extended in two different 
ways by two additional points yielding the $\mP^1$ base of the K3 fibration. 
The Picard--Fuchs equations as well as 
the triple intersections can be mapped onto each other in a similar way as 
above. For the last example this is shown in detail in 
Appendix~\ref{sec:monodr-stu}. We observe that in all these examples the number 
of vertices and points of $\Delta$ is the same, while a single vertex is added 
to the simplex $\Delta^*$.

\section{A selection of other models}
\label{sec:other_models}

In this section we will present a few more codimension two complete 
intersection \CY manifolds, but without going into so much detail as in the 
last section. The selection contains manifolds with the next smallest Hodge 
numbers after (2,30): These are (2,36) and (2,44). The latter is particularly 
interesting since it has realizations as both a simply connected space and a 
free $\mZ_2$ quotient.

In Section~\ref{sec:nef} we mentioned that there are nine polyhedra admitting
nef-partitions giving Hodge numbers (2,44). The CICYs obtained from first 
five polyhedra are         
\begin{eqnarray}
X_1\MN M:272 12 N:8 7 &\sim&\IP\left(\Msize
        \BM 2 & 1 & 1 & \mathbf 1 & 1 & 0 & 0 \cr  
            0 & 1 & 1 & \mathbf 3 & 3 & 2 & 2 \cr \EM\right) 
        \left[\Msize\BM4\cr6\EM\VL\BM2\cr6\EM\right]\\
        \label{eq:X1}
X_2\MN M:294 13 N:9 8 &\sim&\IP\left(\Msize\BM 2&1&1&1&1&0&0&0\cr
        0&1&1&1&3&2&2&0\cr
        0&0&0&0&1&0&0&1\cr\EM\right) 
        \left[\Msize\BM   4\cr4\cr0  \EM\VL\BM   2\cr6\cr2   \EM\right]\\
        \label{eq:X2}
X_3\MN M:298 16 N:9 8   &\sim&\IP\left(\Msize\BM  2&1&1&3&\mathbf1&0&0&0 \cr  
        0&1&1&3&\mathbf3&2&2&0 \cr 0&0&0&1&\bf0&0&0&1 \cr \EM\right) 
        \left[\Msize\BM 6\cr6  \cr2  \EM\VL\BM 2\cr6  \cr0   \EM\right]\\
        \label{eq:X3}
X_4\MN M:232 10 N:9 7   &\sim&\IP\left(\Msize\BM  2&\mathbf1&1&1&1&0&0&0  \cr
        4&\mathbf2&2&2&0&1&1&0  \cr     1&\mathbf0&0&0&0&0&0&1  \cr \EM\right)
        \left[\Msize\BM 4\cr8\cr2  \EM\VL\BM 2\cr4\cr0   \EM\right]
        ~/~ \IZ_2: \hbox{ \Msize 1 1 0 1 0 1 0 0}\\
        \label{eq:X4}
X_5\MN M:232 12 N:9 8 &\sim&\IP\left(\Msize\BM    2&1&\mathbf1&1&1&0&0&0  \cr
        2&1&\mathbf1&0&0&1&1&0  \cr     1&0&\mathbf0&0&0&0&0&1  \cr\EM\right) 
        \left[\Msize\BM 4\cr4\cr2  \EM\VL\BM  2\cr2\cr0   \EM\right]
        ~/~ \IZ_2: \hbox{ \Msize 1 1 0 1 0 1 0 0}
        \label{eq:X5}
\end{eqnarray}

The remaining four 
polytopes yield blow-ups of the ambient spaces of $X_1$, $X_3$, $X_4$ and 
$X_5$, respectively. These blow-ups are obtained by adding a vertex 
$\v_{bl}=-\v_i$, where $\v_i$ is the vertex whose weights are in bold face.
Only the blow-up of the ambient space of $X_3$ 
descends to the complete intersection. 

Some of the nine polytopes admit several nef partitions and/or several 
triangulations. Similar to the $(2,30)$ example, it turns out that for a given 
polytope there is only one topologically inequivalent manifold. This justifies 
the notation in~\eqref{eq:X1} to~\eqref{eq:X5}. A representative will be given
below.

$X_2$, $X_4$, $X_5$, as well as the blow-up of $X_3$, have reflexive hyperplane 
sections of codimension one, and hence admit K3 fibrations (or a $\mZ_2$ 
quotient thereof in the cases of $X_4$ or $X_5$). In addition $X_3$ itself 
also admits a K3 fibration, which however does not come from a toric morphism
in the ambient space. In fact, $X_3$ and its blow-up are related in the same 
manner as $X_{(B)}$ and $X_{(C)}$ discussed in Section~\ref{sec:2,30}. 
  
Two out of the nine polytopes have lattice points which are not vertices. 
These
are related to the ambient spaces of $X_4$ and its blowup. In fact, there is 
only one such lattice point, namely $\v_5=\2(\v_6+\v_7)$. This can be seen by 
subtracting twice the (redundant) first weight vector from the second one. We 
included that point to make the K3 fiber of $X_4$ visible, although it is 
redundant for the characterization of the polytope. 
$X_4$ is thus the same variety as the one given in~\eqref{eq:X4a}. 
Finally, it turns out that 
$X_4$ and $X_5$ are topologically equivalent, for the same reason as 
$X_{(A)}$ and $X_{(B)}$ in Section~\ref{sec:equivalence} were equivalent, cf.
also the discussion below~(\ref{eq:X_C}). Note that $X_4$ and $X_5$ are free
$\mZ_2$ quotients.

We summarize the reduced data for $X_1$ to $X_4$ in the same way as we did for
$X_{(A)}$ in~(\ref{eq:X_A}), together with the intersection numbers and the 
linear forms
\begin{equation*}
  X_1:\qquad
  \begin{footnotesize}
  \begin{array}{ccrrrrrrr|rrcl}
    \multicolumn{9}{c}{ }                        &c^{(1)}&c^{(2)}&&\\
    D_{0,1}&& \um1&\um0& \um0&\um0&\um0&\um0&\um0&     -3&     -1&&\\
    D_{0,2}&&    0&   1&    0&   0&   0&   0&   0&      0&     -2&&\\
    D_1    &&    1&   0&    1&   0&   0&   0&   0&      3&     -1&&\\
    D_2    &&    0&   1&    0&   1&   0&   0&   0&      1&      0&&\\
    D_3    &&    0&   1&    0&   0&   1&   0&   0&      1&      0&&\\
    D_4    &&    1&   0&    0&   0&   0&   1&   0&      0&      1&&\\
    D_5    &&    1&   0&   -2&  -1&  -1&  -1&   0&      0&      1&&\\
    D_6    &&    0&   1&    0&   0&   0&   0&   1&     -1&      1&&\\
    D_7    &&    0&   1&    3&   1&   1&   0&  -1&     -1&      1&&\\
  \end{array}  
  \end{footnotesize}
\end{equation*}
\begin{align*}
  \kappa_{111} &= 1  & \kappa_{112} &= 3  & \kappa_{122} &= 7 & \kappa_{222} &= 11 \\
  &&\ch_2J_1 &= 22 & \ch_2J_2 &= 50
\end{align*}
\begin{equation*}
  X_2:\qquad
  \begin{footnotesize}
  \begin{array}{ccrrrrrrr|rrcl}
    \multicolumn{9}{c}{ }                        &c^{(1)}&c^{(2)}&&\\
    D_{0,1}&& \um1&\um0& \um0&\um0&\um0&\um0&\um0&     -2&     -2&&\\
    D_{0,2}&&    0&   1&    0&   0&   0&   0&   0&     -4&      0&&\\
    D_1    &&    0&   1&    1&   0&   0&   0&   0&      2&     -1&&\\
    D_2    &&    0&   1&    0&   1&   0&   0&   0&      1&      0&&\\
    D_3    &&    0&   1&    0&   0&   1&   0&   0&      1&      0&&\\
    D_4    &&    1&   0&    0&   0&   0&   1&   0&      1&      0&&\\
    D_5    &&    1&   0&   -2&  -1&  -1&  -1&   0&      1&      1&&\\
    D_6    &&    0&   1&    0&   0&   0&   0&   1&      0&      1&&\\
    D_7    &&    1&   0&    3&   1&   1&   1&  -1&      0&      1&&\\
  \end{array}
  \end{footnotesize}  
\end{equation*}
\begin{align*}
  \kappa_{111} &= 2  & \kappa_{112} &= 4  & \kappa_{122} &= 0 & \kappa_{222} &= 0 \\
  &&\ch_2J_1 &= 32 & \ch_2J_2 &= 24
\end{align*}
\begin{equation*}
  X_3:\qquad
  \begin{footnotesize}
  \begin{array}{ccrrrrrrr|rrcl}
    \multicolumn{9}{c}{ }                        &c^{(1)}&c^{(2)}&&\\
    D_{0,1}&& \um1&\um0& \um0&\um0&\um0&\um0&\um0&     -2&     -2&&\\
    D_{0,2}&&    0&   1&    0&   0&   0&   0&   0&     -6&      0&&\\
    D_1    &&    0&   1&    1&   0&   0&   0&   0&      2&     -1&&\\
    D_2    &&    1&   0&    0&   1&   0&   0&   0&      1&      0&&\\
    D_3    &&    1&   0&    0&   0&   1&   0&   0&      1&      0&&\\
    D_4    &&    0&   1&    0&   0&   0&   1&   0&      3&      0&&\\
    D_5    &&    0&   1&   -2&  -1&  -1&  -3&   0&      1&      1&&\\
    D_6    &&    1&   0&    0&   0&   0&   0&   1&      0&      1&&\\
    D_7    &&    1&   0&    3&   1&   1&   3&  -1&      0&      1&&\\
  \end{array}  
  \end{footnotesize}
\end{equation*}
\begin{align*}
  \kappa_{111} &= 1  & \kappa_{112} &= 2  & \kappa_{122} &= 0 & \kappa_{222} &= 0 \\
  &&\ch_2J_1 &= 22 & \ch_2J_2 &= 24
\end{align*}
\begin{equation*}
  X_4:\qquad
  \begin{footnotesize}
  \begin{array}{ccrrrrrrr|rrcl}
    \multicolumn{9}{c}{ }                        &c^{(1)}&c^{(2)}&&\\
    D_{0,1}&& \um1&\um0& \um0&\um0&\um0&\um0&\um0&     -2&      0&&\\
    D_{0,2}&&    0&   1&    0&   0&   0&   0&   0&     -4&      0&&\\
    D_1    &&    0&   1&    1&   0&   0&   0&   0&      2&      0&&\\
    D_2    &&    1&   0&    0&   1&   0&   0&   0&      1&      0&&\\
    D_3    &&    0&   1&    0&   0&   1&   0&   0&      1&      0&&\\
    D_4    &&    1&   0&    0&   0&   0&   1&   0&      1&      0&&\\
    D_5    &&    0&   1&   -2&  -1&  -1&  -1&   0&      1&     -2&&\\
    D_6    &&    0&   1&    1&   1&   0&   1&   2&      0&      1&&\\
    D_7    &&    0&   1&   -5&  -3&  -2&  -3&  -2&      0&      1&&\\
  \end{array}  
  \end{footnotesize}
\end{equation*}
\begin{align*}
  \kappa_{111} &= 4  & \kappa_{112} &= 2  & \kappa_{122} &= 0 & \kappa_{222} &= 0 \\
  &&\ch_2J_1 &= 28 & \ch_2J_2 &= 12
\end{align*}

There are three polyhedra yielding nef-partitions with Hodge numbers (2,36).
They all admit two non-unimodular star triangulations. After going through the 
procedure explained in detail in Section~\ref{sec:X_B,X_C} we can describe 
the (reduced) data for the first polyhedron as follows
\begin{equation*}
  \begin{footnotesize}
  \begin{array}{ccrrrrrrr|rrcl}
    \multicolumn{9}{c}{ }                      &c^{(1)}&c^{(2)}&&\\
    D_{0,1}&&\um1&\um0&\um0&\um0&\um0&\um0&\um0&    -2&     -2&&\\
    D_{0,2}&&    0&   1&    0&   0&   0&   0&   0&     -2&     -2&&\\
    D_1    &&    0&   1&    1&   0&   0&   0&   0&      0&      1&&\\
    D_2    &&    0&   1&    0&   1&   0&   0&   0&     -1&      2&&\\
    D_3    &&    1&   0&    0&   0&   1&   0&   0&      1&      0&&\\
    D_4    &&    0&   1&    0&   0&   0&   1&   0&      1&      1&&\\
    D_5    &&    1&   0&    0&   1&   1&   1&   2&      1&      1&&\\
    D_6    &&    1&   0&   -1&  -3&  -3&  -4&  -4&      0&      1&&\\
    D_7    &&    0&   1&    0&   0&  -1&  -1&  -1&      2&     -2&&\\
  \end{array}  
  \end{footnotesize}
\end{equation*}
\begin{align*}
  \kappa_{111}&=2 & \kappa_{112}&=2 & \kappa_{122}&=2 & \kappa_{222}&=1\\
  &&\ch_2 \cdot J_1&=20 & \ch_2 \cdot J_2&=22
\end{align*}
It turns out that all nef-partitions and triangulations yield topologically
equivalent complete intersections with identical Mori cones.

\section{Toric Calabi--Yau spaces with small Picard numbers}
\label{sec:smallPicard}

In this appendix we compile the Hodge data with $h^{11}\le 3$ that have
been obtained for weighted projective spaces and more general toric 
ambient spaces. The results are complete for hypersurfaces and they are 
probably (at least almost) complete for codimension 2:

\begin{center}
\begin{tabular}{|c|c|c|}\hline
H:(1,$h^{12}$)  & weighted projective   & toric\\\hline
hypersurfaces:  & 101 103 145 149       & 21    \\\hline
codimension 2:  & 61 73 79 89 129       & 25 37 \\\hline
\end{tabular}

\bigskip
\begin{tabular}{|c|c|c|}\hline
H:(2,$h^{12}$)  & weighted projective   & toric\\\hline
hypersurfaces:  & 74 86 95 106 122 128 132 272
                                & 29 38 83 84 90 92 102 116 120 144 \\\hline
codimension 2:  & 62 68 (83 84 90)   & 30 36 44 50 54 56 58 59 60 64 ~~ 
\\&                               & 66 70 72 76 77 78 80 82 100 112 \\\hline
\end{tabular}

\bigskip
\begin{tabular}{|c|c|c|}\hline
H:(3,$h^{12}$)  & weighted projective   & toric\\\hline
hypersurfaces:  & 66 69 75 87 99 103 105 
                        & 43 45 51 57 59 63 65 67 71 72 73 76 77 78 79 81 \\
        & 123 131 165 195 231 243
                & 83 84 85 89 91 93 95 107 111 115 119  127 141\\\hline
codimension 2:  &  47 55 61 87   (45 51 57 67 
        & 23 24 27 29 31 33 35 37 39 41 42 44 48 49 50 
\\& 71 77 81 83 89 91 93 111) 
        & 52 53 54 56 58 60 62 64 68 70 80 101 113      \\\hline
\end{tabular}\end{center}

Complete intersections in products of projective spaces
were enumerated completely for arbitrary codimension many years ago 
\cite{Candelas:1988kf}.  
The relevant Hodge data from \cite{GHL} are 
\\[3pt]\hbox to 13.5cm{~~~~\hfil
        {\bf65} 73 89 101               \hfil}for $h^{11}=1$,    
\\[3pt]\hbox to 13.5cm{~~~~\hfil 
        {\bf 46 47} 50 {\bf55} 56 58 59 62 64 66 68 72 76 77 83 86 
                                        \hfil}for $h^{11}=2$,
\\[3pt]\hbox to 13.5cm{~~~~\hfil 
        27 31 33 35 {\bf 36} 37 {\bf 38} 39 {\bf 40} 41
         43-45 {\bf 46} 47-61 63 66 69 72 75 
                                        \hfil}for $h^{11}=3$.
\\[3pt]
Bold-face numbers are those values of $h^{12}$ that do not occur in the 
above tables. As a check for the completeness of our results we used the 
lists that are available at \cite{CYbonn} to verify that
the missing values indeed require codimension 3 or more. Possible 
representations of minimal codimension are
\[      \IP^7[2~2~2~2]\equiv\IP \left(\Msize
        \BM1~1~1~1~1~1~1~1\cr\EM\right)
        \left[\Msize\BM2~2~2~2\cr\EM\right]_{-128}^{1,65}
\]
i.e. 4 quartics in $\IP^7$ for the example with Picard number 1, and 
\[
        \Msize\BM \IP^2 \cr\IP^5\cr\EM
        \left[\Msize\BM 2~1~0~0 \cr1~1~2~2\EM\right]_{-88}^{2,46}, 
\hspace{7mm}    \Msize\BM \IP^2 \cr\IP^4\cr\EM
        \left[\Msize\BM 2~1~0 \cr1~1~3 \EM\right]_{-90}^{2,47}, 
\hspace{7mm}    \Msize\BM \IP^3 \cr\IP^3\cr\EM
        \left[\Msize\BM 2~1~1 \cr1~2~1 \EM\right]_{-106}^{2,55},
\]\medskip
\[
        \Msize\BM \IP^2\cr\IP^2\cr\IP^2\cr\EM
        \left[\Msize\BM 2~1~0 \cr1~1~1 \cr0~1~2 \EM\right]_{-66}^{3,36},
\hspace{7mm}    \Msize\BM \IP^2\cr\IP^2\cr\IP^3\cr\EM
        \left[\Msize\BM 2~1~0~0 \cr1~0~1~1\cr0~1~2~1\EM\right]_{-70}^{3,38}, 
\hspace{7mm}    \Msize\BM \IP^1\cr\IP^2\cr\IP^3\cr\EM
        \left[\Msize\BM 1~1~0 \cr1~0~2 \cr1~2~1 \EM\right]_{-74}^{3,40},
\hspace{7mm}    \Msize\BM \IP^1\cr\IP^2\cr\IP^3\cr\EM
        \left[\Msize\BM 1~1~0 \cr2~0~1\cr1~2~1 \EM\right]_{-86}^{3,46}
\]
for $h^{11}=2$ and $h^{11}=3$, respectively. 
The data for codimension 2 weight systems and polytopes can be found at~\cite{CICY}.

\section{Free quotients of elliptic K3 fibrations}
\label{sec:FreeQ} 

In this appendix we present the Hodge data and some polytopes that we found 
for free $\IZ_2$ quotients of elliptic K3 fibrations.
More complete data are collected in~\cite{CICY}.

Among all Calabi--Yau hypersurfaces in toric varieties there are 16 polytopes
that correspond to free quotients. This can be compatible with a K3
fibration only for $\IZ_2$ quotients because the action on the
base $\IP^1$ always has fixed points and the K3 fibers only admit a free 
$\IZ_2$ action. 
The well-known example of the free $\IZ_5$ quotient of the quintic 
has no fibration. The two $\IZ_3$ examples are elliptic.
The remaining 13 polytopes have fundamental group $\IZ_2$
and are elliptic and K3 fibrations. The Hodge numbers
of these manifolds and of their double covers are

\begin{tabular}{|c|c|} \hline
$h^{11}$ $h^{12}$ [$\chi$] ~~~&~~~ double cover \\\hline
3 43 [ -80] &  3 \,\,83\, [-160]\\
3 59 [-112] &  3 115 [-224]\\
3 75 [-144] &  4 148 [-288]\\
\hline
\end{tabular}
\begin{tabular}{|c|c|} \hline
$h^{11}$ $h^{12}$ [$\chi$] ~~~&~~~ double cover \\\hline
4 28 [ -48] &  4  52 [ -96]\\
4 36 [ -64] &  4  68 [-128]\\
4 44 [ -80] &  5  85 [-160]\\
\hline
\end{tabular}
\begin{tabular}{|c|c|} \hline
$h^{11}$ $h^{12}$ [$\chi$] ~~~&~~~ double cover \\\hline
5 29 [ -48] &  7  55 [ -96]\\ &\\&\\
\hline
\end{tabular}
\bigskip

At codimension 2 we found 72 polytopes with nef partitions and elliptic K3 
structure that correspond to a free quotient. In 3 cases the lattice quotient
actually corresponds to a $\IZ_4$ quotient, but only the $\IZ_2$ refinement
of the lattice is compatible with the nef partition.


{       
\begin{tabular}{|c|c|} \hline
$h^{11}$ $h^{12}$ [$\chi$] ~~~&~~~ double cover \\\hline
3 23  [-40]   &  3 43   \,[-80]\,   \\
3 27  [-48]   &  3 51   \,[-96]\,  \\
3 29  [-52]   &  3 55  [-104]   \\
3 31  [-56]   &  3 59  [-112]   \\
3 33  [-60]   &  4 64  [-120]   \\
3 39  [-72]   &  4 76  [-144]   \\
\hline\hline
6 24  [-36]   &  10 46  [-72] \,  \\
6 26  [-40]   &  9 49   [-80]   \\      \hline
\end{tabular}
\begin{tabular}{|l|l|} \hline
$h^{11}$ $h^{12}$ [$\chi$] ~~~&~~~ double cover \\\hline
4 22  [-36]   &  5 41   [-72]   \\
4 24  [-40]   &  5 45   [-80]   \\
4 26  [-44]   &  6 50   [-88]   \\
4 36  [-64]   &  6 70  [-128]   \\
4 42  [-76]   &  5 41  [-152]   \\
4 58 [-108]   &  6 114 [-216]   \\\hline\hline
7 19  [-24]   &  11,35  [-48]   \\&\\\hline
\end{tabular}
\begin{tabular}{|l|l|} \hline
$h^{11}$ $h^{12}$ [$\chi$] ~~~&~~~ double cover \\\hline
5 25  [-40]   &  7 47   [-80]   \\
5 27  [-44]   &  8 52   [-88]   \\
5 29  [-48]   &  6 54   [-96]   \\
5 33  [-56]   &  7 63  [-112]   \\
5 35  [-60]   &  7 67  [-120]   \\
5 41  [-72]   &  8 80  [-144]   \\\hline\hline
8 14  [-12]   &  13 25  [-24]   \\&\\\hline
\end{tabular}

\noindent
The hypersurface with Hodge data (3,43) and the complete intersection (4,36) 
are discussed in Section~\ref{sec:IIAhetDual}.

A surprise in view of the Heterotic-Type II anomaly conditions~\cite{Candelas:1996su},~\cite{Aldazabal:1996du}
is the small number of hypermultiplets 
even for small numbers (like $h^{11}=3$) of vectors.
Some of the corresponding polytopes are

\[    
        \IP^{\,3,23}_{-40} \left(\Msize\BM
        \hbox{1 1 1 1 0 0 0 0}\cr
        \hbox{0 0 0 0 1 1 0 0}\cr
        \hbox{0 0 0 0 0 0 1 1}\cr\EM\right)
        \left[\Msize\BM2\cr2\cr0\EM\VL\BM2\cr0\cr2\EM\right]
        ~/~ \IZ_2: \hbox{\Msize 1 1 1 0 1 0 1 0}
\]\[
        \IP^{\,3,27}_{-48} \left(
        \hbox{2 2 1 1 1 1}\right)
        \left[4\VL4\right]
        ~/~ \IZ_4: \hbox{\Msize 0 2 2 1 3 0}
\]\[
\IP^{\,3,29}_{-52} \left(\Msize\BM
        \hbox{2 2 1 0 2 1 0}\cr
        \hbox{0 0 0 1 0 0 1}\cr\EM\right)
        \left[\Msize\BM4\cr2\EM\VL\BM4\cr0\EM\right]
        ~/~ \IZ_2: \hbox{\Msize 1 0 1 1 1 0 0}
\]\[
\IP^{\,3,31}_{-56} \left(\Msize\BM
        \hbox{0 2 1 1 0 2 2}\cr
        \hbox{1 1 0 0 1 1 0}\cr\EM\right)
        \left[\Msize\BM4\cr2\EM\VL\BM4\cr2\EM\right]
        ~/~ \IZ_2: \hbox{\Msize 1 1 1 0 0 0 1}
\]\[
\IP^{\,3,33}_{-60} \left(\Msize\BM
        \hbox{2 2 1 2 0 1 0 0}\cr
        \hbox{1 1 0 2 1 0 1 0}\cr
        \hbox{0 0 0 1 0 0 0 1}\cr\EM\right)
        \left[\Msize\BM4\cr4\cr2\EM\VL\BM4\cr2\cr0\EM\right]
        ~/~ \IZ_2: \hbox{\Msize 1 0 1 1 1 0 0 0}
\]\[
\IP^{\,3,39}_{-72} \left(\Msize\BM
        \hbox{0 2 1 4 2 2 1 0}\cr
        \hbox{1 1 0 1 0 1 0 0}\cr
        \hbox{0 0 0 1 0 0 0 1}\cr\EM\right)
        \left[\Msize\BM8\cr2\cr2\EM\VL\BM4\cr2\cr0\EM\right]
        ~/~ \IZ_2: \hbox{\Msize 1 1 1 0 1 0 0 0}
\]\[
\IP^{\,4,22}_{-36} \left(\Msize\BM
        \hbox{2 2 1 0 1 2 0 0}\cr
        \hbox{0 0 1 2 1 2 2 0}\cr
        \hbox{0 0 0 1 0 0 0 1}\cr\EM\right)
        \left[\Msize\BM4\cr4\cr2\EM\VL\BM4\cr4\cr0\EM\right]
        ~/~ \IZ_2: \hbox{\Msize 0 1 1 1 0 1 0 0}
\]\[
\IP^{\,4,24}_{-40} \left(\Msize\BM
        \hbox{2 2 1 0 0 2 1 0}\cr
        \hbox{1 1 0 1 1 0 0 0}\cr
        \hbox{0 0 0 1 0 0 0 1}\cr\EM\right)
        \left[\Msize\BM4\cr2\cr2\EM\VL\BM4\cr2\cr0\EM\right]
        ~/~ \IZ_2: \hbox{\Msize 1 0 1 1 0 1 0 0}
\]\[
\IP^{\,4,26}_{-44} \left(\Msize\BM
        \hbox{2 2 1 1 0 0 0 2}\cr
        \hbox{1 1 0 0 0 1 1 0}\cr
        \hbox{0 0 0 0 1 0 0 1}\cr\EM\right)
        \left[\Msize\BM4\cr2\cr2\EM\VL\BM4\cr2\cr0\EM\right]
        ~/~ \IZ_2: \hbox{\Msize 0 1 1 0 1 1 0 0}
\]
Note that the second example also has fundamental group $\IZ_2$ 
since $\IP_{2 2 1 1 1 1}
        \left[4\VL4\right]\!/\!\IZ_2:\hbox{\Msize000110}$ is simply connected.
Further examples can be found at \cite{CICY}.

\section{More details on the $STU$ models}
\label{sec:STUdetails}

\subsection{Discriminants and monodromy}
\label{sec:monodr-stu}

We first give the precise relation between the $STU$ models with 
$(12,12)$ and $(10,14)$ instanton embedding. This facilitates the use of
the literature, where certain aspects are discussed in one or the 
other case. The 24 instantons in the gauge bundle on the heterotic K3 are
embedded in both cases into the ${\rm E}_8\times {\rm E}_8$ so
that this group  breaks completely and leads to $3$ vector multiplets, 
apart from the graviphoton, and $243$ hypermultiplets, apart from the type 
II dilaton~\cite{Kachru:1995wm}. On the type II side the symmetric 
embedding $(12,12)$ is identified with the elliptic fibration over the 
Hirzebruch surface $\mF_0$, while the $(10,14)$ embedding corresponds to 
the elliptic fibration over the Hirzebruch surface $\mF_2$. 
The latter is realizable as a degree 24 hypersurface in the weighted 
projective space $\mP^4_{1,1,2,8,12}$. The models describe the same complex 
and symplectic families, but one complex structure deformation (hypermultiplet)
is fixed in the second realization~\cite{Morrison:1996na}. In this case there 
is a unique K3 fibration, whose base is also the base of the rational 
fibration $\mF_2$. Decomposing the K\"ahler class as 
$J = \sum_{i=1}^3 {\tilde t}_i J_i$ with $J_1$, $J_2$, and $J_3$ the classes of the 
elliptic fiber, the fiber of $\mF_2$ and the base, respectively,
we have  
\begin{equation}
\tilde t_1=U, \qquad \tilde t_2=T-U, \qquad  {\rm and} \qquad \tilde t_3=\tilde S \ . 
\label{eq:F0F2t} 
\end{equation}
The identification of $\tilde t_1$ and $ \tilde t_2$ is uniquely defined 
by the identical perturbative heterotic limit. However $S=\tilde S-(T-U)$, which 
reflects that we have no canonical geometric way of identifying the heterotic 
dilaton. All topological data including the topological string amplitudes 
$n^{(g)}_{i,j,k}={\tilde n}^{(g)}_{i,j+k,k}$ are identified by the 
coordinate change in the vector multiplet moduli space.

For the $(12,12)$ and $(10,14)$ models one gets two conifold discriminants
\begin{equation}
\label{eq:STUdiscr}
\begin{array}{rclrcl}
\Delta_1&=&[(1-a)^2-a^2 (b+c)]^2- 4 a^4 bc, &
{\tilde \Delta}_1&=&[(1-{\tilde a})^2- {\tilde a}^2 \tilde b]^2- {\tilde a}^4 {\tilde b}^2 {\tilde c}, \\
\Delta_2&=&(1-(b+c))^2- 4 bc, &
{\tilde \Delta}_2&=&(1-{\tilde b})^2- {\tilde c} {\tilde b}^2,\\
&&&{\tilde \Delta}_s&=&1-\tilde c \ ,
\end{array}
\end{equation}
but the $(10,14)$ has in addition an explicit strong 
coupling discriminant $\tilde \Delta_s$. However, it is easily seen that 
the Yukawa couplings
\begin{equation}
\begin{array}{rclrclrcl}
Y_{111}&=&{8 (1-a)\over a^3 \Delta_1},&  Y_{112}&=&{2 ((1-a)^2+a^2(b-c))\over a^2 b \Delta_1},& 
Y_{222}&=&{(2 a-1)((1-a)^2+(b-c)((1-a)^2+a^2(1-c-3 b)))\over 2 b^2 \Delta_1\Delta_2} \\  
Y_{122}&=&{2(1-a) a\over b \Delta_1},& Y_{123}&=&{(1-a)((1-a)^2- a^2 (b+c))\over abc\Delta_1},&
Y_{223}&=&{(2 a-1)((1-a)^2+(b-c)(b -c)((1-a)^2+a^2(1-c-3b)))\over 2 bc \Delta_1 \Delta_2} \\
\end{array}
\label{eq:yukef0}
\end{equation}
transform into each other by
\begin{equation}
\begin{array}{rcl}
\tilde a&=&a \\
\tilde b&=& b+c \\ [2 mm]
\tilde c&=& \ds{4 bc \over (b+c)^2},
\label{eq:stutrans}
\end{array}
\end{equation}
without change of the gauge of  $\Omega$ (vacuum line bundle~\cite{Bershadsky:1993cx}). 
This is of course reflected by the transformation of the Picard--Fuchs 
operators. 
Here we display the ones for $(12,12)$
\begin{equation}
\label{eq:PF(12,12)}
\begin{array}{rcl}
{\cal L}_1&=&5 - 36 (1-2 (a-b \partial_b-c\partial_c)+a(1-a)\partial_a)\partial_a \\
{\cal L}_2&=&(1+c\partial_c)\partial_c -(1+b\partial_b)\partial_b  \\
{\cal L}_3&=&a(a\partial_a- 4 c \partial_c)\partial_a +
             2 b(3+ 4 c \partial_c+2 b\partial_b-2 a\partial_a)\partial_b+
             2 (3 c + 2(c(c-1)\partial_c-1))\partial_c, \\
\end{array}
\end{equation}
which under (\ref{eq:stutrans}) transform\footnote{Keeping in mind that the 
essential information in ${\cal L}_i$ is the differential 
left ideal they span at a generic point in the moduli space.} into the ones 
listed in~\cite{Hosono:1993qy}
for $\mP^4_{1,1,2,8,12}[24]$. This situation is the same as the one discussed 
in Appendix~\ref{sec:PF} with the only difference that here we do not need to 
change the choice of the section of ${\cal L}^2$.

The transformations (\ref{eq:stutrans}) are useful, because 
now information that
appeared in the literature about the moduli space of either of
these models can be connected. One example is that the 
$\mP^4_{1,1,2,8,12}[24]$ 
realization has an obvious $\mZ_{24}$ symmetric point in the moduli space corresponding to the 
Fermat form of the constraint. This symmetry identifies this point with the Gepner point where an
orbifold of an $N=(2,2)$ minimal tensor product with levels $\vec k=(22,22,10,1)$ is 
conjectured to be the exact $\sigma$-model description. By (\ref{eq:stutrans}) it is 
easily identified in the moduli space of the $E\rightarrow X\rightarrow \mF_0$ description, 
where a $\mZ_2$ subgroup of the $\mZ_{24}$ acts as permutation of the $\mP^1$'s. It is known
that certain complete intersections with one modulus have no Gepner point. 
On the other hand, some complete intersections do have Gepner points related to $DE$ 
nondiagonal invariants~\cite{Fuchs:1989yv}. The above analysis shows that hypersurfaces 
in toric varieties can have Gepner points even if the polynomial cannot be written in 
weighted projective space. The same argument also applies to the other examples listed at the
end of Appendix~\ref{sec:PF}.  
Extending~(\ref{eq:F0F2t}) we note from instanton numbers and intersection numbers 
that $\cF\left(t(\tilde t)\right)=\cF(\tilde t)$ and from this it follows that 
$\Pi_{\mF_0}=B \Pi_{\mF_2}$, where $B$ is a symplectic matrix 
$(F_0=X^0(2 \cF-t^i\partial_i \cF))$
\begin{equation}
\begin{array}{rclrcl}
X^0&=&\tilde X^0, & F_0&=&\tilde F_0,\\
X^1&=&\tilde X^1, & F_1&=&\tilde F_1,  \\
X^2&=&\tilde X^2, & F_2&=&\tilde F_2-\tilde F_3, \\
X^3&=&\tilde X^3 +\tilde X^2, & F_3&=&\tilde F_3 \ .
\end{array}
\end{equation}

We will now discuss the monodromies in the large volume 
base $(X,F)$ of the $\mF_0$ or 
$(12,12)$ model. At the principal conifold discriminants $\Delta_1$ 
and $\Delta_2$ one has 
\begin{equation}
\begin{array}{cc}
\footnotesize
C_1=\left(
\begin{array}{cccccccc}
1&0&0&0&-1&0&0&0\cr
0&1&0&0&0&0&0&0\cr
0&0&1&0&0&0&0&0\cr
0&0&0&1&0&0&0&0\cr
0&0&0&0&1&0&0&0\cr
0&0&0&0&0&1&0&0\cr
0&0&0&0&0&0&1&0\cr
0&0&0&0&0&0&0&1\cr
\end{array}\right)\ ,&
\footnotesize
C_2=\left(
\begin{array}{cccccccc}
1&0&0&0&0&\um 0&\um 0&\um 0\cr
0&1&0&0&0&   -1&  \um 2&\um   2\cr
0&0&1&0&0&\um 2& -4& -4\cr
0&0&0&1&0&\um 2& -4& -4\cr
0&0&0&0&1&\um 0&\um 0&\um 0\cr
0&0&0&0&0&\um 1&\um 0&\um 0\cr
0&0&0&0&0&\um 0&\um 1&\um 0\cr
0&0&0&0&0&\um 0&\um 0&\um 1\cr
\end{array}\right) \ .
\end{array} 
\end{equation}
This follows from the Picard-Lefshetz formula. For the single vanishing cycle 
$\nu$ the monodromy transformation of the cycle $\gamma$ is given by
$$ 
\gamma\rightarrow \gamma + \langle \gamma,\nu\rangle \nu \ .
$$ 
One establishes e.g. by numerical analytic continuation that $\nu_1 =F_0$ 
vanishes 
at $\Delta_1$ and $\nu_2 = F_1-2F_2-2F_3$ vanishes at $\Delta_2$. 

The transformations (\ref{eq:stutrans}) show that the strong coupling 
singularity $\tilde \Delta_s$ is identified with the $\mZ_2$ singularity 
$(b-c)^2$ corresponding to the exchange of the two $\mP^1$ classes. 
This divisor appears in the $\mF_0$ moduli space by blowing up the contact 
order two point between $\Delta_2$ and ${1\over b+c}$ at $a=0$. 
Physically the local expansion of the topological string was identified 
as the large $N$ dual of Chern-Simons theory on a lens space $S^3/\mZ_2$. One 
parameter was identified with the 't Hooft coupling, and the other as the 
non-trivial Wilson line  in $H^1(S^3/\mZ_2)$. This theory can be solved exactly 
by a matrix model~\cite{Aganagic:2002wv}. On the other hand, from the space time 
point of view it is a strong coupling point with an SU(2) $N=4$ spectrum. Due 
to conformal invariance, the Weyl reflection is not augmented by shifts and in 
the large volume basis the monodromy around $(b-c)^2$ generates the 
$t_2\leftrightarrow t_3$ exchange
\begin{equation}
\begin{array}{cc}
\footnotesize
W=\left(\begin{array}{rrrrrrrr}
1&0&0&0&0&0&0&0\cr
0&1&0&0&0&0&0&0\cr
0&0&0&1&0&0&0&0\cr
0&0&1&0&0&0&0&0\cr
0&0&0&0&1&0&0&0\cr
0&0&0&0&0&1&0&0\cr
0&0&0&0&0&0&0&1\cr
0&0&0&0&0&0&1&0\cr
\end{array}\right) \ . 
\end{array}
\end{equation}
Other monodromies are given by the $t_i\rightarrow t_i+1$ shifts from moves
around $a,b,c=0$ and can be read off from the classical intersection data 
\begin{equation}
\begin{array}{cc}
\footnotesize
T_1=\left(\begin{array}{rrrrrrrr}
    1&    0&    0&    0&\um 0&    0&\um 0&\um 0\cr
    1&    1&    0&    0&    0&    0&    0&    0\cr
    0&    0&    1&    0&    0&    0&    0&    0\cr
    0&    0&    0&    1&    0&    0&    0&    0\cr
    9&    4&    1&    1&    1&   -1&    0&    0\cr
   -4&   -8&   -2&   -2&    0&    1&    0&    0\cr
   -1&   -2&    0&   -1&    0&    0&    1&    0\cr
   -1&   -2&   -1&    0&    0&    0&    0&    1\cr
\end{array}\right)&
\footnotesize
T_2=\left(\begin{array}{rrrrrrrr}
1&    0&\um 0&    0&\um 0&\um 0&    0&\um 0\cr
0&    1&    0&    0&    0&    0&    0&    0\cr
1&    0&    1&    0&    0&    0&    0&    0\cr
0&    0&    0&    1&    0&    0&    0&    0\cr
2&    0&    0&    0&    1&    0&   -1&    0\cr
0&   -2&    0&   -1&    0&    1&    0&    0\cr
0&    0&    0&    0&    0&    0&    1&    0\cr
0&   -1&    0&    0&    0&    0&    0&    1\cr
\end{array}\right) \ .
\end{array} 
\end{equation}
and $T_3=WT_2 W$.

Around the Gepner point in the $\mF_2$ realization one has a 
$\mZ_{24}$ symmetry. This is reflected by $24$ rational solutions of the indicial 
equations of the Picard--Fuchs equation corresponding to a branching of order $24$ 
in the solutions. One can choose $b_3=8$ linearly independent solutions, on which 
$\mZ_{24}$ acts by cyclic exchanges of the branches. Analytic continuation to the 
large complex structure points is possible due to Barnes integral representation. 
In contrast to
the connecting matrix from the conifolds to infinity, where the 
non-trivial numbers $\zeta(2) {\int_X\ch_2 J/24}$ and $\zeta(3)\chi$ 
containing the 
topological data arise from the analogue of the Gauss resummation for the 
hypergeometric function $_2F_1$, the connection matrix from the $\mZ_{24}$ 
point to 
infinity is integer~\cite{Candelas:1993dm},~\cite{Candelas:1994hw}. 
A nice open string interpretation of the 
connection matrix was given in~\cite{Brunner:1999jq},~\cite{Mayr:2000as}. 
In particular~\cite{Mayr:2000as} calculates the matrix for the $\mF_2$ case. 
Conjugated to the $\mF_0$ basis it is
\begin{equation}
\begin{array}{cc}
\footnotesize
A=\left(\begin{array}{rrrrrrrr}
 1&\um0& 0& 0& 1& 0& 0& 0\cr
-1&   1& 1& 1&-1& 1&-2&-2\cr
 1&   0&-2&-1& 1&-2& 4& 4\cr
 2&   0&-1&-2& 2&-2& 4& 4\cr
-2&   0& 0& 0&-1& 0& 0& 1\cr
 0&   2& 1& 0& 0& 1& 0& 0\cr
 1&   0&-1&-1& 1&-1& 2& 3\cr
 0&   1& 0& 0& 0& 0& 1& 0\cr
\end{array}\right) \ . 
\end{array}
\end{equation}
The combination $M=T_3 A C A C T_2$ 
\begin{equation}
\begin{array}{cc}
\footnotesize
M=\left(\begin{array}{rrrrrrrr}
1&\um0&0&\um0&\um0&\um0&0&0\cr
0&1&1&0&0&0&0&0\cr
0&0&-1&0&0&0&0&0\cr
0&0&-2&1&0&0&0&0\cr
0&0&0&0&1&0&0&0\cr
0&0&0&0&0&1&0&0\cr
0&0&1&0&0&1&-1&-2\cr
0&0&0&0&0&0&0&1\cr
\end{array}\right) \ . 
\end{array}
\end{equation}
is identified precisely with the mirror symmetry transformation 
$T\leftrightarrow U$ of the two torus $T^2$ of the heterotic 
string including the shift in the quadratic term, which cancels in 
$\cF_0=2 \cF - t_i \partial_i \cF$ and shows up only
in $\partial_2 F$~\footnote{The fact that the $A C A C$ element leaves the
large complex structure limit invariant was already noted in~\cite{Mayr:2000as}.}. 

\subsection{Restriction to Calabi--Yau families with less moduli}
\label{sec:restrict}

We can consider the restriction $b=c$ in the complex structure moduli
space of the mirror of 
$X=\IP\left({\footnotesize\BM 6~4~1~0~1~0\cr6~4~0~1~0~1\cr\EM}\right)
\left[\footnotesize\BM12\cr12\EM\right]$, in other words we identify 
the classes $t_2$ and $t_3$ of the two $\mP^1$'s in~(\ref{eq:classes}).
There is a toric realization of this restriction given by the complete
intersection
\begin{equation}
  \label{eq:STU'}
  X' = \IP\left(\Msize\BM    6&4&0&1&1&1&1  \cr
        3&2&1&0&0&0&0  \cr\EM\right) 
        \left[\Msize\BM 2\cr0\cr  \EM\VL\BM  12\cr6   \EM\right]
\end{equation}
This model has the same Hodge numbers as $X$, $(h^{1,1},h^{2,1})=(3,243)$, 
but only two of the three K\"ahler moduli are realized torically, namely 
the elliptic fiber and the combination of the two $\mP^1$'s. One way to see 
this is to compare the Mori generators $l^{(a)}$ of $X$ with those of $X'$
\begin{eqnarray}
  \label{eq:MoriSTU'}
  l'{}^{(1)}=&l^{(1)}&=(\um0,-6,3,2,\um1,0,0,0,0)\cr
  l'{}^{(2)}=&l^{(2)}+l^{(3)}&=(-2,\um0,0,0,-2,1,1,1,1)\ .
\end{eqnarray}
and to compute the intersection ring
\begin{align}
  \kappa_{111} &= 8,  & \kappa_{112} &= 4,  & \kappa_{122} &= 2, & \kappa_{222} &= 0, \cr
  &&\ch_2J_1 &= 92, & \ch_2J_2 &= 48 \ .&& 
\end{align}
Another way is to look at the Picard--Fuchs operators 
in~(\ref{eq:PF(12,12)}) and set $b=c$ which yields the same 
operators as those obtained from~(\ref{eq:MoriSTU'}) and~(\ref{eq:DL})
with $a=432z_1$ and $b=4z_2$
\begin{align}
  \label{eq:PF2}
  \cL_1 &= \theta_1(\theta_1 - 2\theta_2) - 12(6\theta_1 - 1)(6\theta_1 - 5)z_1\ ,\cr 
  \cL_2 &= \theta_2^3 - 2(2\theta_2 -1)(2\theta_2 - \theta_1 -2)(2\theta_2 -\theta_1 -1)z_2\ .
\end{align}
and the discriminants in~(\ref{eq:STUdiscr}) reduce to
\begin{align}
  \label{eq:discrX'}
  \Delta_1 &=
  (1-432\,z_{{1}})^{2}-2985984\,z_{{1}}^{2}{z_{{2}}}, &
  \Delta_2 &= 1-16\,z_{{2}}, & \Delta_3 &= 1-432\,z_{{1}}  \ .
\end{align}
For the GV invariants we then get a relation of the 
form~(\ref{eq:birat})
\begin{equation}
  n_{i,j}^{(g)}(X') = \sum_{k} n^{(g)}_{i,k,j-k}(X) \ .
\end{equation}

Next, we can consider the sublocus of the complex structure moduli space
of the mirror of $X'$ defined by the limit $z_1 \to 0$, $z_2 \to \infty$ 
such that $z_1^2z_2$ remains fixed. We will now argue that there is a toric
realization of a singular one--parameter family $X''$ parameterized by 
this sublocus. For this purpose, we set  $x=z_1^2z_2$ and $y=z_1$, and get
$\theta_x=\theta_2$ and $\theta_y=\theta_1-2\theta_2$. In these variables, 
the operators in~(\ref{eq:PF2}) become
\begin{align}
  \cL_1 &= \theta_y(2\theta_x+\theta_y) - 12 y(12\theta_x + 6\theta_y +5)(12\theta_x + 6\theta_y +1),\cr
  \cL_2 &= \theta_x^3 - \tfrac{2x}{y^2}(2\theta_x + 1)\theta_y(\theta_y-1)\ .
\end{align}
With the ansatz $\varpi(z_1,z_2) = \sum_{n=0}^\infty a_n(x) y^n$ for the 
fundamental period of $X'$, $\varpi(z_1,z_2) = \sum_{n_1,n_2} 
\frac{(6n_1)!(2n_2)!}{(n_2!)^4(n_1-2n_2)!(2n_1)!(3n_1)!}z_1^{n_1}z_2^{n_2}$
we find
\begin{align}
  \theta_x^3 a_0 - 4 x(2\theta_x + 1)a_2 + O(y) &= 0\cr
  \left((2\theta_x+1)a_1 - 12 (12\theta_x +1)(12\theta_x +5)a_0\right)y+&\cr
  +\left(2(2\theta_x+2)a_2 - 12 (12\theta_x +7)(12\theta_x +11)a_1\right)y^2 + O(y^3) &= 0 
\end{align}
Multiplying the first of these equations with $\theta_x$, multiplying the 
order $y^2$ term with $x(2\theta_x+1)$ and the order $y$ term with 
$- 12 x(12\theta_x +7)(12\theta_x + 11)$ in the second equation, we can 
eliminate $a_1$ and $a_2$, and in the limit $y\to 0$ we find that
\begin{align}
  \label{eq:PF1}
  \cL &= \theta^4- 144x(12\theta_x +1)(12\theta_x +5)(12\theta_x + 7)(12\theta_x + 11)
\end{align}
annihilates $a_0(x)$. This is precisely the Picard--Fuchs operator 
corresponding to the Mori generator
\begin{equation}
  \label{eq:MoriX''}
  l''{}^{(1)}=2l'{}^{(1)}+l'{}^{(2)}=(-2,-12,6,4,0,1,1,1,1)\ ,
\end{equation}
which is the Mori generator for the complete intersection 
$X''=\mP^5_{1,1,1,1,4,6}[2,12]$. Note that due the two weights having
a common factor, $Z''=\mP^5_{1,1,1,1,4,6}$ is singular, and since we do
not blow-up the singularity, $X''$ is a singular Calabi--Yau family.
Another way to see this is to look for toric blow-ups $Z' \to Z''$ such that
the polyhedron $\Delta^*_{Z'}$ is reflexive and admits a nef partition.
There are two minimal such blow-ups, $Z'_1$ and $Z'_2$, where we added
the vertices $\nu^*_{7,1} = (0,0,0,2,3)$ and $\nu^*_{7,2} = (0,0,0,1,2)$,
respectively. $Z'_1$ is exactly the ambient space of the two--parameter
family $X'$ in~(\ref{eq:STU'}). ($Z'_2$ is the ambient space for the 
complete intersection $\IP\left(\Msize\BM 6&4&1&1&1&1&0 \cr 2&1&0&0&0&0&1
\cr\EM\right) \left[\Msize\BM 6\cr2\cr \EM\VL\BM 8\cr2 \EM\right]$ with 
$(h^{1,1},h^{2,1})=(1,149)$.) Since $X'$ intersects $D_{7,1}$, blowing
down $D_{7,1}$ will yield a singular manifold $X''$. 

The fundamental period $a_0(x)$ of $X''$ is the solution  
${}_4F_3(\frac{1}{12},\frac{5}{12},\frac{7}{12},\frac{11}{12};1,1,1;x)$
of a generalized hypergeometric system, and is the $14^{\rm th}$ 
one--parameter system with three regular points and integral 
monodromy~\cite{Chuck}. The discriminant $\Delta = 1-2985984x$ follows 
from~(\ref{eq:PF1}) and agrees with the limit of~(\ref{eq:discrX'}). 
The Gopakumar--Vafa invariants $n^{(g)}_d$ are
\vskip 5 mm
{\vbox{\footnotesize{
$$
\vbox{\offinterlineskip\tabskip=0pt \halign{\strut \vrule#& &\hfil
~$#$~&\hfil ~$#$ &\hfil ~$#$ &\hfil ~$#$ &\hfil ~$#$ &\hfil ~$#$~&~ \vrule#\cr \noalign{\hrule} 
& g & d=1        &2                 &3                      &4                     &5                          &  \cr
\noalign{\hrule}
&   0& 678816&137685060720&69080128815414048&51172489466251340674608&46928387692914781844159094240  & \cr 
&   1& 480&-1191139920&1399124442888000&8310445299962958677280&22083962595341011092128873088       & \cr 
\noalign{\hrule}}\hrule}$$}}}
The ambiguity $f_2$ for the genus two invariants can only be fixed up
to two integer invariants. Furthermore, due to the complicated nature 
of the map $X' \to X''$ there is no simple relation between the 
$n^{(g)}_d(X'')$ and $n_{i,j}^{(g)}(X')$. If we were able to solve
these two problems, we could further constrain the six unknown GV 
invariants of the $STU$ model in Section~\ref{sec:STUprops}.

\end{document}